\documentclass[
aip,
preprint,
amsmath,
amssymb]{revtex4-1}
\usepackage{graphicx}
\usepackage{hyperref}
\hypersetup{
    colorlinks=true,
    linkcolor=blue,
    filecolor=blue,      
    urlcolor=blue,
    citecolor=blue,
}
\usepackage[usenames,dvipsnames]{color}

\begin{document}


\title{Ferromagnetic Materials for Josephson $\pi$ Junctions} 



\author{Norman~O.~Birge}
\email{birge@msu.edu}
\affiliation{Department of Physics and Astronomy, Michigan State University, East Lansing, Michigan 48824, USA}

\author{Nathan~Satchell}
\email{satchell@txstate.edu}
\affiliation{Department of Physics, Texas State University, San Marcos, Texas, 78666, USA}


\date{\today}

\begin{abstract}
The past two decades have seen an explosion of work on Josephson junctions containing ferromagnetic materials.  Such junctions are under consideration for applications in digital superconducting logic and memory.  In the presence of the exchange field, spin-singlet Cooper pairs from conventional superconductors undergo rapid oscillations in phase as they propagate through a ferromagnetic material.  As a result, the ground-state phase difference across a ferromagnetic Josephson junction oscillates between 0 and $\pi$ as a function of the thickness of the ferromagnetic material.  $\pi$-junctions have been proposed as circuit elements in superconducting digital logic and in certain qubit designs for quantum computing.  If a junction contains two or more ferromagnetic layers whose relative magnetization directions can be controlled by a small applied magnetic field, then the junction can serve as the foundation for a memory cell.  Success in all of those applications requires careful choices of ferromagnetic materials.  Often, materials that optimize magnetic properties do not optimize supercurrent propagation, and vice versa.  In this review we discuss the significant progress that has been made in identifying and testing a wide range of ferromagnetic materials in Josephson junctions over the past two decades.  The review concentrates on ferromagnetic metals, partly because eventual industrial applications of ferromagnetic Josephson junctions will most likely start with metallic ferromagnets (either in all metal junctions or junctions containing also an insulating layer). We will briefly mention work on non-metallic barriers, including ferromagnetic insulators, and some of the exciting work on spin-triplet supercurrent in junctions containing noncollinear magnetic inhomogeneity.
\end{abstract}


\maketitle 

\section{Introduction}

It has been over 20 years since Ryazanov and co-workers first demonstrated a magnetic Josephson junction whose ground-state phase difference could be either zero or $\pi$ depending on the sample temperature.\cite{veretennikov_2000,Ryazanov_PRL_2001}  That breakthrough was followed soon after by Kontos \textit{et al.}, who demonstrated the $0 - \pi$ transition as a function of the thickness of the ferromagnetic material in a series of magnetic junctions.\cite{Kontos_PRL_2002}  The evidence for the $0 - \pi$ transition in those pioneering works was indirect in the sense that only the amplitude of the critical current was measured; direct measurements of the junction phase were carried out soon thereafter.\cite{ryazanov_2001,guichard_2003,bauer_2004,frolov_2004} The physical mechanism underlying these phenomena had been understood already 20 years earlier;\cite{bulaevskii_1977,buzdin_1982,buzdin_1991} nevertheless, the laboratory demonstrations of these phenomena sparked widespread interest in magnetic Josephson junctions and in $\pi$-junctions in general.\cite{golubov_2004,lyuksyutov_2005,Buzdin_review_2005} There have been many proposals to use $\pi$-junctions as circuit elements in superconducting digital logic\cite{terzioglu_1998,ustinov_2003} and in certain qubit designs for quantum computing,\cite{ioffe_1999,blatter_2001} and several of these ideas are being tested in the laboratory (see Section \ref{Sec:App} of this review).  

At about the same time as the first $\pi$-junction demonstrations, Bergeret, Volkov and Efetov predicted theoretically that magnetic Josephson junctions made from conventional spin-singlet superconductors could carry long-range spin-triplet supercurrent under specific conditions that include the presence of non-collinear magnetic inhomogeneities inside the junctions.\cite{bergeret_2001b,kadigrobov_2001,volkov_2003,Bergeret_review_2005}  Unlike the demonstration of magnetic $\pi$ junctions, the Bergeret prediction came as a total surprise to the superconductivity community.  The idea that supercurrents could be spin-polarized led to much excitement and the coining of the name ``superconducting spintronics” for this new area of condensed matter physics.\cite{eschrig_review_2011,Eschrig_review_2015,linder_review_2015,melnikov_2022,cai_2023} 

Put together, the two breakthroughs mentioned above catalyzed a huge surge in interest in superconducting/ferromagnetic (S/F) hybrid systems.  This review covers only a small subset of the work that has been done since 2001, focusing mainly on S/F/S $\pi$-junctions and their variants, including S/I/F/S, S/I/s/F/S and S/FI/S junctions (where I = insulator and FI = ferromagnetic insulator).  Our emphasis is on the wide range of magnetic materials that have been used in such devices.  The centerpiece of the review are the Tables in Section IV, which we hope provide a complete list of magnetic materials that have been used inside Josephson junctions, usually with the aim of making $\pi$-junctions.  To our knowledge, the last time such a compilation has been made was in 2006 by Kupriyanov, Golubov, and Siegel.\cite{Kupriyanov_review_2006}  That review covered both experimental and theoretical developments; ours focuses almost entirely on experimental results.

There has been a lot of work on other experimental signatures of S/F physics besides $\pi$-state Josephson junctions, such as the critical temperature and critical field of S/F bilayers and multilayers.  We do not discuss those results here, even though some of that work\cite{wong_1986,wong_1986a,uher_1986,radovic_1988,radovic_1991,jiang_1995,mhge_1996,mhge_1997,chien_1999} predated the experimental demonstrations of $\pi$ Josephson junctions discussed above. 

The structure of this review is as follows: Section~\ref{Sec:Gen} discusses general considerations, including the physical parameters that describe Josephson junctions, ballistic vs diffusive transport regimes, the different types of junctions, and how the junction response to a magnetic field is modified when the junction contain ferromagnetic materials.  Section~\ref{Sec:App} discusses applications of $\pi$-junctions in quantum and classical circuits.  Section~\ref{Sec:Tables} discusses metallic S/F/S $\pi$-junctions and contains the Tables of magnetic materials mentioned above.  It includes a short subsection about ``spin-valve" junctions whose critical current or phase state are controllable by changing the magnetic configuration inside the junction.  Section~\ref{Sec:NonMet} mentions briefly the many different types of $\pi$-junctions that have been created in the past few years in exotic materials other than metals.  Section~\ref{Sec:Triplet} discusses spin-triplet Josephson junctions, including a very brief mention of so-called ``$\phi$ and $\phi_0$-junctions."  Finally, Section~\ref{Sec:Chall} contains a discussion of some open challenges and directions for the future.

\section{General considerations}\label{Sec:Gen}

\subsection{S/F physics, length scales, and theoretical predictions for $I_c R_N$}

The physics of S/F systems has been reviewed many times;\cite{golubov_2004,Buzdin_review_2005,lyuksyutov_2005,Bergeret_review_2005,Eschrig_review_2015} we encourage readers new to the topic to read the excellent pedagogical article by Demler \textit{et al}.\cite{demler_1997}  Superconducting and ferromagnetic materials in contact or in close proximity to each other interact in two distinct ways.  Effects due to the magnetic field produced by F are generally referred to as ``orbital effects"; they are nearly always present in experiments and should not be forgotten, but they are not the main focus of this review. Here we focus on proximity effects, whereby electron pair correlations are induced in a material in contact with a superconductor.  

We start by defining several important length scales.  The superconducting coherence length is defined as $\xi_S^\textit{bal} = \hbar v_F/\Delta$ or $\xi_S^\textit{dif} = (\hbar D_S/\Delta)^{1/2}$ in the ballistic and diffusive limits, respectively, where $\Delta$ is the superconducting gap, and $v_F$ and $D_S$ are the Fermi velocity and diffusion constant in S. The diffusion constant, $D$, is related to the electron mean free path, $l$, via $D = \frac{1}{3} v_F l$.  In an S/N bilayer (N = normal metal), pair correlations penetrate N over a distance called the ``normal-metal coherence length", which has the following forms in the ballistic or diffusive limits: $\xi_N^\textit{bal} = \hbar v_F/(2 \pi k_B T)$ or $\xi_N^\textit{dif} = (\hbar D_N/(2 \pi k_B T))^{1/2}$, where $v_F$ and $D_N$ are now in the normal metal, and $T$ is the temperature. That length can be quite long -- up to several hundred nm in noble metals at dilution refrigerator temperatures.  In contrast, in ferromagnetic materials the pair correlations oscillate in sign due to the exchange splitting between the majority and minority spin bands in F. (Those oscillations are sometimes referred to as ``FFLO oscillations" after Fulde-Ferrell\cite{Fulde_1964} and Larkin-Ovchinnikov,\cite{larkin_1965} but we emphasize that the original FFLO prediction was for a bulk ferromagnetic superconductor in the ballistic limit, rather than a proximity system.) The period of the oscillations is $2 \pi \xi_F$, where $\xi_F$ is sometimes called the ``exchange length" or the ``ferromagnetic coherence length". In the ballistic or diffusive limits with $E_{ex} >> k_BT$, the results are $\xi_F^\textit{bal} = \hbar v_F/(2 E_{ex})$ or $\xi_F^\textit{dif} = (\hbar D_F/E_{ex})^{1/2}$, respectively, where $2 E_{ex}$ is the exchange splitting between the majority and minority spin bands. In addition to the oscillation, the pair correlations decay algebraically in the ballistic limit or exponentially in the diffusive limit.  Due to the large exchange energies in strong F materials, $E_{ex} \approx 1 eV$, $\xi_F$ is less than 1 nm in those materials.  That is one reason why experimental progress in S/F systems lagged behind the theoretical developments for so long. One of the keys to the experimental breakthroughs of 2001 was the use of dilute ferromagnetic alloys: CuNi alloy in the case of the Ryazanov group\cite{Ryazanov_PRL_2001} and PdNi alloy in the case of the Aprili group.\cite{kontos_2001,Kontos_PRL_2002} Diluting the Ni decreased $E_{ex}$ substantially thereby increasing $\xi_F$ to values of several nm. In that limit one must include the contribution of the temperature to the exchange length $\xi_F$; in fact, Ryazanov's first demonstration of a $\pi$-junction was carried out by varying the temperature of a Josephson junction with fixed ferromagnetic layer thickness.\cite{Ryazanov_PRL_2001}

When discussing the critical current, $I_c$, of Josephson junctions, we will quote results for the $I_c R_N$ product, where $R_N$ is the normal-state resistance of the junction, usually determined from the slope of the $I – V$ relation for applied currents $I >> I_c$.  $I_c R_N$ is a useful quantity because it is independent of junction area and because it can be compared with the standard Ambegaokar-Baratoff result for short S/I/S junctions: $I_c R_N =\pi\Delta/(2e)$.\cite{ambegaokar_1963} Calculations of $I_c$ for ferromagnetic Josephson junctions have been carried out in several different regimes.  In addition to the length scales defined above, one must also consider the electron mean free path, $l_F$, and the thickness $d_F$ of the F layer.  The first calculation of $I_c R_N$ was performed for the purely ballistic or ``clean" limit, defined by $d_F << \xi_S << l_F, l_S$ -- i.e. neglecting impurity scattering in both F and S.\cite{buzdin_1982}  The decay and oscillations of $I_cR_N$ are given by the numerical maximum with respect to $\varphi$ of the ballistic limit supercurrent $I_\text{S}(\varphi)$, \cite{buzdin_1982}
\begin{multline}
\label{eq:fullballistic}
I_\text{S}(\varphi)R_N = \frac{\pi \Delta \alpha^2}{2 e} \int_{\alpha}^{\infty}\frac{dy}{y^3}\Bigg[ \sin{}\Bigg(\frac{\varphi-y}{2}\Bigg)\tanh{}\Bigg(\frac{\Delta\cos{}\frac{\varphi-y}{2}}{2k_\text{B}T}\Bigg) \\ + \sin{}\Bigg(\frac{\varphi+y}{2}\Bigg)\tanh{}\Bigg(\frac{\Delta\cos{}\frac{\varphi+y}{2}}{2k_\text{B}T}\Bigg)\Bigg],
\end{multline}
\noindent where $\varphi$ is the phase difference across the junction, and $\alpha \equiv d_F / \xi_F^\textit{bal}$. In the limit of large $\alpha$ and for $T$ near $T_c$, Eqn.~\ref{eq:fullballistic} can be simplified to,\cite{buzdin_1982}
\begin{equation}\label{eq:simpleballistic}
I_cR_N = \frac{\pi \Delta^2}{4 e T} \Bigg|\frac{\sin\big(d_F/\xi_F^\textit{bal}\big)}{\big(d_F/\xi_F^\textit{bal}\big)}\Bigg|.
\end{equation}
\noindent The simplified form of the ballistic limit shows $0-\pi$ oscillations at $d_F/\xi_F^\textit{bal}=n\pi$, but misses the first transition, which occurs at $d_F/\xi_F^\textit{bal}\approx 0.33\pi$. 

The opposite limit is the diffusive or ``dirty" limit, defined by $l_F,l_S << d_F, \xi_S, \xi_F$ -- i.e. the limit where impurity scattering is very strong in both F and S.  In that limit the Usadel equation is valid, and the original calculation produces the result:\cite{buzdin_1991,Buzdin_review_2005} 
\begin{equation}\label{eq:fulldiffusive}
I_cR_N = \frac{\pi \Delta^2}{4eT_c} 2x \bigg| \frac{\cos(x)\sinh(x)+\sin(x)\cosh(x)}{\cosh(2x)-\cos(2x)}\bigg|,
\end{equation}
where $x = d_F/\xi_F^\textit{dif}$.  In the limit $d_F >> \xi_F^\textit{dif}$ and $T$ near $T_c$, that result simplifies to:\cite{buzdin_1991,Buzdin_review_2005} 
\begin{equation}\label{eq:simplediffusive}
I_cR_N = \frac{\sqrt{2} \Delta^2}{4eT_c} \Bigg( \frac{d_F}{\xi_F^\textit{dif}} \Bigg) \exp \Bigg( \frac{-d_F}{\xi_F^\textit{dif}} \Bigg) \Bigg|\sin\Bigg( \frac{d_F}{\xi_F^\textit{dif}}+\frac{\pi}{4} \Bigg)\Bigg|,
\end{equation}
which predicts $0-\pi$ oscillations at $d_F/\xi_F^\textit{dif}=(n+3/4)\pi$.

There are several ways in which real Josephson junctions deviate from the strict assumptions used to calculate the expressions above. We know of only one paper that incorporates realistic models of the Fermi surfaces of the S and F materials into a calculation of $I_c$.\cite{ness_2022}  In addition, strong F materials have different mean free paths for majority and minority electrons, which are not captured in the usual theoretical formulations. In the presence of spin-flip or spin-orbit scattering in F, the length scales governing the decay and oscillations are no longer equal,\cite{faure_2006, pugach_2011} hence one defines independent length scales $\xi_{F1}$ and $\xi_{F2}$ for the two phenomena. In addition, it has been shown that the position of the first $0 - \pi$ transition varies depending on the boundary conditions at the S/F interface, including the presence of an insulating barrier or a normal-metal spacer layer.\cite{buzdin_2003a,pugach_2011,heim_2015}   

It is common to address some of the effects discussed above by utilizing a phenomenological expression for the diffusive regime:
\begin{equation}\label{eq:generaldiffusive}
I_cR_N = I_cR_N (0) \exp \bigg( \frac{-d_F}{\xi_{F1}} \bigg) \bigg|\sin\bigg( \frac{d_F}{\xi_{F2}}+\phi \bigg)\bigg|,
\end{equation}
where $\phi$ is not necessarily equal to $\pi/4$ and $I_cR_N (0)$ is a fictitious zero thickness fit parameter. Spin-dependent scattering generally causes $\xi_{F1}$ to decrease and $\xi_{F2}$ to increase relative to $\xi_F^\textit{dif}$, resulting in the relation $\xi_{F1} < \xi_{F2}$, which is observed in junctions containing weakly-ferromagnetic CuNi alloys, as seen in Table \ref{weak}.  

Several authors have pointed out that the strong elemental F materials, Ni, Fe, and Co, do not satisfy the conditions of validity of the Usadel equation.  In such materials the exchange energy is very large, hence the length $\xi_F^\textit{bal}$ is very short.  Bergeret \textit{et al.}\cite{bergeret_2001a} were the first to discuss the intermediate limit, defined by $\xi_F^\textit{bal} << l_F << \xi_S^\textit{bal}$.  This limit clearly requires that $E_{ex} >> \Delta$, which is true for the strong F materials. The expression for $I_c R_N$ in the intermediate limit involves a sum over Matsubara frequencies, which we do not reproduce here.  (See Eqns. (19) and (20) in Ref. \onlinecite{bergeret_2001a}.)  The dependence of $I_c R_N$ on $d_F$ has a similar form to Eqn. (\ref{eq:generaldiffusive}), except now one finds that the oscillation period is governed by the ballistic expression for the exchange length, $\xi_{F2} = \xi_F^\textit{bal}$, while the decay is governed by the mean free path, $\xi_{F1} \approx l_F$. Thus, in the intermediate limit one expects to observe $\xi_{F1} > \xi_{F2}$, which is indeed the case for many of the materials listed in Table \ref{pure}. 

When it comes to fitting experimental data, one requires data over a wide range of thicknesses to obtain accurate estimates of all the parameters in Eqn. (\ref{eq:generaldiffusive}).  If only a single $0-\pi$ transition is observed in the measured thickness range, then the parameters $\xi_{F2}$ and $\phi$ have large uncertainties and are highly correlated.  To remove that correlation, we prefer to use the following fitting form: 
\begin{equation}\label{eq:generic}
I_cR_N = I_cR_N (0) \exp \bigg( \frac{-d_F}{\xi_{F1}} \bigg) \bigg|\sin\bigg( \frac{d_F - d_{0-\pi}}{\xi_{F2}} \bigg)\bigg|,
\end{equation}
\noindent where $d_{0-\pi}$ is the thickness of the first $0-\pi$ transition, which is often determined quite accurately in the experiments.  Expression (\ref{eq:generic}) is the one we used when analyzing data from other groups, to extract the parameters shown in the Tables in Section IV.

\subsection{Josephson junction geometry: planar vs ``sandwich-style"}

Microfabricated Josephson junctions generally take one of two geometries.  In planar or lateral junctions, the supercurrent flows in the plane of the substrate, while in ``sandwich-style" junctions, the current flows perpendicular to the plane.  Most S/I/S junctions are of the sandwich type due to the necessity of keeping the insulating barrier extremely thin.  In contrast, most S/N/S junctions studied nowadays are planar so that the supercurrent flows over a long distance in the N metal.  Exceptions include junctions containing very high-resistivity metals such as silicides,\cite{chong_2005} or junctions studied long ago on very thick alloy films.\cite{clarke_1969}  Most S/F/S Josephson junctions have the sandwich structure due to the very short length scale over which the critical current oscillates and decays in ferromagnetic materials.  The notable exceptions are junctions based on half-metallic CrO$_2$\cite{anwar_2010} or on single-crystal ferromagnetic nanowires,\cite{wang_2010} where the supercurrent may propagate over very long distances. 

All of the Josephson junctions analyzed in the Tables in Section IV have the sandwich geometry.

\subsection{{S/I/F/S vs S/F/S} $\pi$-junctions}

Josephson junctions are characterized by several parameters. In addition to $\Delta$ and $I_c R_N$, these include the Josephson energy, $E_J = (\hbar /2e)I_c$ and the Stewart-McCumber parameter, $\beta_c = (2e/\hbar)I_cR_N^2C$, where $C$ is the junction capacitance.  The value of $\beta_c$ determines whether the junction dynamics are underdamped ($\beta_c > 1$) or overdamped ($\beta_c < 1$). Without an explicit insulating tunnel barrier, all S/F/S Josephson junctions are overdamped. They can be made underdamped by inserting an insulating barrier to make an S/I/F/S junction,\cite{Kontos_PRL_2002,Weides_APL_2006} or by using a ferromagnetic insulator (FI) as the barrier material: S/FI/S.\cite{tanaka_1997,kawabata_2010,senapati_2011}  The difference between overdamped and underdamped junction dynamics is crucial for all $\pi$-junction applications -- both quantum and classical.

A related issue relevant for applications is whether the $\pi$-junction is being used as an active or passive device.  In the former situation, the phase drop across the junction is an important dynamical variable, whereas in the latter situation the $\pi$-junction acts as a passive phase shifter with the junction phase hardly deviating from $\pi$ during circuit operation.  For passive applications, the $\pi$-junction is invariably paired up with conventional S/I/S junctions whose critical currents are much smaller than that of the $\pi$-junction. That configuration is common in applications that require underdamped junction dynamics or large $I_c R_N$ product -- both of which are difficult to achieve with S/F/S junctions containing a metallic barrier.

\subsection{Dependence of critical current on applied magnetic field}\label{Sec:Fraunhofer}

A common way to assess the quality of a Josephson junction is to measure the critical current vs magnetic field applied in a direction perpendicular to the current flow, $I_c(H)$.  The local supercurrent in a junction depends on the local vector potential; in the case of a uniform field, the vector potential varies linearly along the direction perpendicular to both the current and applied field.\cite{barone1982physics}  For the case of a small rectangular junction of width $w$, with applied magnetic field, $H$, parallel to one of the symmetry axes, the resulting interference pattern follows the ``sync” function familiar from the single-slit diffraction pattern in optics:
\begin{equation} \label{Eqn:Fraunhofer}
    I_c(\Phi) = I_{c0} \left| \frac{\sin(\pi \Phi /\Phi_0)} {(\pi \Phi / \Phi_0)} \right|
\end{equation}
where $\Phi_0 = h/2e = 2.07 \times 10^{-15}$ Tm$^2$ is the flux quantum for electron pairs, and $\Phi = \mu_0 H w (2\lambda_{\mathrm{eff}} + d_\text{barrier})$ is the magnetic flux through the junction, where $\lambda_{\mathrm{eff}}$ is the effective London penetration depth of the S electrodes and $d_\text{barrier}$ is the total thickness of the barrier.  A small junction has $w$ smaller than the Josephson penetration depth and thus uniform current distribution.\cite{barone1982physics}  Because of the similarity with optics, a plot of such $I_c(H)$ data is colloquially referred to as the ``Fraunhofer pattern.”  The period of the Fraunhofer pattern in the applied field $H$ is inversely proportional to the width $w$ of the junction in the direction transverse to the applied field.  For the case of circular junctions or elliptical junctions with the field applied along a symmetry axis, the interference pattern follows an Airy function,\cite{barone1982physics} but we still refer to the data as the Fraunhofer pattern out of habit.  

The introduction of magnetic materials inside the junction leads to complications due to the field produced by the magnetic material; the Fraunhofer pattern is modified and may be severely distorted.  In principle, if one knew the exact configuration of the magnetization $\mathbf{M}$ as a function of $H$, one could calculate the vector potential everywhere and deduce the $I_c(H)$ pattern.\cite{blamire_2013,golovchanskiy_2016}  The converse is not true, however, except in the simplest cases discussed below.  Consider first the worst situation, which is an S/F/S Josephson junction of relatively large lateral size (more than a few $\mu$m) containing a thick layer of a strong F material such as Ni, Fe, Co, or Gd.  Due to the complicated multi-domain magnetic configuration of the F material, the $I_c(H)$ pattern has a random-looking set of peaks and valleys with no discernible period and no clear central peak.\cite{bourgeois_2001,Khaire_PRB_2009,khasawneh_2009}  In such a situation it is neither possible to calculate the domain structure from the $I_c(H)$ data nor to deduce the maximum critical current density $J_c$ inside the junction.  

Fortunately, there exist many situations that produce well-behaved Fraunhofer patterns from which one can extract key junction parameters – in particular the maximum value of $J_c$.  An illustrative example is the early work of Ryazanov on a Josephson junction containing a weakly-ferromagnetic CuNi alloy.\cite{ryazanov_1999} That material has a mild perpendicular magnetic anisotropy (PMA),\cite{veshchunov_2008} which is advantageous because the vector potential corresponding to an out-of-plane magnetization points in the plane and does not affect the Fraunhofer pattern.  In addition, if both the magnetization and domain size are small, then the vector potential effectively undergoes a random walk throughout the sample and never gets very large in magnitude. For those reasons, Ryazanov's CuNi virgin-state junction produced an excellent Fraunhofer pattern centered at zero field.  After the junction was magnetized in a large in-plane field, the Fraunhofer pattern was shifted in field with only minor distortion, indicating that the CuNi alloy had acquired a nearly-uniform in-plane component to its magnetization.  After demagnetizing the sample, the Fraunhofer pattern reverted back to its original shape centered at the origin. 

Even in junctions containing strong F materials, the Fraunhofer pattern is well-behaved if the magnetization of the material is uniform and parallel to the direction of the applied field $H$.  If it stays uniform over the full measured field range, then Eqn.(\ref{Eqn:Fraunhofer}) is still valid but with a total magnetic flux that includes the contribution from the F material: 
\begin{equation} \label{Eqn:Flux}
    \Phi = \mu_0 H w (2\lambda_{\mathrm{eff}} + d_F) + \mu_0 M (1 – N_{demag}) w d_F 
\end{equation}
where $M$ is the magnetization of the F material and $N_{demag}$ is the demagnetizing factor determined by the shape of the F layer.  Most workers using Eqn.(\ref{Eqn:Flux}) ignore the demagnetizing factor, which is small for thin oblong-shaped F layers.  Note that Eqn.(\ref{Eqn:Flux}) neglects any magnetic flux that returns inside the junction – a quantity that is difficult to calculate accurately.  Setting the total flux $\Phi = 0$ shows that the central peak of the Fraunhofer pattern is shifted in field by an amount:
\begin{equation} \label{Eqn:Shift}
    H_\textit{shift} = \frac{-M d_F} {2\lambda_{\mathrm{eff}} + d_F} 
\end{equation}
where we have neglected $N_{demag}$.  Note that the denominator should also include the thicknesses of any other non-superconducting layers inside the junction – e.g. the Cu spacers used in most of the junctions studied by the authors. 

\begin{figure}
  \begin{center}
    \includegraphics[width=0.5\textwidth]{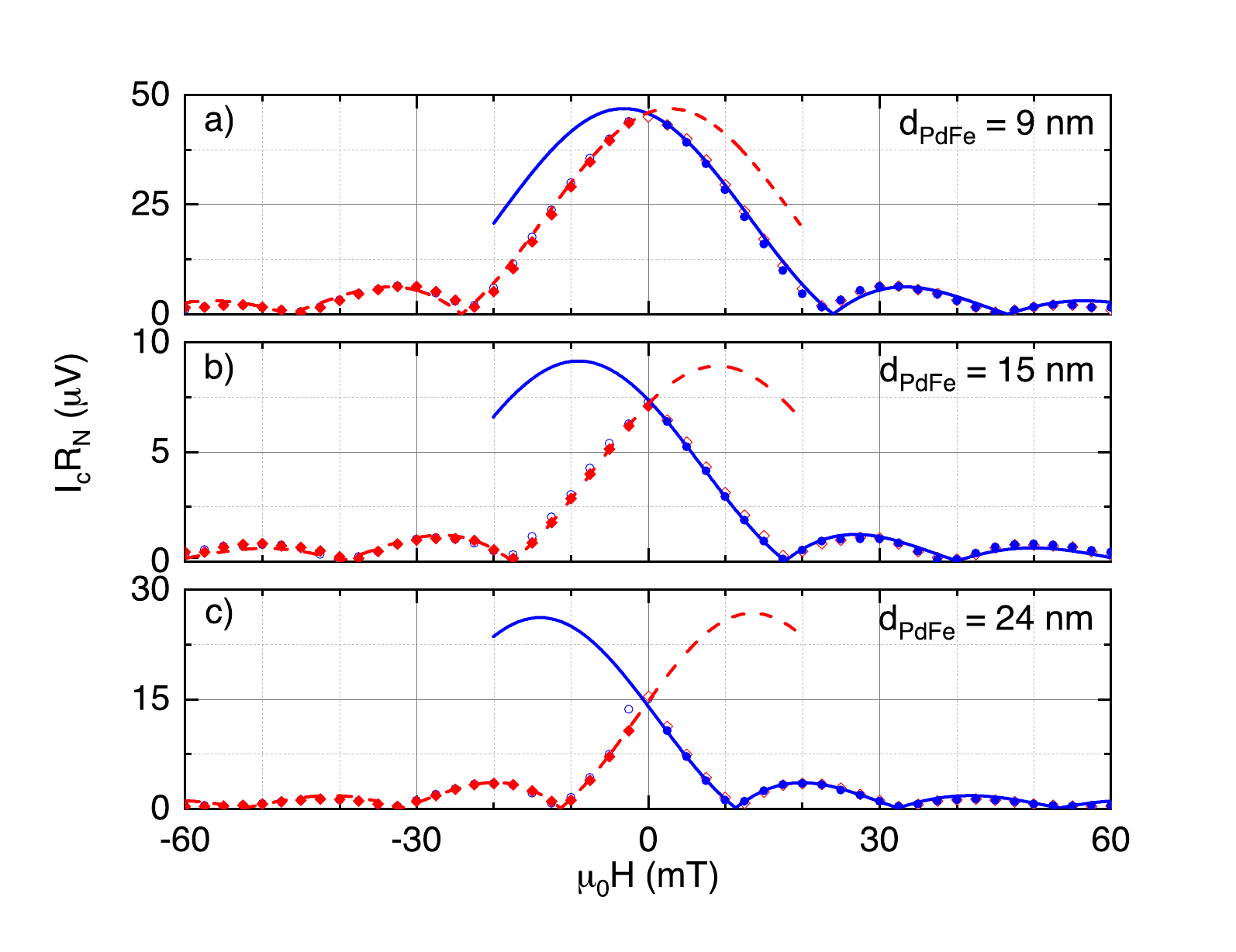}
    \caption{\small $I_c R_N$ vs $\mu_0 H$ for Josephson junctions containing a PdFe alloy with 3\% Fe with thicknesses labeled in the figure.  Junctions have elliptical shape and lateral dimensions of $\approx 1.3 \times 0.55 \mu$m$^2$. The data acquired before the field at which the PdFe magnetization vector reverses direction (solid markers), and the corresponding fits (lines) show good agreement for both the positive (red, dashed) and negative (blue) field sweep directions. The hollow circles are the corresponding data points after the PdFe magnetization switches. These Fraunhofer patterns display magnetic hysteresis and are increasingly shifted with larger $d_F$. Adapted with permission from IEEE Trans. Appl. Supercond. 27, 1800205 (2017).\cite{Glick_IEEE_2017}}
    \label{FigPdFeFraun}
  \end{center}
\end{figure}

\begin{figure}
  \begin{center}
    \includegraphics[width=0.5\textwidth]{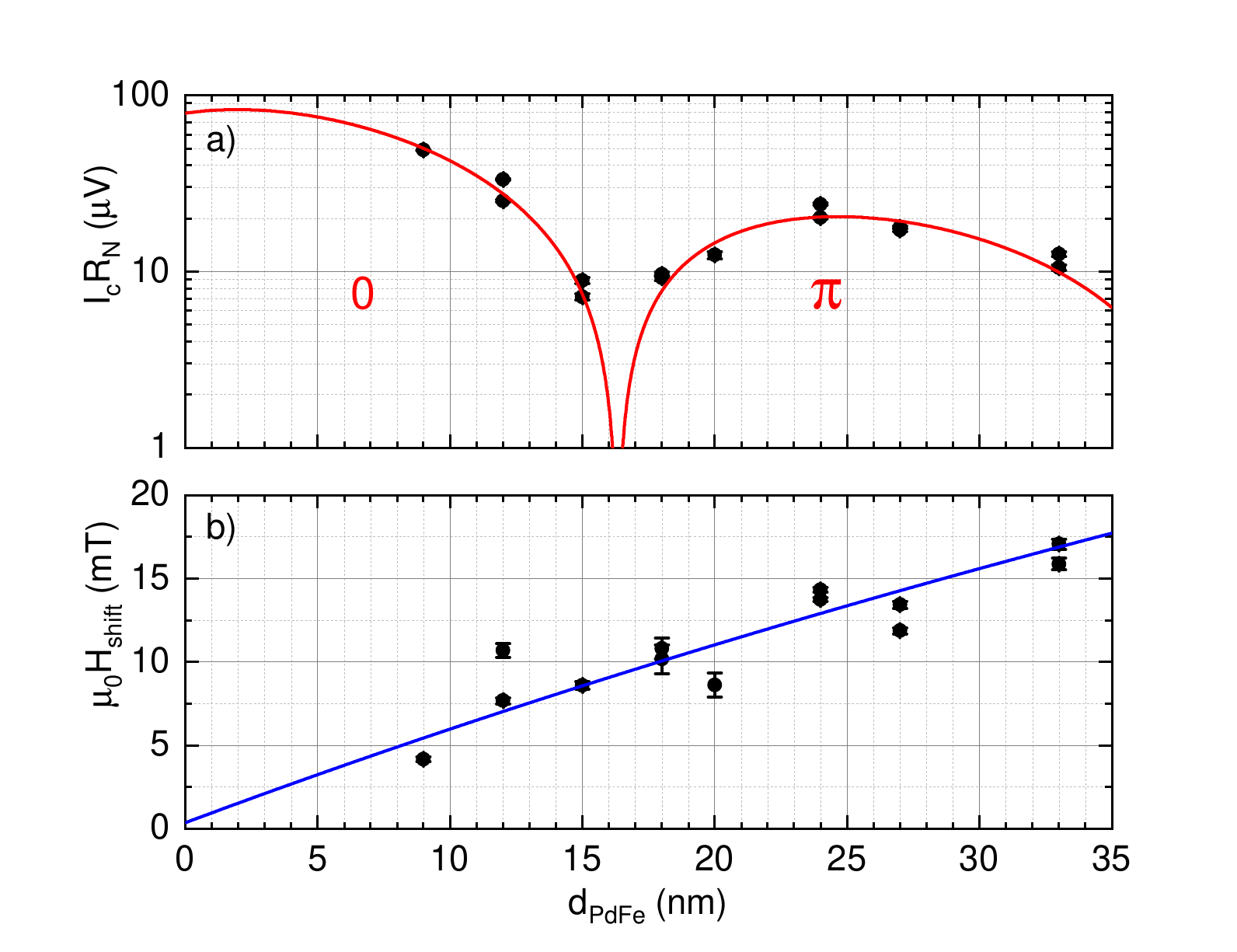}
    \caption{\small Characterization of Josephson junctions containing a weakly ferromagnetic PdFe alloy.  a) The maximal $I_c R_N$ product of the junctions versus PdFe thickness.  The minimum indicates the PdFe thickness at which the junctions transition between the 0 and $\pi$-phase states. The solid line is a fit of the data to Eqn.~\ref{eq:generic}.  b) The field shift of the $I_c(B)$ data (the ``Fraunhofer pattern") vs PdFe thickness.  The field shift increases with $d_{\mathrm{PdFe}}$ due to the increasing total magnetic moment inside the junctions and is fit to Eqn.~\ref{Eqn:Shift}. Reproduced with permission from IEEE Trans. Appl. Supercond. 27, 1800205 (2017).\cite{Glick_IEEE_2017}  Copyright 2017, IEEE.}
    \label{FigPdFeIcRn}
  \end{center}
\end{figure}

We illustrate Eqn. \ref{Eqn:Shift} with data from junctions containing a PdFe alloy with about 3\% Fe, shown in Figs. \ref{FigPdFeFraun} and \ref{FigPdFeIcRn}. We have chosen this data set because this weakly-ferromagnetic alloy has a large value of $\xi_F$ (several nm, see Table \ref{weak}), hence thickness fluctuations are negligible and there is very little scatter in the data of $I_c R_N$ vs $d_F$. Nevertheless, the magnetization is large enough to shift the Fraunhofer patterns considerably.  The PdFe thickness where $I_c R_N$ has its peak value in the $\pi$ state is about 24 nm. At that thickness, the Fraunhofer pattern is shifted by about 14 mT, and the value of $I_c$ measured at zero field is significantly smaller than the peak value extrapolated from the fit, Figure~\ref{FigPdFeFraun} (c). Hence this material would not be ideal for $\pi$-junction applications unless the junctions were considerably narrower in their lateral dimension, with correspondingly wider Fraunhofer patterns.

\section{Applications of $\pi$-junctions}\label{Sec:App}

Several applications of Josephson junctions involve superconducting loops threaded by half a quantum of magnetic flux, which has two degenerate ground states with supercurrent circulating in opposite directions.  The fundamental motivation for using a $\pi$-junction in place of the flux bias is to improve the circuit performance in some way, e.g. by reducing noise, reducing the size of the circuit (e.g. by eliminating the need for flux bias lines or a large geometrical inductance in the loop), or increasing the operating margins.

Even before Ryazanov’s 2001 experimental breakthrough with S/F/S junctions, there already existed experimental realizations of $\pi$-junctions without magnetic materials.  The most famous example involved the demonstration of the \textit{d}-wave character of the high-T$_c$ superconductors, where it was shown that a phase shift of $\pi$ is created by a Pb/YBCO/Pd double-junction based on two orthogonal crystal faces of a YBa$_2$Cu$_3$O$_7$ single crystal,\cite{wollman_1993} or by three YBCO grain-boundary junctions in series.\cite{kirtley_1995} $\pi$-junctions can also be realized in 4-terminal S/N/S junctions driven out of equilibrium by a control current flowing orthogonal to the supercurrent direction\cite{yip_1998,wilhelm_1998,baselmans_2001} or in 3-terminal junctions with the control current injected into the junction.\cite{vanweesbj_1991,huang_2002,crosser_2008} While the nonequilibrium $\pi$-junctions demonstrate fascinating physics, we would argue that they are not likely to be suitable for practical applications. 

\subsection{$\pi$-junctions in superconducting qubits}

The results from \textit{d}-wave superconductors\cite{wollman_1993,kirtley_1995} inspired Ioffe and co-workers to propose using $\pi$-junctions in superconducting phase or flux qubits for quantum computing.\cite{ioffe_1999,blatter_2001}   A conventional flux qubit consists of a superconducting loop containing three Josephson junctions.\cite{mooij_1999,Orlando_1999}  Under an applied flux bias of exactly $\Phi_0/2$, states with clockwise and counterclockwise supercurrent flow are degenerate, and the symmetric and antisymmetric linear combinations of those states can form the two states of a qubit.  Replacing the $\Phi_0/2$ flux bias with a $\pi$-junction is beneficial in several ways.  In systems with many qubits the $\Phi_0/2$ flux bias can be provided either by a global magnetic field or by individual flux lines for each qubit.  The former suffers from field inhomogeneity and fluctuations in loop area among the qubits while the latter suffers from crosstalk between neighboring qubits and the corresponding need to calibrate every qubit flux precisely.  Both suffer from fluctuations in dc supply current.  Replacing the flux bias with $\pi$-junctions alleviates those issues.  In addition, removing the large static flux bias means that the qubits can be made much smaller, thereby reducing sensitivity to magnetic field fluctuations and improving prospects for large-scale integration.  We note that individual flux bias lines are still needed for qubit operations, but they supply only small, pulsed currents; there is no need for a large static dc current.

The idea to use $\pi$-junctions in various types of qubits has been developed further by several authors,\cite{yamashita_2005,yamashita_2006,kawabata_2006,kato_2007,kawabata_2007,feofanov_2010,shcherbakova_2015} and is being pursued to this day.\cite{mori_2021,ahmad_2022,massarotti_2023,kim_2024}  While in the early proposals, the $\pi$-junction was considered to be an active device, \cite{yamashita_2005,yamashita_2006,kawabata_2006} later works emphasized the advantages of using  the $\pi$-junction as a passive phase-shifter,\cite{kato_2007,feofanov_2010,shcherbakova_2015,kim_2024}  where the only constraint on the critical current of the $\pi$-junction is that it be much larger than the critical currents of the conventional S/I/S junctions in the loop.  

A crucial issue confronting all types of qubits is decoherence.  Kato \textit{et al.} published the first calculation of the relaxation and dephasing rates of a flux qubit containing a ferromagnetic $\pi$-junction as a passive phase shifter.\cite{kato_2007}  Those authors state explicitly that only underdamped $\pi$-junctions with very large critical current density have sufficiently long dephasing times to be suitable for qubit applications. Nevertheless, the first demonstration of a ferromagnetic $\pi$-junction in a qubit, carried out by Feofanov \textit{et al.} in 2010,\cite{feofanov_2010} used an overdamped S/F/S $\pi$-junction in series with a standard underdamped S/I/S junction to form a phase qubit. The choice of using a phase qubit rather than a flux qubit was for device simplicity.  Feofanov \textit{et al.} compared the properties of their $\pi$-junction qubit with a reference qubit without the $\pi$-junction. While there was no noticeable decrease in the decoherence time of the $\pi$-junction qubit relative to the reference qubit, both qubits had coherence times of only a few nanoseconds.  Calculation of the expected decoherence time using the theoretical framework of Kato \text{et al.}\cite{kato_2007} gave a result in the nanosecond range, but that theory was not intended to describe the phase qubit of Feofanov \textit{et al}. Later, Shcherbakova \textit{et al.}\cite{shcherbakova_2015} fabricated flux qubits with and without a $\pi$-junction and showed explicitly the shift of the magnetic-field dependence of the qubit energy splitting by exactly half of a flux quantum, but they did not measure the decoherence times of their qubits. 

Very recently, Kim \textit{et al.} fabricated and measured flux qubits with and without an S/F/S $\pi$-junction.\cite{kim_2024}  The $\pi$-junction was used as a passive phase shifter with a critical current much larger than those of the conventional S/I/S junctions in the qubit loop.  The coherence time of the standard qubit was more than 1,000 times larger than that of the phase qubit studied earlier by Feofanov \textit{et al.}\cite{feofanov_2010}, making this a much more stringent test of how the $\pi$-junction affects qubit performance.  Kim \textit{et al.} found that the energy relaxation time ($T_1$) of the flux qubit decreased from about $15 \mu s$ to $1.5 \mu s$ upon insertion of the $\pi$-junction, while the value of $T_2$ decreased from about $11 \mu s$ to $1.5 \mu s$.  The authors attribute the decreases to dissipation in the S/F/S junction due to quasiparticle excitations.  While these results are somewhat disappointing, they are not surprising given the insistence by Kato \textit{et al.} that only underdamped $\pi$-junctions are suitable for use in quantum circuits, even when the $\pi$-junction is used as a passive phase shifter.\cite{kato_2007}  Kim \textit{et al.} suggest using a $\pi$-junction based on a ferromagnetic insulator (FI) rather than a metallic ferromagnet, and we certainly agree.  Unfortunately, the only FI that has been studied extensively in Josephson junctions is GdN, where a number of complex behaviors have been observed.  We delay our discussion of GdN Josephson junctions to Section V of this review.  Another possibility would be to use an S/I/F/S junction, if the critical current density can be made large enough and the damping small enough to satisfy the requirements laid out by Kato \textit{et al.}

Nowadays it is possible to fabricate so-called ``fluxonium” qubits with coherence times as large as a millisecond.\cite{somoroff_2023}  Like traditional flux qubits, fluxonium qubits also have an optimal working point when flux biased at $\Phi_0/2$.\cite{pop_2014}  It would be very informative to repeat the experiment of Kim \textit{et al.} on a fluxonium qubit, but with an underdamped S/FI/S or S/I/F/S $\pi$-junction in place of Kim’s S/F/S $\pi$-junction.

\subsection{$\pi$-junctions in superconducting digital logic}

A more promising application of $\pi$-junctions, in our opinion, is in superconducting digital electronics.\cite{likharev_2012,soloviev2017beyond} There have been numerous proposals for using $\pi$ phase shifters in superconducting digital circuits, starting with the pioneering papers by Terzioglu and Beasley in 1998\cite{terzioglu_1998} and Ustinov and Kaplunenko\cite{ustinov_2003} in 2003, and continuing to the present.\cite{ortlepp_2006,mielke_2008,khabipov_2010,wetzstein_2011,kamiya_2018,yamanashi_2018,hasegawa_2019,arai_2019,yamanashi_2019,yamashita_2021,takeshita_2021,li_2021,li_2023,li_2023a,tanemura_2023,soloviev_2021,soloviev_2022,maksimovskaya_2022,khismatullin_2023}  The initial motivation for introducing $\pi$-junctions into single-flux-quantum (SFQ) digital logic circuits was to eliminate the need for the large geometrical loop inductances needed to store single flux quanta, thereby reducing the circuit size substantially.\cite{ustinov_2003}  Subsequently it was argued that incorporating $\pi$-junctions could also alleviate the need for current bias lines and increase circuit operating margins.\cite{ortlepp_2006}  More recently, entire new families of superconducting digital logic incorporating $\pi$-junctions are being explored, as we will see below. 

The distinction between active and passive roles of the $\pi$-junction is crucial in this context of classical digital circuits, just as it was in qubits. The speed at which a Josephson junction switches into the voltage state is limited by the time $\tau_{switch} = \hbar/(e I_c R_N)$.  Metallic S/F/S $\pi$-junctions without an insulating barrier have very low values of $I_c R_N$, hence they are not suitable as switching elements in high-speed circuits. The easiest way around that problem is to use S/F/S junctions only as passive phase shifters embedded in circuits that contain conventional S/I/S junctions.\cite{feofanov_2010}  The critical current of the S/F/S phase shifter must be considerably larger than that of the S/I/S junctions, so that the S/F/S junction never switches into the voltage state and its phase drop stays close to $\pi$.  An alternative approach is to develop $\pi$-junctions with a large $I_c R_N$ product.  That is not trivial; simply adding an insulating barrier increases the value of $R_N$ by a large factor but simultaneously decreases $I_c$ -- usually by a larger factor.\cite{Weides_APL_2006}  An ingenious solution is to insert a thin superconducting layer between the insulating barrier and the F layer, to get S/I/s/F/S.\cite{Larkin_APL_2012,vernik_2013,bakurskiy_2013a,Parlato_JAP_2020}  Depending on the thickness of the ``s" layer, the S/I/s/F/S behaves either as a single Josephson junction with large $I_c R_N$ product, or as a series combination of a standard S/I/S junction and an S/F/S junction.

The first laboratory demonstration of a single-flux-quantum (SFQ) digital logic circuit containing intrinsic $\pi$ phase shifts was reported by Ortlepp \textit{et al.}\cite{ortlepp_2006} using high-T$_c$/low-T$_c$ hybrid Josephson junctions.\cite{hilgenkamp_2003}  The $\pi$ phase shifts were introduced into superconducting loops by using orthogonal facets of the YBCO a-b plane, as was done in the original measurement of the \textit{d}-wave symmetry in YBCO.\cite{wollman_1993} Ortlepp \textit{et al.}\cite{ortlepp_2006} demonstrated proper functioning of a toggle flip-flop circuit, and they used circuit simulations to compare two identically-functioning circuits with and without $\pi$ phase shifters. The circuit with the phase shifters contained fewer current bias lines, took up considerably less area, and had wider operating margins -- i.e. better stability against variations in circuit parameters.  Shortly thereafter, Balashov \textit{et al.}\cite{balashov_2007} demonstrated an all-niobium toggle flip-flop with a $\pi$ phase shifter based on the ``trapped flux" method.\cite{majer_2002} The $\pi$ phase shifter replaces a large geometrical inductance, hence diminishes the circuit size substantially. Using the same type of trapped-flux $\pi$-shifter, Wetzstein \textit{et al.} compared the bit error rates of toggle flip-flops with and without a $\pi$ phase shifter, and concluded that the former had improved robustness against circuit noise.\cite{wetzstein_2011} 

The first demonstrations of all-Nb SFQ circuits containing ferromagnetic $\pi$-junctions were reported by Feofanov \textit{et al.}\cite{feofanov_2010} and by Khabipov \textit{et al.}\cite{khabipov_2010} in 2010. Those works represent a major milestone because all modern large-scale SFQ circuits are fabricated using only Nb-based superconductors, and because they represent the first example of a scalable process for incorporating $\pi$-junctions into SFQ circuits. In comparison to trapped-flux $\pi$-shifters, S/F/S junctions are more compact and do not require a delicate initialization procedure.  Both groups used the $\pi$-junction as a passive phase shifter in a circuit containing conventional S/I/S switching junctions.  Both groups identified circuit size as the main advantage of using $\pi$-junctions. 

There has been a surge of interest in $\pi$-junctions in the past few years in Japan.  The Yoshikawa group at Yokohama University has shown through simulations that including $\pi$-junctions in certain SFQ circuits reduces the junction count significantly, while the new circuits have wider operating margins and lower static power dissipation that their conventional counterparts.\cite{yamanashi_2018,yamanashi_2019} They and others propose similar advantages using $\pi$-junctions in ultra-low-power adiabatic quantum flux parametron circuits.\cite{arai_2019,soloviev_2022,khismatullin_2023} The Fujimaki group at Nogoya University recently invented a new family of superconducting logic called ``half-flux-quantum" logic, where information propagates in the form of voltage pulses with area equal to $\Phi_0 /2$ rather than $\Phi_0$ as in standard SFQ circuits.\cite{kamiya_2018,hasegawa_2019,li_2021,takeshita_2021,li_2023,li_2023a,tanemura_2023} The basic unit of half-flux quantum logic was originally a $0-\pi$ SQUID, where the 0-junction and $\pi$-junction in the SQUID must have similar critical currents and switching properties. The inventors soon realized, however, that it is more practical to use a $0-0-\pi$ SQUID instead, where the $\pi$-junction is used as a passive phase shifter and never switches into the voltage state. Half-flux-quantum circuits offer several advantages compared to standard SFQ circuits, including smaller bias currents, smaller circuit size, and lower power dissipation.

Another new logic family, called ``phase logic", was introduced very recently by the group at Moscow State University.\cite{soloviev_2021,maksimovskaya_2022} A major selling point of phase logic is the absence of inductors, which take up much of the space in standard SFQ circuits and cause crosstalk between neighboring cells, thereby being a major hindrance to scaling those circuits to smaller sizes. In the original proposal, phase logic circuits were based on bistable Josephson junctions with a dominant second harmonic in their current phase relation: $I_s(\phi)=I_c \textrm{sin}(2\phi)$.  The existence of such junctions has been demonstrated in special situations,\cite{Stoutimore_PRL_2018,pal_2014} and proposed to exist in others,\cite{trifunovic_2011a,trifunovic_2011,melnikov_2012,richard_2013} but we emphasize that they are not available as ``off the shelf" components ready for use in electronic circuits.  Hence it was encouraging when Maksimovskaya \textit{et al.}\cite{maksimovskaya_2022} suggested a way to construct phase logic circuits using standard $\pi$-junctions. Phase logic appears to be a promising avenue for scaling superconducting circuits to higher densities, and is being pursued also by circuit designers in the US.\cite{salameh_2022,jabbari_2023,razmkhah_2024,cong_2023} 

We note that the $0-0-\pi$ SQUIDs that form the basic unit of the Nagoya group's half-flux-quantum logic are equivalent to a single junction with a negative second harmonic current-phase relation: $I_s(\phi)=-I_c \textrm{sin}(2\phi)$.\cite{soloviev_2021} The ground state of such a SQUID is doubly-degenerate, with a phase shift of $\pm\pi/2$ across the SQUID.  In contrast, the basic unit of the Moscow group's original phase logic proposal is a junction with a positive second harmonic current-phase relation, which puts its ground state phase at either 0 or $\pi$.  Those two situations have very different consequences for circuit design.

\subsection{Controllable junctions: superconducting memory}\label{Sec:SVApps}

Josephson junctions with controllable properties are suitable for applications in superconducting memory.  The controllable property can be either the magnitude of $I_c$ or the phase state of the junction, but in the former case the junction must be active, i.e. it must switch into the voltage state when the applied current surpasses $I_c$, which means it must have a large $I_c R_N$ product for high-speed operation, as discussed earlier.  

There are several proposals for controlling the magnitude of $I_c$ in Josephson junctions containing a single magnetic layer. The Ryazanov group developed controllable S/F/S junctions containing a weakly-ferromagnetic PdFe alloy.\cite{bolginov_2012}  In a collaboration with the Hypres Corporation in the US, that idea was extended to controllable S/I/s/F/S junctions with a large $I_c R_N$ product.\cite{Larkin_APL_2012,vernik_2013} Control of $I_c$ was achieved by the orbital effect -- i.e. by shifting the Fraunhofer pattern of the junction via partial magnetization of the PdFe.  A disadvantage of that approach is that it does not scale well as the junction size is reduced.  The junction size can be reduced in one in-plane dimension, but must remain large in the other dimension to achieve adequate Fraunhofer shift.\cite{karelina_2021}  A recent proposal to control $I_c$ via the stray field produced by the magnetic texture of a Co disk also relies on the orbital effect, but the constraints on junction size are less severe because of the large magnetization of the Co.\cite{fermin_2022}

Several other schemes for controlling $I_c$ of Josephson junctions have been proposed, and we do not try to review all of them here.  We mention one based on placing Abrikosov flux vortices in close vicinity to a Josephson junction,\cite{golod_2015} one based on rotating the magnetization in a combination S/F/S and S/F/N/S device,\cite{soloviev_2014} and another based on a multiterminal S/F device.\cite{nevirkovets_2018,nevirkovets_2018a} 

The possibilities become richer when there are two independent F layers inside the junction -- one with fixed magnetization and the other with a magnetization that can be reoriented by application of a small external field. Either the magnitude of $I_c$ or the phase state of the junction can be controlled in such junctions, as predicted in a number of theoretical works.\cite{bergeret_2001,krivoruchko_2001,golubov_2002,barash_2002,chtchelkatchev_2002,zaitsev_2003,blanter_2004,pajovi_2006,crouzy_2007}  The first experimental verification of $I_c$ modulation in such ``spin-valve" junctions was performed by Bell \textit{et al.} in 2004, using Co and Ni$_{.80}$Fe$_{.20}$ (NiFe) as the fixed and free F layer materials, respectively.\cite{Bell_APL_2004}  A similar experiment was revisited a decade later in a much more thorough study by Baek \textit{et al.},\cite{Baek_NComms_2014} and also by Qadar \textit{et al.}.\cite{Qader_APL_2014} Those groups used NiFe-based alloys as the free layer because of their excellent magnetic switching properties, even though they are not so great for supercurrent transmission (see Table \ref{NiFe}). Baek \textit{et al.} used Ni as the fixed layer material since Ni has the best supercurrent transport of any strong F material (see Table \ref{pure} in Section IV). 

Controlling the phase state of the junction rather than the amplitude of $I_c$ has two advantages. First, there is no need for a large $I_c R_N$ product, which is difficult to achieve in all-metal junctions.  Second, the junction phase is intrinsically a digital rather than analog quantity; it can take only one of the two values zero or $\pi$, as long as the F-layer is far from the thickness corresponding to the $0 - \pi$ transition.  The first demonstration of phase control in spin-valve junctions was carried out by one of the authors and his students,\cite{Gingrich_NatPhys_2016,Madden_SUST_2018} using Ni and NiFe for the fixed and free F-layer materials. Similar junctions formed the basis for Northrop Grumman's ``Josephson Magnetic Random Access Memory" or JMRAM, which was demonstrated on a $2 \times 2$ memory array in 2017.\cite{Dayton_IEEE_2017}  A later demonstration on an $8 \times 8$ memory array revealed a difficulty with this approach, namely the inconsistent behavior of the magnetic layers in different memory cells.  The Ni fixed layers are multi-domain at the $\mu$m-scale lateral dimensions of the junctions, and require a very large initialization field to magnetize completely.\cite{Baek_NComms_2014}  And the switching fields of the NiFe free layers varied significantly from cell to cell.  A great deal of materials research would be needed to turn this idea into a successful memory technology. 

Finally, we mention proposals that involve intrinsic bistability of $I_c$ or of the junction phase state without requiring any change in the magnetic configuration.\cite{bakurskiy_2016,bakurskiy_2018} 

\section{Metallic $\pi$-junctions}\label{Sec:Tables}

The Tables in this section are intended for the reader to make quick comparisons between materials and works. We attempt to include as much of the available literature as possible, even where fit parameters could not be extracted from the data; hence there are gaps in the Tables. We present four Tables on single F layer junctions: \ref{pure} -- pure elements, \ref{NiFe} -- NiFe and NiFe-based alloys, \ref{weak} -- weak ferromagnetic alloys, and \ref{strong} -- strong and other ferromagnetic alloys. Within each Table, materials are grouped together, and where there are multiple works on the same material, the works are listed chronologically. 

The methodology for putting together the Tables in this section is as follows. Fit parameters are either extracted from the original publications, or datasets are extracted from the plotted graphs, fit to Eqn.~\ref{eq:generic} and the best fit parameters are reported in the Table. We fit the expression only to the nominal thickness of the ferromagnetic layer, ignoring magnetic dead layers and any role of additional buffer layers.
The reported uncertainties in the original datasets are not taken into consideration for the fitting, so we quote all fit parameters to 2 significant figures without consideration of their uncertainties. For more rigorous data analysis, the reader should refer to the published works, particularly for cases where Eqn.~\ref{eq:generic} does not provide the best fit to the experimental data. We attempt to note in the main text and figure captions as much as possible where subtleties exist in the experimental data which are not accounted for in the fitting here. 

The parameters we report in the Tables include the fit parameters from Eqn.~\ref{eq:generic}, corresponding to the decay and oscillation lengths $\xi_{F1}$ and $\xi_{F2}$, the thickness at which the first 0-$\pi$ transition occurs, $d_{0-\pi}$, and the fictitious zero thickness fit parameter, $I_cR_N (0)$, which is necessary for the reader to reproduce the fits but doesn't have a physical meaning. (Junctions without the ferromagnet will not have the extracted $d_{F}=0$ critical current as there are fewer interfaces in such devices.) We also extract the critical current at the peak of the $\pi$-state, which is a useful parameter for device consideration and for comparing materials. When the range of thicknesses studied is not broad enough to cover multiple 0-$\pi$ oscillations (multiple minima in $I_cR_N$), the $\xi_{F2}$ fitting parameter ambiguous, in which case that parameter will be missing from the Tables.
 
There are two common conventions for reporting the magnitude of critical Josephson current. The first is the product of the critical current and normal state resistance, $I_cR_N$, quoted here in $\mu$V. The second convention is to report the critical current density, $J_c$, quoted here in kAcm$^{-2}$. The main difference between $I_cR_N$ and $J_c$ is that the junction area need not be known to report $I_cR_N$, which may be advantageous for a study where multiple junction sizes are fabricated (intentionally or unintentionally). Where the original publications contain all the information necessary to perform the conversion, we report both quantities at the peak of the $\pi$-state in the Tables. Otherwise, the reader can consider the approximation that for metallic S/F/S junctions the product of area and normal state resistance, $A R_N \approx 10^{-14}~\Omega$m$^2$, which arises primarily from the boundary resistances between the S electrodes and the F layer.\cite{fierz_1990} Hence, for S/F/S junctions, one can estimate $J_c$ by dividing $I_cR_N$ by $A R_N$ (in the Tables, $I_cR_N$ values quoted in $\mu$V can be multiplied by 10 to get $J_c$ in kAcm$^{-2}$ assuming $A R_N \approx 10^{-14}~\Omega$m$^2$). For junctions containing an insulating layer, such as the S/I/F/S geometry, the story is very different: $A R_N$ is typically several orders of magnitude larger, and varies widely depending on the oxidation conditions used to produce the insulating barrier.

\subsection{Pure elements}

\begin{table}[h!]
\resizebox{0.9\columnwidth}{!}{%
\begin{tabular}{l|c|c|c|c|c|c|c|c|c}
\hline
Barrier &
  \begin{tabular}[c]{@{}c@{}}$\xi_{F1}$\\ (nm)\end{tabular} &
  \begin{tabular}[c]{@{}c@{}}$\xi_{F2}$\\ (nm)\end{tabular} &
  \begin{tabular}[c]{@{}c@{}}$d_{0-\pi}$\\ (nm)\end{tabular} &
  \begin{tabular}[c]{@{}c@{}}$I_cR_N (0)$\\ ($\mu$V)\end{tabular} &
  \begin{tabular}[c]{@{}c@{}}$I_cR_N (\pi)$\\ ($\mu$V)\end{tabular} &
        \begin{tabular}[c]{@{}c@{}}$J_c (\pi)$\\ (kAcm$^{-2}$)\end{tabular} &
  \begin{tabular}[c]{@{}c@{}}dead layer\\ (nm)\end{tabular} &
  \multicolumn{1}{c|}{Ref.} &
  Comments \\ \hline
Cu/Ni/Cu       & 3.8  & 0.98 & 2.6  & 2.8  & 1.0  & 10 &     & \onlinecite{Blum_PRL_2002}        &  \\ \hline
Cu(Au)/Ni/Cu(Au) & 1.7 & 0.45 & & & & & &  \onlinecite{Shelukhin_PRB_2006} & $\bigcirc$   \\ \hline
Ni             & 4.5  & 1.2  & 4.0  & 450  & 110 & & 1.3 & \onlinecite{Robinson_PRL_2006}    &   \\ \hline
Al$_2$O$_3$/Cu/Ni    & 0.66 & 0.53 & 3.0  &      & 3.7 & $3.4\times10^{-3}$ & 2.3 & \onlinecite{Bannykh_PRB_2009}     &  \\ \hline
Cu/Ni/Cu             & 2.4  & 0.79  & 0.90  & 330  & $\geq100$ & & 0.7 & \onlinecite{Baek_PRApp_2017}    &   \\ \hline
Cu/Ni/Cu             &   &   & 3.3$^\bigstar$  &   &  & 200 & & \onlinecite{Dayton_IEEE_2017}    & $\bigcirc$, $\dagger$   \\ \hline
Ni             &   &   &   &   & $>30$  & $>300$ & & \onlinecite{Tolpygo_2019}    & $\dagger$   \\ \hline
Ni \& Cu/Ni             &   &   &    &   & & &  & \onlinecite{Kapran_2021}    & $\dagger$   \\ \hline
Cu/Co/Cu             &   &   &    &   &  &  & & \onlinecite{Surgers_2002}    & $\dagger$   \\ \hline
Co             & 2.1  & 0.29 & 1.0  & 130  & 65 &  & 0.8 & \onlinecite{Robinson_PRL_2006}    &  \\ \hline
Rh/Co/Rh       & 0.81 & 0.21 & 0.64 & 260  & 82 & 1600 & 0.0 & \onlinecite{Robinson_APL_2009}    &  \\ \hline
Co/Ru/Co       & 1.2  &      &      & 2.9  &     &   &  &  \onlinecite{khasawneh_2009}  & $\ddagger$\\ \hline
Cu/Co/Ru/Co/Cu & 2.3  &      &      & 0.90 &     &   &  & \onlinecite{khasawneh_2009}  &  $\ddagger$ \\ \hline
Fe             & 5.4  & 0.25 & 0.81 & 170  & 140 & & 1.1 & \onlinecite{Robinson_PRB_2007}    &  \\ \hline
Al/Gd/Al       & 1.2  &      &      & 20   &     &  &   & \onlinecite{bourgeois_2001} & $\ddagger$, 1~K \\ \hline
Ho             & 4.3  &      &      & 2300 &     & & 2.0 & \onlinecite{Witt_PRB_2012}        &  $\ddagger$  \\ \hline
\end{tabular}%
}
\caption{Extracted parameters for Nb based Josephson junctions with pure elements as the ferromagnet in the barriers. The fitting methodology to Eqn~\ref{eq:generic} is given in the main text and best fit values are quoted to 2 significant figures without consideration to their uncertainty so should be treated as approximations. The full barrier layers, including normal metals and insulators, are listed. Unless otherwise commented, the measurements were performed at 4.2~K. Comments: $\bigcirc$ The range of thickness values did not cover the thickness where the first 0-$\pi$ transition was predicted. $\dagger$ The range of thickness values studied was not sufficient for a meaningful fit to theory. $\ddagger$ No 0-$\pi$ oscillations were observed or claimed, and the data are fitted here with only the exponential decay component of Eqn~\ref{eq:generic}. $\bigstar$ This thickness corresponds to the first $\pi$-0 transition.  }
\label{pure}
\end{table}

In Table \ref{pure}, we highlight works studying Josephson junctions with pure elements.

Pure elemental Ni was among the first ferromagnetic barrier layers studied.\cite{Blum_PRL_2002} Ni has the lowest bulk magnetization of the transition metal ferromagnets (550 emu/cm$^3$). Experimentally, it is found that Ni has favorable supercurrent carrying properties - the decay length is relatively long for a strong ferromagnet and the $I_cR_N$ is high. These favourable properties have resulted in a number of reports on Ni Josephson junctions. \cite{Dayton_IEEE_2017, Tolpygo_2019, Kapran_2021, Blum_PRL_2002, Shelukhin_PRB_2006, Robinson_PRL_2006, Robinson_PRB_2007, Robinson_IEEE_2007, Bannykh_PRB_2009, Baek_NComms_2014, Baek_PRApp_2017, Baek_IEEE_2018} It is worthwhile noting that while Table \ref{pure} is obtained by fitting to Eqn.~\ref{eq:generic} which describes diffusive and intermediate limit transport, Ni junctions have been reported in the literature in both the diffusive\cite{Bannykh_PRB_2009} and ballistic\cite{Baek_PRApp_2017} limits. While Ni may have favourable supercurrent carrying properties, the magnetic properties of thin Ni layers are far from ideal. Ni layers are magnetically `hard', requiring a large initialization field to set the magnetization direction and a large switching field. It is well known that Ni films break up into small magnetic domains, which sets a limit on single domain junction size, and can contribute stray fields into the surrounding superconducting layers.  

For determining the position of the $0-\pi$ transition for Ni from Table~\ref{pure}, we guide the reader towards Refs.~\onlinecite{Baek_PRApp_2017} and \onlinecite{Dayton_IEEE_2017}, which have the smallest increments in thickness around the transitions. The data of Baek \textit{et al.} covers both the first $0-\pi$ and $\pi-0$ transitions, which occur at 0.90 and 3.4~nm respectively.\cite{Baek_PRApp_2017} The first $\pi-0$ transition thickness is confirmed by Dayton \textit{et al.} who observe it at 3.3~nm, again taking small increments in thickness around the transition.\cite{Dayton_IEEE_2017}  Most works on Ni do not study the very thin layers of Ni in the range of Baek \textit{et al.}'s observed $0-\pi$ transition, and we note that in some cases it may still be possible to model the data in those works assuming $d_{0-\pi}=0.9$~nm. 

Critical current oscillations have also been reported in pure elements of Co\cite{Robinson_PRL_2006, Robinson_PRB_2007, Robinson_IEEE_2007, Robinson_APL_2009} and Fe,\cite{Robinson_PRB_2007, Piano_EPJB_2007} although they are less studied than Ni. Potential advantages of Co and Fe are that thin films tend to be magnetically softer than Ni, with larger magnetic domains. An unexplored potential is that thin layers of Co can have perpendicular magnetic anisotropy when deposited on Pt or Pd normal metal spacer layers. 

Finally we mention that the rare-earth metals Gd\cite{bourgeois_2001} and Ho\cite{Witt_PRB_2012} have been studied in Josephson junctions, including the thickness series to extract the decay length, however no critical current oscillations attributed to the 0-$\pi$ transition were observed. In the case of Gd, this may be due to the thickness increments (step size) between samples in that early work being too large to observe the oscillation. Ho has been extensively studied as a spin-triplet generator due to its intrinsic magnetic inhomogeneity,\cite{robinson_2010} which may make it unsuitable as a $\pi$-junction. Nb/Gd multilayers have been extensively studied for the observation of critical temperature oscillations,\cite{jiang_1995} which can guide future junction experiments. It is additionally noteworthy that Nb can seed rare-earth ferromagnet growth and that the interface is considered to be sharp with minimal interdiffusion.\cite{Kwo_1986, satchell_2017, khaydukov_2018}

\subsection{NiFe alloys}

\begin{table}[h!]
\resizebox{0.9\columnwidth}{!}{%
\begin{tabular}{l|c|c|c|c|c|c|c|c|c}
\hline
Barrier &
  \begin{tabular}[c]{@{}c@{}}$\xi_{F1}$\\ (nm)\end{tabular} &
  \begin{tabular}[c]{@{}c@{}}$\xi_{F2}$\\ (nm)\end{tabular} &
  \begin{tabular}[c]{@{}c@{}}$d_{0-\pi}$\\ (nm)\end{tabular} &
  \begin{tabular}[c]{@{}c@{}}$I_cR_N (0)$\\ ($\mu$V)\end{tabular} &
  \begin{tabular}[c]{@{}c@{}}$I_cR_N (\pi)$\\ ($\mu$V)\end{tabular} &
          \begin{tabular}[c]{@{}c@{}}$J_c (\pi)$\\ (kAcm$^{-2}$)\end{tabular} &
  \begin{tabular}[c]{@{}c@{}}dead layer\\ (nm)\end{tabular} &
  \multicolumn{1}{c|}{Ref.} &
  Comments \\ \hline
Cu/NiFe/Cu       & 1.0  & 0.76 & 2.4  & 360  & 12   & 170 & 0.7  & \onlinecite{Bell_PRB_2005}        & $\#$ \\ \hline
NiFe       & 1.6  & 0.33 & 1.3  & 240  & 78   & &  0.5  & \onlinecite{Robinson_PRL_2006}        &  \\ \hline
NiFe       & 1.2  & 0.61 &  2.4 & 17  &  1.2  &  &    & \onlinecite{Qader_APL_2014}        &  \\ \hline
Cu/NiFe/Cu       & 1.5  & 0.58 & 1.8  & 69  & 12   & 160 & 0.0  & \onlinecite{Glick_JAP_2017}        &  \\ \hline
Cu/NiFe/Cu       &   &  & 1.66  &   &    & 25  &  & \onlinecite{Dayton_IEEE_2017}        & $\dagger$ \\ \hline
Al/AlO$_\text{x}$/Nb/NiFe       &   &  &   &   &   &  &   & \onlinecite{Parlato_JAP_2020}        & $\dagger$ \\ \hline
NiFe       &   &  &   &   &   &  &   & \onlinecite{Yao_2021}        & $\dagger$ \\ \hline
Cu/NiFe/Cu       & 0.71  & 0.74 &  1.5 & 250  & 10   & 100 &   & \onlinecite{Mishra_PRB_2022}        &  \\ \hline
Cu/Ni$_{.73}$Fe$_{.21}$Mo$_{.06}$/Cu       & 0.48  & 0.96 &  2.3 & 150  & 0.20 & 1.8 &  0.2   & \onlinecite{Niedzielski_SUST_2015}        & \\ \hline
Cu/Ni$_{.65}$Fe$_{.15}$Co$_{.20}$/Cu       & 1.1  & 0.48 & 1.2  & 30  & 5   & 67 & 0.0  & \onlinecite{Glick_JAP_2017}        &  \\ \hline
Cu/Ni/NiFe/Ni/Cu       & 0.64  & 0.78 &  0.92 & 800  & 50   & 530 &   & \onlinecite{Mishra_PRB_2022}        &  \\ \hline

\end{tabular}%
}
\caption{Extracted parameters for Nb based Josephson junctions with NiFe and NiFe-based alloys as the ferromagnet in the barriers. The fitting methodology to Eqn~\ref{eq:generic} is given in the main text and best fit values are quoted to 2 significant figures without consideration to their uncertainty so should be treated as approximations. The full barrier layers, including normal metals and in one case an additional ferromagnetic layer, are listed. Unless otherwise stated in the Table, the composition was Ni$_{.80}$Fe$_{.20}$ and the measurements were performed at 4.2~K. Comments: $\#$ The bottom Nb electrode and barrier layers were epitaxial. $\dagger$ The range of thickness values studied was not sufficient for a meaningful fit to theory. }
\label{NiFe}
\end{table}

In Table \ref{NiFe}, we highlight works studying Josephson junctions with NiFe and NiFe-based alloys.

Ni$_{.80}$Fe$_{.20}$ (hereafter NiFe) is a strong ferromagnetic alloy which is optimized for magnetic switching at small magnetic fields, making it an excellent choice in spintronics research as a soft magnetic layer. In many papers, NiFe is referred to as permalloy or Py. Josephson studies of NiFe were well motivated by the spintronics community, and it has been used as the free layer of choice in superconducting spin valves.\cite{Bell_APL_2004, Gingrich_NatPhys_2016, Dayton_IEEE_2017} Additional functionality offered by NiFe is the ability to set the direction of in-plane magnetic anisotropy by applying a magnetic field during growth.

From studying Table~\ref{NiFe}, the reader will notice two possible positions of $d_{0-\pi}$ for NiFe, either between $1.3-1.8$~nm or at 2.4~nm. We guide the reader towards the work of Dayton \textit{et al.},\cite{Dayton_IEEE_2017} which takes the smallest increments in thickness for NiFe around the $0-\pi$ transition and concludes that $d_{0-\pi}=1.66$~nm, falling in the middle of the $1.3-1.8$~nm range reported by other works.\cite{Dayton_IEEE_2017}  

Although the switching properties of NiFe at room temperature are excellent, the switching degrades at low temperatures relevant to Josephson junctions. To address this limitation, NiFe-based alloys with even lower coercivity have been developed\cite{Boothby_1947, Chen_1991} and tested in Josephson junctions, such as Ni$_{.73}$Fe$_{.21}$Mo$_{.06}$ (supermalloy), which we include also in Table \ref{NiFe}. Unfortunately, the decay length inside NiFe-based alloys is experimentally found to be even shorter than NiFe, which reduces considerably the potential critical Josephson current of a $\pi$-junction based on these materials. This is likely due to the short electron mean free path in these alloys.

Other NiFe-based alloys which have been explored for use in Josephson junctions include Ni$_{.70}$Fe$_{.17}$Nb$_{.13}$, which was used by Baek \textit{et al.} as the free layer in a spin-valve device.\cite{Baek_NComms_2014} In later works, the same group used a more dilute non-magnetic concentration of the alloy as a normal metal spacer layer in junctions containing ferromagnetic Ni.\cite{Baek_PRApp_2017, Baek_IEEE_2018} Qader \textit{et al.} characterized Cu$_\text{1-x}$(Ni$_{.80}$Fe$_{.20}$)$_\text{x}$ as a possible barrier layer, however single layer Josephson junctions with this alloy have not been reported.\cite{Qader_JMMM_2017}  

The work of Bell \textit{et al.} is noteworthy because the bottom Nb electrode and barrier layers were grown epitaxially.\cite{Bell_PRB_2005} This work may, therefore, not be directly comparable to the other works in Table \ref{NiFe}. As far as we are aware, this is the only work listed in any of the Tables to study an epitaxial barrier. Epitaxy plays a role in improving the mean free path of electrons through metals, which may improve the supercurrent carrying properties of junctions with epitaxial barriers. Epitaxy has not yet been systematically studied for $\pi$-junctions and may provide fertile ground for future exploration. 

\subsection{Weak ferromagnetic alloys}

\begin{table}[]
\resizebox{0.85\columnwidth}{!}{%
\begin{tabular}{l|c|c|c|c|c|c|c|c|c}
\hline
Barrier &
  \begin{tabular}[c]{@{}c@{}}$\xi_{F1}$\\ (nm)\end{tabular} &
  \begin{tabular}[c]{@{}c@{}}$\xi_{F2}$\\ (nm)\end{tabular} &
  \begin{tabular}[c]{@{}c@{}}$d_{0-\pi}$\\ (nm)\end{tabular} &
  \begin{tabular}[c]{@{}c@{}}$I_cR_N (0)$\\ ($\mu$V)\end{tabular} &
  \begin{tabular}[c]{@{}c@{}}$I_cR_N (\pi)$\\ ($\mu$V)\end{tabular} &
      \begin{tabular}[c]{@{}c@{}}$J_c (\pi)$\\ (kAcm$^{-2}$)\end{tabular} &
  \begin{tabular}[c]{@{}c@{}}dead layer\\ (nm)\end{tabular} &
  \multicolumn{1}{c|}{Ref.} &
  Comments \\ \hline
Cu$_{.48}$Ni$_{.52}$/Cu       &   &  &   &   & &   &    & \onlinecite{Ryazanov_PRL_2001}        & $\bowtie$ \\ \hline
Cu$_{.52}$Ni$_{.48}$       &   &  &   &   &  &  &    & \onlinecite{Sellier_PRB_2003}        & $\bowtie$ \\ \hline
Cu$_{.47}$Ni$_{.53}$/Cu       & 1.3  & 3.7 & 11  & 2200  & 0.15 & 1 &   4.3  & \onlinecite{Oboznov_PRL_2006}        &  \\ \hline
Al$_2$O$_3$/Cu$_{.40}$Ni$_{.60}$       &  0.78 & 1.35 & 5.2  &   & 400 & $5.0\times10^{-3}$ &   3.1  & \onlinecite{Weides_APL_2006}        & 2.11~K \\ \hline
AlO$_\text{x}$/Cu$_{.40}$Ni$_{.60}$       & 1.2  &  & 5.8  &   &  & $1.5\times10^{-2}$ &    &  \onlinecite{Ruppelt_2015}        &  \\ \hline
AlO$_\text{x}$/Nb/Cu$_{.40}$Ni$_{.60}$       & 1.2  &  &  7.2 &   &  & $6.5\times10^{-2}$ &    &  \onlinecite{Ruppelt_2015}        &  \\ \hline
Cu$_{.47}$Ni$_{.53}$       & 1.3  & 4.3 &  7.4 &  3200 & 2.0 & 20 &     & \onlinecite{Bolginov_JLTP_2018}        &  \\ \hline
NbN/Cu$_{.40}$Ni$_{.60}$/NbN       &  1.8 & 2.0 & 5.6  & 160  & 2.2 & 8.8 &    &  \onlinecite{Yamashita_2017}        &  \\ \hline
Cu$_{.48}$Ni$_{.52}$/Al       &  1.3 & 3.2 &  15$^\bigstar$ &   & $>4.2$ & $>60$ &    &  \onlinecite{zeng_2022}        & $\bigcirc$ \\ \hline
Al/Al$_2$O$_3$/Pd$_{.88}$Ni$_{.12}$       &  2.9 & 2.3 & 6.5  & 450  & 18 &  &    & \onlinecite{Kontos_PRL_2002}        & 1.5~K \\ \hline
Pd$_{.88}$Ni$_{.12}$       &  7.7 & 4.4 & 7.4$^\ast$  & 100  &  &  &    & \onlinecite{Khaire_PRB_2009}        & $\bigcirc$ \\ \hline
NbN/Pd$_{.89}$Ni$_{.11}$/NbN       & 4.4  & 3.6 & 8.5  &  540 & $\geq18$ & $\geq71$ &    & \onlinecite{Pham_2022}        &  \\ \hline
Pd$_{.89}$Ni$_{.11}$/Al/AlO$_\text{x}$       &   &  &   &   & 22 & $2.2\times10^{-2}$ &    & \onlinecite{li_2023a}        & $\dagger$ \\ \hline
Al/AlO$_\text{x}$/Nb/Pd$_{.99}$Fe$_{.01}$       &   &  &   &   &  &  &    & \onlinecite{Larkin_APL_2012}        & $\dagger$ \\ \hline
Pd$_{.97}$Fe$_{.03}$       &  16.2 & 7.2 & 16  & 100  & 21 & 190 & 2.8   & \onlinecite{Glick_IEEE_2017}        &  \\ \hline
Pt$_\text{x}$Ni$_\text{1-x}$       &   &  &   &   & &   &    & \onlinecite{Kapran_2021}        & $\bowtie$ \\ \hline
\end{tabular}%
}
\caption{Extracted parameters for Nb based Josephson junctions with weak ferromagnetic alloy barriers. In Refs.~\onlinecite{Yamashita_2017} and \onlinecite{Pham_2022}, the superconductor was NbN, as indicated in the Table. The fitting methodology to Eqn~\ref{eq:generic} is given in the main text and best fit values are quoted to 2 significant figures without consideration to their uncertainty so should be treated as approximations. The full barrier layers, including normal metals and insulators, are listed. Unless otherwise commented, measurements were performed at 4.2~K. Comments: $\bowtie$ The main focus was to study the temperature and/or compositional variation of the Josephson effect instead of the thickness dependence. $\dagger$ The range of thickness values studied was not sufficient for a meaningful fit to theory. $\ast$ This value was extrapolated from measurements on junctions outside of this thickness. $\bigcirc$ The range of thickness values did not cover the thickness where the first 0-$\pi$ transition was predicted. $\bigstar$ This thickness corresponds to the first $\pi$-0 transition.}
\label{weak}
\end{table}

In Table \ref{weak}, we highlight works studying Josephson junctions with weak ferromagnetic alloys.

In the context of ferromagnetic Josephson junctions, weak ferromagnet refers to materials with a low Curie temperature, and therefore low exchange energy. The structural and magnetic properties of weak ferromagnetic alloys was reviewed in detail by Kupriyanov, Golubov, and Siegel.\cite{Kupriyanov_review_2006} 

From Table~\ref{weak}, the reader will notice that although CuNi alloys have been widely studied, they are far from an ideal material as a $\pi$-junction. The CuNi alloys are firmly in the diffusive limit where $\xi_{F1} < \xi_{F2}$, which in comparison to other materials presented in the Tables is somewhat of a rare case. Superconducting junctions with weak ferromagnets are a system where the theoretical Usadel equations are valid, and therefore it is possible in these systems to directly test predictions of that theoretical framework. Oboznov \textit{et al.} apply the Usadel equations to model their data, considering that their junctions showed strong spin dependent scattering, attributed to magnetic inhomogeneities in the barrier caused by the formation of Ni-rich clusters.\cite{Oboznov_PRL_2006} Further work studying Nb/CuNi bilayers using scanning tunneling spectroscopy support this idea, finding significant spatial variation in the thickness and composition of Ni in the CuNi layer.\cite{Stolyarov_2022} In comparison to CuNi, the Pd-based alloys appear much more promising as potential $\pi$-junctions due to the larger critical Josephson currents observed in the $\pi$-state. 

Other works on weak ferromagnetic alloys that are not included in Table~\ref{weak} but which we wish to highlight include an experiment where CuNi junctions were fabricated where the barrier layer has a thickness step, such that junction has both 0 and $\pi$ parts.\cite{Weides_PRL_2006} Further work on CuNi includes observation of a positive second harmonic in the current-phase relation in junctions with a CuNi thickness close to the $0-\pi$ transition.\cite{Stoutimore_PRL_2018}  

\subsection{Strong and other ferromagnetic alloys}

\begin{table}[]
\resizebox{0.9\columnwidth}{!}{%
\begin{tabular}{l|c|c|c|c|c|c|c|c|c}
\hline
Barrier &
  \begin{tabular}[c]{@{}c@{}}$\xi_{F1}$\\ (nm)\end{tabular} &
  \begin{tabular}[c]{@{}c@{}}$\xi_{F2}$\\ (nm)\end{tabular} &
  \begin{tabular}[c]{@{}c@{}}$d_{0-\pi}$\\ (nm)\end{tabular} &
  \begin{tabular}[c]{@{}c@{}}$I_cR_N (0)$\\ ($\mu$V)\end{tabular} &
  \begin{tabular}[c]{@{}c@{}}$I_cR_N (\pi)$\\ ($\mu$V)\end{tabular} &
    \begin{tabular}[c]{@{}c@{}}$J_c (\pi)$\\ (kAcm$^{-2}$)\end{tabular} &
  \begin{tabular}[c]{@{}c@{}}dead layer\\ (nm)\end{tabular} &
  \multicolumn{1}{c|}{Ref.} &
  Comments \\ \hline
Al/Al$_2$O$_3$/Ni$_{3}$Al       & 4.6  & 0.45 &   &   &  & $\geq1.5^\diamond$ &  5 - 8  & \onlinecite{Born_PRB_2006}  & $\bigcirc$ \\ \hline
Al/AlO$_\text{x}$/Cu/Fe$_{.75}$Co$_{.25}$  & 0.22 & 0.79 & 1.9$^\ast$  &  $>1000$ & $<0.10$ &  &  0.6  & \onlinecite{Sprungmann_2009}  & $\bigcirc$ \\ \hline
Pt/Co$_{.68}$B$_{.32}$/Pt       &  0.28  & 0.20  & 0.30  & 56  & 7  & 35 &  0.0  & \onlinecite{Satchell_SciRep_2021}        & 1.8~K \\ \hline
Cu/Co$_{.56}$Fe$_{.24}$B$_{.20}$/Cu       & 0.10  &  & 0.80  &  & &  19  &     & \onlinecite{Loving_arXiv_2021}        &  \\ \hline
[Nb/Cu]$_\text{ML}$/Co$_{.56}$Fe$_{.24}$B$_{.20}$/Cu       & 0.13  &  & 0.80  & &  &  9.6  &     & \onlinecite{Loving_arXiv_2021}        &  \\ \hline
Co$_{.40}$Fe$_{.40}$B$_{.20}$       &  0.93 &  &   & 440  &  &  &     & \onlinecite{Komori_PRApp_2022}        & $\ddagger$, 1.6~K \\ \hline
Co$_{.47}$Gd$_{.53}$       & 0.16  & 0.21  & 1.2  &   &  1.9  &  &   & \onlinecite{madden_thesis_2022}        &   \\ \hline
NbN/GdN/NbN       & 0.40  &   &   & 80000  &    &  &   & \onlinecite{caruso_2019}        & $\ddagger$, 0.3~K  \\ \hline
 NbSe$_2$/Cr$_2$Ge$_2$Te$_6$/NbSe$_2$      & 2.5  & 2.4  & 8.4  & 940  &  11  & $4.0\times10^{-3}$ &   & \onlinecite{kang_2022}        & 1.6~K  \\ \hline
\end{tabular}%
}
\caption{Extracted parameters for Nb based Josephson junctions with strong and other ferromagnetic alloy barriers. In Ref.~\onlinecite{caruso_2019}, the superconductor was NbN and in Ref.~\onlinecite{kang_2022}, NbSe$_2$, as indicated in the Table. The fitting methodology to Eqn~\ref{eq:generic} is given in the main text and best fit values are quoted to 2 significant figures without consideration to their uncertainty so should be treated as approximations. The full barrier layers, including normal metals and insulators, are listed. Unless otherwise commented, measurements were performed at 4.2~K. Comments: $\bigcirc$ The range of thickness values did not cover the thickness where the first 0-$\pi$ transition was predicted. $\ddagger$ No 0-$\pi$ oscillations were observed or claimed, and the data are fitted here with only the exponential decay component of Eqn~\ref{eq:generic}. $\diamond$ This lower limit value is taken at the first observed oscillation, which is unlikely the first $\pi$-state due to the thickness range studied. $\ast$ This value was extrapolated from measurements on junctions outside of this thickness.}
\label{strong}
\end{table}

In Table \ref{strong}, we highlight works studying Josephson junctions with strong and other ferromagnetic alloys.

It is noteworthy that while most alloy barriers we have included in the Tables show supercurrent transport in the diffusive or intermediate limits, the Ni$_{3}$Al barrier was found to be in the ballistic limit.\cite{Born_PRB_2006} The long $\xi_{F1}$ decay length and short $\xi_{F2}$ oscillation lengths meant that multiple 0-$\pi$ oscillations in this material were accessible in the thickness range 10-18 nm. The thickness range studied was unlikely to cover the first 0-$\pi$ oscillation, so we do not report a $d_{0-\pi}$ parameter for this material, and the reported $J_c (\pi)$ corresponds to the peak of the first oscillation in the measured thickness range, so a higher $J_c (\pi)$ may be possible in thinner barriers. However, a potential downside of Ni$_{3}$Al is that the magnetic dead layer in this material was very large compared to others. 

The amorphous Co-based alloys, CoFeB and CoB, are popular in spintronics and magnetic memory research due to their lack of crystalline anisotropy and weaker pinning of magnetic domain walls due to the reduced density of grain boundaries.\cite{Satchell_SciRep_2021, Lavrijsen_2010} For application in MRAM, they have been found to have large values of TMR when placed in magnetic tunnel junctions.\cite{Ikeda_2008} In addition, thin layers can have perpendicular magnetic anisotropy when placed on spacer layers such as Pt or Pd.\cite{Satchell_SciRep_2021, Lavrijsen_2010} CoGd is an amorphous ferrimagnet, where the composition can be tuned to have zero net magnetization while retaining spin-polarized transport.\cite{Naylor2012} In Josephson junction studies, the general conclusion from Table \ref{strong} is that the amorphous alloys have very short $\xi_{F1}$ decay length, most likely related to the short electron mean free path in amorphous metals. This limitation may make them unsuitable as $\pi$-junctions. Komori \textit{et al.} showed that $J_c$ can be enhanced in CoFeB by thermally annealing the junctions - which has been shown to crystallize the layer.\cite{Komori_PRApp_2022, Jimbo1997}

\subsection{Advantages of weak vs strong ferromagnets (length scales)}

An important consideration for constructing a $\pi$-junction is to consider the ratio of $d_{0-\pi}/\xi_{F1}$. Since the supercurrent decays exponentially with $\xi_{F1}$, if the 0-$\pi$ transition occurs at too great of a thickness with respect to $\xi_{F1}$, the resulting $\pi$-junctions will have small critical Josephson currents.

The materials that we can describe as strong ferromagnets in Tables~\ref{pure}, \ref{NiFe}, and \ref{strong} have $d_{0-\pi}/\xi_{F1}$ ratios which can be $\leq1$. While this is great for optimizing the critical Josephson current, one of the difficulties of fabricating $\pi$-junctions with strong ferromagnetic barriers is that the length scales $\xi_{F1}$ and $\xi_{F2}$ are both short -- particularly in the amorphous alloys in Table~\ref{strong}, where $\xi_{F1}$ and $\xi_{F2}$ are $<1$~nm. As a result, very precise control over film thickness during deposition is needed to create a thickness series of samples with a small enough increments (step size) to observe the 0-$\pi$ oscillations. For integrating junctions into an industrial process, the layer thickness margin of error for reproducibility of the $\pi$-junction will be small, although sub-nm thick magnetic layers are routinely used in MRAM, so while difficult, this level of control on an industrial scale is not impossible.

Weak ferromagnets offer the advantage that $\xi_{F1}$ and $\xi_{F2}$ can be much longer, up to several nm in the Pd-based weak alloys (Table~\ref{weak}), which similarly offer the $d_{0-\pi}/\xi_{F1}$ ratio $\leq1$. Longer length scales in the Josephson junctions offer the advantage that the parameters of a $\pi$-junction, such as $J_c$, should in principle be more robust against small sample-to-sample variations in the thickness of the barrier. In the weak ferromagnets, CuNi is somewhat of an outlier having a particularly large ratio of $d_{0-\pi}/\xi_{F1}$, causing the critical Josephson current in the $\pi$ state to be small.

\subsection{Magnetic dead layers and normal metal spacer layers}

Ferromagnetic materials are often modelled as slabs with uniform magnetization, however in reality they are rarely so simple. At the interface of a thin magnetic film, it is often observed that the moments of the first few atomic layers may not align, reducing the observed magnetization from the expected one. These are commonly referred to as magnetic dead layers. Furthermore, the presence of such magnetic dead layers can degrade key properties such as the coercive field of the layer. Nb/ferromagnet interfaces are well known for having magnetic dead layers\cite{Robinson_PRB_2007} and much effort has been dedicated to studying the related interfacial roughness at Nb/ferromagnet interfaces.\cite{Vecchione_2011, Khaydukov_2015} Where possible, we have included the thickness of magnetic dead layers in the Tables.  

The most reliable method to determine the presence of dead layers is to measure several samples of varying ferromagnet thickness. The magnetometer returns a measurement of the total magnetic moment of the sample which, so long as due care has been taken in the handling of the sample and the data corrections, can be used to report the area normalized magnetic moment.\cite{Garcia_2009} If the moment/area is linear with ferromagnet thickness, then the total dead layer thickness ($d_i$) can be determined according to moment/area $=M(d_F-d_i)$.

Additional normal metal layers between the Nb and ferromagnet layer are often introduced in the junctions, and we have included such layers as part of the ``barrier" in the Tables. These additional layers are commonly referred to in the literature as either buffers, seeds, spacers, or interlayers. Here we will use the term spacer layers to describe any additional normal metal layers. The motivation for adding the spacers is often to provide better lattice matching for the ferromagnet to be deposited. 

Considering the example of Co, the fcc spacer layers Cu, Pt and Rh have been studied. The predicted mechanism is that the fcc spacer layers promote fcc growth of Co, which otherwise may have a mixed fcc/hcp phase. The two desirable benefits that Co grown on the fcc spacer layers gain compared to Co grown directly on Nb is that the magnetic dead layer is reduced or removed completely, and that the improved Co crystal structure can improve supercurrent carrying properties.\cite{Robinson_APL_2009, khasawneh_2009, satchellSOC2018} 

As a counter example, the inclusion of Cu spacer layers in junctions containing Ni and CuNi alloys may be detrimental. Cu and Ni alloy very readily which can cause the Ni to interdiffuse out of the intended layer and form interfacial alloy layers. Bolginov \textit{et al.}\cite{Bolginov_JLTP_2018} argue that early works on CuNi, Refs. \onlinecite{ryazanov_2001, Oboznov_PRL_2006}, contain large magnetic dead layers because of interdiffusion of the Ni out of the CuNi layer into an adjacent Cu layer (which was added as a necessary step in the fabrication process). By improving the fabrication process to exclude the extra Cu layer, the resulting junctions without interdiffusion are shown in Table~\ref{weak} to have an order of magnitude improvement in $J_c (\pi)$.\cite{Bolginov_JLTP_2018} Kapran \textit{et al.} describe a similar effect comparing single layer Ni and bilayer Cu/Ni junctions, where the former junctions have an order of magnitude higher $J_c$ for the same thickness of Ni.\cite{Kapran_2021} In addition to this Ni specific issue, a known downside of adding normal metal spacer layers is that every interface in the barrier may reduce the critical Josephson current. In particular, normal metals with known large interface resistances should be avoided.\cite{bass_2007}

A further motivation for adding spacer layers is that they can provide a smoother surface for growing the very thin ferromagnetic layers on compared to Nb. Thin film Nb has a columnar growth of grains which can result in a surface roughness comparable to the thickness of the ferromagnetic barrier layer in a $\pi$-junction. Certain metals including Al, Au, and Cu have been shown to act as planarization layers which significantly reduce the surface roughness. We highlight two methods making use of this property as they are employed in works featured in the Tables. The first is to replace the bottom Nb electrode with a superlattice of Nb and a thin normal metal, where the superlattice will have lower surface roughness.\cite{Thomas_1998, Wang_2012, quarterman2020distortions} Several works in the Tables use this method, however they are not distinguished in the Tables on the assumption that the use of a thick superlattice is unlikely to affect the reported junction parameters. The second approach is to place a much thinner superlattice on the surface of the single layer Nb electrode in place of the normal metal spacer layer, see Ref.~\onlinecite{Loving_arXiv_2021} in Table~\ref{strong}.

\subsection{Spin-valve junctions}\label{Sec:SVTables}

\begin{table}[]
\resizebox{0.9\columnwidth}{!}
{%
\begin{tabular}{l|c|c|c|c|c|c}
\hline
Barrier &
  \begin{tabular}[c]{@{}c@{}}$I_cR_N$ (max)\\ ($\mu$V)\end{tabular} &
  \begin{tabular}[c]{@{}c@{}}$J_c$ (max)\\ (kAcm$^{-2}$)\end{tabular} &
  \begin{tabular}[c]{@{}c@{}}$\Delta I_c$ (P vs AP)\\ (\%)\end{tabular} &
  \begin{tabular}[c]{@{}c@{}}Phase\\ detection\end{tabular} &
  Ref. &
  Comment \\ \hline
Co/Cu/NiFe & 2 & 20 & 50 &  &   \onlinecite{Bell_APL_2004} & \\ \hline
NiFe/Cu/Co & 3.5 &  & 25 & &   \onlinecite{Bell_IEEE_2005} & \\ \hline
Cu/Ni$_{.70}$Fe$_{.17}$Nb$_{.13}$/Cu/Ni/Cu & 4 & 50 & 500 &  &   \onlinecite{Baek_NComms_2014} & \\ \hline
Cu/Cu$_{.70}$(NiFe)$_{.30}$/Cu/NiFe/Cu & 1 & 2.5 & 380 &  &   \onlinecite{Qader_APL_2014} & \\ \hline
Ni/Cu/Ni & 66 & 860 & 210 &  &   \onlinecite{Baek_PRApp_2015} & \\ \hline
Cu/NiFe/Cu/Ni/Cu & 11 & 110 & 93 & yes &  \onlinecite{Gingrich_NatPhys_2016} &  \\ \hline
NiFe/Cu/Ni &  &  &  & yes & \onlinecite{Dayton_IEEE_2017} & $\mathsection$\\ \hline
Cu/NiFe/Cu/Ni/Cu & 9 & 90 & 800 & &  \onlinecite{Niedzielski_PRB_2018} & \\ \hline
Cu/NiFe/Cu/Ni/Cu & 6.5 & 65 & 340 & yes &  \onlinecite{Madden_SUST_2018} &  \\ \hline
Pt/Co/Pt/Co$_{.68}$B$_{.32}$/Pt & 0.5 & 5 & 60 & &  \onlinecite{Satchell_APL_2020} & 1.8~K \\ \hline
\end{tabular}%
}
\caption{Summary of literature on Nb based spin-valve Josephson junctions. The full barrier layers, including normal metal spacer layers are listed. For single junction measurements, $I_cR_N$ (max) corresponds to the largest critical Josephson current reported and $\Delta I_c$ (P vs AP) is (the difference in $I_c$ between P and AP)/(the lesser of $I_c$ in P or AP) expressed as a percentage, where P is the parallel and AP the antiparallel magnetic alignment. Unless otherwise commented, measurements were performed at 4.2~K. Comments: $\mathsection$ The magnetic junctions were passive and therefore the switching parameters were not directly measured.}
\label{spinvalves}
\end{table}

In Table~\ref{spinvalves} we highlight works studying controllable spin-valve junctions.

For spin-valve junctions we extract the parameters that may be useful for potential applications, including the maximum critical Josephson currents, $I_cR_N$ and $J_c$, achieved in the work, and what was the largest observed difference in critical current between the parallel and antiparallel magnetic alignments. The first quantity is an important consideration where the junction is passive and does not switch (as the critical current must be larger than the active junctions in the circuit) and the second is useful for memory schemes where readout is achieved by measuring directly the state of the spin-valve junction (superconducting or normal) or any other application where large differences between the magnetic alignments is desirable.

It is possible that for the entries in Table~\ref{spinvalves} the two magnetic alignments correspond also to a change in phase difference between 0 and $\pi$. This can be inferred based on the $0-\pi$ transition of single layer junctions reported in the other Tables. Direct detection of the phase was achieved in spin-valve junctions by placing the junctions into a phase sensitive SQUID circuit.\cite{Gingrich_NatPhys_2016, Dayton_IEEE_2017, Madden_SUST_2018}

The highest reported critical current in Table~\ref{spinvalves} was for a junction where both magnetic layers were pure Ni, which is not surprising from studying the presented Tables for single layer junctions. While using NiFe as the free layer is magnetically beneficial (NiFe has much improved switching properties compared to Ni) the resulting junctions have a much reduced critical current.

\section{Non-metallic and exotic $\pi$-junctions}\label{Sec:NonMet}

Ferromagnetic insulators (FIs) have been predicted to show 0-$\pi$ transitions.\cite{Kulik_1966, kawabata_2006, kawabata_2010} Progress in this area has been more difficult due to the limited availability of FI materials, where GdN and Eu chalcogenides have been the most studied for potential integration into superconducting devices. Of the available studies, the only FI material to show a Josephson current in S/FI/S junctions is GdN.\cite{senapati_2011} GdN Josephson junctions have shown a range of interesting phenomena, suggesting that the physics inside these junctions is complex.\cite{senapati_2011, pal_2014, massarotti_2015, massarotti_2017, cascales_2019, caruso_2019, Ahmad_2020, ahmad_2022, Ahmad_2022b} Complex behaviors of GdN junctions includes observation of a pure second harmonic current-phase relation.\cite{pal_2014} We also note an experiment on hybrid junctions containing InAs barriers coated in the FI EuS, where switching between 0 and $\pi$ coupling is attributed to the magnetic domain structure of the EuS.\cite{razmadze_2023}

Caruso \textit{et al}. studied the thickness and temperature dependence of supercurrent in GdN junctions.\cite{caruso_2019} The thickness dependence at 0.3~K resulted in an exponential decay, where the fit parameters are shown in Table~\ref{strong}. By studying the temperature dependence of the junctions, the authors characterize an incomplete 0-$\pi$ transition with temperature accompanied by spin-triplet physics.\cite{caruso_2019} Follow up work reports on the possibility to quantitatively describe and eventually control spin-triplet transport through external magnetic fields which is correlated to an ``extended" $0-\pi$ transition in temperature in these GdN FI barriers.\cite{Ahmad_2022b} 

The van der Waals family of materials have only recently been explored in the context of $\pi$-junctions. In this case, both the superconductor and magnetic layer must be compatible with the van der Waals fabrication method, so typically NbSe$_2$, an $s$-wave superconductor, is used. Reports on junctions and SQUIDs with a barrier of Cr$_2$Ge$_2$Te$_6$ (which is described as either a ferromagnetic semiconductor or ferromagnetic insulator) showed a coexistence of 0 and $\pi$ phase in the junction region.\cite{ai_2021, idzuchi_2021} Another work studied the thickness dependence of S/F/S Josephson junctions with the Cr$_2$Ge$_2$Te$_6$ barrier showing a pronounced minimum in supercurrent at a thickness of 8.4 nm, attributed to the $0-\pi$ transition.\cite{kang_2022} Fitting Eqn~\ref{eq:generic} to the thickness dependence yields the fit parameters shown on the final line of Table~\ref{strong}. In a planar junction, a long ranged supercurrent was shown across a Fe$_3$GeTe$_2$ van der Waals barrier.\cite{hu_2023}

The $0-\pi$ transition can occur in Josephson junctions without magnetic barriers if the geometry of the junction allows for Zeeman splitting in a large enough magnetic field. The idea is that in a S/N/S junction, the normal metal in a large enough magnetic field will resemble a weak ferromagnet as the up and down spin bands are displaced by the Zeeman energy.\cite{heikkila_2000, Yip_2000} An important factor here is that the applied magnetic field must be parallel to the direction of supercurrent propagation, such that it does not contribute to the Fraunhofer pattern discussed previously. Although an early attempt to observe a so-called Zeeman $\pi$-junction with the normal metal Ag by one of the authors was unsuccessful,\cite{crosser_2008} later works with non-metallic barriers were successful. Zeeman $\pi$-junctions have been realized with barriers of the Dirac semimetal Bi$_\text{1-x}$Sn$_\text{x}$,\cite{Li_2019} the 2D material graphene,\cite{Dvir_2021} the topological crystalline insulator SnTe,\cite{schnle_2019} the two-dimensional electron gas system InSb,\cite{ke_2019, haxell_2023} and in a quantum dot.\cite{Whiticar_2021} In some of the successful implementations, the large g-factors and spin-orbit coupling in these materials contributed along with the Zeeman effect to the observation of 0-$\pi$ transitions. In the InSb junctions, it was possible to apply a gate voltage which could induce the 0-$\pi$ transition at a set applied magnetic field.\cite{ke_2019}

Electron transport across a quantum dot is heavily influenced by the strong Coulomb interaction, including the possibility of observing single electron tunneling.\cite{likharev_1999} A Josephson junction can be formed across a quantum dot when pairs can coherently tunnel. In a Josephson quantum dot junction, $0-\pi$ transitions are possible by tuning the occupancy levels. Experimental realizations of $\pi$-junctions in quantum dots were achieved in an InAs nanowire\cite{van_2006} and in carbon nanotubes.\cite{jorgensen_2007, Maurand_2012, Delagrange_2016}

Several examples in this section explore planar junctions, as opposed to the sandwich geometry of all the works presented in the Tables. Planar junctions open additional possibilities for experimental control, such as adding voltage gates and manipulating strain by use of piezoelectric substrate. Another possibility in planar junctions is to control the $0-\pi$ transition using only the geometry of the junction by introducing curvature.\cite{Salamone_2021, Skarpeid_2023}

\section{Spin-triplet junctions}\label{Sec:Triplet}

The data in the Tables show clearly that the length scales governing the decay and period of the 0-$\pi$ oscillations, $\xi_{F1}$ and $\xi_{F2}$, are both very short – typically less than 1 nm in the strong F materials and up to several nm in the weak F materials.  In the late 1990’s, however, there were some experimental hints of a long-range proximity effect in S/F systems based on measurements of the electrical resistance of long F wires connected to an S electrode.\cite{giroud_1998,lawrence_1999,petrashov_1999}  Those results foreshadowed the 2001 theoretical breakthrough by Bergeret \textit{et al.} mentioned in the Introduction.  Further experimental support for the theory appeared in 2006,\cite{keizer_2006,Sosnin_2006} but the experimental breakthroughs that finally convinced the skeptics in the community did not appear until 2010.\cite{khaire_2010,robinson_2010,anwar_2010,sprungmann_2010}  Those works all showed that the supercurrent in an S/F/S junction decayed over a distance much longer than $\xi_F$ if the F layer contained the appropriate kind of noncollinear magnetization needed to convert spin-singlet Cooper pairs from the S electrodes into spin-triplet pairs in the central region of the junction.  In the case of the ``half-metal” CrO$_2$, a supercurrent can propagate hundreds of nanometers through the F material.\cite{anwar_2010,anwar_2012}  Since that time, there have been several reviews of spin-triplet proximity effects in S/F systems;\cite{Bergeret_review_2005,eschrig_review_2011,Eschrig_review_2015,linder_review_2015,Birge_review_2018} we do not review that large body of work here.   

\subsection{Controllable $\pi$-junctions}

Very few experimental studies of spin-triplet supercurrents address the issue of the ground-state phase difference across the junction.  The theoretical papers, however, make specific predictions about this issue.\cite{volkov_2003,houzet_2007,volkov_2010,trifunovic_2010,volkov_2006,halsz_2009}  Here we discuss only the sample geometry suggested by Houzet and Buzdin in 2007,\cite{houzet_2007} which was realized in many experiments.  In that theoretical work, the sequence of layers in the junction is S/F$^{\prime}$/F/F$^{\prime\prime}$/S, where the directions of the magnetizations of the three F layers can be controlled independently.  (In experimental devices, normal spacer layers are placed between adjacent F layers to prevent exchange coupling between them.)  In particular, the magnetizations $\mathbf{M}$ of any two adjacent layers must be non-collinear; let us choose adjacent layers to have orthogonal magnetizations to maximize generation of spin-triplet supercurrents.  If we define the direction of $\mathbf{M^{\prime}}$ as the z-direction and the direction of $\mathbf{M}$ as the x-direction, and if we keep all three magnetizations coplanar, then the direction of $\mathbf{M^{\prime\prime}}$ may be either plus or minus z.  According to theory, the former case produces a $\pi$-junction whereas the latter produces a 0-junction.\cite{volkov_2003,houzet_2007,volkov_2010,trifunovic_2010}  The only experimental work that tests this prediction, to our knowledge, is the one by Glick \textit{et al.} in 2018.\cite{Glick_SciAdv_2018}  To achieve the required orthogonal magnetizations, those authors used a Pd/Co multilayer with perpendicular anisotropy as the central F layer.  (Actually, two such multilayers were placed back-to-back separated by a Ru spacer layer to achieve a synthetic antiferromagnet.  That does not change the $0-\pi$ physics, according to theory.\cite{volkov_2010,trifunovic_2010})  Those authors used Ni as the fixed F$^{\prime\prime}$ layer and NiFe as the free F$^{\prime}$ layer.  Seven out of eight junctions showed robust $\pi$ phase switching when the direction of the NiFe magnetization was reversed by a small applied field. 

From a practical perspective, spin-triplet junctions appear less attractive than the simpler spin-valve junctions discussed in Section~\ref{Sec:SVApps}.  They do have one advantage, however: the constraints on the thicknesses of the F layers needed for switching between the 0 and $\pi$ states of the junctions are less stringest in the case of spin-triplet junctions than they are for spin-valve junctions.\cite{Birge_IEEE_2019}

\subsection{$\phi$ and $\phi_0$-junctions}

The spin-singlet physics discussed in the majority of this article would appear to allow only junctions with ground-state phase differences of 0 or $\pi$.  That is not true, however.  A spatially-extended junction containing only a single F layer with varying thickness, such that some parts of the junction prefer the 0-state while others prefer the $\pi$-state, can have degenerate ground states at phases $\pm \phi$.\cite{buzdin_2003,goldobin_2007,pugach_2010,sickinger_2012,bakurskiy_2013}  Such junctions are called ``$\phi$-junctions," where the phase $\phi$ can take any value between 0 and $\pi$.  Such junctions still obey the conventional time-reversal symmetry relation, $I_s(-\varphi) = -I_s(\varphi)$, where $\varphi$ is the phase difference between the superconducting condensates in the two S electrodes.  

Spin-triplet junctions containing three F-layers have an additional possibility, if the three magnetizations are non-coplanar.  In that case, time-reversal symmetry is broken, and cannot be restored by a simple rotation of the entire spin system.\cite{silaev_2017}  Such junctions may have a ground-state phase difference of $\phi_0$, which can take any value between 0 and $2\pi$, and is nondegenerate on that interval. Such junctions are now called $\phi_0$-junctions, to distinguish them from the $\phi$-junctions discussed in the previous paragraph. The existence of $\phi_0$-junctions was predicted in a number of theoretical works starting in 2007,\cite{braude_2007,buzdin_2008,galaktionov_2008,grein_2009,Margaris_2010,liu_2010,alidoust_2013,kulagina_2014,konschelle_2015,silaev_2017,pal_2019} and two reviews have appeared recently.\cite{shukrinov_2022,bobkova_2022} Because of the broken symmetries, $\phi_0$-junctions do not obey $I_s(-\varphi) = -I_s(\varphi)$, hence this phenomenon is also referred to as the ``anomalous Josephson effect", or AJE.  Since 2016, the AJE has been experimentally observed in a wide variety of exotic systems with strong spin-orbit interaction\cite{szombati_2016,murani_2017,assouline_2019,ren_2019,Mayer_2020,strambini_2020,idzuchi_2021,baumgartner_2022} and in planar out-of-equilibrium devices,\cite{Margineda_2023} but so far it has not been observed in metallic sandwich junctions.

\section{Open challenges and outlook}\label{Sec:Chall}

We hope that the presented Tables act as both a record of current literature and inspiration for the future direction of the field. While magnetic $\pi$-junctions are somewhat of a mature field, we feel from the presented Tables that a Goldilocks junction material that passes a large supercurrent in the $\pi$-state while having well controlled magnetic properties is still to be discovered. Such a material is particularly needed to realize practical spin-valve junctions. In this section, we outline some open challenges and provide an outlook on the field. 

\subsection{Spread in junction parameters and influence of fabrication method}

From examination of the Tables in this work, the reader may find themselves overwhelmed with seemingly contradictory information on the same material - a thickness of ferromagnet that provides a $\pi$-state from one work may not directly translate to the $\pi$-state in another work. For Ni and NiFe, the two most studied barriers, the authors have directed the reader towards what they consider to be the most reliable value for the thickness $d_{0-\pi}$.

In some cases, the explanation for spread in the Tables may be as simple as differences in the way the ultra-thin layer thicknesses were calibrated, or that the range of thicknesses studied unintentionally missed a $0-\pi$ transition. In other cases, the explanation may be more complex. The decay length inside the ferromagnet, $\xi_{F1}$, is shown in the Tables to vary considerably between works. This parameter is sensitive to the exact layer morphology, which itself is influenced by the presence or absence of normal metal spacer layers. We also must consider what role, if any, does the junction fabrication process have on the measured parameters in the Tables. Bolginov \textit{et al.} provide a detailed account of improvements made to their fabrication process, which resulted in a significant difference to the parameters, particularly the thickness of $d_{0-\pi}$, in junctions measured with CuNi barriers.\cite{Bolginov_JLTP_2018}

Two common fabrication methods are based around either the ion beam etching or focused ion beam techniques, which are described in detail elsewhere.\cite{Bell_Nano_2003, Wang_2012} Here, we highlight a significant difference in the two methods which may affect junction parameters. The ion beam etching process typically requires that the top Nb electrode be deposited separately to the bottom Nb electrode and barrier layers, post-fabrication and etching. To protect the barrier layers from oxidation during fabrication, a normal metal capping layer must be added above the barrier layers. In focused ion beam fabrication, both bottom and top electrodes as well as the barrier layers can be deposited in the same vacuum cycle, meaning no additional capping layers are required by the processing. We speculate that this difference may account for the generally higher critical Josephson currents observed in focused ion beam junctions, although there has been no systematic study comparing the two techniques to date.  

Based on the discussion in this section, the authors highly recommend that while the Tables in this work can be used to help choose materials to implement as $\pi$-junctions in an application, that researchers don't assume that parameters provided in the Tables will exactly reproduce in their own process. It is therefore necessary to conduct a full systematic study of a material using the deposition and fabrication methods needed for the ultimate application to determine the exact parameters for junctions produced by that process.  

\subsection{Magnetic initialization}

In the as-grown state, the direction of the F layer's magnetization in a junction may be uncontrolled and could have broken up into magnetic domains. In a single F layer junction, the physics driving the $0-\pi$ oscillations should not depend on the direction of magnetization. However, for in-plane magnetization, the Fraunhofer pattern of an individual junction can be strongly influenced by the direction and domain state of the F layer, as discussed in Section~\ref{Sec:Fraunhofer}. Since determining the critical Josephson current in the junction relies on interpretation of the Fraunhofer response, it may be preferable to initialize the magnetization to a known direction and to minimize domains. Another case where setting the magnetization direction of the ferromagnet in the junction is crucial are spin-valve junctions, where it is essential to know and control the relative orientations of the magnetic layers. For junctions with perpendicular magnetic anisotropy, the magnetization is perpendicular to the applied field so the Fraunhofer pattern should be less influenced and therefore magnetic initialization may not be necessary.

To aid control of the in-plane magnetization direction in a junction, it is usual to create a shape anisotropy by defining the junction area as an ellipse, where it is favourable for the magnetization to point in the long axis. Even with a shape anisotropy, for a single domain F layer, there are now two directions that the magnetization could point. When the magnetic domains are small the situation is worse, as the magnetic layer may break up into domains pointing in many different directions, and stray magnetic fields may emerge from the domain walls. 

The standard procedure, to set the direction of magnetization and minimize the number of domains, is to initialize the magnetic layer in a large magnetic field. The magnetic field must be applied while the circuit is far below the Curie temperature – which might be below room temperature in the case of very thin magnetic layers – but above the superconducting critical temperature of Nb to avoid flux trapping. For Ni, where the small domain size is particularly problematic, it is typical to apply a global initialization field of $\geq 350$~mT to saturate the Ni magnetization.\cite{Baek_PRApp_2017} A superconducting digital logic circuit is unlikely to be implemented in a cryogenic environment which includes the ability to apply a large global field, hence it would be favourable to have a magnetic layer which forms a single magnetic domain and does not require an initialization field procedure. 

In some materials, such as NiFe, magnetic anisotropy can be set by a growth field during deposition, which may be enough to set the direction of magnetization without the need for additional initialization fields. Another mechanism which can set the magnetization is the exchange bias effect between a ferromagnetic and antiferromagnetic material.\cite{Meiklejohn_1957} However, it has been found that supercurrent is severely suppressed in Josephson junctions containing metallic antiferromagnetic barriers.\cite{Bell_2003, Weides_PRB_2009, Klaes_2023}

\subsection{Materials compatibility with Nb or CMOS fab (MRAM example)}

MRAM is an example where integrating the magnetic junctions into the traditional CMOS wafer fabrication has been extremely difficult. Generally, the magnetic layers of the MRAM are incompatible with the CMOS process. One of the reasons for this is that elevated temperatures used for the CMOS process would cause the thin magnetic layers to interdiffuse. The solution is that the standard CMOS wafer is manufactured in one front-end-of-the-line facility, then shipped to a separate facility for backend-of-the-line processing, where the MRAM is built. This approach has several technological disadvantages, including severely limiting circuit design possibilities.

For superconducting electronics, although a multi-facility approach is possible without compromising the overall transport properties of individual junctions,\cite{khabipov_2010,Larkin_APL_2012,vernik_2013,Vettoliere_2022a,Vettoliere_2022b} in our opinion it would be best to avoid the multi-facility approach so that circuit design is not limited. In 2019, MIT Lincoln Laboratory announced a new fabrication process for superconductor electronics called PSE2, which integrates Ni $\pi$-junctions - an important step towards using $\pi$-junctions in logic circuits.\cite{Tolpygo_2019} But we are not aware of any published circuit demonstrations using that process. As we have previously motivated, pure Ni may not be the ideal choice of material due to the high initialization field required and small magnetic domains. 

The Tables presented here might motivate the reader to revisit pure Fe or to pursue other Fe containing alloys (NiFe, PdFe, etc.) as potential $\pi$-junction materials. However, we would urge caution on exploring further materials that include Fe when considering incorporation into an integrated fabrication process. Fe may be incompatible with scaled fabrication processes due to the potential detrimental effect Fe contamination would have on the rest of the process. 

Our conclusion from a materials compatibility perspective is that Ni based alloys are well motivated for further work, particularly those that are in the limit Kapran \textit{et al.} describe as ``strong-but-clean", meaning strong ferromagnetism with ballistic transport properties.\cite{Kapran_2021} We also mention that it may be possible to explore alloys and compounds for magnetic barriers where the component materials are otherwise non-magnetic, such as PdMn,\cite{Kupriyanov_review_2006} some Heuslers,\cite{sprungmann_2010, Tavares_2023} and $L1_0$-ordered MnAl\cite{Kono_1958} or MnGa.\cite{Bither_2004} Such an approach may negate any concerns about ferromagnetic particle contamination in a Nb process. 

\subsection{Materials and read/write operations for spin-valves}

We identify some outstanding challenges specific to spin-valve junctions. Firstly, to achieve a spin-valve junction with high critical current, it seems necessary to replace the NiFe free layer. Secondly, we must remove the requirement for global magnetic field for read/write operations and move towards on-chip operations.

The materials for spin-valve junctions have several requirements and important considerations: the fixed layer must be robust enough against the switching of the free layer, both layers should be single domain so that the supercurrent and phase difference across the junction is uniform, both layers must be sufficiently thin to allow a significant supercurrent to propagate through the junction, magnetic dead layers (which can dominate in very thin layers) must be minimized, and the $0-\pi$ oscillations of each component layer must be well characterized. It is not obvious from the Tables presented in this work that there is an immediate replacement for either NiFe as the free layer or Ni as the fixed layer. 

On the prospect of on-chip writing, we note that there are adventurous proposals for switching magnetic layers in the superconducting state of S-F systems,\cite{Waintal_2002, Zhao_2008, Konschelle_2009, Linder_2011} although there is not yet experimental realization. We therefore suggest that established mechanisms used in the field of spintronics and technologies for MRAM can also be applied in the normal state of a superconducting spin-valve. Spin transfer torque switching, where a current pulse is passed through the junction and induced magnetization dynamics causes the free layer to switch, was demonstrated by Baek \textit{et al.} by applying a current density to their junction which greatly exceeded $J_c$.\cite{Baek_PRApp_2015} Spin orbit torque switching, where a current pulse is passed through a layer with large spin Hall angle adjacent to the free layer causing it to switch, has shown promising results in room-temperature magnetic tunnel junctions, and promises faster switching speeds than spin transfer torque.\cite{Shao_2021} Raytheon BBN Technologies have proposed a low temperature superconductor-ferromagnet hybrid memory (not based on the Josephson effect) where magnetic switching is achieved by spin Hall effect.\cite{nguyen_2020} 

\subsection{Altermagnets}

Solids with an intrinsic magnetic phase are traditionally classed as having either ferromagnetic or antiferromagnetic ordering. Recently, there has been great interest in the fundamental study and potential applications of an emerging third class of magnetic solid, where the magnetic phase is known as altermagnetism.\cite{smejkal_2022} Altermagnets have properties that blur the lines between the traditional ferromagnetic and antiferromagnetic orderings. On the one hand, these materials have electronic properties consistent with ferromagnetism, they show an anomalous Hall effect and have spin-polarized conduction bands. On the other hand, altermagnetic materials have antiparallel magnetic crystal order and zero net magnetization, consistent with antiferromagnetism. Altermagnets are expected to be abundant in nature. \cite{smejkal_2022}

The presence of a spin-polarized conduction band suggests that much of the rich physics present in ferromagnet-superconductor hybrid systems are also present when the ferromagnet is replaced by an altermagnet, along with additional unique physics in the new altermagnet system. Notably in the altermagnet, the finite pair momentum appears in the absence of a net magnetization. \cite{zhang_2023} Zero net magnetic moment junctions offers the significant advantages over ferromagnet systems that the junctions produce either zero or negligibly small stray magnetic fields, and they don't require any magnetic initialization.

Of particular interest to the topic of this review are predictions of 0-$\pi$ oscillations in Josephson junctions containing altermagnetic barriers.\cite{ouassou_2023, zhang_2023, beenakker_2023, papaj_2023} To date, there are no experimental studies of altermagnet-superconductor hybrid systems. The most studied candidate altermagnet is RuO$_2$, a metal with a bulk room-temperature resistivity of $\rho \approx 35~\mu \Omega$-cm.\cite{feng_2022}

\subsection{Neuromorphic and novel analogue computing}

Magnetic Josephson junctions have been proposed as artificial synapses for neuromorphic computing schemes.\cite{Crotty_2010, Schneider_2022, Schegolev_2022} Neuromorphic computing is seen as a potential solution to the problem of poor energy efficiency when running neural networks on traditional computing architectures. The Tutorial of Schneider \textit{et al.} covers the topic of neuromorphic systems based on magnetic Josephson junctions in detail.\cite{schneider_2018} Here we focus on the materials and requirements for this application.

The requirements in this potential application is for the junction to provide a continuously tunable (analogue) response, provide a degree of plasticity/memory, and be scalable and energy efficient enough for the architecture required to interconnect many such devices needed for neuromorphic application. 

So far, synapse behavior in magnetic Josephson junctions has been achieved in magnetic nanoclusters. The magnetic clusters do not spontaneously order, but can be aligned by subsequent current pulses. The magnetic order can be tuned continuously from the disordered to ordered magnetic states, and the degree of order affects the critical current of the junction.\cite{schneider_2018a, schneider_2018, jue_2022, jue_2022a} Similar to the $\pi$-junctions presented in the Tables, there will be a wide variation of performance in metrics such as the $I_cR_N$ of barriers with different magnetic nanoclusters. We are aware of two magnetic nanocluster barriers which have been studied for this application, Mn clusters embedded in Si\cite{schneider_2018a} and Fe clusters embedded in Ge.\cite{jue_2022} Current limitations are that the $I_cR_N$ products in these barriers are quite a low and the Mn clusters in Si required an annealing step. A potential alternative to nanoclusters are ultra-thin magnetic layers, sometimes called dusting layers, which can also be close to the superparamagnetic transition.

\section{Conclusions}\label{Sec:Conc}

Readers new to this field may be overwhelmed by the sheer number of materials listed in the Tables in Section~\ref{Sec:Tables}, as well as the wide variability in parameters extracted by different research groups. To avoid that parting impression, we finish the review by expressing a few of our own opinions.

First, if we had to choose a material from the Tables to make highly-reproducible $\pi$-junctions that can tolerate small variations in thickness, we would choose a PdNi alloy with a Ni concentration in the range of 10 - 15\%.  Pham \textit{et al.} report values of $J_c$ and $I_c R_N$ in the $\pi$-state as large as 70 kA/cm${^2}$ and 18 $\mu$V, respectively, in Pd$_{89}$Ni$_{11}$ junctions with NbN superconducting electrodes.\cite{Pham_2022}  Because of its mild perpendicular anisotropy, junctions containing PdNi may not need magnetic initialization. The only drawback of this material is the cost of Pd; but we remind readers that only very small amounts of the material are needed.

Second, if one wants the highest possible critical current density in a $\pi$-junction, then Ni is the obvious choice. But Ni junctions require magnetic initialization in a rather large magnetic field to establish a reproducible magnetized initial state.  Without that initialization, the Fraunhofer pattern will vary randomly from sample to sample, yielding non-reproducible values of $J_c$.

Third, if one wants a magnetically-soft free layer for a spin-valve junction, then NiFe is the best choice so far.  Our own attempts to dilute NiFe to lower its magnetization (and hence the magnetic switching energy) have been largely unsuccessful; in the case of doping with Mo or Cr, the resulting material has a short mean free path, leading to a steep decay of $J_c$ with thickness.

And finally, we lament the heavy use of CuNi alloy by groups new to the field.  We acknowledge the many ``firsts" achieved by the Ryazanov group and their collaborators using junctions containing CuNi, but the high rate of spin-flip scattering in the material causes $J_c$ to decay very steeply with thickness, leading to very small values of $J_c$ and $I_c R_N$ in the $\pi$-state. The main advantages of CuNi are: i) $\xi_F$ is long due to the small magnetization, so thickness fluctuations are tolerated; ii) the perpendicular anisotropy avoids distortions of the Fraunhofer pattern; and iii) the material is less expensive than PdNi. We would recommend using that material only when junction size is not an issue, so that the small value of $J_c$ is tolerated.  But we much prefer PdNi alloy.

In summary, we believe that $\pi$-junctions have an important role to play in the development of superconducting digital electronics.  We hope that this review will help workers navigate the vast literature as they explore new materials for $\pi$-junctions.

\begin{acknowledgments}

N.O.B. wishes to thank all the students who have worked with him on superconducting/ferromagnetic hybrid systems over the past two decades. Special thanks also go to B. Bi, D. Edmunds, R. Loloee, and W.P. Pratt, Jr., without whom none of this work would have been possible, and to the JMRAM team at Northrop Grumman Corporation, especially A.Y. Herr, D.L. Miller, O. Naaman, N. Rizzo, T.F. Ambrose, and M.G. Loving.
N.S. wishes to thank G. Burnell and everyone who has worked with him on this topic, and acknowledges support from new faculty startup funding made available by Texas State University. 
We acknowledge helpful suggestions from F.S. Bergeret, A.I. Buzdin, A.A. Golubov, S. Jacobsen, M. Kupriyanov, O. Mukhanov, Z. Radovic, V.V. Ryazanov, I.I. Soloviev, and A.F. Volkov.
 
\end{acknowledgments}

\section*{AUTHOR DECLARATIONS}

\subsection*{Conflict of Interest}
The authors have no conflicts to disclose.

\subsection*{Author Contributions}

\noindent\textbf{N.O. Birge}: writing - original draft (equal) \textbf{N. Satchell}: writing - original draft (equal) 

\section*{DATA AVAILABILITY}

Data sharing is not applicable to this article as no new data were created or analyzed in this study. 


\bibliography{APLMatRev,APLMatRev_adds}

\begin{thebibliography}{305}%
\makeatletter
\providecommand \@ifxundefined [1]{%
 \@ifx{#1\undefined}
}%
\providecommand \@ifnum [1]{%
 \ifnum #1\expandafter \@firstoftwo
 \else \expandafter \@secondoftwo
 \fi
}%
\providecommand \@ifx [1]{%
 \ifx #1\expandafter \@firstoftwo
 \else \expandafter \@secondoftwo
 \fi
}%
\providecommand \natexlab [1]{#1}%
\providecommand \enquote  [1]{``#1''}%
\providecommand \bibnamefont  [1]{#1}%
\providecommand \bibfnamefont [1]{#1}%
\providecommand \citenamefont [1]{#1}%
\providecommand \href@noop [0]{\@secondoftwo}%
\providecommand \href [0]{\begingroup \@sanitize@url \@href}%
\providecommand \@href[1]{\@@startlink{#1}\@@href}%
\providecommand \@@href[1]{\endgroup#1\@@endlink}%
\providecommand \@sanitize@url [0]{\catcode `\\12\catcode `\$12\catcode `\&12\catcode `\#12\catcode `\^12\catcode `\_12\catcode `\%12\relax}%
\providecommand \@@startlink[1]{}%
\providecommand \@@endlink[0]{}%
\providecommand \url  [0]{\begingroup\@sanitize@url \@url }%
\providecommand \@url [1]{\endgroup\@href {#1}{\urlprefix }}%
\providecommand \urlprefix  [0]{URL }%
\providecommand \Eprint [0]{\href }%
\providecommand \doibase [0]{http://dx.doi.org/}%
\providecommand \selectlanguage [0]{\@gobble}%
\providecommand \bibinfo  [0]{\@secondoftwo}%
\providecommand \bibfield  [0]{\@secondoftwo}%
\providecommand \translation [1]{[#1]}%
\providecommand \BibitemOpen [0]{}%
\providecommand \bibitemStop [0]{}%
\providecommand \bibitemNoStop [0]{.\EOS\space}%
\providecommand \EOS [0]{\spacefactor3000\relax}%
\providecommand \BibitemShut  [1]{\csname bibitem#1\endcsname}%
\let\auto@bib@innerbib\@empty
\bibitem [{\citenamefont {Veretennikov}(2000)}]{veretennikov_2000}%
  \BibitemOpen
  \bibfield  {author} {\bibinfo {author} {\bibfnamefont {A.}~\bibnamefont {Veretennikov}},\ }\bibfield  {title} {\enquote {\bibinfo {title} {Supercurrents through the superconductor–ferromagnet–superconductor ({SFS}) junctions},}\ }\href {\doibase 10.1016/S0921-4526(99)02086-4} {\bibfield  {journal} {\bibinfo  {journal} {Physica B Condens. Matter.}\ }\textbf {\bibinfo {volume} {284-288}},\ \bibinfo {pages} {495--496} (\bibinfo {year} {2000})}\BibitemShut {NoStop}%
\bibitem [{\citenamefont {Ryazanov}\ \emph {et~al.}(2001{\natexlab{a}})\citenamefont {Ryazanov}, \citenamefont {Oboznov}, \citenamefont {Rusanov}, \citenamefont {Veretennikov}, \citenamefont {Golubov},\ and\ \citenamefont {Aarts}}]{Ryazanov_PRL_2001}%
  \BibitemOpen
  \bibfield  {author} {\bibinfo {author} {\bibfnamefont {V.~V.}\ \bibnamefont {Ryazanov}}, \bibinfo {author} {\bibfnamefont {V.~A.}\ \bibnamefont {Oboznov}}, \bibinfo {author} {\bibfnamefont {A.~Y.}\ \bibnamefont {Rusanov}}, \bibinfo {author} {\bibfnamefont {A.~V.}\ \bibnamefont {Veretennikov}}, \bibinfo {author} {\bibfnamefont {A.~A.}\ \bibnamefont {Golubov}}, \ and\ \bibinfo {author} {\bibfnamefont {J.}~\bibnamefont {Aarts}},\ }\bibfield  {title} {\enquote {\bibinfo {title} {Coupling of {T}wo {S}uperconductors through a {F}erromagnet: {E}vidence for a $\ensuremath{\pi}$ {J}unction},}\ }\href {\doibase 10.1103/PhysRevLett.86.2427} {\bibfield  {journal} {\bibinfo  {journal} {Phys. Rev. Lett.}\ }\textbf {\bibinfo {volume} {86}},\ \bibinfo {pages} {2427--2430} (\bibinfo {year} {2001}{\natexlab{a}})}\BibitemShut {NoStop}%
\bibitem [{\citenamefont {Kontos}\ \emph {et~al.}(2002)\citenamefont {Kontos}, \citenamefont {Aprili}, \citenamefont {Lesueur}, \citenamefont {Gen\^et}, \citenamefont {Stephanidis},\ and\ \citenamefont {Boursier}}]{Kontos_PRL_2002}%
  \BibitemOpen
  \bibfield  {author} {\bibinfo {author} {\bibfnamefont {T.}~\bibnamefont {Kontos}}, \bibinfo {author} {\bibfnamefont {M.}~\bibnamefont {Aprili}}, \bibinfo {author} {\bibfnamefont {J.}~\bibnamefont {Lesueur}}, \bibinfo {author} {\bibfnamefont {F.}~\bibnamefont {Gen\^et}}, \bibinfo {author} {\bibfnamefont {B.}~\bibnamefont {Stephanidis}}, \ and\ \bibinfo {author} {\bibfnamefont {R.}~\bibnamefont {Boursier}},\ }\bibfield  {title} {\enquote {\bibinfo {title} {Josephson {J}unction through a {T}hin {F}erromagnetic {L}ayer: {N}egative {C}oupling},}\ }\href {\doibase 10.1103/PhysRevLett.89.137007} {\bibfield  {journal} {\bibinfo  {journal} {Phys. Rev. Lett.}\ }\textbf {\bibinfo {volume} {89}},\ \bibinfo {pages} {137007} (\bibinfo {year} {2002})}\BibitemShut {NoStop}%
\bibitem [{\citenamefont {Ryazanov}\ \emph {et~al.}(2001{\natexlab{b}})\citenamefont {Ryazanov}, \citenamefont {Oboznov}, \citenamefont {Veretennikov},\ and\ \citenamefont {Rusanov}}]{ryazanov_2001}%
  \BibitemOpen
  \bibfield  {author} {\bibinfo {author} {\bibfnamefont {V.~V.}\ \bibnamefont {Ryazanov}}, \bibinfo {author} {\bibfnamefont {V.~A.}\ \bibnamefont {Oboznov}}, \bibinfo {author} {\bibfnamefont {A.~V.}\ \bibnamefont {Veretennikov}}, \ and\ \bibinfo {author} {\bibfnamefont {A.~Y.}\ \bibnamefont {Rusanov}},\ }\bibfield  {title} {\enquote {\bibinfo {title} {Intrinsically frustrated superconducting array of superconductor-ferromagnet-superconductor $\pi$ junctions},}\ }\href {\doibase 10.1103/PhysRevB.65.020501} {\bibfield  {journal} {\bibinfo  {journal} {Phys. Rev. B}\ }\textbf {\bibinfo {volume} {65}},\ \bibinfo {pages} {020501} (\bibinfo {year} {2001}{\natexlab{b}})}\BibitemShut {NoStop}%
\bibitem [{\citenamefont {Guichard}\ \emph {et~al.}(2003)\citenamefont {Guichard}, \citenamefont {Aprili}, \citenamefont {Bourgeois}, \citenamefont {Kontos}, \citenamefont {Lesueur},\ and\ \citenamefont {Gandit}}]{guichard_2003}%
  \BibitemOpen
  \bibfield  {author} {\bibinfo {author} {\bibfnamefont {W.}~\bibnamefont {Guichard}}, \bibinfo {author} {\bibfnamefont {M.}~\bibnamefont {Aprili}}, \bibinfo {author} {\bibfnamefont {O.}~\bibnamefont {Bourgeois}}, \bibinfo {author} {\bibfnamefont {T.}~\bibnamefont {Kontos}}, \bibinfo {author} {\bibfnamefont {J.}~\bibnamefont {Lesueur}}, \ and\ \bibinfo {author} {\bibfnamefont {P.}~\bibnamefont {Gandit}},\ }\bibfield  {title} {\enquote {\bibinfo {title} {Phase {Sensitive Experiments in F}erromagnetic-{B}ased {Josephson} {J}unctions},}\ }\href {\doibase 10.1103/PhysRevLett.90.167001} {\bibfield  {journal} {\bibinfo  {journal} {Phys. Rev. Lett.}\ }\textbf {\bibinfo {volume} {90}},\ \bibinfo {pages} {167001} (\bibinfo {year} {2003})}\BibitemShut {NoStop}%
\bibitem [{\citenamefont {Bauer}\ \emph {et~al.}(2004)\citenamefont {Bauer}, \citenamefont {Bentner}, \citenamefont {Aprili}, \citenamefont {Della~Rocca}, \citenamefont {Reinwald}, \citenamefont {Wegscheider},\ and\ \citenamefont {Strunk}}]{bauer_2004}%
  \BibitemOpen
  \bibfield  {author} {\bibinfo {author} {\bibfnamefont {A.}~\bibnamefont {Bauer}}, \bibinfo {author} {\bibfnamefont {J.}~\bibnamefont {Bentner}}, \bibinfo {author} {\bibfnamefont {M.}~\bibnamefont {Aprili}}, \bibinfo {author} {\bibfnamefont {M.~L.}\ \bibnamefont {Della~Rocca}}, \bibinfo {author} {\bibfnamefont {M.}~\bibnamefont {Reinwald}}, \bibinfo {author} {\bibfnamefont {W.}~\bibnamefont {Wegscheider}}, \ and\ \bibinfo {author} {\bibfnamefont {C.}~\bibnamefont {Strunk}},\ }\bibfield  {title} {\enquote {\bibinfo {title} {Spontaneous supercurrent induced by ferromagnetic $\pi$ junctions},}\ }\href {\doibase 10.1103/PhysRevLett.92.217001} {\bibfield  {journal} {\bibinfo  {journal} {Phys. Rev. Lett.}\ }\textbf {\bibinfo {volume} {92}},\ \bibinfo {pages} {217001} (\bibinfo {year} {2004})}\BibitemShut {NoStop}%
\bibitem [{\citenamefont {Frolov}\ \emph {et~al.}(2004)\citenamefont {Frolov}, \citenamefont {Van~Harlingen}, \citenamefont {Oboznov}, \citenamefont {Bolginov},\ and\ \citenamefont {Ryazanov}}]{frolov_2004}%
  \BibitemOpen
  \bibfield  {author} {\bibinfo {author} {\bibfnamefont {S.~M.}\ \bibnamefont {Frolov}}, \bibinfo {author} {\bibfnamefont {D.~J.}\ \bibnamefont {Van~Harlingen}}, \bibinfo {author} {\bibfnamefont {V.~A.}\ \bibnamefont {Oboznov}}, \bibinfo {author} {\bibfnamefont {V.~V.}\ \bibnamefont {Bolginov}}, \ and\ \bibinfo {author} {\bibfnamefont {V.~V.}\ \bibnamefont {Ryazanov}},\ }\bibfield  {title} {\enquote {\bibinfo {title} {Measurement of the current-phase relation of superconductor/ferromagnet/superconductor $\pi$ {J}osephson junctions},}\ }\href {\doibase 10.1103/PhysRevB.70.144505} {\bibfield  {journal} {\bibinfo  {journal} {Phys. Rev. B}\ }\textbf {\bibinfo {volume} {70}},\ \bibinfo {pages} {144505} (\bibinfo {year} {2004})}\BibitemShut {NoStop}%
\bibitem [{\citenamefont {Bulaevskii}, \citenamefont {Kuzii},\ and\ \citenamefont {Sobyanin}(1977)}]{bulaevskii_1977}%
  \BibitemOpen
  \bibfield  {author} {\bibinfo {author} {\bibfnamefont {L.~N.}\ \bibnamefont {Bulaevskii}}, \bibinfo {author} {\bibfnamefont {V.~V.}\ \bibnamefont {Kuzii}}, \ and\ \bibinfo {author} {\bibfnamefont {A.~A.}\ \bibnamefont {Sobyanin}},\ }\bibfield  {title} {\enquote {\bibinfo {title} {Superconducting system with weak coupling with a current in the ground state},}\ }\href@noop {} {\bibfield  {journal} {\bibinfo  {journal} {Pis'ma Zh. Eksp. Teor. Fiz.}\ }\textbf {\bibinfo {volume} {25}},\ \bibinfo {pages} {314 -- 318} (\bibinfo {year} {1977})}\BibitemShut {NoStop}%
\bibitem [{\citenamefont {Buzdin}, \citenamefont {Bulaevskii},\ and\ \citenamefont {Panyukov}(1982)}]{buzdin_1982}%
  \BibitemOpen
  \bibfield  {author} {\bibinfo {author} {\bibfnamefont {A.~I.}\ \bibnamefont {Buzdin}}, \bibinfo {author} {\bibfnamefont {L.~N.}\ \bibnamefont {Bulaevskii}}, \ and\ \bibinfo {author} {\bibfnamefont {S.~V.}\ \bibnamefont {Panyukov}},\ }\bibfield  {title} {\enquote {\bibinfo {title} {Critical-current oscillations as a function of the exchange field and thickness of the ferromagnetic metal ({F}) in an {SFS} {J}osephson junction},}\ }\href@noop {} {\bibfield  {journal} {\bibinfo  {journal} {JETP Lett}\ }\textbf {\bibinfo {volume} {35}},\ \bibinfo {pages} {178--180} (\bibinfo {year} {1982})}\BibitemShut {NoStop}%
\bibitem [{\citenamefont {Buzdin}\ and\ \citenamefont {Kupriyanov}(1991)}]{buzdin_1991}%
  \BibitemOpen
  \bibfield  {author} {\bibinfo {author} {\bibfnamefont {A.}~\bibnamefont {Buzdin}}\ and\ \bibinfo {author} {\bibfnamefont {M.}~\bibnamefont {Kupriyanov}},\ }\bibfield  {title} {\enquote {\bibinfo {title} {Josephson junction with a ferromagnetic layer},}\ }\href@noop {} {\bibfield  {journal} {\bibinfo  {journal} {{JETP} Letters}\ }\textbf {\bibinfo {volume} {53}},\ \bibinfo {pages} {321--326} (\bibinfo {year} {1991})}\BibitemShut {NoStop}%
\bibitem [{\citenamefont {Golubov}, \citenamefont {Kupriyanov},\ and\ \citenamefont {Il'ichev}(2004)}]{golubov_2004}%
  \BibitemOpen
  \bibfield  {author} {\bibinfo {author} {\bibfnamefont {A.~A.}\ \bibnamefont {Golubov}}, \bibinfo {author} {\bibfnamefont {M.~Y.}\ \bibnamefont {Kupriyanov}}, \ and\ \bibinfo {author} {\bibfnamefont {E.}~\bibnamefont {Il'ichev}},\ }\bibfield  {title} {\enquote {\bibinfo {title} {The current-phase relation in {Josephson} junctions},}\ }\href {\doibase 10.1103/RevModPhys.76.411} {\bibfield  {journal} {\bibinfo  {journal} {Rev. Mod. Phys.}\ }\textbf {\bibinfo {volume} {76}},\ \bibinfo {pages} {411--469} (\bibinfo {year} {2004})}\BibitemShut {NoStop}%
\bibitem [{\citenamefont {Lyuksyutov}\ and\ \citenamefont {Pokrovsky}(2005)}]{lyuksyutov_2005}%
  \BibitemOpen
  \bibfield  {author} {\bibinfo {author} {\bibfnamefont {I.~F.}\ \bibnamefont {Lyuksyutov}}\ and\ \bibinfo {author} {\bibfnamefont {V.~L.}\ \bibnamefont {Pokrovsky}},\ }\bibfield  {title} {\enquote {\bibinfo {title} {Ferromagnet–superconductor hybrids},}\ }\href {\doibase 10.1080/00018730500057536} {\bibfield  {journal} {\bibinfo  {journal} {Adv. Phys.}\ }\textbf {\bibinfo {volume} {54}},\ \bibinfo {pages} {67--136} (\bibinfo {year} {2005})}\BibitemShut {NoStop}%
\bibitem [{\citenamefont {Buzdin}(2005)}]{Buzdin_review_2005}%
  \BibitemOpen
  \bibfield  {author} {\bibinfo {author} {\bibfnamefont {A.~I.}\ \bibnamefont {Buzdin}},\ }\bibfield  {title} {\enquote {\bibinfo {title} {Proximity effects in superconductor-ferromagnet heterostructures},}\ }\href {\doibase 10.1103/RevModPhys.77.935} {\bibfield  {journal} {\bibinfo  {journal} {Rev. Mod. Phys.}\ }\textbf {\bibinfo {volume} {77}},\ \bibinfo {pages} {935--976} (\bibinfo {year} {2005})}\BibitemShut {NoStop}%
\bibitem [{\citenamefont {Terzioglu}\ and\ \citenamefont {Beasley}(1998)}]{terzioglu_1998}%
  \BibitemOpen
  \bibfield  {author} {\bibinfo {author} {\bibfnamefont {E.}~\bibnamefont {Terzioglu}}\ and\ \bibinfo {author} {\bibfnamefont {M.}~\bibnamefont {Beasley}},\ }\bibfield  {title} {\enquote {\bibinfo {title} {Complementary {Josephson} junction devices and circuits: a possible new approach to superconducting electronics},}\ }\href {\doibase 10.1109/77.678441} {\bibfield  {journal} {\bibinfo  {journal} {{IEEE} Trans. Appl. Supercond.}\ }\textbf {\bibinfo {volume} {8}},\ \bibinfo {pages} {48--53} (\bibinfo {year} {1998})}\BibitemShut {NoStop}%
\bibitem [{\citenamefont {Ustinov}\ and\ \citenamefont {Kaplunenko}(2003)}]{ustinov_2003}%
  \BibitemOpen
  \bibfield  {author} {\bibinfo {author} {\bibfnamefont {A.~V.}\ \bibnamefont {Ustinov}}\ and\ \bibinfo {author} {\bibfnamefont {V.~K.}\ \bibnamefont {Kaplunenko}},\ }\bibfield  {title} {\enquote {\bibinfo {title} {Rapid single-flux quantum logic using $\pi$-shifters},}\ }\href {\doibase 10.1063/1.1604964} {\bibfield  {journal} {\bibinfo  {journal} {J. Appl. Phys.}\ }\textbf {\bibinfo {volume} {94}},\ \bibinfo {pages} {5405} (\bibinfo {year} {2003})}\BibitemShut {NoStop}%
\bibitem [{\citenamefont {Ioffe}\ \emph {et~al.}(1999)\citenamefont {Ioffe}, \citenamefont {Geshkenbein}, \citenamefont {Feigel'man}, \citenamefont {Fauchère},\ and\ \citenamefont {Blatter}}]{ioffe_1999}%
  \BibitemOpen
  \bibfield  {author} {\bibinfo {author} {\bibfnamefont {L.~B.}\ \bibnamefont {Ioffe}}, \bibinfo {author} {\bibfnamefont {V.~B.}\ \bibnamefont {Geshkenbein}}, \bibinfo {author} {\bibfnamefont {M.~V.}\ \bibnamefont {Feigel'man}}, \bibinfo {author} {\bibfnamefont {A.~L.}\ \bibnamefont {Fauchère}}, \ and\ \bibinfo {author} {\bibfnamefont {G.}~\bibnamefont {Blatter}},\ }\bibfield  {title} {\enquote {\bibinfo {title} {Environmentally decoupled $sds$-wave {Josephson} junctions for quantum computing},}\ }\href {\doibase 10.1038/19464} {\bibfield  {journal} {\bibinfo  {journal} {Nature}\ }\textbf {\bibinfo {volume} {398}},\ \bibinfo {pages} {679--681} (\bibinfo {year} {1999})}\BibitemShut {NoStop}%
\bibitem [{\citenamefont {Blatter}, \citenamefont {Geshkenbein},\ and\ \citenamefont {Ioffe}(2001)}]{blatter_2001}%
  \BibitemOpen
  \bibfield  {author} {\bibinfo {author} {\bibfnamefont {G.}~\bibnamefont {Blatter}}, \bibinfo {author} {\bibfnamefont {V.~B.}\ \bibnamefont {Geshkenbein}}, \ and\ \bibinfo {author} {\bibfnamefont {L.~B.}\ \bibnamefont {Ioffe}},\ }\bibfield  {title} {\enquote {\bibinfo {title} {Design aspects of superconducting-phase quantum bits},}\ }\href {\doibase 10.1103/PhysRevB.63.174511} {\bibfield  {journal} {\bibinfo  {journal} {Phys. Rev. B}\ }\textbf {\bibinfo {volume} {63}},\ \bibinfo {pages} {174511} (\bibinfo {year} {2001})}\BibitemShut {NoStop}%
\bibitem [{\citenamefont {Bergeret}, \citenamefont {Volkov},\ and\ \citenamefont {Efetov}(2001{\natexlab{a}})}]{bergeret_2001b}%
  \BibitemOpen
  \bibfield  {author} {\bibinfo {author} {\bibfnamefont {F.~S.}\ \bibnamefont {Bergeret}}, \bibinfo {author} {\bibfnamefont {A.~F.}\ \bibnamefont {Volkov}}, \ and\ \bibinfo {author} {\bibfnamefont {K.~B.}\ \bibnamefont {Efetov}},\ }\bibfield  {title} {\enquote {\bibinfo {title} {Long-range proximity effects in superconductor-ferromagnet structures.}}\ }\href {\doibase 10.1103/{PhysRevLett}.86.4096} {\bibfield  {journal} {\bibinfo  {journal} {Phys Rev Lett}\ }\textbf {\bibinfo {volume} {86}},\ \bibinfo {pages} {4096--4099} (\bibinfo {year} {2001}{\natexlab{a}})}\BibitemShut {NoStop}%
\bibitem [{\citenamefont {Kadigrobov}, \citenamefont {Shekhter},\ and\ \citenamefont {Jonson}(2001)}]{kadigrobov_2001}%
  \BibitemOpen
  \bibfield  {author} {\bibinfo {author} {\bibfnamefont {A.}~\bibnamefont {Kadigrobov}}, \bibinfo {author} {\bibfnamefont {R.~I.}\ \bibnamefont {Shekhter}}, \ and\ \bibinfo {author} {\bibfnamefont {M.}~\bibnamefont {Jonson}},\ }\bibfield  {title} {\enquote {\bibinfo {title} {Quantum spin fluctuations as a source of long-range proximity effects in diffusive ferromagnet-superconductor structures},}\ }\href {\doibase 10.1209/epl/i2001-00107-2} {\bibfield  {journal} {\bibinfo  {journal} {Europhys. Lett.}\ }\textbf {\bibinfo {volume} {54}},\ \bibinfo {pages} {394--400} (\bibinfo {year} {2001})}\BibitemShut {NoStop}%
\bibitem [{\citenamefont {Volkov}, \citenamefont {Bergeret},\ and\ \citenamefont {Efetov}(2003)}]{volkov_2003}%
  \BibitemOpen
  \bibfield  {author} {\bibinfo {author} {\bibfnamefont {A.~F.}\ \bibnamefont {Volkov}}, \bibinfo {author} {\bibfnamefont {F.~S.}\ \bibnamefont {Bergeret}}, \ and\ \bibinfo {author} {\bibfnamefont {K.~B.}\ \bibnamefont {Efetov}},\ }\bibfield  {title} {\enquote {\bibinfo {title} {Odd triplet superconductivity in superconductor-ferromagnet multilayered structures},}\ }\href {\doibase 10.1103/PhysRevLett.90.117006} {\bibfield  {journal} {\bibinfo  {journal} {Phys. Rev. Lett.}\ }\textbf {\bibinfo {volume} {90}},\ \bibinfo {pages} {117006} (\bibinfo {year} {2003})}\BibitemShut {NoStop}%
\bibitem [{\citenamefont {Bergeret}, \citenamefont {Volkov},\ and\ \citenamefont {Efetov}(2005)}]{Bergeret_review_2005}%
  \BibitemOpen
  \bibfield  {author} {\bibinfo {author} {\bibfnamefont {F.~S.}\ \bibnamefont {Bergeret}}, \bibinfo {author} {\bibfnamefont {A.~F.}\ \bibnamefont {Volkov}}, \ and\ \bibinfo {author} {\bibfnamefont {K.~B.}\ \bibnamefont {Efetov}},\ }\bibfield  {title} {\enquote {\bibinfo {title} {Odd triplet superconductivity and related phenomena in superconductor-ferromagnet structures},}\ }\href {\doibase 10.1103/RevModPhys.77.1321} {\bibfield  {journal} {\bibinfo  {journal} {Rev. Mod. Phys.}\ }\textbf {\bibinfo {volume} {77}},\ \bibinfo {pages} {1321--1373} (\bibinfo {year} {2005})}\BibitemShut {NoStop}%
\bibitem [{\citenamefont {Eschrig}(2011)}]{eschrig_review_2011}%
  \BibitemOpen
  \bibfield  {author} {\bibinfo {author} {\bibfnamefont {M.}~\bibnamefont {Eschrig}},\ }\bibfield  {title} {\enquote {\bibinfo {title} {Spin-polarized supercurrents for spintronics},}\ }\href {\doibase 10.1063/1.3541944} {\bibfield  {journal} {\bibinfo  {journal} {Phys. Today}\ }\textbf {\bibinfo {volume} {64}},\ \bibinfo {pages} {43--49} (\bibinfo {year} {2011})}\BibitemShut {NoStop}%
\bibitem [{\citenamefont {Eschrig}(2015)}]{Eschrig_review_2015}%
  \BibitemOpen
  \bibfield  {author} {\bibinfo {author} {\bibfnamefont {M.}~\bibnamefont {Eschrig}},\ }\bibfield  {title} {\enquote {\bibinfo {title} {Spin-polarized supercurrents for spintronics: a review of current progress},}\ }\href {\doibase 10.1088/0034-4885/78/10/104501} {\bibfield  {journal} {\bibinfo  {journal} {Rep. Prog. Phys}\ }\textbf {\bibinfo {volume} {78}},\ \bibinfo {pages} {104501} (\bibinfo {year} {2015})}\BibitemShut {NoStop}%
\bibitem [{\citenamefont {Linder}\ and\ \citenamefont {Robinson}(2015)}]{linder_review_2015}%
  \BibitemOpen
  \bibfield  {author} {\bibinfo {author} {\bibfnamefont {J.}~\bibnamefont {Linder}}\ and\ \bibinfo {author} {\bibfnamefont {J.~W.~A.}\ \bibnamefont {Robinson}},\ }\bibfield  {title} {\enquote {\bibinfo {title} {Superconducting spintronics},}\ }\href {\doibase 10.1038/nphys3242} {\bibfield  {journal} {\bibinfo  {journal} {Nat. Phys.}\ }\textbf {\bibinfo {volume} {11}},\ \bibinfo {pages} {307--315} (\bibinfo {year} {2015})}\BibitemShut {NoStop}%
\bibitem [{\citenamefont {Mel'nikov}\ \emph {et~al.}(2022)\citenamefont {Mel'nikov}, \citenamefont {Mironov}, \citenamefont {Samokhvalov},\ and\ \citenamefont {Buzdin}}]{melnikov_2022}%
  \BibitemOpen
  \bibfield  {author} {\bibinfo {author} {\bibfnamefont {A.~S.}\ \bibnamefont {Mel'nikov}}, \bibinfo {author} {\bibfnamefont {S.~V.}\ \bibnamefont {Mironov}}, \bibinfo {author} {\bibfnamefont {A.~V.}\ \bibnamefont {Samokhvalov}}, \ and\ \bibinfo {author} {\bibfnamefont {A.~I.}\ \bibnamefont {Buzdin}},\ }\bibfield  {title} {\enquote {\bibinfo {title} {Superconducting spintronics: state of the art and prospects},}\ }\href {\doibase 10.3367/{UFNe}.2021.07.039020} {\bibfield  {journal} {\bibinfo  {journal} {Phys.-Usp.}\ }\textbf {\bibinfo {volume} {65}},\ \bibinfo {pages} {1248--1289} (\bibinfo {year} {2022})}\BibitemShut {NoStop}%
\bibitem [{\citenamefont {Cai}, \citenamefont {Žutić},\ and\ \citenamefont {Han}(2023)}]{cai_2023}%
  \BibitemOpen
  \bibfield  {author} {\bibinfo {author} {\bibfnamefont {R.}~\bibnamefont {Cai}}, \bibinfo {author} {\bibfnamefont {I.}~\bibnamefont {Žutić}}, \ and\ \bibinfo {author} {\bibfnamefont {W.}~\bibnamefont {Han}},\ }\bibfield  {title} {\enquote {\bibinfo {title} {Superconductor/ferromagnet heterostructures: {A} platform for superconducting spintronics and quantum computation},}\ }\href {\doibase 10.1002/qute.202200080} {\bibfield  {journal} {\bibinfo  {journal} {Adv. Quantum Technol.}\ }\textbf {\bibinfo {volume} {6}},\ \bibinfo {pages} {2200080} (\bibinfo {year} {2023})}\BibitemShut {NoStop}%
\bibitem [{\citenamefont {Kupriyanov}, \citenamefont {Golubov},\ and\ \citenamefont {Siegel}(2006)}]{Kupriyanov_review_2006}%
  \BibitemOpen
  \bibfield  {author} {\bibinfo {author} {\bibfnamefont {M.~Y.}\ \bibnamefont {Kupriyanov}}, \bibinfo {author} {\bibfnamefont {A.~A.}\ \bibnamefont {Golubov}}, \ and\ \bibinfo {author} {\bibfnamefont {M.}~\bibnamefont {Siegel}},\ }\bibfield  {title} {\enquote {\bibinfo {title} {{Josephson junctions with ferromagnetic materials}},}\ }in\ \href {\doibase 10.1117/12.681774} {\emph {\bibinfo {booktitle} {Micro- and Nanoelectronics 2005}}},\ Vol.\ \bibinfo {volume} {6260},\ \bibinfo {editor} {edited by\ \bibinfo {editor} {\bibfnamefont {K.~A.}\ \bibnamefont {Valiev}}\ and\ \bibinfo {editor} {\bibfnamefont {A.~A.}\ \bibnamefont {Orlikovsky}}},\ \bibinfo {organization} {International Society for Optics and Photonics}\ (\bibinfo  {publisher} {SPIE},\ \bibinfo {year} {2006})\ p.\ \bibinfo {pages} {62600S}\BibitemShut {NoStop}%
\bibitem [{\citenamefont {Wong}\ and\ \citenamefont {Ketterson}(1986)}]{wong_1986}%
  \BibitemOpen
  \bibfield  {author} {\bibinfo {author} {\bibfnamefont {H.~K.}\ \bibnamefont {Wong}}\ and\ \bibinfo {author} {\bibfnamefont {J.~B.}\ \bibnamefont {Ketterson}},\ }\bibfield  {title} {\enquote {\bibinfo {title} {Superconducting properties of {Fe/V/Fe} sandwiches},}\ }\href {\doibase 10.1007/{BF00682067}} {\bibfield  {journal} {\bibinfo  {journal} {J Low Temp Phys}\ }\textbf {\bibinfo {volume} {63}},\ \bibinfo {pages} {139--150} (\bibinfo {year} {1986})}\BibitemShut {NoStop}%
\bibitem [{\citenamefont {Wong}\ \emph {et~al.}(1986)\citenamefont {Wong}, \citenamefont {Jin}, \citenamefont {Yang}, \citenamefont {Ketterson},\ and\ \citenamefont {Hilliard}}]{wong_1986a}%
  \BibitemOpen
  \bibfield  {author} {\bibinfo {author} {\bibfnamefont {H.~K.}\ \bibnamefont {Wong}}, \bibinfo {author} {\bibfnamefont {B.~Y.}\ \bibnamefont {Jin}}, \bibinfo {author} {\bibfnamefont {H.~Q.}\ \bibnamefont {Yang}}, \bibinfo {author} {\bibfnamefont {J.~B.}\ \bibnamefont {Ketterson}}, \ and\ \bibinfo {author} {\bibfnamefont {J.~E.}\ \bibnamefont {Hilliard}},\ }\bibfield  {title} {\enquote {\bibinfo {title} {Superconducting properties of {V/Fe} superlattices},}\ }\href {\doibase 10.1007/{BF00683770}} {\bibfield  {journal} {\bibinfo  {journal} {J Low Temp Phys}\ }\textbf {\bibinfo {volume} {63}},\ \bibinfo {pages} {307--315} (\bibinfo {year} {1986})}\BibitemShut {NoStop}%
\bibitem [{\citenamefont {Uher}, \citenamefont {Cohn},\ and\ \citenamefont {Schuller}(1986)}]{uher_1986}%
  \BibitemOpen
  \bibfield  {author} {\bibinfo {author} {\bibfnamefont {C.}~\bibnamefont {Uher}}, \bibinfo {author} {\bibfnamefont {J.}~\bibnamefont {Cohn}}, \ and\ \bibinfo {author} {\bibfnamefont {I.}~\bibnamefont {Schuller}},\ }\bibfield  {title} {\enquote {\bibinfo {title} {Upper critical field in anisotropic superconductors.}}\ }\href {\doibase 10.1103/physrevb.34.4906} {\bibfield  {journal} {\bibinfo  {journal} {Phys Rev, B Condens Matter}\ }\textbf {\bibinfo {volume} {34}},\ \bibinfo {pages} {4906--4908} (\bibinfo {year} {1986})}\BibitemShut {NoStop}%
\bibitem [{\citenamefont {Radovic}\ \emph {et~al.}(1988)\citenamefont {Radovic}, \citenamefont {Dobrosavljevic-Grujic}, \citenamefont {Buzdin},\ and\ \citenamefont {Clem}}]{radovic_1988}%
  \BibitemOpen
  \bibfield  {author} {\bibinfo {author} {\bibfnamefont {Z.}~\bibnamefont {Radovic}}, \bibinfo {author} {\bibfnamefont {L.}~\bibnamefont {Dobrosavljevic-Grujic}}, \bibinfo {author} {\bibfnamefont {A.}~\bibnamefont {Buzdin}}, \ and\ \bibinfo {author} {\bibfnamefont {J.}~\bibnamefont {Clem}},\ }\bibfield  {title} {\enquote {\bibinfo {title} {Upper critical fields of superconductor-ferromagnet multilayers.}}\ }\href {\doibase 10.1103/physrevb.38.2388} {\bibfield  {journal} {\bibinfo  {journal} {Phys Rev, B Condens Matter}\ }\textbf {\bibinfo {volume} {38}},\ \bibinfo {pages} {2388--2393} (\bibinfo {year} {1988})}\BibitemShut {NoStop}%
\bibitem [{\citenamefont {Radovic}\ \emph {et~al.}(1991)\citenamefont {Radovic}, \citenamefont {Ledvij}, \citenamefont {Dobrosavljevic-Grujic}, \citenamefont {Buzdin},\ and\ \citenamefont {Clem}}]{radovic_1991}%
  \BibitemOpen
  \bibfield  {author} {\bibinfo {author} {\bibfnamefont {Z.}~\bibnamefont {Radovic}}, \bibinfo {author} {\bibfnamefont {M.}~\bibnamefont {Ledvij}}, \bibinfo {author} {\bibfnamefont {L.}~\bibnamefont {Dobrosavljevic-Grujic}}, \bibinfo {author} {\bibfnamefont {A.}~\bibnamefont {Buzdin}}, \ and\ \bibinfo {author} {\bibfnamefont {J.}~\bibnamefont {Clem}},\ }\bibfield  {title} {\enquote {\bibinfo {title} {Transition temperatures of superconductor-ferromagnet superlattices.}}\ }\href {\doibase 10.1103/physrevb.44.759} {\bibfield  {journal} {\bibinfo  {journal} {Phys Rev, B Condens Matter}\ }\textbf {\bibinfo {volume} {44}},\ \bibinfo {pages} {759--764} (\bibinfo {year} {1991})}\BibitemShut {NoStop}%
\bibitem [{\citenamefont {Jiang}\ \emph {et~al.}(1995)\citenamefont {Jiang}, \citenamefont {Davidovic}, \citenamefont {Reich},\ and\ \citenamefont {Chien}}]{jiang_1995}%
  \BibitemOpen
  \bibfield  {author} {\bibinfo {author} {\bibfnamefont {J.}~\bibnamefont {Jiang}}, \bibinfo {author} {\bibfnamefont {D.}~\bibnamefont {Davidovic}}, \bibinfo {author} {\bibfnamefont {D.}~\bibnamefont {Reich}}, \ and\ \bibinfo {author} {\bibfnamefont {C.}~\bibnamefont {Chien}},\ }\bibfield  {title} {\enquote {\bibinfo {title} {Oscillatory {S}uperconducting {T}ransition {T}emperature in {Nb/Gd} {M}ultilayers},}\ }\href {\doibase 10.1103/PhysRevLett.74.314} {\bibfield  {journal} {\bibinfo  {journal} {Phys. Rev. Lett.}\ }\textbf {\bibinfo {volume} {74}},\ \bibinfo {pages} {314--317} (\bibinfo {year} {1995})}\BibitemShut {NoStop}%
\bibitem [{\citenamefont {Mühge}\ \emph {et~al.}(1996)\citenamefont {Mühge}, \citenamefont {Garif'yanov}, \citenamefont {Goryunov}, \citenamefont {Khaliullin}, \citenamefont {Tagirov}, \citenamefont {Westerholt}, \citenamefont {Garifullin},\ and\ \citenamefont {Zabel}}]{mhge_1996}%
  \BibitemOpen
  \bibfield  {author} {\bibinfo {author} {\bibfnamefont {T.}~\bibnamefont {Mühge}}, \bibinfo {author} {\bibfnamefont {N.}~\bibnamefont {Garif'yanov}}, \bibinfo {author} {\bibfnamefont {Y.}~\bibnamefont {Goryunov}}, \bibinfo {author} {\bibfnamefont {G.}~\bibnamefont {Khaliullin}}, \bibinfo {author} {\bibfnamefont {L.}~\bibnamefont {Tagirov}}, \bibinfo {author} {\bibfnamefont {K.}~\bibnamefont {Westerholt}}, \bibinfo {author} {\bibfnamefont {I.}~\bibnamefont {Garifullin}}, \ and\ \bibinfo {author} {\bibfnamefont {H.}~\bibnamefont {Zabel}},\ }\bibfield  {title} {\enquote {\bibinfo {title} {Possible origin for oscillatory superconducting transition temperature in superconductor/ferromagnet multilayers},}\ }\href {\doibase 10.1103/PhysRevLett.77.1857} {\bibfield  {journal} {\bibinfo  {journal} {Phys. Rev. Lett.}\ }\textbf {\bibinfo {volume} {77}},\ \bibinfo {pages} {1857--1860} (\bibinfo {year} {1996})}\BibitemShut {NoStop}%
\bibitem [{\citenamefont {Mühge}\ \emph {et~al.}(1997)\citenamefont {Mühge}, \citenamefont {Westerholt}, \citenamefont {Zabel}, \citenamefont {Garif'yanov}, \citenamefont {Goryunov}, \citenamefont {Garifullin},\ and\ \citenamefont {Khaliullin}}]{mhge_1997}%
  \BibitemOpen
  \bibfield  {author} {\bibinfo {author} {\bibfnamefont {T.}~\bibnamefont {Mühge}}, \bibinfo {author} {\bibfnamefont {K.}~\bibnamefont {Westerholt}}, \bibinfo {author} {\bibfnamefont {H.}~\bibnamefont {Zabel}}, \bibinfo {author} {\bibfnamefont {N.~N.}\ \bibnamefont {Garif'yanov}}, \bibinfo {author} {\bibfnamefont {Y.~V.}\ \bibnamefont {Goryunov}}, \bibinfo {author} {\bibfnamefont {I.~A.}\ \bibnamefont {Garifullin}}, \ and\ \bibinfo {author} {\bibfnamefont {G.~G.}\ \bibnamefont {Khaliullin}},\ }\bibfield  {title} {\enquote {\bibinfo {title} {Magnetism and superconductivity of {Fe/Nb/Fe} trilayers},}\ }\href {\doibase 10.1103/PhysRevB.55.8945} {\bibfield  {journal} {\bibinfo  {journal} {Phys. Rev. B}\ }\textbf {\bibinfo {volume} {55}},\ \bibinfo {pages} {8945--8954} (\bibinfo {year} {1997})}\BibitemShut {NoStop}%
\bibitem [{\citenamefont {Chien}\ and\ \citenamefont {Reich}(1999)}]{chien_1999}%
  \BibitemOpen
  \bibfield  {author} {\bibinfo {author} {\bibfnamefont {C.}~\bibnamefont {Chien}}\ and\ \bibinfo {author} {\bibfnamefont {D.~H.}\ \bibnamefont {Reich}},\ }\bibfield  {title} {\enquote {\bibinfo {title} {Proximity effects in superconducting/magnetic multilayers},}\ }\href {\doibase 10.1016/S0304-8853(99)00318-2} {\bibfield  {journal} {\bibinfo  {journal} {J. Magn. Magn. Mater.}\ }\textbf {\bibinfo {volume} {200}},\ \bibinfo {pages} {83--94} (\bibinfo {year} {1999})}\BibitemShut {NoStop}%
\bibitem [{\citenamefont {Demler}, \citenamefont {Arnold},\ and\ \citenamefont {Beasley}(1997)}]{demler_1997}%
  \BibitemOpen
  \bibfield  {author} {\bibinfo {author} {\bibfnamefont {E.~A.}\ \bibnamefont {Demler}}, \bibinfo {author} {\bibfnamefont {G.~B.}\ \bibnamefont {Arnold}}, \ and\ \bibinfo {author} {\bibfnamefont {M.~R.}\ \bibnamefont {Beasley}},\ }\bibfield  {title} {\enquote {\bibinfo {title} {Superconducting proximity effects in magnetic metals},}\ }\href {\doibase 10.1103/PhysRevB.55.15174} {\bibfield  {journal} {\bibinfo  {journal} {Phys. Rev. B}\ }\textbf {\bibinfo {volume} {55}},\ \bibinfo {pages} {15174--15182} (\bibinfo {year} {1997})}\BibitemShut {NoStop}%
\bibitem [{\citenamefont {Fulde}\ and\ \citenamefont {Ferrell}(1964)}]{Fulde_1964}%
  \BibitemOpen
  \bibfield  {author} {\bibinfo {author} {\bibfnamefont {P.}~\bibnamefont {Fulde}}\ and\ \bibinfo {author} {\bibfnamefont {R.~A.}\ \bibnamefont {Ferrell}},\ }\bibfield  {title} {\enquote {\bibinfo {title} {Superconductivity in a strong spin-exchange field},}\ }\href {\doibase 10.1103/PhysRev.135.A550} {\bibfield  {journal} {\bibinfo  {journal} {Phys. Rev.}\ }\textbf {\bibinfo {volume} {135}},\ \bibinfo {pages} {A550--A563} (\bibinfo {year} {1964})}\BibitemShut {NoStop}%
\bibitem [{\citenamefont {Larkin}\ and\ \citenamefont {Ovchinnikov}(1965)}]{larkin_1965}%
  \BibitemOpen
  \bibfield  {author} {\bibinfo {author} {\bibfnamefont {A.}~\bibnamefont {Larkin}}\ and\ \bibinfo {author} {\bibfnamefont {Y.}~\bibnamefont {Ovchinnikov}},\ }\bibfield  {title} {\enquote {\bibinfo {title} {Nonuniform state of superconductors},}\ }\href@noop {} {\bibfield  {journal} {\bibinfo  {journal} {Sov. Phys. JETP}\ }\textbf {\bibinfo {volume} {20}} (\bibinfo {year} {1965})}\BibitemShut {NoStop}%
\bibitem [{\citenamefont {Kontos}\ \emph {et~al.}(2001)\citenamefont {Kontos}, \citenamefont {Aprili}, \citenamefont {Lesueur},\ and\ \citenamefont {Grison}}]{kontos_2001}%
  \BibitemOpen
  \bibfield  {author} {\bibinfo {author} {\bibfnamefont {T.}~\bibnamefont {Kontos}}, \bibinfo {author} {\bibfnamefont {M.}~\bibnamefont {Aprili}}, \bibinfo {author} {\bibfnamefont {J.}~\bibnamefont {Lesueur}}, \ and\ \bibinfo {author} {\bibfnamefont {X.}~\bibnamefont {Grison}},\ }\bibfield  {title} {\enquote {\bibinfo {title} {Inhomogeneous superconductivity induced in a ferromagnet by proximity effect},}\ }\href {\doibase 10.1103/PhysRevLett.86.304} {\bibfield  {journal} {\bibinfo  {journal} {Phys. Rev. Lett.}\ }\textbf {\bibinfo {volume} {86}},\ \bibinfo {pages} {304--307} (\bibinfo {year} {2001})}\BibitemShut {NoStop}%
\bibitem [{\citenamefont {Ambegaokar}\ and\ \citenamefont {Baratoff}(1963)}]{ambegaokar_1963}%
  \BibitemOpen
  \bibfield  {author} {\bibinfo {author} {\bibfnamefont {V.}~\bibnamefont {Ambegaokar}}\ and\ \bibinfo {author} {\bibfnamefont {A.}~\bibnamefont {Baratoff}},\ }\bibfield  {title} {\enquote {\bibinfo {title} {Tunneling {Between S}uperconductors},}\ }\href {\doibase 10.1103/PhysRevLett.10.486} {\bibfield  {journal} {\bibinfo  {journal} {Phys. Rev. Lett.}\ }\textbf {\bibinfo {volume} {10}},\ \bibinfo {pages} {486--489} (\bibinfo {year} {1963})}\BibitemShut {NoStop}%
\bibitem [{\citenamefont {Ness}\ \emph {et~al.}(2022)\citenamefont {Ness}, \citenamefont {Sadovskyy}, \citenamefont {Antipov}, \citenamefont {van Schilfgaarde},\ and\ \citenamefont {Lutchyn}}]{ness_2022}%
  \BibitemOpen
  \bibfield  {author} {\bibinfo {author} {\bibfnamefont {H.}~\bibnamefont {Ness}}, \bibinfo {author} {\bibfnamefont {I.~A.}\ \bibnamefont {Sadovskyy}}, \bibinfo {author} {\bibfnamefont {A.~E.}\ \bibnamefont {Antipov}}, \bibinfo {author} {\bibfnamefont {M.}~\bibnamefont {van Schilfgaarde}}, \ and\ \bibinfo {author} {\bibfnamefont {R.~M.}\ \bibnamefont {Lutchyn}},\ }\bibfield  {title} {\enquote {\bibinfo {title} {Supercurrent decay in ballistic magnetic {Josephson} junctions},}\ }\href {\doibase 10.1038/s41524-021-00694-3} {\bibfield  {journal} {\bibinfo  {journal} {npj Comput. Mater.}\ }\textbf {\bibinfo {volume} {8}},\ \bibinfo {pages} {23} (\bibinfo {year} {2022})}\BibitemShut {NoStop}%
\bibitem [{\citenamefont {Faur\'e}\ \emph {et~al.}(2006)\citenamefont {Faur\'e}, \citenamefont {Buzdin}, \citenamefont {Golubov},\ and\ \citenamefont {Kupriyanov}}]{faure_2006}%
  \BibitemOpen
  \bibfield  {author} {\bibinfo {author} {\bibfnamefont {M.}~\bibnamefont {Faur\'e}}, \bibinfo {author} {\bibfnamefont {A.~I.}\ \bibnamefont {Buzdin}}, \bibinfo {author} {\bibfnamefont {A.~A.}\ \bibnamefont {Golubov}}, \ and\ \bibinfo {author} {\bibfnamefont {M.~Y.}\ \bibnamefont {Kupriyanov}},\ }\bibfield  {title} {\enquote {\bibinfo {title} {Properties of superconductor/ferromagnet structures with spin-dependent scattering},}\ }\href {\doibase 10.1103/PhysRevB.73.064505} {\bibfield  {journal} {\bibinfo  {journal} {Phys. Rev. B}\ }\textbf {\bibinfo {volume} {73}},\ \bibinfo {pages} {064505} (\bibinfo {year} {2006})}\BibitemShut {NoStop}%
\bibitem [{\citenamefont {Pugach}\ \emph {et~al.}(2011)\citenamefont {Pugach}, \citenamefont {Kupriyanov}, \citenamefont {Goldobin}, \citenamefont {Kleiner},\ and\ \citenamefont {Koelle}}]{pugach_2011}%
  \BibitemOpen
  \bibfield  {author} {\bibinfo {author} {\bibfnamefont {N.~G.}\ \bibnamefont {Pugach}}, \bibinfo {author} {\bibfnamefont {M.~Y.}\ \bibnamefont {Kupriyanov}}, \bibinfo {author} {\bibfnamefont {E.}~\bibnamefont {Goldobin}}, \bibinfo {author} {\bibfnamefont {R.}~\bibnamefont {Kleiner}}, \ and\ \bibinfo {author} {\bibfnamefont {D.}~\bibnamefont {Koelle}},\ }\bibfield  {title} {\enquote {\bibinfo {title} {Superconductor-insulator-ferromagnet-superconductor {J}osephson junction: {F}rom the dirty to the clean limit},}\ }\href {\doibase 10.1103/PhysRevB.84.144513} {\bibfield  {journal} {\bibinfo  {journal} {Phys. Rev. B}\ }\textbf {\bibinfo {volume} {84}},\ \bibinfo {pages} {144513} (\bibinfo {year} {2011})}\BibitemShut {NoStop}%
\bibitem [{\citenamefont {Buzdin}(2003)}]{buzdin_2003a}%
  \BibitemOpen
  \bibfield  {author} {\bibinfo {author} {\bibfnamefont {A.}~\bibnamefont {Buzdin}},\ }\bibfield  {title} {\enquote {\bibinfo {title} {$\pi$-junction realization due to tunneling through a thin ferromagnetic layer},}\ }\href {\doibase 10.1134/1.1641489} {\bibfield  {journal} {\bibinfo  {journal} {Jetp Lett.}\ }\textbf {\bibinfo {volume} {78}},\ \bibinfo {pages} {583--586} (\bibinfo {year} {2003})}\BibitemShut {NoStop}%
\bibitem [{\citenamefont {Heim}\ \emph {et~al.}(2015)\citenamefont {Heim}, \citenamefont {Pugach}, \citenamefont {Kupriyanov}, \citenamefont {Goldobin}, \citenamefont {Koelle}, \citenamefont {Kleiner}, \citenamefont {Ruppelt}, \citenamefont {Weides},\ and\ \citenamefont {Kohlstedt}}]{heim_2015}%
  \BibitemOpen
  \bibfield  {author} {\bibinfo {author} {\bibfnamefont {D.~M.}\ \bibnamefont {Heim}}, \bibinfo {author} {\bibfnamefont {N.~G.}\ \bibnamefont {Pugach}}, \bibinfo {author} {\bibfnamefont {M.~Y.}\ \bibnamefont {Kupriyanov}}, \bibinfo {author} {\bibfnamefont {E.}~\bibnamefont {Goldobin}}, \bibinfo {author} {\bibfnamefont {D.}~\bibnamefont {Koelle}}, \bibinfo {author} {\bibfnamefont {R.}~\bibnamefont {Kleiner}}, \bibinfo {author} {\bibfnamefont {N.}~\bibnamefont {Ruppelt}}, \bibinfo {author} {\bibfnamefont {M.}~\bibnamefont {Weides}}, \ and\ \bibinfo {author} {\bibfnamefont {H.}~\bibnamefont {Kohlstedt}},\ }\bibfield  {title} {\enquote {\bibinfo {title} {The effect of normal and insulating layers on 0-$\pi$ transitions in {Josephson} junctions with a ferromagnetic barrier},}\ }\href {\doibase 10.1088/1367-2630/17/11/113022} {\bibfield  {journal} {\bibinfo  {journal} {New J. Phys.}\ }\textbf {\bibinfo {volume} {17}},\ \bibinfo {pages} {113022} (\bibinfo {year} {2015})}\BibitemShut {NoStop}%
\bibitem [{\citenamefont {Bergeret}, \citenamefont {Volkov},\ and\ \citenamefont {Efetov}(2001{\natexlab{b}})}]{bergeret_2001a}%
  \BibitemOpen
  \bibfield  {author} {\bibinfo {author} {\bibfnamefont {F.~S.}\ \bibnamefont {Bergeret}}, \bibinfo {author} {\bibfnamefont {A.~F.}\ \bibnamefont {Volkov}}, \ and\ \bibinfo {author} {\bibfnamefont {K.~B.}\ \bibnamefont {Efetov}},\ }\bibfield  {title} {\enquote {\bibinfo {title} {Josephson current in superconductor-ferromagnet structures with a nonhomogeneous magnetization},}\ }\href {\doibase 10.1103/PhysRevB.64.134506} {\bibfield  {journal} {\bibinfo  {journal} {Phys. Rev. B}\ }\textbf {\bibinfo {volume} {64}},\ \bibinfo {pages} {134506} (\bibinfo {year} {2001}{\natexlab{b}})}\BibitemShut {NoStop}%
\bibitem [{\citenamefont {Chong}, \citenamefont {Dresselhaus},\ and\ \citenamefont {Benz}(2005)}]{chong_2005}%
  \BibitemOpen
  \bibfield  {author} {\bibinfo {author} {\bibfnamefont {Y.}~\bibnamefont {Chong}}, \bibinfo {author} {\bibfnamefont {P.~D.}\ \bibnamefont {Dresselhaus}}, \ and\ \bibinfo {author} {\bibfnamefont {S.~P.}\ \bibnamefont {Benz}},\ }\bibfield  {title} {\enquote {\bibinfo {title} {Electrical properties of {Nb–MoSi2–Nb} {Josephson} junctions},}\ }\href {\doibase 10.1063/1.1947386} {\bibfield  {journal} {\bibinfo  {journal} {Appl. Phys. Lett}\ }\textbf {\bibinfo {volume} {86}},\ \bibinfo {pages} {232505} (\bibinfo {year} {2005})}\BibitemShut {NoStop}%
\bibitem [{\citenamefont {Clarke}(1969)}]{clarke_1969}%
  \BibitemOpen
  \bibfield  {author} {\bibinfo {author} {\bibfnamefont {J.}~\bibnamefont {Clarke}},\ }\bibfield  {title} {\enquote {\bibinfo {title} {Supercurrents in lead-copper-lead sandwiches},}\ }\href {\doibase 10.1098/rspa.1969.0020} {\bibfield  {journal} {\bibinfo  {journal} {Proc. R. Soc. Lond. A}\ }\textbf {\bibinfo {volume} {308}},\ \bibinfo {pages} {447--471} (\bibinfo {year} {1969})}\BibitemShut {NoStop}%
\bibitem [{\citenamefont {Anwar}\ \emph {et~al.}(2010)\citenamefont {Anwar}, \citenamefont {Czeschka}, \citenamefont {Hesselberth}, \citenamefont {Porcu},\ and\ \citenamefont {Aarts}}]{anwar_2010}%
  \BibitemOpen
  \bibfield  {author} {\bibinfo {author} {\bibfnamefont {M.~S.}\ \bibnamefont {Anwar}}, \bibinfo {author} {\bibfnamefont {F.}~\bibnamefont {Czeschka}}, \bibinfo {author} {\bibfnamefont {M.}~\bibnamefont {Hesselberth}}, \bibinfo {author} {\bibfnamefont {M.}~\bibnamefont {Porcu}}, \ and\ \bibinfo {author} {\bibfnamefont {J.}~\bibnamefont {Aarts}},\ }\bibfield  {title} {\enquote {\bibinfo {title} {Long-range supercurrents through half-metallic ferromagnetic {CrO}$_2$},}\ }\href {\doibase 10.1103/PhysRevB.82.100501} {\bibfield  {journal} {\bibinfo  {journal} {Phys. Rev. B}\ }\textbf {\bibinfo {volume} {82}},\ \bibinfo {pages} {100501} (\bibinfo {year} {2010})}\BibitemShut {NoStop}%
\bibitem [{\citenamefont {Wang}\ \emph {et~al.}(2010)\citenamefont {Wang}, \citenamefont {Singh}, \citenamefont {Tian}, \citenamefont {Kumar}, \citenamefont {Liu}, \citenamefont {Shi}, \citenamefont {Jain}, \citenamefont {Samarth}, \citenamefont {Mallouk},\ and\ \citenamefont {Chan}}]{wang_2010}%
  \BibitemOpen
  \bibfield  {author} {\bibinfo {author} {\bibfnamefont {J.}~\bibnamefont {Wang}}, \bibinfo {author} {\bibfnamefont {M.}~\bibnamefont {Singh}}, \bibinfo {author} {\bibfnamefont {M.}~\bibnamefont {Tian}}, \bibinfo {author} {\bibfnamefont {N.}~\bibnamefont {Kumar}}, \bibinfo {author} {\bibfnamefont {B.}~\bibnamefont {Liu}}, \bibinfo {author} {\bibfnamefont {C.}~\bibnamefont {Shi}}, \bibinfo {author} {\bibfnamefont {J.~K.}\ \bibnamefont {Jain}}, \bibinfo {author} {\bibfnamefont {N.}~\bibnamefont {Samarth}}, \bibinfo {author} {\bibfnamefont {T.~E.}\ \bibnamefont {Mallouk}}, \ and\ \bibinfo {author} {\bibfnamefont {M.~H.~W.}\ \bibnamefont {Chan}},\ }\bibfield  {title} {\enquote {\bibinfo {title} {Interplay between superconductivity and ferromagnetism in crystalline nanowires},}\ }\href {\doibase 10.1038/nphys1621} {\bibfield  {journal} {\bibinfo  {journal} {Nat. Phys.}\ }\textbf {\bibinfo {volume} {6}},\ \bibinfo {pages} {389--394} (\bibinfo {year} {2010})}\BibitemShut {NoStop}%
\bibitem [{\citenamefont {Weides}\ \emph {et~al.}(2006{\natexlab{a}})\citenamefont {Weides}, \citenamefont {Kemmler}, \citenamefont {Goldobin}, \citenamefont {Koelle}, \citenamefont {Kleiner}, \citenamefont {Kohlstedt},\ and\ \citenamefont {Buzdin}}]{Weides_APL_2006}%
  \BibitemOpen
  \bibfield  {author} {\bibinfo {author} {\bibfnamefont {M.}~\bibnamefont {Weides}}, \bibinfo {author} {\bibfnamefont {M.}~\bibnamefont {Kemmler}}, \bibinfo {author} {\bibfnamefont {E.}~\bibnamefont {Goldobin}}, \bibinfo {author} {\bibfnamefont {D.}~\bibnamefont {Koelle}}, \bibinfo {author} {\bibfnamefont {R.}~\bibnamefont {Kleiner}}, \bibinfo {author} {\bibfnamefont {H.}~\bibnamefont {Kohlstedt}}, \ and\ \bibinfo {author} {\bibfnamefont {A.~I.}\ \bibnamefont {Buzdin}},\ }\bibfield  {title} {\enquote {\bibinfo {title} {High quality ferromagnetic 0 and $\pi$ {J}osephson tunnel junctions},}\ }\href {\doibase 10.1063/1.2356104} {\bibfield  {journal} {\bibinfo  {journal} {Appl. Phys. Lett.}\ }\textbf {\bibinfo {volume} {89}},\ \bibinfo {pages} {122511} (\bibinfo {year} {2006}{\natexlab{a}})}\BibitemShut {NoStop}%
\bibitem [{\citenamefont {Tanaka}\ and\ \citenamefont {Kashiwaya}(1997)}]{tanaka_1997}%
  \BibitemOpen
  \bibfield  {author} {\bibinfo {author} {\bibfnamefont {Y.}~\bibnamefont {Tanaka}}\ and\ \bibinfo {author} {\bibfnamefont {S.}~\bibnamefont {Kashiwaya}},\ }\bibfield  {title} {\enquote {\bibinfo {title} {Theory of {Josephson} effect in superconductor-ferromagnetic-insulator-superconductor junction},}\ }\href {\doibase 10.1016/S0921-4534(97)80002-0} {\bibfield  {journal} {\bibinfo  {journal} {Physica C: Supercond.}\ }\textbf {\bibinfo {volume} {274}},\ \bibinfo {pages} {357--363} (\bibinfo {year} {1997})}\BibitemShut {NoStop}%
\bibitem [{\citenamefont {Kawabata}\ \emph {et~al.}(2010)\citenamefont {Kawabata}, \citenamefont {Asano}, \citenamefont {Tanaka}, \citenamefont {Golubov},\ and\ \citenamefont {Kashiwaya}}]{kawabata_2010}%
  \BibitemOpen
  \bibfield  {author} {\bibinfo {author} {\bibfnamefont {S.}~\bibnamefont {Kawabata}}, \bibinfo {author} {\bibfnamefont {Y.}~\bibnamefont {Asano}}, \bibinfo {author} {\bibfnamefont {Y.}~\bibnamefont {Tanaka}}, \bibinfo {author} {\bibfnamefont {A.~A.}\ \bibnamefont {Golubov}}, \ and\ \bibinfo {author} {\bibfnamefont {S.}~\bibnamefont {Kashiwaya}},\ }\bibfield  {title} {\enquote {\bibinfo {title} {Josephson $\pi$ state in a ferromagnetic insulator},}\ }\href {\doibase 10.1103/PhysRevLett.104.117002} {\bibfield  {journal} {\bibinfo  {journal} {Phys. Rev. Lett.}\ }\textbf {\bibinfo {volume} {104}},\ \bibinfo {pages} {117002} (\bibinfo {year} {2010})}\BibitemShut {NoStop}%
\bibitem [{\citenamefont {Senapati}, \citenamefont {Blamire},\ and\ \citenamefont {Barber}(2011)}]{senapati_2011}%
  \BibitemOpen
  \bibfield  {author} {\bibinfo {author} {\bibfnamefont {K.}~\bibnamefont {Senapati}}, \bibinfo {author} {\bibfnamefont {M.~G.}\ \bibnamefont {Blamire}}, \ and\ \bibinfo {author} {\bibfnamefont {Z.~H.}\ \bibnamefont {Barber}},\ }\bibfield  {title} {\enquote {\bibinfo {title} {Spin-filter {Josephson} junctions},}\ }\href {\doibase 10.1038/nmat3116} {\bibfield  {journal} {\bibinfo  {journal} {Nat. Mater.}\ }\textbf {\bibinfo {volume} {10}},\ \bibinfo {pages} {849--852} (\bibinfo {year} {2011})}\BibitemShut {NoStop}%
\bibitem [{\citenamefont {Barone}\ and\ \citenamefont {Patern\`{o}}(1982)}]{barone1982physics}%
  \BibitemOpen
  \bibfield  {author} {\bibinfo {author} {\bibfnamefont {A.}~\bibnamefont {Barone}}\ and\ \bibinfo {author} {\bibfnamefont {G.}~\bibnamefont {Patern\`{o}}},\ }\href {\doibase 10.1002/352760278X} {\emph {\bibinfo {title} {Physics and {A}pplications of the {J}osephson {E}ffect}}}\ (\bibinfo  {publisher} {John Wiley \& Sons, New York},\ \bibinfo {year} {1982})\BibitemShut {NoStop}%
\bibitem [{\citenamefont {Blamire}\ \emph {et~al.}(2013)\citenamefont {Blamire}, \citenamefont {Smiet}, \citenamefont {Banerjee},\ and\ \citenamefont {Robinson}}]{blamire_2013}%
  \BibitemOpen
  \bibfield  {author} {\bibinfo {author} {\bibfnamefont {M.~G.}\ \bibnamefont {Blamire}}, \bibinfo {author} {\bibfnamefont {C.~B.}\ \bibnamefont {Smiet}}, \bibinfo {author} {\bibfnamefont {N.}~\bibnamefont {Banerjee}}, \ and\ \bibinfo {author} {\bibfnamefont {J.~W.~A.}\ \bibnamefont {Robinson}},\ }\bibfield  {title} {\enquote {\bibinfo {title} {Field modulation of the critical current in magnetic {Josephson} junctions},}\ }\href {\doibase 10.1088/0953-2048/26/5/055017} {\bibfield  {journal} {\bibinfo  {journal} {Supercond. Sci. Technol.}\ }\textbf {\bibinfo {volume} {26}},\ \bibinfo {pages} {055017} (\bibinfo {year} {2013})}\BibitemShut {NoStop}%
\bibitem [{\citenamefont {Golovchanskiy}\ \emph {et~al.}(2016)\citenamefont {Golovchanskiy}, \citenamefont {Bol'ginov}, \citenamefont {Stolyarov}, \citenamefont {Abramov}, \citenamefont {Ben~Hamida}, \citenamefont {Emelyanova}, \citenamefont {Stolyarov}, \citenamefont {Kupriyanov}, \citenamefont {Golubov},\ and\ \citenamefont {Ryazanov}}]{golovchanskiy_2016}%
  \BibitemOpen
  \bibfield  {author} {\bibinfo {author} {\bibfnamefont {I.~A.}\ \bibnamefont {Golovchanskiy}}, \bibinfo {author} {\bibfnamefont {V.~V.}\ \bibnamefont {Bol'ginov}}, \bibinfo {author} {\bibfnamefont {V.~S.}\ \bibnamefont {Stolyarov}}, \bibinfo {author} {\bibfnamefont {N.~N.}\ \bibnamefont {Abramov}}, \bibinfo {author} {\bibfnamefont {A.}~\bibnamefont {Ben~Hamida}}, \bibinfo {author} {\bibfnamefont {O.~V.}\ \bibnamefont {Emelyanova}}, \bibinfo {author} {\bibfnamefont {B.~S.}\ \bibnamefont {Stolyarov}}, \bibinfo {author} {\bibfnamefont {M.~Y.}\ \bibnamefont {Kupriyanov}}, \bibinfo {author} {\bibfnamefont {A.~A.}\ \bibnamefont {Golubov}}, \ and\ \bibinfo {author} {\bibfnamefont {V.~V.}\ \bibnamefont {Ryazanov}},\ }\bibfield  {title} {\enquote {\bibinfo {title} {Micromagnetic modeling of critical current oscillations in magnetic {Josephson} junctions},}\ }\href {\doibase 10.1103/PhysRevB.94.214514} {\bibfield  {journal} {\bibinfo  {journal} {Phys. Rev. B}\ }\textbf {\bibinfo {volume} {94}},\ \bibinfo {pages}
  {214514} (\bibinfo {year} {2016})}\BibitemShut {NoStop}%
\bibitem [{\citenamefont {Bourgeois}\ \emph {et~al.}(2001)\citenamefont {Bourgeois}, \citenamefont {Gandit}, \citenamefont {Lesueur}, \citenamefont {Sulpice}, \citenamefont {Grison},\ and\ \citenamefont {Chaussy}}]{bourgeois_2001}%
  \BibitemOpen
  \bibfield  {author} {\bibinfo {author} {\bibfnamefont {O.}~\bibnamefont {Bourgeois}}, \bibinfo {author} {\bibfnamefont {P.}~\bibnamefont {Gandit}}, \bibinfo {author} {\bibfnamefont {J.}~\bibnamefont {Lesueur}}, \bibinfo {author} {\bibfnamefont {A.}~\bibnamefont {Sulpice}}, \bibinfo {author} {\bibfnamefont {X.}~\bibnamefont {Grison}}, \ and\ \bibinfo {author} {\bibfnamefont {J.}~\bibnamefont {Chaussy}},\ }\bibfield  {title} {\enquote {\bibinfo {title} {Josephson effect through a ferromagnetic layer},}\ }\href {\doibase 10.1007/s100510170215} {\bibfield  {journal} {\bibinfo  {journal} {Eur. Phys. J. B}\ }\textbf {\bibinfo {volume} {21}},\ \bibinfo {pages} {75--80} (\bibinfo {year} {2001})}\BibitemShut {NoStop}%
\bibitem [{\citenamefont {Khaire}, \citenamefont {Pratt},\ and\ \citenamefont {Birge}(2009)}]{Khaire_PRB_2009}%
  \BibitemOpen
  \bibfield  {author} {\bibinfo {author} {\bibfnamefont {T.~S.}\ \bibnamefont {Khaire}}, \bibinfo {author} {\bibfnamefont {W.~P.}\ \bibnamefont {Pratt}}, \ and\ \bibinfo {author} {\bibfnamefont {N.~O.}\ \bibnamefont {Birge}},\ }\bibfield  {title} {\enquote {\bibinfo {title} {Critical current behavior in {J}osephson junctions with the weak ferromagnet {PdNi}},}\ }\href {\doibase 10.1103/PhysRevB.79.094523} {\bibfield  {journal} {\bibinfo  {journal} {Phys. Rev. B}\ }\textbf {\bibinfo {volume} {79}},\ \bibinfo {pages} {094523} (\bibinfo {year} {2009})}\BibitemShut {NoStop}%
\bibitem [{\citenamefont {Khasawneh}, \citenamefont {Pratt},\ and\ \citenamefont {Birge}(2009)}]{khasawneh_2009}%
  \BibitemOpen
  \bibfield  {author} {\bibinfo {author} {\bibfnamefont {M.~A.}\ \bibnamefont {Khasawneh}}, \bibinfo {author} {\bibfnamefont {W.~P.}\ \bibnamefont {Pratt}}, \ and\ \bibinfo {author} {\bibfnamefont {N.~O.}\ \bibnamefont {Birge}},\ }\bibfield  {title} {\enquote {\bibinfo {title} {Josephson junctions with a synthetic antiferromagnetic interlayer},}\ }\href {\doibase 10.1103/PhysRevB.80.020506} {\bibfield  {journal} {\bibinfo  {journal} {Phys. Rev. B}\ }\textbf {\bibinfo {volume} {80}},\ \bibinfo {pages} {020506} (\bibinfo {year} {2009})}\BibitemShut {NoStop}%
\bibitem [{\citenamefont {Ryazanov}(1999)}]{ryazanov_1999}%
  \BibitemOpen
  \bibfield  {author} {\bibinfo {author} {\bibfnamefont {V.~V.}\ \bibnamefont {Ryazanov}},\ }\bibfield  {title} {\enquote {\bibinfo {title} {Josephson superconductor - ferromagnet - superconductor - $\pi$-contact as an element of a quantum bit (experiment)},}\ }\href@noop {} {\bibfield  {journal} {\bibinfo  {journal} {Phys.-Uspekhi}\ }\textbf {\bibinfo {volume} {42}},\ \bibinfo {pages} {825 -- 827} (\bibinfo {year} {1999})}\BibitemShut {NoStop}%
\bibitem [{\citenamefont {Veshchunov}\ \emph {et~al.}(2008)\citenamefont {Veshchunov}, \citenamefont {Oboznov}, \citenamefont {Rossolenko}, \citenamefont {Prokokiev}, \citenamefont {Vinnikov}, \citenamefont {Rusanov},\ and\ \citenamefont {Matveev}}]{veshchunov_2008}%
  \BibitemOpen
  \bibfield  {author} {\bibinfo {author} {\bibfnamefont {I.~S.}\ \bibnamefont {Veshchunov}}, \bibinfo {author} {\bibfnamefont {V.~A.}\ \bibnamefont {Oboznov}}, \bibinfo {author} {\bibfnamefont {A.~N.}\ \bibnamefont {Rossolenko}}, \bibinfo {author} {\bibfnamefont {A.~S.}\ \bibnamefont {Prokokiev}}, \bibinfo {author} {\bibfnamefont {L.~Y.}\ \bibnamefont {Vinnikov}}, \bibinfo {author} {\bibfnamefont {A.~Y.}\ \bibnamefont {Rusanov}}, \ and\ \bibinfo {author} {\bibfnamefont {D.~V.}\ \bibnamefont {Matveev}},\ }\bibfield  {title} {\enquote {\bibinfo {title} {Observation of the magnetic domain structure in {Cu}$_{0.47}${Ni}$_{0.53}$ thin films at low temperature},}\ }\href@noop {} {\bibfield  {journal} {\bibinfo  {journal} {Pis'ma v. {ZhETF}}\ }\textbf {\bibinfo {volume} {88}},\ \bibinfo {pages} {873--876} (\bibinfo {year} {2008})}\BibitemShut {NoStop}%
\bibitem [{\citenamefont {{Glick}}\ \emph {et~al.}(2017)\citenamefont {{Glick}}, \citenamefont {{Loloee}}, \citenamefont {{Pratt}},\ and\ \citenamefont {{Birge}}}]{Glick_IEEE_2017}%
  \BibitemOpen
  \bibfield  {author} {\bibinfo {author} {\bibfnamefont {J.~A.}\ \bibnamefont {{Glick}}}, \bibinfo {author} {\bibfnamefont {R.}~\bibnamefont {{Loloee}}}, \bibinfo {author} {\bibfnamefont {W.~P.}\ \bibnamefont {{Pratt}}}, \ and\ \bibinfo {author} {\bibfnamefont {N.~O.}\ \bibnamefont {{Birge}}},\ }\bibfield  {title} {\enquote {\bibinfo {title} {{Critical Current Oscillations of Josephson Junctions Containing PdFe Nanomagnets}},}\ }\href {\doibase 10.1109/TASC.2016.2630024} {\bibfield  {journal} {\bibinfo  {journal} {IEEE Trans. Appl. Supercond.}\ }\textbf {\bibinfo {volume} {27}},\ \bibinfo {pages} {1800205} (\bibinfo {year} {2017})}\BibitemShut {NoStop}%
\bibitem [{\citenamefont {Wollman}\ \emph {et~al.}(1993)\citenamefont {Wollman}, \citenamefont {Van~Harlingen}, \citenamefont {Lee}, \citenamefont {Ginsberg},\ and\ \citenamefont {Leggett}}]{wollman_1993}%
  \BibitemOpen
  \bibfield  {author} {\bibinfo {author} {\bibfnamefont {D.~A.}\ \bibnamefont {Wollman}}, \bibinfo {author} {\bibfnamefont {D.~J.}\ \bibnamefont {Van~Harlingen}}, \bibinfo {author} {\bibfnamefont {W.~C.}\ \bibnamefont {Lee}}, \bibinfo {author} {\bibfnamefont {D.~M.}\ \bibnamefont {Ginsberg}}, \ and\ \bibinfo {author} {\bibfnamefont {A.~J.}\ \bibnamefont {Leggett}},\ }\bibfield  {title} {\enquote {\bibinfo {title} {Experimental determination of the superconducting pairing state in {YBCO} from the phase coherence of {YBCO-Pb} dc {SQUIDs}},}\ }\href {\doibase 10.1103/PhysRevLett.71.2134} {\bibfield  {journal} {\bibinfo  {journal} {Phys. Rev. Lett.}\ }\textbf {\bibinfo {volume} {71}},\ \bibinfo {pages} {2134--2137} (\bibinfo {year} {1993})}\BibitemShut {NoStop}%
\bibitem [{\citenamefont {Kirtley}\ \emph {et~al.}(1995)\citenamefont {Kirtley}, \citenamefont {Tsuei}, \citenamefont {Sun}, \citenamefont {Chi}, \citenamefont {Yu-Jahnes}, \citenamefont {Gupta}, \citenamefont {Rupp},\ and\ \citenamefont {Ketchen}}]{kirtley_1995}%
  \BibitemOpen
  \bibfield  {author} {\bibinfo {author} {\bibfnamefont {J.~R.}\ \bibnamefont {Kirtley}}, \bibinfo {author} {\bibfnamefont {C.~C.}\ \bibnamefont {Tsuei}}, \bibinfo {author} {\bibfnamefont {J.~Z.}\ \bibnamefont {Sun}}, \bibinfo {author} {\bibfnamefont {C.~C.}\ \bibnamefont {Chi}}, \bibinfo {author} {\bibfnamefont {L.~S.}\ \bibnamefont {Yu-Jahnes}}, \bibinfo {author} {\bibfnamefont {A.}~\bibnamefont {Gupta}}, \bibinfo {author} {\bibfnamefont {M.}~\bibnamefont {Rupp}}, \ and\ \bibinfo {author} {\bibfnamefont {M.~B.}\ \bibnamefont {Ketchen}},\ }\bibfield  {title} {\enquote {\bibinfo {title} {Symmetry of the order parameter in the high-{T}$_c$ superconductor {YBa}$_2${Cu}$_3${O}$_{7-\delta}$},}\ }\href {\doibase 10.1038/373225a0} {\bibfield  {journal} {\bibinfo  {journal} {Nature}\ }\textbf {\bibinfo {volume} {373}},\ \bibinfo {pages} {225--228} (\bibinfo {year} {1995})}\BibitemShut {NoStop}%
\bibitem [{\citenamefont {Yip}(1998)}]{yip_1998}%
  \BibitemOpen
  \bibfield  {author} {\bibinfo {author} {\bibfnamefont {S.-K.}\ \bibnamefont {Yip}},\ }\bibfield  {title} {\enquote {\bibinfo {title} {Energy-resolved supercurrent between two superconductors},}\ }\href {\doibase 10.1103/PhysRevB.58.5803} {\bibfield  {journal} {\bibinfo  {journal} {Phys. Rev. B}\ }\textbf {\bibinfo {volume} {58}},\ \bibinfo {pages} {5803--5807} (\bibinfo {year} {1998})}\BibitemShut {NoStop}%
\bibitem [{\citenamefont {Wilhelm}, \citenamefont {Schön},\ and\ \citenamefont {Zaikin}(1998)}]{wilhelm_1998}%
  \BibitemOpen
  \bibfield  {author} {\bibinfo {author} {\bibfnamefont {F.~K.}\ \bibnamefont {Wilhelm}}, \bibinfo {author} {\bibfnamefont {G.}~\bibnamefont {Schön}}, \ and\ \bibinfo {author} {\bibfnamefont {A.~D.}\ \bibnamefont {Zaikin}},\ }\bibfield  {title} {\enquote {\bibinfo {title} {Mesoscopic {Supercon}ducting–{Normal M}etal–{Superconducting T}ransistor},}\ }\href {\doibase 10.1103/PhysRevLett.81.1682} {\bibfield  {journal} {\bibinfo  {journal} {Phys. Rev. Lett.}\ }\textbf {\bibinfo {volume} {81}},\ \bibinfo {pages} {1682--1685} (\bibinfo {year} {1998})}\BibitemShut {NoStop}%
\bibitem [{\citenamefont {Baselmans}, \citenamefont {van Wees},\ and\ \citenamefont {Klapwijk}(2001)}]{baselmans_2001}%
  \BibitemOpen
  \bibfield  {author} {\bibinfo {author} {\bibfnamefont {J.~J.~A.}\ \bibnamefont {Baselmans}}, \bibinfo {author} {\bibfnamefont {B.~J.}\ \bibnamefont {van Wees}}, \ and\ \bibinfo {author} {\bibfnamefont {T.~M.}\ \bibnamefont {Klapwijk}},\ }\bibfield  {title} {\enquote {\bibinfo {title} {Nonequilibrium supercurrent transport in controllable superconductor–normal-metal–superconductor junctions},}\ }\href {\doibase 10.1103/PhysRevB.63.094504} {\bibfield  {journal} {\bibinfo  {journal} {Phys. Rev. B}\ }\textbf {\bibinfo {volume} {63}},\ \bibinfo {pages} {094504} (\bibinfo {year} {2001})}\BibitemShut {NoStop}%
\bibitem [{\citenamefont {van Wees}, \citenamefont {Lenssen},\ and\ \citenamefont {Harmans}(1991)}]{vanweesbj_1991}%
  \BibitemOpen
  \bibfield  {author} {\bibinfo {author} {\bibfnamefont {B.~J.}\ \bibnamefont {van Wees}}, \bibinfo {author} {\bibfnamefont {K.}~\bibnamefont {Lenssen}}, \ and\ \bibinfo {author} {\bibfnamefont {C.}~\bibnamefont {Harmans}},\ }\bibfield  {title} {\enquote {\bibinfo {title} {Transmission formalism for supercurrent flow in multiprobe superconductor-semiconductor-superconductor devices},}\ }\href {\doibase 10.1103/physrevb.44.470} {\bibfield  {journal} {\bibinfo  {journal} {Phys. Rev. B}\ }\textbf {\bibinfo {volume} {44}},\ \bibinfo {pages} {470--473} (\bibinfo {year} {1991})}\BibitemShut {NoStop}%
\bibitem [{\citenamefont {Huang}\ \emph {et~al.}(2002)\citenamefont {Huang}, \citenamefont {Pierre}, \citenamefont {Heikkilä}, \citenamefont {Wilhelm},\ and\ \citenamefont {Birge}}]{huang_2002}%
  \BibitemOpen
  \bibfield  {author} {\bibinfo {author} {\bibfnamefont {J.}~\bibnamefont {Huang}}, \bibinfo {author} {\bibfnamefont {F.}~\bibnamefont {Pierre}}, \bibinfo {author} {\bibfnamefont {T.~T.}\ \bibnamefont {Heikkilä}}, \bibinfo {author} {\bibfnamefont {F.~K.}\ \bibnamefont {Wilhelm}}, \ and\ \bibinfo {author} {\bibfnamefont {N.~O.}\ \bibnamefont {Birge}},\ }\bibfield  {title} {\enquote {\bibinfo {title} {Observation of a controllable $\pi$ junction in a 3-terminal {Josephson} device},}\ }\href {\doibase 10.1103/PhysRevB.66.020507} {\bibfield  {journal} {\bibinfo  {journal} {Phys. Rev. B}\ }\textbf {\bibinfo {volume} {66}},\ \bibinfo {pages} {020507} (\bibinfo {year} {2002})}\BibitemShut {NoStop}%
\bibitem [{\citenamefont {Crosser}\ \emph {et~al.}(2008)\citenamefont {Crosser}, \citenamefont {Huang}, \citenamefont {Pierre}, \citenamefont {Virtanen}, \citenamefont {Heikkilä}, \citenamefont {Wilhelm},\ and\ \citenamefont {Birge}}]{crosser_2008}%
  \BibitemOpen
  \bibfield  {author} {\bibinfo {author} {\bibfnamefont {M.~S.}\ \bibnamefont {Crosser}}, \bibinfo {author} {\bibfnamefont {J.}~\bibnamefont {Huang}}, \bibinfo {author} {\bibfnamefont {F.}~\bibnamefont {Pierre}}, \bibinfo {author} {\bibfnamefont {P.}~\bibnamefont {Virtanen}}, \bibinfo {author} {\bibfnamefont {T.~T.}\ \bibnamefont {Heikkilä}}, \bibinfo {author} {\bibfnamefont {F.~K.}\ \bibnamefont {Wilhelm}}, \ and\ \bibinfo {author} {\bibfnamefont {N.~O.}\ \bibnamefont {Birge}},\ }\bibfield  {title} {\enquote {\bibinfo {title} {Nonequilibrium transport in mesoscopic multi-terminal {SNS} {Josephson} junctions},}\ }\href {\doibase 10.1103/PhysRevB.77.014528} {\bibfield  {journal} {\bibinfo  {journal} {Phys. Rev. B}\ }\textbf {\bibinfo {volume} {77}},\ \bibinfo {pages} {014528} (\bibinfo {year} {2008})}\BibitemShut {NoStop}%
\bibitem [{\citenamefont {Mooij}\ \emph {et~al.}(1999)\citenamefont {Mooij}, \citenamefont {Orlando}, \citenamefont {Levitov}, \citenamefont {Tian}, \citenamefont {van~der Wal C~H},\ and\ \citenamefont {Lloyd}}]{mooij_1999}%
  \BibitemOpen
  \bibfield  {author} {\bibinfo {author} {\bibfnamefont {J.~E.}\ \bibnamefont {Mooij}}, \bibinfo {author} {\bibfnamefont {T.~P.}\ \bibnamefont {Orlando}}, \bibinfo {author} {\bibfnamefont {L.}~\bibnamefont {Levitov}}, \bibinfo {author} {\bibfnamefont {L.}~\bibnamefont {Tian}}, \bibinfo {author} {\bibnamefont {van~der Wal C~H}}, \ and\ \bibinfo {author} {\bibfnamefont {S.}~\bibnamefont {Lloyd}},\ }\bibfield  {title} {\enquote {\bibinfo {title} {Josephson persistent-current qubit.}}\ }\href {\doibase 10.1126/science.285.5430.1036} {\bibfield  {journal} {\bibinfo  {journal} {Science}\ }\textbf {\bibinfo {volume} {285}},\ \bibinfo {pages} {1036--1039} (\bibinfo {year} {1999})}\BibitemShut {NoStop}%
\bibitem [{\citenamefont {Orlando}\ \emph {et~al.}(1999)\citenamefont {Orlando}, \citenamefont {Mooij}, \citenamefont {Tian}, \citenamefont {van~der Wal}, \citenamefont {Levitov}, \citenamefont {Lloyd},\ and\ \citenamefont {Mazo}}]{Orlando_1999}%
  \BibitemOpen
  \bibfield  {author} {\bibinfo {author} {\bibfnamefont {T.~P.}\ \bibnamefont {Orlando}}, \bibinfo {author} {\bibfnamefont {J.~E.}\ \bibnamefont {Mooij}}, \bibinfo {author} {\bibfnamefont {L.}~\bibnamefont {Tian}}, \bibinfo {author} {\bibfnamefont {C.~H.}\ \bibnamefont {van~der Wal}}, \bibinfo {author} {\bibfnamefont {L.~S.}\ \bibnamefont {Levitov}}, \bibinfo {author} {\bibfnamefont {S.}~\bibnamefont {Lloyd}}, \ and\ \bibinfo {author} {\bibfnamefont {J.~J.}\ \bibnamefont {Mazo}},\ }\bibfield  {title} {\enquote {\bibinfo {title} {Superconducting persistent-current qubit},}\ }\href {\doibase 10.1103/PhysRevB.60.15398} {\bibfield  {journal} {\bibinfo  {journal} {Phys. Rev. B}\ }\textbf {\bibinfo {volume} {60}},\ \bibinfo {pages} {15398--15413} (\bibinfo {year} {1999})}\BibitemShut {NoStop}%
\bibitem [{\citenamefont {Yamashita}\ \emph {et~al.}(2005)\citenamefont {Yamashita}, \citenamefont {Tanikawa}, \citenamefont {Takahashi},\ and\ \citenamefont {Maekawa}}]{yamashita_2005}%
  \BibitemOpen
  \bibfield  {author} {\bibinfo {author} {\bibfnamefont {T.}~\bibnamefont {Yamashita}}, \bibinfo {author} {\bibfnamefont {K.}~\bibnamefont {Tanikawa}}, \bibinfo {author} {\bibfnamefont {S.}~\bibnamefont {Takahashi}}, \ and\ \bibinfo {author} {\bibfnamefont {S.}~\bibnamefont {Maekawa}},\ }\bibfield  {title} {\enquote {\bibinfo {title} {Superconducting $\pi$ qubit with a ferromagnetic {Josephson} junction},}\ }\href {\doibase 10.1103/PhysRevLett.95.097001} {\bibfield  {journal} {\bibinfo  {journal} {Phys. Rev. Lett.}\ }\textbf {\bibinfo {volume} {95}},\ \bibinfo {pages} {097001} (\bibinfo {year} {2005})}\BibitemShut {NoStop}%
\bibitem [{\citenamefont {Yamashita}, \citenamefont {Takahashi},\ and\ \citenamefont {Maekawa}(2006)}]{yamashita_2006}%
  \BibitemOpen
  \bibfield  {author} {\bibinfo {author} {\bibfnamefont {T.}~\bibnamefont {Yamashita}}, \bibinfo {author} {\bibfnamefont {S.}~\bibnamefont {Takahashi}}, \ and\ \bibinfo {author} {\bibfnamefont {S.}~\bibnamefont {Maekawa}},\ }\bibfield  {title} {\enquote {\bibinfo {title} {Superconducting $\pi$ qubit with three {Josephson} junctions},}\ }\href {\doibase 10.1063/1.2189191} {\bibfield  {journal} {\bibinfo  {journal} {Appl. Phys. Lett}\ }\textbf {\bibinfo {volume} {88}},\ \bibinfo {pages} {132501} (\bibinfo {year} {2006})}\BibitemShut {NoStop}%
\bibitem [{\citenamefont {Kawabata}\ \emph {et~al.}(2006)\citenamefont {Kawabata}, \citenamefont {Kashiwaya}, \citenamefont {Asano}, \citenamefont {Tanaka},\ and\ \citenamefont {Golubov}}]{kawabata_2006}%
  \BibitemOpen
  \bibfield  {author} {\bibinfo {author} {\bibfnamefont {S.}~\bibnamefont {Kawabata}}, \bibinfo {author} {\bibfnamefont {S.}~\bibnamefont {Kashiwaya}}, \bibinfo {author} {\bibfnamefont {Y.}~\bibnamefont {Asano}}, \bibinfo {author} {\bibfnamefont {Y.}~\bibnamefont {Tanaka}}, \ and\ \bibinfo {author} {\bibfnamefont {A.~A.}\ \bibnamefont {Golubov}},\ }\bibfield  {title} {\enquote {\bibinfo {title} {Macroscopic quantum dynamics of $\pi$ junctions with ferromagnetic insulators},}\ }\href {\doibase 10.1103/PhysRevB.74.180502} {\bibfield  {journal} {\bibinfo  {journal} {Phys. Rev. B}\ }\textbf {\bibinfo {volume} {74}},\ \bibinfo {pages} {180502} (\bibinfo {year} {2006})}\BibitemShut {NoStop}%
\bibitem [{\citenamefont {Kato}, \citenamefont {Golubov},\ and\ \citenamefont {Nakamura}(2007)}]{kato_2007}%
  \BibitemOpen
  \bibfield  {author} {\bibinfo {author} {\bibfnamefont {T.}~\bibnamefont {Kato}}, \bibinfo {author} {\bibfnamefont {A.~A.}\ \bibnamefont {Golubov}}, \ and\ \bibinfo {author} {\bibfnamefont {Y.}~\bibnamefont {Nakamura}},\ }\bibfield  {title} {\enquote {\bibinfo {title} {Decoherence in a superconducting flux qubit with a $\pi$-junction},}\ }\href {\doibase 10.1103/PhysRevB.76.172502} {\bibfield  {journal} {\bibinfo  {journal} {Phys. Rev. B}\ }\textbf {\bibinfo {volume} {76}},\ \bibinfo {pages} {172502} (\bibinfo {year} {2007})}\BibitemShut {NoStop}%
\bibitem [{\citenamefont {Kawabata}\ \emph {et~al.}(2007)\citenamefont {Kawabata}, \citenamefont {Golubov}, \citenamefont {{Ariando}}, \citenamefont {Verwijs}, \citenamefont {Hilgenkamp},\ and\ \citenamefont {Kirtley}}]{kawabata_2007}%
  \BibitemOpen
  \bibfield  {author} {\bibinfo {author} {\bibfnamefont {S.}~\bibnamefont {Kawabata}}, \bibinfo {author} {\bibfnamefont {A.~A.}\ \bibnamefont {Golubov}}, \bibinfo {author} {\bibnamefont {{Ariando}}}, \bibinfo {author} {\bibfnamefont {C.~J.~M.}\ \bibnamefont {Verwijs}}, \bibinfo {author} {\bibfnamefont {H.}~\bibnamefont {Hilgenkamp}}, \ and\ \bibinfo {author} {\bibfnamefont {J.~R.}\ \bibnamefont {Kirtley}},\ }\bibfield  {title} {\enquote {\bibinfo {title} {Macroscopic quantum tunneling and quasiparticle-tunneling blockade effect in $s$-wave/$d$-wave hybrid junctions},}\ }\href {\doibase 10.1103/PhysRevB.76.064505} {\bibfield  {journal} {\bibinfo  {journal} {Phys. Rev. B}\ }\textbf {\bibinfo {volume} {76}},\ \bibinfo {pages} {064505} (\bibinfo {year} {2007})}\BibitemShut {NoStop}%
\bibitem [{\citenamefont {Feofanov}\ \emph {et~al.}(2010)\citenamefont {Feofanov}, \citenamefont {Oboznov}, \citenamefont {Bol'ginov}, \citenamefont {Lisenfeld}, \citenamefont {Poletto}, \citenamefont {Ryazanov}, \citenamefont {Rossolenko}, \citenamefont {Khabipov}, \citenamefont {Balashov}, \citenamefont {Zorin}, \citenamefont {Dmitriev}, \citenamefont {Koshelets},\ and\ \citenamefont {Ustinov}}]{feofanov_2010}%
  \BibitemOpen
  \bibfield  {author} {\bibinfo {author} {\bibfnamefont {A.~K.}\ \bibnamefont {Feofanov}}, \bibinfo {author} {\bibfnamefont {V.~A.}\ \bibnamefont {Oboznov}}, \bibinfo {author} {\bibfnamefont {V.~V.}\ \bibnamefont {Bol'ginov}}, \bibinfo {author} {\bibfnamefont {J.}~\bibnamefont {Lisenfeld}}, \bibinfo {author} {\bibfnamefont {S.}~\bibnamefont {Poletto}}, \bibinfo {author} {\bibfnamefont {V.~V.}\ \bibnamefont {Ryazanov}}, \bibinfo {author} {\bibfnamefont {A.~N.}\ \bibnamefont {Rossolenko}}, \bibinfo {author} {\bibfnamefont {M.}~\bibnamefont {Khabipov}}, \bibinfo {author} {\bibfnamefont {D.}~\bibnamefont {Balashov}}, \bibinfo {author} {\bibfnamefont {A.~B.}\ \bibnamefont {Zorin}}, \bibinfo {author} {\bibfnamefont {P.~N.}\ \bibnamefont {Dmitriev}}, \bibinfo {author} {\bibfnamefont {V.~P.}\ \bibnamefont {Koshelets}}, \ and\ \bibinfo {author} {\bibfnamefont {A.~V.}\ \bibnamefont {Ustinov}},\ }\bibfield  {title} {\enquote {\bibinfo {title} {Implementation of superconductor/ferromagnet/ superconductor $\pi$-shifters
  in superconducting digital and quantum circuits},}\ }\href {\doibase 10.1038/nphys1700} {\bibfield  {journal} {\bibinfo  {journal} {Nat. Phys.}\ }\textbf {\bibinfo {volume} {6}},\ \bibinfo {pages} {593--597} (\bibinfo {year} {2010})}\BibitemShut {NoStop}%
\bibitem [{\citenamefont {Shcherbakova}\ \emph {et~al.}(2015)\citenamefont {Shcherbakova}, \citenamefont {Fedorov}, \citenamefont {Shulga}, \citenamefont {Ryazanov}, \citenamefont {Bolginov}, \citenamefont {Oboznov}, \citenamefont {Egorov}, \citenamefont {Shkolnikov}, \citenamefont {Wolf}, \citenamefont {Beckmann},\ and\ \citenamefont {Ustinov}}]{shcherbakova_2015}%
  \BibitemOpen
  \bibfield  {author} {\bibinfo {author} {\bibfnamefont {A.~V.}\ \bibnamefont {Shcherbakova}}, \bibinfo {author} {\bibfnamefont {K.~G.}\ \bibnamefont {Fedorov}}, \bibinfo {author} {\bibfnamefont {K.~V.}\ \bibnamefont {Shulga}}, \bibinfo {author} {\bibfnamefont {V.~V.}\ \bibnamefont {Ryazanov}}, \bibinfo {author} {\bibfnamefont {V.~V.}\ \bibnamefont {Bolginov}}, \bibinfo {author} {\bibfnamefont {V.~A.}\ \bibnamefont {Oboznov}}, \bibinfo {author} {\bibfnamefont {S.~V.}\ \bibnamefont {Egorov}}, \bibinfo {author} {\bibfnamefont {V.~O.}\ \bibnamefont {Shkolnikov}}, \bibinfo {author} {\bibfnamefont {M.~J.}\ \bibnamefont {Wolf}}, \bibinfo {author} {\bibfnamefont {D.}~\bibnamefont {Beckmann}}, \ and\ \bibinfo {author} {\bibfnamefont {A.~V.}\ \bibnamefont {Ustinov}},\ }\bibfield  {title} {\enquote {\bibinfo {title} {Fabrication and measurements of hybrid {Nb/Al} {Josephson} junctions and flux qubits with $\pi$-shifters},}\ }\href {\doibase 10.1088/0953-2048/28/2/025009} {\bibfield  {journal} {\bibinfo  {journal}
  {Supercond. Sci. Technol.}\ }\textbf {\bibinfo {volume} {28}},\ \bibinfo {pages} {025009} (\bibinfo {year} {2015})}\BibitemShut {NoStop}%
\bibitem [{\citenamefont {Mori}\ and\ \citenamefont {Maekawa}(2021)}]{mori_2021}%
  \BibitemOpen
  \bibfield  {author} {\bibinfo {author} {\bibfnamefont {M.}~\bibnamefont {Mori}}\ and\ \bibinfo {author} {\bibfnamefont {S.}~\bibnamefont {Maekawa}},\ }\bibfield  {title} {\enquote {\bibinfo {title} {Half-integer {S}hapiro-steps in superconducting qubit with a $\pi$-{Josephson} junction},}\ }\href {\doibase 10.35848/1882-0786/ac211d} {\bibfield  {journal} {\bibinfo  {journal} {Appl. Phys. Express}\ }\textbf {\bibinfo {volume} {14}},\ \bibinfo {pages} {103001} (\bibinfo {year} {2021})}\BibitemShut {NoStop}%
\bibitem [{\citenamefont {Ahmad}\ \emph {et~al.}(2022{\natexlab{a}})\citenamefont {Ahmad}, \citenamefont {Brosco}, \citenamefont {Miano}, \citenamefont {Di~Palma}, \citenamefont {Arzeo}, \citenamefont {Montemurro}, \citenamefont {Lucignano}, \citenamefont {Pepe}, \citenamefont {Tafuri}, \citenamefont {Fazio},\ and\ \citenamefont {Massarotti}}]{ahmad_2022}%
  \BibitemOpen
  \bibfield  {author} {\bibinfo {author} {\bibfnamefont {H.~G.}\ \bibnamefont {Ahmad}}, \bibinfo {author} {\bibfnamefont {V.}~\bibnamefont {Brosco}}, \bibinfo {author} {\bibfnamefont {A.}~\bibnamefont {Miano}}, \bibinfo {author} {\bibfnamefont {L.}~\bibnamefont {Di~Palma}}, \bibinfo {author} {\bibfnamefont {M.}~\bibnamefont {Arzeo}}, \bibinfo {author} {\bibfnamefont {D.}~\bibnamefont {Montemurro}}, \bibinfo {author} {\bibfnamefont {P.}~\bibnamefont {Lucignano}}, \bibinfo {author} {\bibfnamefont {G.~P.}\ \bibnamefont {Pepe}}, \bibinfo {author} {\bibfnamefont {F.}~\bibnamefont {Tafuri}}, \bibinfo {author} {\bibfnamefont {R.}~\bibnamefont {Fazio}}, \ and\ \bibinfo {author} {\bibfnamefont {D.}~\bibnamefont {Massarotti}},\ }\bibfield  {title} {\enquote {\bibinfo {title} {Hybrid ferromagnetic transmon qubit: {C}ircuit design, feasibility, and detection protocols for magnetic fluctuations},}\ }\href {\doibase 10.1103/PhysRevB.105.214522} {\bibfield  {journal} {\bibinfo  {journal} {Phys. Rev. B}\ }\textbf {\bibinfo
  {volume} {105}},\ \bibinfo {pages} {214522} (\bibinfo {year} {2022}{\natexlab{a}})}\BibitemShut {NoStop}%
\bibitem [{\citenamefont {Massarotti}\ \emph {et~al.}(2023)\citenamefont {Massarotti}, \citenamefont {Ahmad}, \citenamefont {Satariano}, \citenamefont {Ferraiuolo}, \citenamefont {Di~Palma}, \citenamefont {Mastrovito}, \citenamefont {Serpico}, \citenamefont {Levochkina}, \citenamefont {Caruso}, \citenamefont {Miano}, \citenamefont {Arzeo}, \citenamefont {Ausanio}, \citenamefont {Granata}, \citenamefont {Lucignano}, \citenamefont {Montemurro}, \citenamefont {Parlato}, \citenamefont {Vettoliere}, \citenamefont {Fazio}, \citenamefont {Mukhanov}, \citenamefont {Pepe},\ and\ \citenamefont {Tafuri}}]{massarotti_2023}%
  \BibitemOpen
  \bibfield  {author} {\bibinfo {author} {\bibfnamefont {D.}~\bibnamefont {Massarotti}}, \bibinfo {author} {\bibfnamefont {H.~G.}\ \bibnamefont {Ahmad}}, \bibinfo {author} {\bibfnamefont {R.}~\bibnamefont {Satariano}}, \bibinfo {author} {\bibfnamefont {R.}~\bibnamefont {Ferraiuolo}}, \bibinfo {author} {\bibfnamefont {L.}~\bibnamefont {Di~Palma}}, \bibinfo {author} {\bibfnamefont {P.}~\bibnamefont {Mastrovito}}, \bibinfo {author} {\bibfnamefont {G.}~\bibnamefont {Serpico}}, \bibinfo {author} {\bibfnamefont {A.}~\bibnamefont {Levochkina}}, \bibinfo {author} {\bibfnamefont {R.}~\bibnamefont {Caruso}}, \bibinfo {author} {\bibfnamefont {A.}~\bibnamefont {Miano}}, \bibinfo {author} {\bibfnamefont {M.}~\bibnamefont {Arzeo}}, \bibinfo {author} {\bibfnamefont {G.}~\bibnamefont {Ausanio}}, \bibinfo {author} {\bibfnamefont {C.}~\bibnamefont {Granata}}, \bibinfo {author} {\bibfnamefont {P.}~\bibnamefont {Lucignano}}, \bibinfo {author} {\bibfnamefont {D.}~\bibnamefont {Montemurro}}, \bibinfo {author} {\bibfnamefont
  {L.}~\bibnamefont {Parlato}}, \bibinfo {author} {\bibfnamefont {A.}~\bibnamefont {Vettoliere}}, \bibinfo {author} {\bibfnamefont {R.}~\bibnamefont {Fazio}}, \bibinfo {author} {\bibfnamefont {O.}~\bibnamefont {Mukhanov}}, \bibinfo {author} {\bibfnamefont {G.~P.}\ \bibnamefont {Pepe}}, \ and\ \bibinfo {author} {\bibfnamefont {F.}~\bibnamefont {Tafuri}},\ }\bibfield  {title} {\enquote {\bibinfo {title} {A feasible path for the use of ferromagnetic {Josephson} junctions in quantum circuits: The ferro-transmon},}\ }\href {\doibase 10.1063/10.0019690} {\bibfield  {journal} {\bibinfo  {journal} {Low Temp. Phys.}\ }\textbf {\bibinfo {volume} {49}},\ \bibinfo {pages} {794--802} (\bibinfo {year} {2023})}\BibitemShut {NoStop}%
\bibitem [{\citenamefont {Kim}\ \emph {et~al.}(2024)\citenamefont {Kim}, \citenamefont {Abdurakhimov}, \citenamefont {Pham}, \citenamefont {Qiu}, \citenamefont {Terai}, \citenamefont {Ashhab}, \citenamefont {Saito}, \citenamefont {Yamashita},\ and\ \citenamefont {Semba}}]{kim_2024}%
  \BibitemOpen
  \bibfield  {author} {\bibinfo {author} {\bibfnamefont {S.}~\bibnamefont {Kim}}, \bibinfo {author} {\bibfnamefont {L.~V.}\ \bibnamefont {Abdurakhimov}}, \bibinfo {author} {\bibfnamefont {D.}~\bibnamefont {Pham}}, \bibinfo {author} {\bibfnamefont {W.}~\bibnamefont {Qiu}}, \bibinfo {author} {\bibfnamefont {H.}~\bibnamefont {Terai}}, \bibinfo {author} {\bibfnamefont {S.}~\bibnamefont {Ashhab}}, \bibinfo {author} {\bibfnamefont {S.}~\bibnamefont {Saito}}, \bibinfo {author} {\bibfnamefont {T.}~\bibnamefont {Yamashita}}, \ and\ \bibinfo {author} {\bibfnamefont {K.}~\bibnamefont {Semba}},\ }\bibfield  {title} {\enquote {\bibinfo {title} {Superconducting flux qubit with ferromagnetic {J}osephson $\pi$ junction operating at zero magnetic field},}\ }\href {https://arxiv.org/abs/2401.14597} {\bibfield  {journal} {\bibinfo  {journal} {arXiv:2401.14597}\ } (\bibinfo {year} {2024})}\BibitemShut {NoStop}%
\bibitem [{\citenamefont {Somoroff}\ \emph {et~al.}(2023)\citenamefont {Somoroff}, \citenamefont {Ficheux}, \citenamefont {Mencia}, \citenamefont {Xiong}, \citenamefont {Kuzmin},\ and\ \citenamefont {Manucharyan}}]{somoroff_2023}%
  \BibitemOpen
  \bibfield  {author} {\bibinfo {author} {\bibfnamefont {A.}~\bibnamefont {Somoroff}}, \bibinfo {author} {\bibfnamefont {Q.}~\bibnamefont {Ficheux}}, \bibinfo {author} {\bibfnamefont {R.~A.}\ \bibnamefont {Mencia}}, \bibinfo {author} {\bibfnamefont {H.}~\bibnamefont {Xiong}}, \bibinfo {author} {\bibfnamefont {R.}~\bibnamefont {Kuzmin}}, \ and\ \bibinfo {author} {\bibfnamefont {V.~E.}\ \bibnamefont {Manucharyan}},\ }\bibfield  {title} {\enquote {\bibinfo {title} {Millisecond coherence in a superconducting qubit},}\ }\href {\doibase 10.1103/PhysRevLett.130.267001} {\bibfield  {journal} {\bibinfo  {journal} {Phys. Rev. Lett.}\ }\textbf {\bibinfo {volume} {130}},\ \bibinfo {pages} {267001} (\bibinfo {year} {2023})}\BibitemShut {NoStop}%
\bibitem [{\citenamefont {Pop}\ \emph {et~al.}(2014)\citenamefont {Pop}, \citenamefont {Geerlings}, \citenamefont {Catelani}, \citenamefont {Schoelkopf}, \citenamefont {Glazman},\ and\ \citenamefont {Devoret}}]{pop_2014}%
  \BibitemOpen
  \bibfield  {author} {\bibinfo {author} {\bibfnamefont {I.~M.}\ \bibnamefont {Pop}}, \bibinfo {author} {\bibfnamefont {K.}~\bibnamefont {Geerlings}}, \bibinfo {author} {\bibfnamefont {G.}~\bibnamefont {Catelani}}, \bibinfo {author} {\bibfnamefont {R.~J.}\ \bibnamefont {Schoelkopf}}, \bibinfo {author} {\bibfnamefont {L.~I.}\ \bibnamefont {Glazman}}, \ and\ \bibinfo {author} {\bibfnamefont {M.~H.}\ \bibnamefont {Devoret}},\ }\bibfield  {title} {\enquote {\bibinfo {title} {Coherent suppression of electromagnetic dissipation due to superconducting quasiparticles.}}\ }\href {\doibase 10.1038/nature13017} {\bibfield  {journal} {\bibinfo  {journal} {Nature}\ }\textbf {\bibinfo {volume} {508}},\ \bibinfo {pages} {369--372} (\bibinfo {year} {2014})}\BibitemShut {NoStop}%
\bibitem [{\citenamefont {Likharev}(2012)}]{likharev_2012}%
  \BibitemOpen
  \bibfield  {author} {\bibinfo {author} {\bibfnamefont {K.~K.}\ \bibnamefont {Likharev}},\ }\bibfield  {title} {\enquote {\bibinfo {title} {Superconductor digital electronics},}\ }\href {\doibase 10.1016/j.physc.2012.05.016} {\bibfield  {journal} {\bibinfo  {journal} {Physica C: Supercond.}\ }\textbf {\bibinfo {volume} {482}},\ \bibinfo {pages} {6--18} (\bibinfo {year} {2012})}\BibitemShut {NoStop}%
\bibitem [{\citenamefont {Soloviev}\ \emph {et~al.}(2017)\citenamefont {Soloviev}, \citenamefont {Klenov}, \citenamefont {Bakurskiy}, \citenamefont {Kupriyanov}, \citenamefont {Gudkov},\ and\ \citenamefont {Sidorenko}}]{soloviev2017beyond}%
  \BibitemOpen
  \bibfield  {author} {\bibinfo {author} {\bibfnamefont {I.~I.}\ \bibnamefont {Soloviev}}, \bibinfo {author} {\bibfnamefont {N.~V.}\ \bibnamefont {Klenov}}, \bibinfo {author} {\bibfnamefont {S.~V.}\ \bibnamefont {Bakurskiy}}, \bibinfo {author} {\bibfnamefont {M.~Y.}\ \bibnamefont {Kupriyanov}}, \bibinfo {author} {\bibfnamefont {A.~L.}\ \bibnamefont {Gudkov}}, \ and\ \bibinfo {author} {\bibfnamefont {A.~S.}\ \bibnamefont {Sidorenko}},\ }\bibfield  {title} {\enquote {\bibinfo {title} {Beyond {M}oore’s technologies: operation principles of a superconductor alternative},}\ }\href {\doibase 10.3762/bjnano.8.269} {\bibfield  {journal} {\bibinfo  {journal} {Beilstein J. Nanotechnol}\ }\textbf {\bibinfo {volume} {8}},\ \bibinfo {pages} {2689--2710} (\bibinfo {year} {2017})}\BibitemShut {NoStop}%
\bibitem [{\citenamefont {Ortlepp}\ \emph {et~al.}(2006)\citenamefont {Ortlepp}, \citenamefont {{Ariando}}, \citenamefont {Mielke}, \citenamefont {Verwijs}, \citenamefont {Foo}, \citenamefont {Rogalla}, \citenamefont {Uhlmann},\ and\ \citenamefont {Hilgenkamp}}]{ortlepp_2006}%
  \BibitemOpen
  \bibfield  {author} {\bibinfo {author} {\bibfnamefont {T.}~\bibnamefont {Ortlepp}}, \bibinfo {author} {\bibnamefont {{Ariando}}}, \bibinfo {author} {\bibfnamefont {O.}~\bibnamefont {Mielke}}, \bibinfo {author} {\bibfnamefont {C.~J.~M.}\ \bibnamefont {Verwijs}}, \bibinfo {author} {\bibfnamefont {K.~F.~K.}\ \bibnamefont {Foo}}, \bibinfo {author} {\bibfnamefont {H.}~\bibnamefont {Rogalla}}, \bibinfo {author} {\bibfnamefont {F.~H.}\ \bibnamefont {Uhlmann}}, \ and\ \bibinfo {author} {\bibfnamefont {H.}~\bibnamefont {Hilgenkamp}},\ }\bibfield  {title} {\enquote {\bibinfo {title} {Flip-flopping fractional flux quanta},}\ }\href {\doibase 10.1126/science.1126041} {\bibfield  {journal} {\bibinfo  {journal} {Science}\ }\textbf {\bibinfo {volume} {312}},\ \bibinfo {pages} {1495--1497} (\bibinfo {year} {2006})}\BibitemShut {NoStop}%
\bibitem [{\citenamefont {Mielke}\ \emph {et~al.}(2008)\citenamefont {Mielke}, \citenamefont {Ortlepp}, \citenamefont {Dimov},\ and\ \citenamefont {Uhlmann}}]{mielke_2008}%
  \BibitemOpen
  \bibfield  {author} {\bibinfo {author} {\bibfnamefont {O.}~\bibnamefont {Mielke}}, \bibinfo {author} {\bibfnamefont {T.}~\bibnamefont {Ortlepp}}, \bibinfo {author} {\bibfnamefont {B.}~\bibnamefont {Dimov}}, \ and\ \bibinfo {author} {\bibfnamefont {F.~H.}\ \bibnamefont {Uhlmann}},\ }\bibfield  {title} {\enquote {\bibinfo {title} {Phase engineering techniques in superconducting quantum electronics},}\ }\href {\doibase 10.1088/1742-6596/97/1/012196} {\bibfield  {journal} {\bibinfo  {journal} {J. Phys.: Conf. Ser.}\ }\textbf {\bibinfo {volume} {97}},\ \bibinfo {pages} {012196} (\bibinfo {year} {2008})}\BibitemShut {NoStop}%
\bibitem [{\citenamefont {Khabipov}\ \emph {et~al.}(2010)\citenamefont {Khabipov}, \citenamefont {Balashov}, \citenamefont {Maibaum}, \citenamefont {Zorin}, \citenamefont {Oboznov}, \citenamefont {Bol'ginov}, \citenamefont {Rossolenko},\ and\ \citenamefont {Ryazanov}}]{khabipov_2010}%
  \BibitemOpen
  \bibfield  {author} {\bibinfo {author} {\bibfnamefont {M.~I.}\ \bibnamefont {Khabipov}}, \bibinfo {author} {\bibfnamefont {D.~V.}\ \bibnamefont {Balashov}}, \bibinfo {author} {\bibfnamefont {F.}~\bibnamefont {Maibaum}}, \bibinfo {author} {\bibfnamefont {A.~B.}\ \bibnamefont {Zorin}}, \bibinfo {author} {\bibfnamefont {V.~A.}\ \bibnamefont {Oboznov}}, \bibinfo {author} {\bibfnamefont {V.~V.}\ \bibnamefont {Bol'ginov}}, \bibinfo {author} {\bibfnamefont {A.~N.}\ \bibnamefont {Rossolenko}}, \ and\ \bibinfo {author} {\bibfnamefont {V.~V.}\ \bibnamefont {Ryazanov}},\ }\bibfield  {title} {\enquote {\bibinfo {title} {A single flux quantum circuit with a ferromagnet-based {Josephson} $\pi$-junction},}\ }\href {\doibase 10.1088/0953-2048/23/4/045032} {\bibfield  {journal} {\bibinfo  {journal} {Supercond. Sci. Technol.}\ }\textbf {\bibinfo {volume} {23}},\ \bibinfo {pages} {045032} (\bibinfo {year} {2010})}\BibitemShut {NoStop}%
\bibitem [{\citenamefont {Wetzstein}\ \emph {et~al.}(2011)\citenamefont {Wetzstein}, \citenamefont {Ortlepp}, \citenamefont {Stolz}, \citenamefont {Kunert}, \citenamefont {Meyer},\ and\ \citenamefont {Toepfer}}]{wetzstein_2011}%
  \BibitemOpen
  \bibfield  {author} {\bibinfo {author} {\bibfnamefont {O.}~\bibnamefont {Wetzstein}}, \bibinfo {author} {\bibfnamefont {T.}~\bibnamefont {Ortlepp}}, \bibinfo {author} {\bibfnamefont {R.}~\bibnamefont {Stolz}}, \bibinfo {author} {\bibfnamefont {J.}~\bibnamefont {Kunert}}, \bibinfo {author} {\bibfnamefont {H.-G.}\ \bibnamefont {Meyer}}, \ and\ \bibinfo {author} {\bibfnamefont {H.}~\bibnamefont {Toepfer}},\ }\bibfield  {title} {\enquote {\bibinfo {title} {Comparison of {RSFQ} logic cells with and without phase shifting elements by means of {BER} measurements},}\ }\href {\doibase 10.1109/TASC.2010.2102998} {\bibfield  {journal} {\bibinfo  {journal} {{IEEE} Trans. Appl. Supercond.}\ }\textbf {\bibinfo {volume} {21}},\ \bibinfo {pages} {814--817} (\bibinfo {year} {2011})}\BibitemShut {NoStop}%
\bibitem [{\citenamefont {Kamiya}\ \emph {et~al.}(2018)\citenamefont {Kamiya}, \citenamefont {Tanaka}, \citenamefont {Sano},\ and\ \citenamefont {Fujimaki}}]{kamiya_2018}%
  \BibitemOpen
  \bibfield  {author} {\bibinfo {author} {\bibfnamefont {T.}~\bibnamefont {Kamiya}}, \bibinfo {author} {\bibfnamefont {M.}~\bibnamefont {Tanaka}}, \bibinfo {author} {\bibfnamefont {K.}~\bibnamefont {Sano}}, \ and\ \bibinfo {author} {\bibfnamefont {A.}~\bibnamefont {Fujimaki}},\ }\bibfield  {title} {\enquote {\bibinfo {title} {Energy/{S}pace-{Efficient Rapid S}ingle-{F}lux-{Q}uantum {Circuits by U}sing $\pi$-{S}hifted {Josephson} {J}unctions},}\ }\href {\doibase 10.1587/transele.E101.C.385} {\bibfield  {journal} {\bibinfo  {journal} {{IEICE} Trans. Electron.}\ }\textbf {\bibinfo {volume} {E101.C}},\ \bibinfo {pages} {385--390} (\bibinfo {year} {2018})}\BibitemShut {NoStop}%
\bibitem [{\citenamefont {Yamanashi}\ \emph {et~al.}(2018)\citenamefont {Yamanashi}, \citenamefont {Nakaishi}, \citenamefont {Sugiyama}, \citenamefont {Takeuchi},\ and\ \citenamefont {Yoshikawa}}]{yamanashi_2018}%
  \BibitemOpen
  \bibfield  {author} {\bibinfo {author} {\bibfnamefont {Y.}~\bibnamefont {Yamanashi}}, \bibinfo {author} {\bibfnamefont {S.}~\bibnamefont {Nakaishi}}, \bibinfo {author} {\bibfnamefont {A.}~\bibnamefont {Sugiyama}}, \bibinfo {author} {\bibfnamefont {N.}~\bibnamefont {Takeuchi}}, \ and\ \bibinfo {author} {\bibfnamefont {N.}~\bibnamefont {Yoshikawa}},\ }\bibfield  {title} {\enquote {\bibinfo {title} {Design methodology of single-flux-quantum flip-flops composed of both 0- and $\pi$-shifted {Josephson} junctions},}\ }\href {\doibase 10.1088/1361-6668/aad78d} {\bibfield  {journal} {\bibinfo  {journal} {Supercond. Sci. Technol.}\ }\textbf {\bibinfo {volume} {31}},\ \bibinfo {pages} {105003} (\bibinfo {year} {2018})}\BibitemShut {NoStop}%
\bibitem [{\citenamefont {Hasegawa}\ \emph {et~al.}(2019)\citenamefont {Hasegawa}, \citenamefont {Takeshita}, \citenamefont {Sano}, \citenamefont {Tanaka}, \citenamefont {Fujimaki},\ and\ \citenamefont {Yamashita}}]{hasegawa_2019}%
  \BibitemOpen
  \bibfield  {author} {\bibinfo {author} {\bibfnamefont {D.}~\bibnamefont {Hasegawa}}, \bibinfo {author} {\bibfnamefont {Y.}~\bibnamefont {Takeshita}}, \bibinfo {author} {\bibfnamefont {K.}~\bibnamefont {Sano}}, \bibinfo {author} {\bibfnamefont {M.}~\bibnamefont {Tanaka}}, \bibinfo {author} {\bibfnamefont {A.}~\bibnamefont {Fujimaki}}, \ and\ \bibinfo {author} {\bibfnamefont {T.}~\bibnamefont {Yamashita}},\ }\bibfield  {title} {\enquote {\bibinfo {title} {Magnetic {J}osephson junctions on {Nb} four-layer structure for half flux quantum circuits},}\ }in\ \href {\doibase 10.1109/{ISEC46533}.2019.8990908} {\emph {\bibinfo {booktitle} {2019 {IEEE} International Superconductive Electronics Conference ({ISEC})}}}\ (\bibinfo  {publisher} {IEEE},\ \bibinfo {year} {2019})\ pp.\ \bibinfo {pages} {1--3}\BibitemShut {NoStop}%
\bibitem [{\citenamefont {Arai}\ \emph {et~al.}(2019)\citenamefont {Arai}, \citenamefont {Takeuchi}, \citenamefont {Yamashita},\ and\ \citenamefont {Yoshikawa}}]{arai_2019}%
  \BibitemOpen
  \bibfield  {author} {\bibinfo {author} {\bibfnamefont {K.}~\bibnamefont {Arai}}, \bibinfo {author} {\bibfnamefont {N.}~\bibnamefont {Takeuchi}}, \bibinfo {author} {\bibfnamefont {T.}~\bibnamefont {Yamashita}}, \ and\ \bibinfo {author} {\bibfnamefont {N.}~\bibnamefont {Yoshikawa}},\ }\bibfield  {title} {\enquote {\bibinfo {title} {Adiabatic quantum-flux-parametron with $\pi$ {Josephson} junctions},}\ }\href {\doibase 10.1063/1.5080467} {\bibfield  {journal} {\bibinfo  {journal} {J. Appl. Phys.}\ }\textbf {\bibinfo {volume} {125}},\ \bibinfo {pages} {093901} (\bibinfo {year} {2019})}\BibitemShut {NoStop}%
\bibitem [{\citenamefont {Yamanashi}, \citenamefont {Nakaishi},\ and\ \citenamefont {Yoshikawa}(2019)}]{yamanashi_2019}%
  \BibitemOpen
  \bibfield  {author} {\bibinfo {author} {\bibfnamefont {Y.}~\bibnamefont {Yamanashi}}, \bibinfo {author} {\bibfnamefont {S.}~\bibnamefont {Nakaishi}}, \ and\ \bibinfo {author} {\bibfnamefont {N.}~\bibnamefont {Yoshikawa}},\ }\bibfield  {title} {\enquote {\bibinfo {title} {Simulation of the {Margins in Single Flux Quantum Circuits C}ontaining $\pi$-{S}hifted {Josephson} {J}unctions},}\ }\href {\doibase 10.1109/TASC.2019.2904700} {\bibfield  {journal} {\bibinfo  {journal} {{IEEE} Trans. Appl. Supercond.}\ }\textbf {\bibinfo {volume} {29}},\ \bibinfo {pages} {1301805} (\bibinfo {year} {2019})}\BibitemShut {NoStop}%
\bibitem [{\citenamefont {Yamashita}(2021)}]{yamashita_2021}%
  \BibitemOpen
  \bibfield  {author} {\bibinfo {author} {\bibfnamefont {T.}~\bibnamefont {Yamashita}},\ }\bibfield  {title} {\enquote {\bibinfo {title} {Magnetic {Josephson} {Junctions: New Phenomena and Physics with Diluted Alloy, Conventional Ferromagnet, and Multilayer B}arriers},}\ }\href {\doibase 10.1587/transele.{2020SUI0004}} {\bibfield  {journal} {\bibinfo  {journal} {{IEICE} Trans. Electron.}\ }\textbf {\bibinfo {volume} {E104.C}},\ \bibinfo {pages} {422--428} (\bibinfo {year} {2021})}\BibitemShut {NoStop}%
\bibitem [{\citenamefont {Takeshita}\ \emph {et~al.}(2021)\citenamefont {Takeshita}, \citenamefont {Li}, \citenamefont {Hasegawa}, \citenamefont {Sano}, \citenamefont {Tanaka}, \citenamefont {Yamashita},\ and\ \citenamefont {Fujimaki}}]{takeshita_2021}%
  \BibitemOpen
  \bibfield  {author} {\bibinfo {author} {\bibfnamefont {Y.}~\bibnamefont {Takeshita}}, \bibinfo {author} {\bibfnamefont {F.}~\bibnamefont {Li}}, \bibinfo {author} {\bibfnamefont {D.}~\bibnamefont {Hasegawa}}, \bibinfo {author} {\bibfnamefont {K.}~\bibnamefont {Sano}}, \bibinfo {author} {\bibfnamefont {M.}~\bibnamefont {Tanaka}}, \bibinfo {author} {\bibfnamefont {T.}~\bibnamefont {Yamashita}}, \ and\ \bibinfo {author} {\bibfnamefont {A.}~\bibnamefont {Fujimaki}},\ }\bibfield  {title} {\enquote {\bibinfo {title} {High-{Speed Memory D}riven by {SFQ} {Pulses B}ased on 0-$\pi$ {SQUID}},}\ }\href {\doibase 10.1109/TASC.2021.3060351} {\bibfield  {journal} {\bibinfo  {journal} {IEEE Trans. Appl. Supercond.}\ }\textbf {\bibinfo {volume} {31}},\ \bibinfo {pages} {1100906} (\bibinfo {year} {2021})}\BibitemShut {NoStop}%
\bibitem [{\citenamefont {Li}\ \emph {et~al.}(2021)\citenamefont {Li}, \citenamefont {Takeshita}, \citenamefont {Hasegawa}, \citenamefont {Tanaka}, \citenamefont {Yamashita},\ and\ \citenamefont {Fujimaki}}]{li_2021}%
  \BibitemOpen
  \bibfield  {author} {\bibinfo {author} {\bibfnamefont {F.}~\bibnamefont {Li}}, \bibinfo {author} {\bibfnamefont {Y.}~\bibnamefont {Takeshita}}, \bibinfo {author} {\bibfnamefont {D.}~\bibnamefont {Hasegawa}}, \bibinfo {author} {\bibfnamefont {M.}~\bibnamefont {Tanaka}}, \bibinfo {author} {\bibfnamefont {T.}~\bibnamefont {Yamashita}}, \ and\ \bibinfo {author} {\bibfnamefont {A.}~\bibnamefont {Fujimaki}},\ }\bibfield  {title} {\enquote {\bibinfo {title} {Low-power high-speed half-flux-quantum circuits driven by low bias voltages},}\ }\href {\doibase 10.1088/1361-6668/abcaac} {\bibfield  {journal} {\bibinfo  {journal} {Supercond. Sci. Technol.}\ }\textbf {\bibinfo {volume} {34}},\ \bibinfo {pages} {025013} (\bibinfo {year} {2021})}\BibitemShut {NoStop}%
\bibitem [{\citenamefont {Li}\ \emph {et~al.}(2023{\natexlab{a}})\citenamefont {Li}, \citenamefont {Pham}, \citenamefont {Takeshita}, \citenamefont {Higashi}, \citenamefont {Yamashita}, \citenamefont {Tanaka},\ and\ \citenamefont {Fujimaki}}]{li_2023}%
  \BibitemOpen
  \bibfield  {author} {\bibinfo {author} {\bibfnamefont {F.}~\bibnamefont {Li}}, \bibinfo {author} {\bibfnamefont {D.}~\bibnamefont {Pham}}, \bibinfo {author} {\bibfnamefont {Y.}~\bibnamefont {Takeshita}}, \bibinfo {author} {\bibfnamefont {M.}~\bibnamefont {Higashi}}, \bibinfo {author} {\bibfnamefont {T.}~\bibnamefont {Yamashita}}, \bibinfo {author} {\bibfnamefont {M.}~\bibnamefont {Tanaka}}, \ and\ \bibinfo {author} {\bibfnamefont {A.}~\bibnamefont {Fujimaki}},\ }\bibfield  {title} {\enquote {\bibinfo {title} {Energy efficient half-flux-quantum circuit aiming at milli-kelvin stage operation},}\ }\href {\doibase 10.1088/1361-6668/acf0f2} {\bibfield  {journal} {\bibinfo  {journal} {Supercond. Sci. Technol.}\ }\textbf {\bibinfo {volume} {36}},\ \bibinfo {pages} {105006} (\bibinfo {year} {2023}{\natexlab{a}})}\BibitemShut {NoStop}%
\bibitem [{\citenamefont {Li}\ \emph {et~al.}(2023{\natexlab{b}})\citenamefont {Li}, \citenamefont {Takeshita}, \citenamefont {Tanaka},\ and\ \citenamefont {Fujimaki}}]{li_2023a}%
  \BibitemOpen
  \bibfield  {author} {\bibinfo {author} {\bibfnamefont {F.}~\bibnamefont {Li}}, \bibinfo {author} {\bibfnamefont {Y.}~\bibnamefont {Takeshita}}, \bibinfo {author} {\bibfnamefont {M.}~\bibnamefont {Tanaka}}, \ and\ \bibinfo {author} {\bibfnamefont {A.}~\bibnamefont {Fujimaki}},\ }\bibfield  {title} {\enquote {\bibinfo {title} {Superconductor digital circuits with $\pi$ junctions alone},}\ }\href {\doibase 10.1063/5.0144604} {\bibfield  {journal} {\bibinfo  {journal} {Appl. Phys. Lett}\ }\textbf {\bibinfo {volume} {122}},\ \bibinfo {pages} {162601} (\bibinfo {year} {2023}{\natexlab{b}})}\BibitemShut {NoStop}%
\bibitem [{\citenamefont {Tanemura}\ \emph {et~al.}(2023)\citenamefont {Tanemura}, \citenamefont {Takeshita}, \citenamefont {Li}, \citenamefont {Nakayama}, \citenamefont {Tanaka},\ and\ \citenamefont {Fujimaki}}]{tanemura_2023}%
  \BibitemOpen
  \bibfield  {author} {\bibinfo {author} {\bibfnamefont {S.}~\bibnamefont {Tanemura}}, \bibinfo {author} {\bibfnamefont {Y.}~\bibnamefont {Takeshita}}, \bibinfo {author} {\bibfnamefont {F.}~\bibnamefont {Li}}, \bibinfo {author} {\bibfnamefont {T.}~\bibnamefont {Nakayama}}, \bibinfo {author} {\bibfnamefont {M.}~\bibnamefont {Tanaka}}, \ and\ \bibinfo {author} {\bibfnamefont {A.}~\bibnamefont {Fujimaki}},\ }\bibfield  {title} {\enquote {\bibinfo {title} {Optimization of {Half-Flux-Quantum Circuits C}omposed of $\pi$-{Shift and C}onventional {Josephson} {J}unctions},}\ }\href {\doibase 10.1109/TASC.2023.3258374} {\bibfield  {journal} {\bibinfo  {journal} {{IEEE} Trans. Appl. Supercond.}\ }\textbf {\bibinfo {volume} {33}},\ \bibinfo {pages} {1701305} (\bibinfo {year} {2023})}\BibitemShut {NoStop}%
\bibitem [{\citenamefont {Soloviev}\ \emph {et~al.}(2021)\citenamefont {Soloviev}, \citenamefont {Ruzhickiy}, \citenamefont {Bakurskiy}, \citenamefont {Klenov}, \citenamefont {Kupriyanov}, \citenamefont {Golubov}, \citenamefont {Skryabina},\ and\ \citenamefont {Stolyarov}}]{soloviev_2021}%
  \BibitemOpen
  \bibfield  {author} {\bibinfo {author} {\bibfnamefont {I.~I.}\ \bibnamefont {Soloviev}}, \bibinfo {author} {\bibfnamefont {V.~I.}\ \bibnamefont {Ruzhickiy}}, \bibinfo {author} {\bibfnamefont {S.~V.}\ \bibnamefont {Bakurskiy}}, \bibinfo {author} {\bibfnamefont {N.~V.}\ \bibnamefont {Klenov}}, \bibinfo {author} {\bibfnamefont {M.~Y.}\ \bibnamefont {Kupriyanov}}, \bibinfo {author} {\bibfnamefont {A.~A.}\ \bibnamefont {Golubov}}, \bibinfo {author} {\bibfnamefont {O.~V.}\ \bibnamefont {Skryabina}}, \ and\ \bibinfo {author} {\bibfnamefont {V.~S.}\ \bibnamefont {Stolyarov}},\ }\bibfield  {title} {\enquote {\bibinfo {title} {Superconducting {Circuits without Inductors Based on B}istable {Josephson} {J}unctions},}\ }\href {\doibase 10.1103/PhysRevApplied.16.014052} {\bibfield  {journal} {\bibinfo  {journal} {Phys. Rev. Appl.}\ }\textbf {\bibinfo {volume} {16}},\ \bibinfo {pages} {014052} (\bibinfo {year} {2021})}\BibitemShut {NoStop}%
\bibitem [{\citenamefont {Soloviev}\ \emph {et~al.}(2022)\citenamefont {Soloviev}, \citenamefont {Khismatullin}, \citenamefont {Klenov},\ and\ \citenamefont {Schegolev}}]{soloviev_2022}%
  \BibitemOpen
  \bibfield  {author} {\bibinfo {author} {\bibfnamefont {I.~I.}\ \bibnamefont {Soloviev}}, \bibinfo {author} {\bibfnamefont {G.~S.}\ \bibnamefont {Khismatullin}}, \bibinfo {author} {\bibfnamefont {N.~V.}\ \bibnamefont {Klenov}}, \ and\ \bibinfo {author} {\bibfnamefont {A.~E.}\ \bibnamefont {Schegolev}},\ }\bibfield  {title} {\enquote {\bibinfo {title} {$\pi$ junctions in adiabatic superconductor logic cells},}\ }\href {\doibase 10.1134/{S106422692212021X}} {\bibfield  {journal} {\bibinfo  {journal} {J. Commun. Technol. Electron.}\ }\textbf {\bibinfo {volume} {67}},\ \bibinfo {pages} {1479--1491} (\bibinfo {year} {2022})}\BibitemShut {NoStop}%
\bibitem [{\citenamefont {Maksimovskaya}\ \emph {et~al.}(2022)\citenamefont {Maksimovskaya}, \citenamefont {Ruzhickiy}, \citenamefont {Klenov}, \citenamefont {Bakurskiy}, \citenamefont {Kupriyanov},\ and\ \citenamefont {Soloviev}}]{maksimovskaya_2022}%
  \BibitemOpen
  \bibfield  {author} {\bibinfo {author} {\bibfnamefont {A.~A.}\ \bibnamefont {Maksimovskaya}}, \bibinfo {author} {\bibfnamefont {V.~I.}\ \bibnamefont {Ruzhickiy}}, \bibinfo {author} {\bibfnamefont {N.~V.}\ \bibnamefont {Klenov}}, \bibinfo {author} {\bibfnamefont {S.~V.}\ \bibnamefont {Bakurskiy}}, \bibinfo {author} {\bibfnamefont {M.~Y.}\ \bibnamefont {Kupriyanov}}, \ and\ \bibinfo {author} {\bibfnamefont {I.~I.}\ \bibnamefont {Soloviev}},\ }\bibfield  {title} {\enquote {\bibinfo {title} {Phase logic based on $\pi$ {Josephson} junctions},}\ }\href {\doibase 10.1134/S0021364022600884} {\bibfield  {journal} {\bibinfo  {journal} {Jetp Lett.}\ }\textbf {\bibinfo {volume} {115}},\ \bibinfo {pages} {735--741} (\bibinfo {year} {2022})}\BibitemShut {NoStop}%
\bibitem [{\citenamefont {Khismatullin}, \citenamefont {Klenov},\ and\ \citenamefont {Soloviev}(2023)}]{khismatullin_2023}%
  \BibitemOpen
  \bibfield  {author} {\bibinfo {author} {\bibfnamefont {G.~S.}\ \bibnamefont {Khismatullin}}, \bibinfo {author} {\bibfnamefont {N.~V.}\ \bibnamefont {Klenov}}, \ and\ \bibinfo {author} {\bibfnamefont {I.~I.}\ \bibnamefont {Soloviev}},\ }\bibfield  {title} {\enquote {\bibinfo {title} {Optimization of adiabatic superconducting logic cells by using $\pi$ {J}osephson junctions},}\ }\href {\doibase 10.1134/S0021364023601331} {\bibfield  {journal} {\bibinfo  {journal} {JETP Lett.}\ }\textbf {\bibinfo {volume} {118}},\ \bibinfo {pages} {220--229} (\bibinfo {year} {2023})}\BibitemShut {NoStop}%
\bibitem [{\citenamefont {Larkin}\ \emph {et~al.}(2012)\citenamefont {Larkin}, \citenamefont {Bol’ginov}, \citenamefont {Stolyarov}, \citenamefont {Ryazanov}, \citenamefont {Vernik}, \citenamefont {Tolpygo},\ and\ \citenamefont {Mukhanov}}]{Larkin_APL_2012}%
  \BibitemOpen
  \bibfield  {author} {\bibinfo {author} {\bibfnamefont {T.~I.}\ \bibnamefont {Larkin}}, \bibinfo {author} {\bibfnamefont {V.~V.}\ \bibnamefont {Bol’ginov}}, \bibinfo {author} {\bibfnamefont {V.~S.}\ \bibnamefont {Stolyarov}}, \bibinfo {author} {\bibfnamefont {V.~V.}\ \bibnamefont {Ryazanov}}, \bibinfo {author} {\bibfnamefont {I.~V.}\ \bibnamefont {Vernik}}, \bibinfo {author} {\bibfnamefont {S.~K.}\ \bibnamefont {Tolpygo}}, \ and\ \bibinfo {author} {\bibfnamefont {O.~A.}\ \bibnamefont {Mukhanov}},\ }\bibfield  {title} {\enquote {\bibinfo {title} {{Ferromagnetic {J}osephson switching device with high characteristic voltage}},}\ }\href {\doibase 10.1063/1.4723576} {\bibfield  {journal} {\bibinfo  {journal} {Appl. Phys. Lett.}\ }\textbf {\bibinfo {volume} {100}},\ \bibinfo {pages} {222601} (\bibinfo {year} {2012})}\BibitemShut {NoStop}%
\bibitem [{\citenamefont {Vernik}\ \emph {et~al.}(2013)\citenamefont {Vernik}, \citenamefont {Bol'ginov}, \citenamefont {Bakurskiy}, \citenamefont {Golubov}, \citenamefont {Kupriyanov}, \citenamefont {Ryazanov},\ and\ \citenamefont {Mukhanov}}]{vernik_2013}%
  \BibitemOpen
  \bibfield  {author} {\bibinfo {author} {\bibfnamefont {I.~V.}\ \bibnamefont {Vernik}}, \bibinfo {author} {\bibfnamefont {V.~V.}\ \bibnamefont {Bol'ginov}}, \bibinfo {author} {\bibfnamefont {S.~V.}\ \bibnamefont {Bakurskiy}}, \bibinfo {author} {\bibfnamefont {A.~A.}\ \bibnamefont {Golubov}}, \bibinfo {author} {\bibfnamefont {M.~Y.}\ \bibnamefont {Kupriyanov}}, \bibinfo {author} {\bibfnamefont {V.~V.}\ \bibnamefont {Ryazanov}}, \ and\ \bibinfo {author} {\bibfnamefont {O.~A.}\ \bibnamefont {Mukhanov}},\ }\bibfield  {title} {\enquote {\bibinfo {title} {Magnetic {Josephson} junctions with superconducting interlayer for cryogenic memory},}\ }\href {\doibase 10.1109/TASC.2012.2233270} {\bibfield  {journal} {\bibinfo  {journal} {{IEEE} Trans. Appl. Supercond.}\ }\textbf {\bibinfo {volume} {23}},\ \bibinfo {pages} {1701208} (\bibinfo {year} {2013})}\BibitemShut {NoStop}%
\bibitem [{\citenamefont {Bakurskiy}\ \emph {et~al.}(2013{\natexlab{a}})\citenamefont {Bakurskiy}, \citenamefont {Klenov}, \citenamefont {Soloviev}, \citenamefont {Bol'ginov}, \citenamefont {Ryazanov}, \citenamefont {Vernik}, \citenamefont {Mukhanov}, \citenamefont {Kupriyanov},\ and\ \citenamefont {Golubov}}]{bakurskiy_2013a}%
  \BibitemOpen
  \bibfield  {author} {\bibinfo {author} {\bibfnamefont {S.~V.}\ \bibnamefont {Bakurskiy}}, \bibinfo {author} {\bibfnamefont {N.~V.}\ \bibnamefont {Klenov}}, \bibinfo {author} {\bibfnamefont {I.~I.}\ \bibnamefont {Soloviev}}, \bibinfo {author} {\bibfnamefont {V.~V.}\ \bibnamefont {Bol'ginov}}, \bibinfo {author} {\bibfnamefont {V.~V.}\ \bibnamefont {Ryazanov}}, \bibinfo {author} {\bibfnamefont {I.~V.}\ \bibnamefont {Vernik}}, \bibinfo {author} {\bibfnamefont {O.~A.}\ \bibnamefont {Mukhanov}}, \bibinfo {author} {\bibfnamefont {M.~Y.}\ \bibnamefont {Kupriyanov}}, \ and\ \bibinfo {author} {\bibfnamefont {A.~A.}\ \bibnamefont {Golubov}},\ }\bibfield  {title} {\enquote {\bibinfo {title} {Theoretical model of superconducting spintronic {SIsFS} devices},}\ }\href {\doibase 10.1063/1.4805032} {\bibfield  {journal} {\bibinfo  {journal} {Appl. Phys. Lett}\ }\textbf {\bibinfo {volume} {102}},\ \bibinfo {pages} {192603} (\bibinfo {year} {2013}{\natexlab{a}})}\BibitemShut {NoStop}%
\bibitem [{\citenamefont {Parlato}\ \emph {et~al.}(2020)\citenamefont {Parlato}, \citenamefont {Caruso}, \citenamefont {Vettoliere}, \citenamefont {Satariano}, \citenamefont {Ahmad}, \citenamefont {Miano}, \citenamefont {Montemurro}, \citenamefont {Salvoni}, \citenamefont {Ausanio}, \citenamefont {Tafuri}, \citenamefont {Pepe}, \citenamefont {Massarotti},\ and\ \citenamefont {Granata}}]{Parlato_JAP_2020}%
  \BibitemOpen
  \bibfield  {author} {\bibinfo {author} {\bibfnamefont {L.}~\bibnamefont {Parlato}}, \bibinfo {author} {\bibfnamefont {R.}~\bibnamefont {Caruso}}, \bibinfo {author} {\bibfnamefont {A.}~\bibnamefont {Vettoliere}}, \bibinfo {author} {\bibfnamefont {R.}~\bibnamefont {Satariano}}, \bibinfo {author} {\bibfnamefont {H.~G.}\ \bibnamefont {Ahmad}}, \bibinfo {author} {\bibfnamefont {A.}~\bibnamefont {Miano}}, \bibinfo {author} {\bibfnamefont {D.}~\bibnamefont {Montemurro}}, \bibinfo {author} {\bibfnamefont {D.}~\bibnamefont {Salvoni}}, \bibinfo {author} {\bibfnamefont {G.}~\bibnamefont {Ausanio}}, \bibinfo {author} {\bibfnamefont {F.}~\bibnamefont {Tafuri}}, \bibinfo {author} {\bibfnamefont {G.~P.}\ \bibnamefont {Pepe}}, \bibinfo {author} {\bibfnamefont {D.}~\bibnamefont {Massarotti}}, \ and\ \bibinfo {author} {\bibfnamefont {C.}~\bibnamefont {Granata}},\ }\bibfield  {title} {\enquote {\bibinfo {title} {{Characterization of scalable Josephson memory element containing a strong ferromagnet}},}\ }\href {\doibase
  10.1063/5.0004554} {\bibfield  {journal} {\bibinfo  {journal} {J. Appl. Phys.}\ }\textbf {\bibinfo {volume} {127}},\ \bibinfo {pages} {193901} (\bibinfo {year} {2020})}\BibitemShut {NoStop}%
\bibitem [{\citenamefont {Hilgenkamp}\ \emph {et~al.}(2003)\citenamefont {Hilgenkamp}, \citenamefont {{Ariando}}, \citenamefont {Smilde}, \citenamefont {Blank}, \citenamefont {Rijnders}, \citenamefont {Rogalla}, \citenamefont {Kirtley},\ and\ \citenamefont {Tsuei}}]{hilgenkamp_2003}%
  \BibitemOpen
  \bibfield  {author} {\bibinfo {author} {\bibfnamefont {H.}~\bibnamefont {Hilgenkamp}}, \bibinfo {author} {\bibnamefont {{Ariando}}}, \bibinfo {author} {\bibfnamefont {H.-J.~H.}\ \bibnamefont {Smilde}}, \bibinfo {author} {\bibfnamefont {D.~H.~A.}\ \bibnamefont {Blank}}, \bibinfo {author} {\bibfnamefont {G.}~\bibnamefont {Rijnders}}, \bibinfo {author} {\bibfnamefont {H.}~\bibnamefont {Rogalla}}, \bibinfo {author} {\bibfnamefont {J.~R.}\ \bibnamefont {Kirtley}}, \ and\ \bibinfo {author} {\bibfnamefont {C.~C.}\ \bibnamefont {Tsuei}},\ }\bibfield  {title} {\enquote {\bibinfo {title} {Ordering and manipulation of the magnetic moments in large-scale superconducting $\pi$-loop arrays},}\ }\href {\doibase 10.1038/nature01442} {\bibfield  {journal} {\bibinfo  {journal} {Nature}\ }\textbf {\bibinfo {volume} {422}},\ \bibinfo {pages} {50--53} (\bibinfo {year} {2003})}\BibitemShut {NoStop}%
\bibitem [{\citenamefont {Balashov}\ \emph {et~al.}(2007)\citenamefont {Balashov}, \citenamefont {Dimov}, \citenamefont {Khabipov}, \citenamefont {Ortlepp}, \citenamefont {Hagedorn}, \citenamefont {Zorin}, \citenamefont {Buchholz}, \citenamefont {Uhlmann},\ and\ \citenamefont {Niemeyer}}]{balashov_2007}%
  \BibitemOpen
  \bibfield  {author} {\bibinfo {author} {\bibfnamefont {D.}~\bibnamefont {Balashov}}, \bibinfo {author} {\bibfnamefont {B.}~\bibnamefont {Dimov}}, \bibinfo {author} {\bibfnamefont {M.}~\bibnamefont {Khabipov}}, \bibinfo {author} {\bibfnamefont {T.}~\bibnamefont {Ortlepp}}, \bibinfo {author} {\bibfnamefont {D.}~\bibnamefont {Hagedorn}}, \bibinfo {author} {\bibfnamefont {A.}~\bibnamefont {Zorin}}, \bibinfo {author} {\bibfnamefont {F.-I.}\ \bibnamefont {Buchholz}}, \bibinfo {author} {\bibfnamefont {F.}~\bibnamefont {Uhlmann}}, \ and\ \bibinfo {author} {\bibfnamefont {J.}~\bibnamefont {Niemeyer}},\ }\bibfield  {title} {\enquote {\bibinfo {title} {Passive phase shifter for superconducting {Josephson} circuits},}\ }\href {\doibase 10.1109/TASC.2007.897382} {\bibfield  {journal} {\bibinfo  {journal} {{IEEE} Trans. Appl. Supercond.}\ }\textbf {\bibinfo {volume} {17}},\ \bibinfo {pages} {142--145} (\bibinfo {year} {2007})}\BibitemShut {NoStop}%
\bibitem [{\citenamefont {Majer}, \citenamefont {Butcher},\ and\ \citenamefont {Mooij}(2002)}]{majer_2002}%
  \BibitemOpen
  \bibfield  {author} {\bibinfo {author} {\bibfnamefont {J.~B.}\ \bibnamefont {Majer}}, \bibinfo {author} {\bibfnamefont {J.~R.}\ \bibnamefont {Butcher}}, \ and\ \bibinfo {author} {\bibfnamefont {J.~E.}\ \bibnamefont {Mooij}},\ }\bibfield  {title} {\enquote {\bibinfo {title} {Simple phase bias for superconducting circuits},}\ }\href {\doibase 10.1063/1.1478150} {\bibfield  {journal} {\bibinfo  {journal} {Appl. Phys. Lett}\ }\textbf {\bibinfo {volume} {80}},\ \bibinfo {pages} {3638--3640} (\bibinfo {year} {2002})}\BibitemShut {NoStop}%
\bibitem [{\citenamefont {Stoutimore}\ \emph {et~al.}(2018)\citenamefont {Stoutimore}, \citenamefont {Rossolenko}, \citenamefont {Bolginov}, \citenamefont {Oboznov}, \citenamefont {Rusanov}, \citenamefont {Baranov}, \citenamefont {Pugach}, \citenamefont {Frolov}, \citenamefont {Ryazanov},\ and\ \citenamefont {Van~Harlingen}}]{Stoutimore_PRL_2018}%
  \BibitemOpen
  \bibfield  {author} {\bibinfo {author} {\bibfnamefont {M.~J.~A.}\ \bibnamefont {Stoutimore}}, \bibinfo {author} {\bibfnamefont {A.~N.}\ \bibnamefont {Rossolenko}}, \bibinfo {author} {\bibfnamefont {V.~V.}\ \bibnamefont {Bolginov}}, \bibinfo {author} {\bibfnamefont {V.~A.}\ \bibnamefont {Oboznov}}, \bibinfo {author} {\bibfnamefont {A.~Y.}\ \bibnamefont {Rusanov}}, \bibinfo {author} {\bibfnamefont {D.~S.}\ \bibnamefont {Baranov}}, \bibinfo {author} {\bibfnamefont {N.}~\bibnamefont {Pugach}}, \bibinfo {author} {\bibfnamefont {S.~M.}\ \bibnamefont {Frolov}}, \bibinfo {author} {\bibfnamefont {V.~V.}\ \bibnamefont {Ryazanov}}, \ and\ \bibinfo {author} {\bibfnamefont {D.~J.}\ \bibnamefont {Van~Harlingen}},\ }\bibfield  {title} {\enquote {\bibinfo {title} {{Second-Harmonic Current-Phase Relation in Josephson Junctions with Ferromagnetic Barriers}},}\ }\href {\doibase 10.1103/PhysRevLett.121.177702} {\bibfield  {journal} {\bibinfo  {journal} {Phys. Rev. Lett.}\ }\textbf {\bibinfo {volume} {121}},\ \bibinfo {pages}
  {177702} (\bibinfo {year} {2018})}\BibitemShut {NoStop}%
\bibitem [{\citenamefont {Pal}\ \emph {et~al.}(2014)\citenamefont {Pal}, \citenamefont {Barber}, \citenamefont {Robinson},\ and\ \citenamefont {Blamire}}]{pal_2014}%
  \BibitemOpen
  \bibfield  {author} {\bibinfo {author} {\bibfnamefont {A.}~\bibnamefont {Pal}}, \bibinfo {author} {\bibfnamefont {Z.}~\bibnamefont {Barber}}, \bibinfo {author} {\bibfnamefont {J.}~\bibnamefont {Robinson}}, \ and\ \bibinfo {author} {\bibfnamefont {M.}~\bibnamefont {Blamire}},\ }\bibfield  {title} {\enquote {\bibinfo {title} {Pure second harmonic current-phase relation in spin-filter {J}osephson junctions},}\ }\href {\doibase 10.1038/ncomms4340} {\bibfield  {journal} {\bibinfo  {journal} {Nat. Commun.}\ }\textbf {\bibinfo {volume} {5}},\ \bibinfo {pages} {3340} (\bibinfo {year} {2014})}\BibitemShut {NoStop}%
\bibitem [{\citenamefont {Trifunovic}(2011)}]{trifunovic_2011a}%
  \BibitemOpen
  \bibfield  {author} {\bibinfo {author} {\bibfnamefont {L.}~\bibnamefont {Trifunovic}},\ }\bibfield  {title} {\enquote {\bibinfo {title} {Long-range superharmonic {J}osephson current.}}\ }\href {\doibase 10.1103/{PhysRevLett}.107.047001} {\bibfield  {journal} {\bibinfo  {journal} {Phys Rev Lett}\ }\textbf {\bibinfo {volume} {107}},\ \bibinfo {pages} {047001} (\bibinfo {year} {2011})}\BibitemShut {NoStop}%
\bibitem [{\citenamefont {Trifunovic}, \citenamefont {Popović},\ and\ \citenamefont {Radović}(2011)}]{trifunovic_2011}%
  \BibitemOpen
  \bibfield  {author} {\bibinfo {author} {\bibfnamefont {L.}~\bibnamefont {Trifunovic}}, \bibinfo {author} {\bibfnamefont {Z.}~\bibnamefont {Popović}}, \ and\ \bibinfo {author} {\bibfnamefont {Z.}~\bibnamefont {Radović}},\ }\bibfield  {title} {\enquote {\bibinfo {title} {Josephson effect and spin-triplet pairing correlations in {SFFS} junctions},}\ }\href {\doibase 10.1103/{PhysRevB}.84.064511} {\bibfield  {journal} {\bibinfo  {journal} {Phys. Rev. B}\ }\textbf {\bibinfo {volume} {84}},\ \bibinfo {pages} {064511} (\bibinfo {year} {2011})}\BibitemShut {NoStop}%
\bibitem [{\citenamefont {Mel'nikov}\ \emph {et~al.}(2012)\citenamefont {Mel'nikov}, \citenamefont {Samokhvalov}, \citenamefont {Kuznetsova},\ and\ \citenamefont {Buzdin}}]{melnikov_2012}%
  \BibitemOpen
  \bibfield  {author} {\bibinfo {author} {\bibfnamefont {A.~S.}\ \bibnamefont {Mel'nikov}}, \bibinfo {author} {\bibfnamefont {A.~V.}\ \bibnamefont {Samokhvalov}}, \bibinfo {author} {\bibfnamefont {S.~M.}\ \bibnamefont {Kuznetsova}}, \ and\ \bibinfo {author} {\bibfnamefont {A.~I.}\ \bibnamefont {Buzdin}},\ }\bibfield  {title} {\enquote {\bibinfo {title} {Interference phenomena and long-range proximity effect in clean superconductor-ferromagnet systems.}}\ }\href {\doibase 10.1103/{PhysRevLett}.109.237006} {\bibfield  {journal} {\bibinfo  {journal} {Phys Rev Lett}\ }\textbf {\bibinfo {volume} {109}},\ \bibinfo {pages} {237006} (\bibinfo {year} {2012})}\BibitemShut {NoStop}%
\bibitem [{\citenamefont {Richard}, \citenamefont {Houzet},\ and\ \citenamefont {Meyer}(2013)}]{richard_2013}%
  \BibitemOpen
  \bibfield  {author} {\bibinfo {author} {\bibfnamefont {C.}~\bibnamefont {Richard}}, \bibinfo {author} {\bibfnamefont {M.}~\bibnamefont {Houzet}}, \ and\ \bibinfo {author} {\bibfnamefont {J.~S.}\ \bibnamefont {Meyer}},\ }\bibfield  {title} {\enquote {\bibinfo {title} {Superharmonic long-range triplet current in a diffusive {J}osephson junction.}}\ }\href {\doibase 10.1103/{PhysRevLett}.110.217004} {\bibfield  {journal} {\bibinfo  {journal} {Phys Rev Lett}\ }\textbf {\bibinfo {volume} {110}},\ \bibinfo {pages} {217004} (\bibinfo {year} {2013})}\BibitemShut {NoStop}%
\bibitem [{\citenamefont {Salameh}, \citenamefont {Friedman},\ and\ \citenamefont {Kvatinsky}(2022)}]{salameh_2022}%
  \BibitemOpen
  \bibfield  {author} {\bibinfo {author} {\bibfnamefont {I.}~\bibnamefont {Salameh}}, \bibinfo {author} {\bibfnamefont {E.~G.}\ \bibnamefont {Friedman}}, \ and\ \bibinfo {author} {\bibfnamefont {S.}~\bibnamefont {Kvatinsky}},\ }\bibfield  {title} {\enquote {\bibinfo {title} {Superconductive logic using 2$\phi$ {J}osephson junctions with half flux quantum pulses},}\ }\href {\doibase 10.1109/TCSII.2022.3162723} {\bibfield  {journal} {\bibinfo  {journal} {{IEEE} Trans. Circuits Syst. {II}}\ }\textbf {\bibinfo {volume} {69}},\ \bibinfo {pages} {2533--2537} (\bibinfo {year} {2022})}\BibitemShut {NoStop}%
\bibitem [{\citenamefont {Jabbari}, \citenamefont {Bocko},\ and\ \citenamefont {Friedman}(2023)}]{jabbari_2023}%
  \BibitemOpen
  \bibfield  {author} {\bibinfo {author} {\bibfnamefont {T.}~\bibnamefont {Jabbari}}, \bibinfo {author} {\bibfnamefont {M.}~\bibnamefont {Bocko}}, \ and\ \bibinfo {author} {\bibfnamefont {E.~G.}\ \bibnamefont {Friedman}},\ }\bibfield  {title} {\enquote {\bibinfo {title} {All-{JJ} logic based on bistable {JJs}},}\ }\href {\doibase 10.1109/{TASC}.2023.3260774} {\bibfield  {journal} {\bibinfo  {journal} {{IEEE} Trans. Appl. Supercond.}\ }\textbf {\bibinfo {volume} {33}},\ \bibinfo {pages} {1--7} (\bibinfo {year} {2023})}\BibitemShut {NoStop}%
\bibitem [{\citenamefont {Razmkhah}\ and\ \citenamefont {Pedram}(2024)}]{razmkhah_2024}%
  \BibitemOpen
  \bibfield  {author} {\bibinfo {author} {\bibfnamefont {S.}~\bibnamefont {Razmkhah}}\ and\ \bibinfo {author} {\bibfnamefont {M.}~\bibnamefont {Pedram}},\ }\bibfield  {title} {\enquote {\bibinfo {title} {High-density superconductive logic circuits utilizing 0 and $\pi$ {J}osephson junctions},}\ }\href {\doibase 10.1088/2631-8695/ad27f5} {\bibfield  {journal} {\bibinfo  {journal} {Eng. Res. Express}\ }\textbf {\bibinfo {volume} {6}},\ \bibinfo {pages} {015307} (\bibinfo {year} {2024})}\BibitemShut {NoStop}%
\bibitem [{\citenamefont {Cong}\ \emph {et~al.}(2023)\citenamefont {Cong}, \citenamefont {Razmkhah}, \citenamefont {Karamuftuoglu},\ and\ \citenamefont {Pedram}}]{cong_2023}%
  \BibitemOpen
  \bibfield  {author} {\bibinfo {author} {\bibfnamefont {H.}~\bibnamefont {Cong}}, \bibinfo {author} {\bibfnamefont {S.}~\bibnamefont {Razmkhah}}, \bibinfo {author} {\bibfnamefont {M.~A.}\ \bibnamefont {Karamuftuoglu}}, \ and\ \bibinfo {author} {\bibfnamefont {M.}~\bibnamefont {Pedram}},\ }\bibfield  {title} {\enquote {\bibinfo {title} {Superconductor logic implementation with all-{JJ} inductor-free cell library},}\ }\href {https://arxiv.org/abs/2310.13857} {\bibfield  {journal} {\bibinfo  {journal} {arXiv:2310.13857}\ } (\bibinfo {year} {2023})}\BibitemShut {NoStop}%
\bibitem [{\citenamefont {Bol'ginov}\ \emph {et~al.}(2012)\citenamefont {Bol'ginov}, \citenamefont {Stolyarov}, \citenamefont {Sobanin}, \citenamefont {Karpovich},\ and\ \citenamefont {Ryazanov}}]{bolginov_2012}%
  \BibitemOpen
  \bibfield  {author} {\bibinfo {author} {\bibfnamefont {V.~V.}\ \bibnamefont {Bol'ginov}}, \bibinfo {author} {\bibfnamefont {V.~S.}\ \bibnamefont {Stolyarov}}, \bibinfo {author} {\bibfnamefont {D.~S.}\ \bibnamefont {Sobanin}}, \bibinfo {author} {\bibfnamefont {A.~L.}\ \bibnamefont {Karpovich}}, \ and\ \bibinfo {author} {\bibfnamefont {V.~V.}\ \bibnamefont {Ryazanov}},\ }\bibfield  {title} {\enquote {\bibinfo {title} {Magnetic switches based on {Nb-PdFe-Nb} {Josephson} junctions with a magnetically soft ferromagnetic interlayer},}\ }\href {\doibase 10.1134/S0021364012070028} {\bibfield  {journal} {\bibinfo  {journal} {JETP Lett.}\ }\textbf {\bibinfo {volume} {95}},\ \bibinfo {pages} {366--371} (\bibinfo {year} {2012})}\BibitemShut {NoStop}%
\bibitem [{\citenamefont {Karelina}\ \emph {et~al.}(2021)\citenamefont {Karelina}, \citenamefont {Hovhannisyan}, \citenamefont {Golovchanskiy}, \citenamefont {Chichkov}, \citenamefont {Ben~Hamida}, \citenamefont {Stolyarov}, \citenamefont {Uspenskaya}, \citenamefont {Erkenov}, \citenamefont {Bolginov},\ and\ \citenamefont {Ryazanov}}]{karelina_2021}%
  \BibitemOpen
  \bibfield  {author} {\bibinfo {author} {\bibfnamefont {L.~N.}\ \bibnamefont {Karelina}}, \bibinfo {author} {\bibfnamefont {R.~A.}\ \bibnamefont {Hovhannisyan}}, \bibinfo {author} {\bibfnamefont {I.~A.}\ \bibnamefont {Golovchanskiy}}, \bibinfo {author} {\bibfnamefont {V.~I.}\ \bibnamefont {Chichkov}}, \bibinfo {author} {\bibfnamefont {A.}~\bibnamefont {Ben~Hamida}}, \bibinfo {author} {\bibfnamefont {V.~S.}\ \bibnamefont {Stolyarov}}, \bibinfo {author} {\bibfnamefont {L.~S.}\ \bibnamefont {Uspenskaya}}, \bibinfo {author} {\bibfnamefont {S.~A.}\ \bibnamefont {Erkenov}}, \bibinfo {author} {\bibfnamefont {V.~V.}\ \bibnamefont {Bolginov}}, \ and\ \bibinfo {author} {\bibfnamefont {V.~V.}\ \bibnamefont {Ryazanov}},\ }\bibfield  {title} {\enquote {\bibinfo {title} {Scalable memory elements based on rectangular {SIsFS} junctions},}\ }\href {\doibase 10.1063/5.0063274} {\bibfield  {journal} {\bibinfo  {journal} {J. Appl. Phys.}\ }\textbf {\bibinfo {volume} {130}},\ \bibinfo {pages} {173901} (\bibinfo {year}
  {2021})}\BibitemShut {NoStop}%
\bibitem [{\citenamefont {Fermin}\ \emph {et~al.}(2022)\citenamefont {Fermin}, \citenamefont {Scheinowitz}, \citenamefont {Aarts},\ and\ \citenamefont {Lahabi}}]{fermin_2022}%
  \BibitemOpen
  \bibfield  {author} {\bibinfo {author} {\bibfnamefont {R.}~\bibnamefont {Fermin}}, \bibinfo {author} {\bibfnamefont {N.~M.~A.}\ \bibnamefont {Scheinowitz}}, \bibinfo {author} {\bibfnamefont {J.}~\bibnamefont {Aarts}}, \ and\ \bibinfo {author} {\bibfnamefont {K.}~\bibnamefont {Lahabi}},\ }\bibfield  {title} {\enquote {\bibinfo {title} {Mesoscopic superconducting memory based on bistable magnetic textures},}\ }\href {\doibase 10.1103/PhysRevResearch.4.033136} {\bibfield  {journal} {\bibinfo  {journal} {Phys. Rev. Research}\ }\textbf {\bibinfo {volume} {4}},\ \bibinfo {pages} {033136} (\bibinfo {year} {2022})}\BibitemShut {NoStop}%
\bibitem [{\citenamefont {Golod}, \citenamefont {Iovan},\ and\ \citenamefont {Krasnov}(2015)}]{golod_2015}%
  \BibitemOpen
  \bibfield  {author} {\bibinfo {author} {\bibfnamefont {T.}~\bibnamefont {Golod}}, \bibinfo {author} {\bibfnamefont {A.}~\bibnamefont {Iovan}}, \ and\ \bibinfo {author} {\bibfnamefont {V.~M.}\ \bibnamefont {Krasnov}},\ }\bibfield  {title} {\enquote {\bibinfo {title} {Single {A}brikosov vortices as quantized information bits.}}\ }\href {\doibase 10.1038/ncomms9628} {\bibfield  {journal} {\bibinfo  {journal} {Nat. Commun.}\ }\textbf {\bibinfo {volume} {6}},\ \bibinfo {pages} {8628} (\bibinfo {year} {2015})}\BibitemShut {NoStop}%
\bibitem [{\citenamefont {Soloviev}\ \emph {et~al.}(2014)\citenamefont {Soloviev}, \citenamefont {Klenov}, \citenamefont {Bakurskiy}, \citenamefont {Bol'ginov}, \citenamefont {Ryazanov}, \citenamefont {Kupriyanov},\ and\ \citenamefont {Golubov}}]{soloviev_2014}%
  \BibitemOpen
  \bibfield  {author} {\bibinfo {author} {\bibfnamefont {I.~I.}\ \bibnamefont {Soloviev}}, \bibinfo {author} {\bibfnamefont {N.~V.}\ \bibnamefont {Klenov}}, \bibinfo {author} {\bibfnamefont {S.~V.}\ \bibnamefont {Bakurskiy}}, \bibinfo {author} {\bibfnamefont {V.~V.}\ \bibnamefont {Bol'ginov}}, \bibinfo {author} {\bibfnamefont {V.~V.}\ \bibnamefont {Ryazanov}}, \bibinfo {author} {\bibfnamefont {M.~Y.}\ \bibnamefont {Kupriyanov}}, \ and\ \bibinfo {author} {\bibfnamefont {A.~A.}\ \bibnamefont {Golubov}},\ }\bibfield  {title} {\enquote {\bibinfo {title} {{Josephson} magnetic rotary valve},}\ }\href {\doibase 10.1063/1.4904012} {\bibfield  {journal} {\bibinfo  {journal} {Appl. Phys. Lett.}\ }\textbf {\bibinfo {volume} {105}},\ \bibinfo {pages} {242601} (\bibinfo {year} {2014})}\BibitemShut {NoStop}%
\bibitem [{\citenamefont {Nevirkovets}, \citenamefont {Shafraniuk},\ and\ \citenamefont {Mukhanov}(2018)}]{nevirkovets_2018}%
  \BibitemOpen
  \bibfield  {author} {\bibinfo {author} {\bibfnamefont {I.~P.}\ \bibnamefont {Nevirkovets}}, \bibinfo {author} {\bibfnamefont {S.~E.}\ \bibnamefont {Shafraniuk}}, \ and\ \bibinfo {author} {\bibfnamefont {O.~A.}\ \bibnamefont {Mukhanov}},\ }\bibfield  {title} {\enquote {\bibinfo {title} {Multiterminal {Superconducting-Ferromagnetic Device with Magnetically Tunable Supercurrent for Memory A}pplication},}\ }\href {\doibase 10.1109/TASC.2018.2836938} {\bibfield  {journal} {\bibinfo  {journal} {{IEEE} Trans. Appl. Supercond.}\ }\textbf {\bibinfo {volume} {28}},\ \bibinfo {pages} {1800904} (\bibinfo {year} {2018})}\BibitemShut {NoStop}%
\bibitem [{\citenamefont {Nevirkovets}\ and\ \citenamefont {Mukhanov}(2018)}]{nevirkovets_2018a}%
  \BibitemOpen
  \bibfield  {author} {\bibinfo {author} {\bibfnamefont {I.~P.}\ \bibnamefont {Nevirkovets}}\ and\ \bibinfo {author} {\bibfnamefont {O.~A.}\ \bibnamefont {Mukhanov}},\ }\bibfield  {title} {\enquote {\bibinfo {title} {Memory {Cell for High-Density Arrays Based on a Multiterminal Superconducting-Ferromagnetic D}evice},}\ }\href {\doibase 10.1103/PhysRevApplied.10.034013} {\bibfield  {journal} {\bibinfo  {journal} {Phys. Rev. Appl.}\ }\textbf {\bibinfo {volume} {10}},\ \bibinfo {pages} {034013} (\bibinfo {year} {2018})}\BibitemShut {NoStop}%
\bibitem [{\citenamefont {Bergeret}, \citenamefont {Volkov},\ and\ \citenamefont {Efetov}(2001{\natexlab{c}})}]{bergeret_2001}%
  \BibitemOpen
  \bibfield  {author} {\bibinfo {author} {\bibfnamefont {F.~S.}\ \bibnamefont {Bergeret}}, \bibinfo {author} {\bibfnamefont {A.~F.}\ \bibnamefont {Volkov}}, \ and\ \bibinfo {author} {\bibfnamefont {K.~B.}\ \bibnamefont {Efetov}},\ }\bibfield  {title} {\enquote {\bibinfo {title} {Enhancement of the {Josephson Current by an Exchange Field in Su}perconductor-{Ferromagnet S}tructures},}\ }\href {\doibase 10.1103/PhysRevLett.86.3140} {\bibfield  {journal} {\bibinfo  {journal} {Phys. Rev. Lett.}\ }\textbf {\bibinfo {volume} {86}},\ \bibinfo {pages} {3140--3143} (\bibinfo {year} {2001}{\natexlab{c}})}\BibitemShut {NoStop}%
\bibitem [{\citenamefont {Krivoruchko}\ and\ \citenamefont {Koshina}(2001)}]{krivoruchko_2001}%
  \BibitemOpen
  \bibfield  {author} {\bibinfo {author} {\bibfnamefont {V.~N.}\ \bibnamefont {Krivoruchko}}\ and\ \bibinfo {author} {\bibfnamefont {E.~A.}\ \bibnamefont {Koshina}},\ }\bibfield  {title} {\enquote {\bibinfo {title} {From inversion to enhancement of the dc {Josephson} current in ${S/F}-{I}-{F/S}$ tunnel structures},}\ }\href {\doibase 10.1103/PhysRevB.64.172511} {\bibfield  {journal} {\bibinfo  {journal} {Phys. Rev. B}\ }\textbf {\bibinfo {volume} {64}},\ \bibinfo {pages} {172511} (\bibinfo {year} {2001})}\BibitemShut {NoStop}%
\bibitem [{\citenamefont {Golubov}, \citenamefont {Kupriyanov},\ and\ \citenamefont {Fominov}(2002)}]{golubov_2002}%
  \BibitemOpen
  \bibfield  {author} {\bibinfo {author} {\bibfnamefont {A.~A.}\ \bibnamefont {Golubov}}, \bibinfo {author} {\bibfnamefont {M.~Y.}\ \bibnamefont {Kupriyanov}}, \ and\ \bibinfo {author} {\bibfnamefont {Y.~V.}\ \bibnamefont {Fominov}},\ }\bibfield  {title} {\enquote {\bibinfo {title} {Critical current in {SFIFS} junctions},}\ }\href {\doibase 10.1134/1.1475721} {\bibfield  {journal} {\bibinfo  {journal} {Jetp Lett.}\ }\textbf {\bibinfo {volume} {75}},\ \bibinfo {pages} {190--194} (\bibinfo {year} {2002})}\BibitemShut {NoStop}%
\bibitem [{\citenamefont {Barash}, \citenamefont {Bobkova},\ and\ \citenamefont {Kopp}(2002)}]{barash_2002}%
  \BibitemOpen
  \bibfield  {author} {\bibinfo {author} {\bibfnamefont {Y.~S.}\ \bibnamefont {Barash}}, \bibinfo {author} {\bibfnamefont {I.~V.}\ \bibnamefont {Bobkova}}, \ and\ \bibinfo {author} {\bibfnamefont {T.}~\bibnamefont {Kopp}},\ }\bibfield  {title} {\enquote {\bibinfo {title} {Josephson current in {S-FIF-S} junctions: {N}onmonotonic dependence on misorientation angle},}\ }\href {\doibase 10.1103/PhysRevB.66.140503} {\bibfield  {journal} {\bibinfo  {journal} {Phys. Rev. B}\ }\textbf {\bibinfo {volume} {66}},\ \bibinfo {pages} {140503} (\bibinfo {year} {2002})}\BibitemShut {NoStop}%
\bibitem [{\citenamefont {Chtchelkatchev}, \citenamefont {Belzig},\ and\ \citenamefont {Bruder}(2002)}]{chtchelkatchev_2002}%
  \BibitemOpen
  \bibfield  {author} {\bibinfo {author} {\bibfnamefont {N.~M.}\ \bibnamefont {Chtchelkatchev}}, \bibinfo {author} {\bibfnamefont {W.}~\bibnamefont {Belzig}}, \ and\ \bibinfo {author} {\bibfnamefont {C.}~\bibnamefont {Bruder}},\ }\bibfield  {title} {\enquote {\bibinfo {title} {Josephson effect in {SFXSF} junctions},}\ }\href@noop {} {\bibfield  {journal} {\bibinfo  {journal} {{JETP} Lett.}\ }\textbf {\bibinfo {volume} {75}},\ \bibinfo {pages} {772--776} (\bibinfo {year} {2002})}\BibitemShut {NoStop}%
\bibitem [{\citenamefont {Zaitsev}(2003)}]{zaitsev_2003}%
  \BibitemOpen
  \bibfield  {author} {\bibinfo {author} {\bibfnamefont {A.}~\bibnamefont {Zaitsev}},\ }\bibfield  {title} {\enquote {\bibinfo {title} {Josephson effect in superconducting junctions with different types of magnetic barrier between the superconductors},}\ }\href {\doibase 10.1016/S0921-4526(02)02407-9} {\bibfield  {journal} {\bibinfo  {journal} {Physica B Condens. Matter.}\ }\textbf {\bibinfo {volume} {329-333}},\ \bibinfo {pages} {1498--1499} (\bibinfo {year} {2003})}\BibitemShut {NoStop}%
\bibitem [{\citenamefont {Blanter}\ and\ \citenamefont {Hekking}(2004)}]{blanter_2004}%
  \BibitemOpen
  \bibfield  {author} {\bibinfo {author} {\bibfnamefont {Y.~M.}\ \bibnamefont {Blanter}}\ and\ \bibinfo {author} {\bibfnamefont {F.~W.~J.}\ \bibnamefont {Hekking}},\ }\bibfield  {title} {\enquote {\bibinfo {title} {Supercurrent in long {SFFS} junctions with antiparallel domain configuration},}\ }\href {\doibase 10.1103/PhysRevB.69.024525} {\bibfield  {journal} {\bibinfo  {journal} {Phys. Rev. B}\ }\textbf {\bibinfo {volume} {69}},\ \bibinfo {pages} {024525} (\bibinfo {year} {2004})}\BibitemShut {NoStop}%
\bibitem [{\citenamefont {Pajović}\ \emph {et~al.}(2006)\citenamefont {Pajović}, \citenamefont {Božović}, \citenamefont {Radović}, \citenamefont {Cayssol},\ and\ \citenamefont {Buzdin}}]{pajovi_2006}%
  \BibitemOpen
  \bibfield  {author} {\bibinfo {author} {\bibfnamefont {Z.}~\bibnamefont {Pajović}}, \bibinfo {author} {\bibfnamefont {M.}~\bibnamefont {Božović}}, \bibinfo {author} {\bibfnamefont {Z.}~\bibnamefont {Radović}}, \bibinfo {author} {\bibfnamefont {J.}~\bibnamefont {Cayssol}}, \ and\ \bibinfo {author} {\bibfnamefont {A.}~\bibnamefont {Buzdin}},\ }\bibfield  {title} {\enquote {\bibinfo {title} {Josephson coupling through ferromagnetic heterojunctions with noncollinear magnetizations},}\ }\href {\doibase 10.1103/PhysRevB.74.184509} {\bibfield  {journal} {\bibinfo  {journal} {Phys. Rev. B}\ }\textbf {\bibinfo {volume} {74}},\ \bibinfo {pages} {184509} (\bibinfo {year} {2006})}\BibitemShut {NoStop}%
\bibitem [{\citenamefont {Crouzy}, \citenamefont {Tollis},\ and\ \citenamefont {Ivanov}(2007)}]{crouzy_2007}%
  \BibitemOpen
  \bibfield  {author} {\bibinfo {author} {\bibfnamefont {B.}~\bibnamefont {Crouzy}}, \bibinfo {author} {\bibfnamefont {S.}~\bibnamefont {Tollis}}, \ and\ \bibinfo {author} {\bibfnamefont {D.~A.}\ \bibnamefont {Ivanov}},\ }\bibfield  {title} {\enquote {\bibinfo {title} {Josephson current in a superconductor-ferromagnet junction with two noncollinear magnetic domains},}\ }\href {\doibase 10.1103/PhysRevB.75.054503} {\bibfield  {journal} {\bibinfo  {journal} {Phys. Rev. B}\ }\textbf {\bibinfo {volume} {75}},\ \bibinfo {pages} {054503} (\bibinfo {year} {2007})}\BibitemShut {NoStop}%
\bibitem [{\citenamefont {Bell}\ \emph {et~al.}(2004)\citenamefont {Bell}, \citenamefont {Burnell}, \citenamefont {Leung}, \citenamefont {Tarte}, \citenamefont {Kang},\ and\ \citenamefont {Blamire}}]{Bell_APL_2004}%
  \BibitemOpen
  \bibfield  {author} {\bibinfo {author} {\bibfnamefont {C.}~\bibnamefont {Bell}}, \bibinfo {author} {\bibfnamefont {G.}~\bibnamefont {Burnell}}, \bibinfo {author} {\bibfnamefont {C.~W.}\ \bibnamefont {Leung}}, \bibinfo {author} {\bibfnamefont {E.~J.}\ \bibnamefont {Tarte}}, \bibinfo {author} {\bibfnamefont {D.-J.}\ \bibnamefont {Kang}}, \ and\ \bibinfo {author} {\bibfnamefont {M.~G.}\ \bibnamefont {Blamire}},\ }\bibfield  {title} {\enquote {\bibinfo {title} {Controllable {Josephson} current through a pseudospin-valve structure},}\ }\href {\doibase http://dx.doi.org/10.1063/1.1646217} {\bibfield  {journal} {\bibinfo  {journal} {Appl. Phys. Lett.}\ }\textbf {\bibinfo {volume} {84}},\ \bibinfo {pages} {1153--1155} (\bibinfo {year} {2004})}\BibitemShut {NoStop}%
\bibitem [{\citenamefont {Baek}\ \emph {et~al.}(2014)\citenamefont {Baek}, \citenamefont {Rippard}, \citenamefont {Benz}, \citenamefont {Russek},\ and\ \citenamefont {Dresselhaus}}]{Baek_NComms_2014}%
  \BibitemOpen
  \bibfield  {author} {\bibinfo {author} {\bibfnamefont {B.}~\bibnamefont {Baek}}, \bibinfo {author} {\bibfnamefont {W.~H.}\ \bibnamefont {Rippard}}, \bibinfo {author} {\bibfnamefont {S.~P.}\ \bibnamefont {Benz}}, \bibinfo {author} {\bibfnamefont {S.~E.}\ \bibnamefont {Russek}}, \ and\ \bibinfo {author} {\bibfnamefont {P.~D.}\ \bibnamefont {Dresselhaus}},\ }\bibfield  {title} {\enquote {\bibinfo {title} {Hybrid superconducting-magnetic memory device using competing order parameters},}\ }\href {\doibase 10.1038/ncomms4888} {\bibfield  {journal} {\bibinfo  {journal} {Nat. Commun.}\ }\textbf {\bibinfo {volume} {5}},\ \bibinfo {pages} {3888} (\bibinfo {year} {2014})}\BibitemShut {NoStop}%
\bibitem [{\citenamefont {Abd El~Qader}\ \emph {et~al.}(2014)\citenamefont {Abd El~Qader}, \citenamefont {Singh}, \citenamefont {Galvin}, \citenamefont {Yu}, \citenamefont {Rowell},\ and\ \citenamefont {Newman}}]{Qader_APL_2014}%
  \BibitemOpen
  \bibfield  {author} {\bibinfo {author} {\bibfnamefont {M.}~\bibnamefont {Abd El~Qader}}, \bibinfo {author} {\bibfnamefont {R.~K.}\ \bibnamefont {Singh}}, \bibinfo {author} {\bibfnamefont {S.~N.}\ \bibnamefont {Galvin}}, \bibinfo {author} {\bibfnamefont {L.}~\bibnamefont {Yu}}, \bibinfo {author} {\bibfnamefont {J.~M.}\ \bibnamefont {Rowell}}, \ and\ \bibinfo {author} {\bibfnamefont {N.}~\bibnamefont {Newman}},\ }\bibfield  {title} {\enquote {\bibinfo {title} {Switching at small magnetic fields in {J}osephson junctions fabricated with ferromagnetic barrier layers},}\ }\href {\doibase 10.1063/1.4862195} {\bibfield  {journal} {\bibinfo  {journal} {Appl. Phys. Lett.}\ }\textbf {\bibinfo {volume} {104}},\ \bibinfo {pages} {022602} (\bibinfo {year} {2014})}\BibitemShut {NoStop}%
\bibitem [{\citenamefont {Gingrich}\ \emph {et~al.}(2016)\citenamefont {Gingrich}, \citenamefont {Niedzielski}, \citenamefont {Glick}, \citenamefont {Wang}, \citenamefont {Miller}, \citenamefont {Loloee}, \citenamefont {Pratt},\ and\ \citenamefont {Birge}}]{Gingrich_NatPhys_2016}%
  \BibitemOpen
  \bibfield  {author} {\bibinfo {author} {\bibfnamefont {E.~C.}\ \bibnamefont {Gingrich}}, \bibinfo {author} {\bibfnamefont {B.~M.}\ \bibnamefont {Niedzielski}}, \bibinfo {author} {\bibfnamefont {J.~A.}\ \bibnamefont {Glick}}, \bibinfo {author} {\bibfnamefont {Y.}~\bibnamefont {Wang}}, \bibinfo {author} {\bibfnamefont {D.~L.}\ \bibnamefont {Miller}}, \bibinfo {author} {\bibfnamefont {R.}~\bibnamefont {Loloee}}, \bibinfo {author} {\bibfnamefont {W.~P.}\ \bibnamefont {Pratt}}, \ and\ \bibinfo {author} {\bibfnamefont {N.~O.}\ \bibnamefont {Birge}},\ }\bibfield  {title} {\enquote {\bibinfo {title} {Controllable 0-$\pi$ {Josephson} junctions containing a ferromagnetic spin valve.}}\ }\href {\doibase doi:10.1038/nphys3681} {\bibfield  {journal} {\bibinfo  {journal} {Nat. Phys.}\ }\textbf {\bibinfo {volume} {12}},\ \bibinfo {pages} {564--567} (\bibinfo {year} {2016})}\BibitemShut {NoStop}%
\bibitem [{\citenamefont {Madden}\ \emph {et~al.}(2018)\citenamefont {Madden}, \citenamefont {Willard}, \citenamefont {Loloee},\ and\ \citenamefont {Birge}}]{Madden_SUST_2018}%
  \BibitemOpen
  \bibfield  {author} {\bibinfo {author} {\bibfnamefont {A.~E.}\ \bibnamefont {Madden}}, \bibinfo {author} {\bibfnamefont {J.~C.}\ \bibnamefont {Willard}}, \bibinfo {author} {\bibfnamefont {R.}~\bibnamefont {Loloee}}, \ and\ \bibinfo {author} {\bibfnamefont {N.~O.}\ \bibnamefont {Birge}},\ }\bibfield  {title} {\enquote {\bibinfo {title} {Phase controllable {J}osephson junctions for cryogenic memory},}\ }\href {\doibase 10.1088/1361-6668/aae8cf} {\bibfield  {journal} {\bibinfo  {journal} {Supercond. Sci. Technol.}\ }\textbf {\bibinfo {volume} {32}},\ \bibinfo {pages} {015001} (\bibinfo {year} {2018})}\BibitemShut {NoStop}%
\bibitem [{\citenamefont {Dayton}\ \emph {et~al.}(2018)\citenamefont {Dayton}, \citenamefont {Sage}, \citenamefont {Gingrich}, \citenamefont {Loving}, \citenamefont {Ambrose}, \citenamefont {Siwak}, \citenamefont {Keebaugh}, \citenamefont {Kirby}, \citenamefont {Miller}, \citenamefont {Herr}, \citenamefont {Herr},\ and\ \citenamefont {Naaman}}]{Dayton_IEEE_2017}%
  \BibitemOpen
  \bibfield  {author} {\bibinfo {author} {\bibfnamefont {I.~M.}\ \bibnamefont {Dayton}}, \bibinfo {author} {\bibfnamefont {T.}~\bibnamefont {Sage}}, \bibinfo {author} {\bibfnamefont {E.~C.}\ \bibnamefont {Gingrich}}, \bibinfo {author} {\bibfnamefont {M.~G.}\ \bibnamefont {Loving}}, \bibinfo {author} {\bibfnamefont {T.~F.}\ \bibnamefont {Ambrose}}, \bibinfo {author} {\bibfnamefont {N.~P.}\ \bibnamefont {Siwak}}, \bibinfo {author} {\bibfnamefont {S.}~\bibnamefont {Keebaugh}}, \bibinfo {author} {\bibfnamefont {C.}~\bibnamefont {Kirby}}, \bibinfo {author} {\bibfnamefont {D.~L.}\ \bibnamefont {Miller}}, \bibinfo {author} {\bibfnamefont {A.~Y.}\ \bibnamefont {Herr}}, \bibinfo {author} {\bibfnamefont {Q.~P.}\ \bibnamefont {Herr}}, \ and\ \bibinfo {author} {\bibfnamefont {O.}~\bibnamefont {Naaman}},\ }\bibfield  {title} {\enquote {\bibinfo {title} {Experimental demonstration of a {J}osephson magnetic memory cell with a programmable $\pi$-junction},}\ }\href {\doibase 10.1109/LMAG.2018.2801820} {\bibfield  {journal}
  {\bibinfo  {journal} {IEEE Magn. Lett.}\ }\textbf {\bibinfo {volume} {9}},\ \bibinfo {pages} {3301905} (\bibinfo {year} {2018})}\BibitemShut {NoStop}%
\bibitem [{\citenamefont {Bakurskiy}\ \emph {et~al.}(2016)\citenamefont {Bakurskiy}, \citenamefont {Klenov}, \citenamefont {Soloviev}, \citenamefont {Kupriyanov},\ and\ \citenamefont {Golubov}}]{bakurskiy_2016}%
  \BibitemOpen
  \bibfield  {author} {\bibinfo {author} {\bibfnamefont {S.~V.}\ \bibnamefont {Bakurskiy}}, \bibinfo {author} {\bibfnamefont {N.~V.}\ \bibnamefont {Klenov}}, \bibinfo {author} {\bibfnamefont {I.~I.}\ \bibnamefont {Soloviev}}, \bibinfo {author} {\bibfnamefont {M.~Y.}\ \bibnamefont {Kupriyanov}}, \ and\ \bibinfo {author} {\bibfnamefont {A.~A.}\ \bibnamefont {Golubov}},\ }\bibfield  {title} {\enquote {\bibinfo {title} {Superconducting phase domains for memory applications},}\ }\href {\doibase 10.1063/1.4940440} {\bibfield  {journal} {\bibinfo  {journal} {Appl. Phys. Lett}\ }\textbf {\bibinfo {volume} {108}},\ \bibinfo {pages} {042602} (\bibinfo {year} {2016})}\BibitemShut {NoStop}%
\bibitem [{\citenamefont {Bakurskiy}\ \emph {et~al.}(2018)\citenamefont {Bakurskiy}, \citenamefont {Klenov}, \citenamefont {Soloviev}, \citenamefont {Pugach}, \citenamefont {Kupriyanov},\ and\ \citenamefont {Golubov}}]{bakurskiy_2018}%
  \BibitemOpen
  \bibfield  {author} {\bibinfo {author} {\bibfnamefont {S.~V.}\ \bibnamefont {Bakurskiy}}, \bibinfo {author} {\bibfnamefont {N.~V.}\ \bibnamefont {Klenov}}, \bibinfo {author} {\bibfnamefont {I.~I.}\ \bibnamefont {Soloviev}}, \bibinfo {author} {\bibfnamefont {N.~G.}\ \bibnamefont {Pugach}}, \bibinfo {author} {\bibfnamefont {M.~Y.}\ \bibnamefont {Kupriyanov}}, \ and\ \bibinfo {author} {\bibfnamefont {A.~A.}\ \bibnamefont {Golubov}},\ }\bibfield  {title} {\enquote {\bibinfo {title} {Protected 0-$\pi$ states in {SIsFS} junctions for {Josephson} memory and logic},}\ }\href {\doibase 10.1063/1.5045490} {\bibfield  {journal} {\bibinfo  {journal} {Appl. Phys. Lett}\ }\textbf {\bibinfo {volume} {113}},\ \bibinfo {pages} {082602} (\bibinfo {year} {2018})}\BibitemShut {NoStop}%
\bibitem [{\citenamefont {Fierz}\ \emph {et~al.}(1990)\citenamefont {Fierz}, \citenamefont {Lee}, \citenamefont {Bass}, \citenamefont {Pratt},\ and\ \citenamefont {Schroeder}}]{fierz_1990}%
  \BibitemOpen
  \bibfield  {author} {\bibinfo {author} {\bibfnamefont {C.}~\bibnamefont {Fierz}}, \bibinfo {author} {\bibfnamefont {S.~F.}\ \bibnamefont {Lee}}, \bibinfo {author} {\bibfnamefont {J.}~\bibnamefont {Bass}}, \bibinfo {author} {\bibfnamefont {W.~P.}\ \bibnamefont {Pratt}}, \ and\ \bibinfo {author} {\bibfnamefont {P.~A.}\ \bibnamefont {Schroeder}},\ }\bibfield  {title} {\enquote {\bibinfo {title} {Superconductor/ferromagnet boundary resistances},}\ }\href {\doibase 10.1088/0953-8984/2/48/024} {\bibfield  {journal} {\bibinfo  {journal} {J. Phys.: Condens. Matter}\ }\textbf {\bibinfo {volume} {2}},\ \bibinfo {pages} {9701--9706} (\bibinfo {year} {1990})}\BibitemShut {NoStop}%
\bibitem [{\citenamefont {Blum}\ \emph {et~al.}(2002)\citenamefont {Blum}, \citenamefont {Tsukernik}, \citenamefont {Karpovski},\ and\ \citenamefont {Palevski}}]{Blum_PRL_2002}%
  \BibitemOpen
  \bibfield  {author} {\bibinfo {author} {\bibfnamefont {Y.}~\bibnamefont {Blum}}, \bibinfo {author} {\bibfnamefont {A.}~\bibnamefont {Tsukernik}}, \bibinfo {author} {\bibfnamefont {M.}~\bibnamefont {Karpovski}}, \ and\ \bibinfo {author} {\bibfnamefont {A.}~\bibnamefont {Palevski}},\ }\bibfield  {title} {\enquote {\bibinfo {title} {{Oscillations of the Superconducting Critical Current in Nb-Cu-Ni-Cu-Nb Junctions}},}\ }\href {\doibase 10.1103/PhysRevLett.89.187004} {\bibfield  {journal} {\bibinfo  {journal} {Phys. Rev. Lett.}\ }\textbf {\bibinfo {volume} {89}},\ \bibinfo {pages} {187004} (\bibinfo {year} {2002})}\BibitemShut {NoStop}%
\bibitem [{\citenamefont {Shelukhin}\ \emph {et~al.}(2006)\citenamefont {Shelukhin}, \citenamefont {Tsukernik}, \citenamefont {Karpovski}, \citenamefont {Blum}, \citenamefont {Efetov}, \citenamefont {Volkov}, \citenamefont {Champel}, \citenamefont {Eschrig}, \citenamefont {L\"ofwander}, \citenamefont {Sch\"on},\ and\ \citenamefont {Palevski}}]{Shelukhin_PRB_2006}%
  \BibitemOpen
  \bibfield  {author} {\bibinfo {author} {\bibfnamefont {V.}~\bibnamefont {Shelukhin}}, \bibinfo {author} {\bibfnamefont {A.}~\bibnamefont {Tsukernik}}, \bibinfo {author} {\bibfnamefont {M.}~\bibnamefont {Karpovski}}, \bibinfo {author} {\bibfnamefont {Y.}~\bibnamefont {Blum}}, \bibinfo {author} {\bibfnamefont {K.~B.}\ \bibnamefont {Efetov}}, \bibinfo {author} {\bibfnamefont {A.~F.}\ \bibnamefont {Volkov}}, \bibinfo {author} {\bibfnamefont {T.}~\bibnamefont {Champel}}, \bibinfo {author} {\bibfnamefont {M.}~\bibnamefont {Eschrig}}, \bibinfo {author} {\bibfnamefont {T.}~\bibnamefont {L\"ofwander}}, \bibinfo {author} {\bibfnamefont {G.}~\bibnamefont {Sch\"on}}, \ and\ \bibinfo {author} {\bibfnamefont {A.}~\bibnamefont {Palevski}},\ }\bibfield  {title} {\enquote {\bibinfo {title} {Observation of periodic $\ensuremath{\pi}$-phase shifts in ferromagnet-superconductor multilayers},}\ }\href {\doibase 10.1103/PhysRevB.73.174506} {\bibfield  {journal} {\bibinfo  {journal} {Phys. Rev. B}\ }\textbf {\bibinfo {volume}
  {73}},\ \bibinfo {pages} {174506} (\bibinfo {year} {2006})}\BibitemShut {NoStop}%
\bibitem [{\citenamefont {Robinson}\ \emph {et~al.}(2006)\citenamefont {Robinson}, \citenamefont {Piano}, \citenamefont {Burnell}, \citenamefont {Bell},\ and\ \citenamefont {Blamire}}]{Robinson_PRL_2006}%
  \BibitemOpen
  \bibfield  {author} {\bibinfo {author} {\bibfnamefont {J.~W.~A.}\ \bibnamefont {Robinson}}, \bibinfo {author} {\bibfnamefont {S.}~\bibnamefont {Piano}}, \bibinfo {author} {\bibfnamefont {G.}~\bibnamefont {Burnell}}, \bibinfo {author} {\bibfnamefont {C.}~\bibnamefont {Bell}}, \ and\ \bibinfo {author} {\bibfnamefont {M.~G.}\ \bibnamefont {Blamire}},\ }\bibfield  {title} {\enquote {\bibinfo {title} {Critical {Current Oscillations in Strong F}erromagnetic $\ensuremath{\pi}$ {J}unctions},}\ }\href {\doibase 10.1103/PhysRevLett.97.177003} {\bibfield  {journal} {\bibinfo  {journal} {Phys. Rev. Lett.}\ }\textbf {\bibinfo {volume} {97}},\ \bibinfo {pages} {177003} (\bibinfo {year} {2006})}\BibitemShut {NoStop}%
\bibitem [{\citenamefont {Bannykh}\ \emph {et~al.}(2009)\citenamefont {Bannykh}, \citenamefont {Pfeiffer}, \citenamefont {Stolyarov}, \citenamefont {Batov}, \citenamefont {Ryazanov},\ and\ \citenamefont {Weides}}]{Bannykh_PRB_2009}%
  \BibitemOpen
  \bibfield  {author} {\bibinfo {author} {\bibfnamefont {A.~A.}\ \bibnamefont {Bannykh}}, \bibinfo {author} {\bibfnamefont {J.}~\bibnamefont {Pfeiffer}}, \bibinfo {author} {\bibfnamefont {V.~S.}\ \bibnamefont {Stolyarov}}, \bibinfo {author} {\bibfnamefont {I.~E.}\ \bibnamefont {Batov}}, \bibinfo {author} {\bibfnamefont {V.~V.}\ \bibnamefont {Ryazanov}}, \ and\ \bibinfo {author} {\bibfnamefont {M.}~\bibnamefont {Weides}},\ }\bibfield  {title} {\enquote {\bibinfo {title} {Josephson tunnel junctions with a strong ferromagnetic interlayer},}\ }\href {\doibase 10.1103/PhysRevB.79.054501} {\bibfield  {journal} {\bibinfo  {journal} {Phys. Rev. B}\ }\textbf {\bibinfo {volume} {79}},\ \bibinfo {pages} {054501} (\bibinfo {year} {2009})}\BibitemShut {NoStop}%
\bibitem [{\citenamefont {Baek}\ \emph {et~al.}(2017)\citenamefont {Baek}, \citenamefont {Schneider}, \citenamefont {Pufall},\ and\ \citenamefont {Rippard}}]{Baek_PRApp_2017}%
  \BibitemOpen
  \bibfield  {author} {\bibinfo {author} {\bibfnamefont {B.}~\bibnamefont {Baek}}, \bibinfo {author} {\bibfnamefont {M.~L.}\ \bibnamefont {Schneider}}, \bibinfo {author} {\bibfnamefont {M.~R.}\ \bibnamefont {Pufall}}, \ and\ \bibinfo {author} {\bibfnamefont {W.~H.}\ \bibnamefont {Rippard}},\ }\bibfield  {title} {\enquote {\bibinfo {title} {{Phase Offsets in the Critical-Current Oscillations of Josephson Junctions Based on Ni and Ni}-($\text{Ni}_{81}\text{Fe}_{19})_{x}\text{Nb}_{y}$ {B}arriers},}\ }\href {\doibase 10.1103/PhysRevApplied.7.064013} {\bibfield  {journal} {\bibinfo  {journal} {Phys. Rev. Applied}\ }\textbf {\bibinfo {volume} {7}},\ \bibinfo {pages} {064013} (\bibinfo {year} {2017})}\BibitemShut {NoStop}%
\bibitem [{\citenamefont {Tolpygo}\ \emph {et~al.}(2019)\citenamefont {Tolpygo}, \citenamefont {Bolkhovsky}, \citenamefont {Rastogi}, \citenamefont {Zarr}, \citenamefont {Day}, \citenamefont {Golden}, \citenamefont {Weir}, \citenamefont {Wynn},\ and\ \citenamefont {Johnson}}]{Tolpygo_2019}%
  \BibitemOpen
  \bibfield  {author} {\bibinfo {author} {\bibfnamefont {S.~K.}\ \bibnamefont {Tolpygo}}, \bibinfo {author} {\bibfnamefont {V.}~\bibnamefont {Bolkhovsky}}, \bibinfo {author} {\bibfnamefont {R.}~\bibnamefont {Rastogi}}, \bibinfo {author} {\bibfnamefont {S.}~\bibnamefont {Zarr}}, \bibinfo {author} {\bibfnamefont {A.~L.}\ \bibnamefont {Day}}, \bibinfo {author} {\bibfnamefont {E.}~\bibnamefont {Golden}}, \bibinfo {author} {\bibfnamefont {T.~J.}\ \bibnamefont {Weir}}, \bibinfo {author} {\bibfnamefont {A.}~\bibnamefont {Wynn}}, \ and\ \bibinfo {author} {\bibfnamefont {L.~M.}\ \bibnamefont {Johnson}},\ }\bibfield  {title} {\enquote {\bibinfo {title} {Planarized {Fabrication Process With Two L}ayers of {SIS} {Josephson} {Junctions and I}ntegration of {SIS} and {SFS} $\pi$-{J}unctions},}\ }\href {\doibase 10.1109/TASC.2019.2901709} {\bibfield  {journal} {\bibinfo  {journal} {IEEE Trans. Appl. Supercond.}\ }\textbf {\bibinfo {volume} {29}},\ \bibinfo {pages} {1101208} (\bibinfo {year} {2019})}\BibitemShut {NoStop}%
\bibitem [{\citenamefont {Kapran}\ \emph {et~al.}(2021)\citenamefont {Kapran}, \citenamefont {Golod}, \citenamefont {Iovan}, \citenamefont {Sidorenko}, \citenamefont {Golubov},\ and\ \citenamefont {Krasnov}}]{Kapran_2021}%
  \BibitemOpen
  \bibfield  {author} {\bibinfo {author} {\bibfnamefont {O.~M.}\ \bibnamefont {Kapran}}, \bibinfo {author} {\bibfnamefont {T.}~\bibnamefont {Golod}}, \bibinfo {author} {\bibfnamefont {A.}~\bibnamefont {Iovan}}, \bibinfo {author} {\bibfnamefont {A.~S.}\ \bibnamefont {Sidorenko}}, \bibinfo {author} {\bibfnamefont {A.~A.}\ \bibnamefont {Golubov}}, \ and\ \bibinfo {author} {\bibfnamefont {V.~M.}\ \bibnamefont {Krasnov}},\ }\bibfield  {title} {\enquote {\bibinfo {title} {Crossover between short- and long-range proximity effects in superconductor/ferromagnet/superconductor junctions with {N}i-based ferromagnets},}\ }\href {\doibase 10.1103/PhysRevB.103.094509} {\bibfield  {journal} {\bibinfo  {journal} {Phys. Rev. B}\ }\textbf {\bibinfo {volume} {103}},\ \bibinfo {pages} {094509} (\bibinfo {year} {2021})}\BibitemShut {NoStop}%
\bibitem [{\citenamefont {Sürgers}\ \emph {et~al.}(2002)\citenamefont {Sürgers}, \citenamefont {Hoss}, \citenamefont {Schönenberger},\ and\ \citenamefont {Strunk}}]{Surgers_2002}%
  \BibitemOpen
  \bibfield  {author} {\bibinfo {author} {\bibfnamefont {C.}~\bibnamefont {Sürgers}}, \bibinfo {author} {\bibfnamefont {T.}~\bibnamefont {Hoss}}, \bibinfo {author} {\bibfnamefont {C.}~\bibnamefont {Schönenberger}}, \ and\ \bibinfo {author} {\bibfnamefont {C.}~\bibnamefont {Strunk}},\ }\bibfield  {title} {\enquote {\bibinfo {title} {Fabrication and superconducting properties of nanostructured {SFS} contacts},}\ }\href {\doibase https://doi.org/10.1016/S0304-8853(01)00862-9} {\bibfield  {journal} {\bibinfo  {journal} {J. Magn. Magn. Mater.}\ }\textbf {\bibinfo {volume} {240}},\ \bibinfo {pages} {598--600} (\bibinfo {year} {2002})},\ \bibinfo {note} {4th International Symposium on Metallic Multilayers}\BibitemShut {NoStop}%
\bibitem [{\citenamefont {Robinson}, \citenamefont {Barber},\ and\ \citenamefont {Blamire}(2009)}]{Robinson_APL_2009}%
  \BibitemOpen
  \bibfield  {author} {\bibinfo {author} {\bibfnamefont {J.~W.~A.}\ \bibnamefont {Robinson}}, \bibinfo {author} {\bibfnamefont {Z.~H.}\ \bibnamefont {Barber}}, \ and\ \bibinfo {author} {\bibfnamefont {M.~G.}\ \bibnamefont {Blamire}},\ }\bibfield  {title} {\enquote {\bibinfo {title} {Strong ferromagnetic {J}osephson devices with optimized magnetism},}\ }\href {\doibase 10.1063/1.3262969} {\bibfield  {journal} {\bibinfo  {journal} {Appl. Phys. Lett.}\ }\textbf {\bibinfo {volume} {95}},\ \bibinfo {pages} {192509} (\bibinfo {year} {2009})}\BibitemShut {NoStop}%
\bibitem [{\citenamefont {Robinson}\ \emph {et~al.}(2007)\citenamefont {Robinson}, \citenamefont {Piano}, \citenamefont {Burnell}, \citenamefont {Bell},\ and\ \citenamefont {Blamire}}]{Robinson_PRB_2007}%
  \BibitemOpen
  \bibfield  {author} {\bibinfo {author} {\bibfnamefont {J.~W.~A.}\ \bibnamefont {Robinson}}, \bibinfo {author} {\bibfnamefont {S.}~\bibnamefont {Piano}}, \bibinfo {author} {\bibfnamefont {G.}~\bibnamefont {Burnell}}, \bibinfo {author} {\bibfnamefont {C.}~\bibnamefont {Bell}}, \ and\ \bibinfo {author} {\bibfnamefont {M.~G.}\ \bibnamefont {Blamire}},\ }\bibfield  {title} {\enquote {\bibinfo {title} {Zero to $\ensuremath{\pi}$ transition in superconductor-ferromagnet-superconductor junctions},}\ }\href {\doibase 10.1103/PhysRevB.76.094522} {\bibfield  {journal} {\bibinfo  {journal} {Phys. Rev. B}\ }\textbf {\bibinfo {volume} {76}},\ \bibinfo {pages} {094522} (\bibinfo {year} {2007})}\BibitemShut {NoStop}%
\bibitem [{\citenamefont {Witt}, \citenamefont {Robinson},\ and\ \citenamefont {Blamire}(2012)}]{Witt_PRB_2012}%
  \BibitemOpen
  \bibfield  {author} {\bibinfo {author} {\bibfnamefont {J.~D.~S.}\ \bibnamefont {Witt}}, \bibinfo {author} {\bibfnamefont {J.~W.~A.}\ \bibnamefont {Robinson}}, \ and\ \bibinfo {author} {\bibfnamefont {M.~G.}\ \bibnamefont {Blamire}},\ }\bibfield  {title} {\enquote {\bibinfo {title} {Josephson junctions incorporating a conical magnetic holmium interlayer},}\ }\href {\doibase 10.1103/PhysRevB.85.184526} {\bibfield  {journal} {\bibinfo  {journal} {Phys. Rev. B}\ }\textbf {\bibinfo {volume} {85}},\ \bibinfo {pages} {184526} (\bibinfo {year} {2012})}\BibitemShut {NoStop}%
\bibitem [{\citenamefont {{Robinson}}\ \emph {et~al.}(2007)\citenamefont {{Robinson}}, \citenamefont {{Piano}}, \citenamefont {{Burnell}}, \citenamefont {{Bell}},\ and\ \citenamefont {{Blamire}}}]{Robinson_IEEE_2007}%
  \BibitemOpen
  \bibfield  {author} {\bibinfo {author} {\bibfnamefont {J.~W.~A.}\ \bibnamefont {{Robinson}}}, \bibinfo {author} {\bibfnamefont {S.}~\bibnamefont {{Piano}}}, \bibinfo {author} {\bibfnamefont {G.}~\bibnamefont {{Burnell}}}, \bibinfo {author} {\bibfnamefont {C.}~\bibnamefont {{Bell}}}, \ and\ \bibinfo {author} {\bibfnamefont {M.~G.}\ \bibnamefont {{Blamire}}},\ }\bibfield  {title} {\enquote {\bibinfo {title} {Transport and {M}agnetic {P}roperties of {S}trong {F}erromagnetic {P}i-{J}unctions},}\ }\href {\doibase 10.1109/TASC.2007.898720} {\bibfield  {journal} {\bibinfo  {journal} {IEEE Trans. Appl. Supercond.}\ }\textbf {\bibinfo {volume} {17}},\ \bibinfo {pages} {641--644} (\bibinfo {year} {2007})}\BibitemShut {NoStop}%
\bibitem [{\citenamefont {{Baek}}\ \emph {et~al.}(2018)\citenamefont {{Baek}}, \citenamefont {{Schneider}}, \citenamefont {{Pufall}},\ and\ \citenamefont {{Rippard}}}]{Baek_IEEE_2018}%
  \BibitemOpen
  \bibfield  {author} {\bibinfo {author} {\bibfnamefont {B.}~\bibnamefont {{Baek}}}, \bibinfo {author} {\bibfnamefont {M.~L.}\ \bibnamefont {{Schneider}}}, \bibinfo {author} {\bibfnamefont {M.~R.}\ \bibnamefont {{Pufall}}}, \ and\ \bibinfo {author} {\bibfnamefont {W.~H.}\ \bibnamefont {{Rippard}}},\ }\bibfield  {title} {\enquote {\bibinfo {title} {{Anomalous Supercurrent Modulation in Josephson Junctions With Ni-Based Barriers}},}\ }\href {\doibase 10.1109/TASC.2018.2836961} {\bibfield  {journal} {\bibinfo  {journal} {IEEE Trans. Appl. Supercond.}\ }\textbf {\bibinfo {volume} {28}},\ \bibinfo {pages} {1800705} (\bibinfo {year} {2018})}\BibitemShut {NoStop}%
\bibitem [{\citenamefont {Piano}\ \emph {et~al.}(2007)\citenamefont {Piano}, \citenamefont {Robinson}, \citenamefont {Burnell},\ and\ \citenamefont {Blamire}}]{Piano_EPJB_2007}%
  \BibitemOpen
  \bibfield  {author} {\bibinfo {author} {\bibfnamefont {S.}~\bibnamefont {Piano}}, \bibinfo {author} {\bibfnamefont {J.~W.~A.}\ \bibnamefont {Robinson}}, \bibinfo {author} {\bibfnamefont {G.}~\bibnamefont {Burnell}}, \ and\ \bibinfo {author} {\bibfnamefont {M.~G.}\ \bibnamefont {Blamire}},\ }\bibfield  {title} {\enquote {\bibinfo {title} {0-$\pi$ oscillations in nanostructured {Nb/Fe/Nb J}osephson junctions},}\ }\href {\doibase 10.1140/epjb/e2007-00210-8} {\bibfield  {journal} {\bibinfo  {journal} {Eur. Phys. J. B}\ }\textbf {\bibinfo {volume} {58}},\ \bibinfo {pages} {123--126} (\bibinfo {year} {2007})}\BibitemShut {NoStop}%
\bibitem [{\citenamefont {Robinson}, \citenamefont {Witt},\ and\ \citenamefont {Blamire}(2010)}]{robinson_2010}%
  \BibitemOpen
  \bibfield  {author} {\bibinfo {author} {\bibfnamefont {J.~W.~A.}\ \bibnamefont {Robinson}}, \bibinfo {author} {\bibfnamefont {J.~D.~S.}\ \bibnamefont {Witt}}, \ and\ \bibinfo {author} {\bibfnamefont {M.~G.}\ \bibnamefont {Blamire}},\ }\bibfield  {title} {\enquote {\bibinfo {title} {Controlled injection of spin-triplet supercurrents into a strong ferromagnet},}\ }\href {\doibase 10.1126/science.1189246} {\bibfield  {journal} {\bibinfo  {journal} {Science}\ }\textbf {\bibinfo {volume} {329}},\ \bibinfo {pages} {59--61} (\bibinfo {year} {2010})}\BibitemShut {NoStop}%
\bibitem [{\citenamefont {Kwo}, \citenamefont {Hong},\ and\ \citenamefont {Nakahara}(1986)}]{Kwo_1986}%
  \BibitemOpen
  \bibfield  {author} {\bibinfo {author} {\bibfnamefont {J.}~\bibnamefont {Kwo}}, \bibinfo {author} {\bibfnamefont {M.}~\bibnamefont {Hong}}, \ and\ \bibinfo {author} {\bibfnamefont {S.}~\bibnamefont {Nakahara}},\ }\bibfield  {title} {\enquote {\bibinfo {title} {{Growth of rare‐earth single crystals by molecular beam epitaxy: The epitaxial relationship between hcp rare earth and bcc niobium}},}\ }\href {\doibase 10.1063/1.97155} {\bibfield  {journal} {\bibinfo  {journal} {Appl. Phys. Lett.}\ }\textbf {\bibinfo {volume} {49}},\ \bibinfo {pages} {319--321} (\bibinfo {year} {1986})}\BibitemShut {NoStop}%
\bibitem [{\citenamefont {Satchell}\ \emph {et~al.}(2017)\citenamefont {Satchell}, \citenamefont {Witt}, \citenamefont {Burnell}, \citenamefont {Curran}, \citenamefont {Kinane}, \citenamefont {Charlton}, \citenamefont {Langridge},\ and\ \citenamefont {Cooper}}]{satchell_2017}%
  \BibitemOpen
  \bibfield  {author} {\bibinfo {author} {\bibfnamefont {N.}~\bibnamefont {Satchell}}, \bibinfo {author} {\bibfnamefont {J.~D.~S.}\ \bibnamefont {Witt}}, \bibinfo {author} {\bibfnamefont {G.}~\bibnamefont {Burnell}}, \bibinfo {author} {\bibfnamefont {P.~J.}\ \bibnamefont {Curran}}, \bibinfo {author} {\bibfnamefont {C.~J.}\ \bibnamefont {Kinane}}, \bibinfo {author} {\bibfnamefont {T.~R.}\ \bibnamefont {Charlton}}, \bibinfo {author} {\bibfnamefont {S.}~\bibnamefont {Langridge}}, \ and\ \bibinfo {author} {\bibfnamefont {J.~F.~K.}\ \bibnamefont {Cooper}},\ }\bibfield  {title} {\enquote {\bibinfo {title} {Probing the spiral magnetic phase in 6 nm textured erbium using polarised neutron reflectometry},}\ }\href {\doibase 10.1088/1361-648X/29/5/055801} {\bibfield  {journal} {\bibinfo  {journal} {J. Condens. Matter Phys.}\ }\textbf {\bibinfo {volume} {29}},\ \bibinfo {pages} {055801} (\bibinfo {year} {2017})}\BibitemShut {NoStop}%
\bibitem [{\citenamefont {Khaydukov}\ \emph {et~al.}(2018)\citenamefont {Khaydukov}, \citenamefont {Vasenko}, \citenamefont {Kravtsov}, \citenamefont {Progliado}, \citenamefont {Zhaketov}, \citenamefont {Csik}, \citenamefont {Nikitenko}, \citenamefont {Petrenko}, \citenamefont {Keller}, \citenamefont {Golubov}, \citenamefont {Kupriyanov}, \citenamefont {Ustinov}, \citenamefont {Aksenov},\ and\ \citenamefont {Keimer}}]{khaydukov_2018}%
  \BibitemOpen
  \bibfield  {author} {\bibinfo {author} {\bibfnamefont {Y.~N.}\ \bibnamefont {Khaydukov}}, \bibinfo {author} {\bibfnamefont {A.~S.}\ \bibnamefont {Vasenko}}, \bibinfo {author} {\bibfnamefont {E.~A.}\ \bibnamefont {Kravtsov}}, \bibinfo {author} {\bibfnamefont {V.~V.}\ \bibnamefont {Progliado}}, \bibinfo {author} {\bibfnamefont {V.~D.}\ \bibnamefont {Zhaketov}}, \bibinfo {author} {\bibfnamefont {A.}~\bibnamefont {Csik}}, \bibinfo {author} {\bibfnamefont {Y.~V.}\ \bibnamefont {Nikitenko}}, \bibinfo {author} {\bibfnamefont {A.~V.}\ \bibnamefont {Petrenko}}, \bibinfo {author} {\bibfnamefont {T.}~\bibnamefont {Keller}}, \bibinfo {author} {\bibfnamefont {A.~A.}\ \bibnamefont {Golubov}}, \bibinfo {author} {\bibfnamefont {M.~Y.}\ \bibnamefont {Kupriyanov}}, \bibinfo {author} {\bibfnamefont {V.~V.}\ \bibnamefont {Ustinov}}, \bibinfo {author} {\bibfnamefont {V.~L.}\ \bibnamefont {Aksenov}}, \ and\ \bibinfo {author} {\bibfnamefont {B.}~\bibnamefont {Keimer}},\ }\bibfield  {title} {\enquote {\bibinfo {title} {Magnetic
  and superconducting phase diagram of {Nb/Gd/Nb} trilayers},}\ }\href {\doibase 10.1103/{PhysRevB}.97.144511} {\bibfield  {journal} {\bibinfo  {journal} {Phys. Rev. B}\ }\textbf {\bibinfo {volume} {97}},\ \bibinfo {pages} {144511} (\bibinfo {year} {2018})}\BibitemShut {NoStop}%
\bibitem [{\citenamefont {Bell}\ \emph {et~al.}(2005)\citenamefont {Bell}, \citenamefont {Loloee}, \citenamefont {Burnell},\ and\ \citenamefont {Blamire}}]{Bell_PRB_2005}%
  \BibitemOpen
  \bibfield  {author} {\bibinfo {author} {\bibfnamefont {C.}~\bibnamefont {Bell}}, \bibinfo {author} {\bibfnamefont {R.}~\bibnamefont {Loloee}}, \bibinfo {author} {\bibfnamefont {G.}~\bibnamefont {Burnell}}, \ and\ \bibinfo {author} {\bibfnamefont {M.~G.}\ \bibnamefont {Blamire}},\ }\bibfield  {title} {\enquote {\bibinfo {title} {Characteristics of strong ferromagnetic {J}osephson junctions with epitaxial barriers},}\ }\href {\doibase 10.1103/PhysRevB.71.180501} {\bibfield  {journal} {\bibinfo  {journal} {Phys. Rev. B}\ }\textbf {\bibinfo {volume} {71}},\ \bibinfo {pages} {180501(R)} (\bibinfo {year} {2005})}\BibitemShut {NoStop}%
\bibitem [{\citenamefont {Glick}\ \emph {et~al.}(2017)\citenamefont {Glick}, \citenamefont {Khasawneh}, \citenamefont {Niedzielski}, \citenamefont {Loloee}, \citenamefont {Pratt}, \citenamefont {Birge}, \citenamefont {Gingrich}, \citenamefont {Kotula},\ and\ \citenamefont {Missert}}]{Glick_JAP_2017}%
  \BibitemOpen
  \bibfield  {author} {\bibinfo {author} {\bibfnamefont {J.~A.}\ \bibnamefont {Glick}}, \bibinfo {author} {\bibfnamefont {M.~A.}\ \bibnamefont {Khasawneh}}, \bibinfo {author} {\bibfnamefont {B.~M.}\ \bibnamefont {Niedzielski}}, \bibinfo {author} {\bibfnamefont {R.}~\bibnamefont {Loloee}}, \bibinfo {author} {\bibfnamefont {W.~P.}\ \bibnamefont {Pratt}}, \bibinfo {author} {\bibfnamefont {N.~O.}\ \bibnamefont {Birge}}, \bibinfo {author} {\bibfnamefont {E.~C.}\ \bibnamefont {Gingrich}}, \bibinfo {author} {\bibfnamefont {P.~G.}\ \bibnamefont {Kotula}}, \ and\ \bibinfo {author} {\bibfnamefont {N.}~\bibnamefont {Missert}},\ }\bibfield  {title} {\enquote {\bibinfo {title} {Critical current oscillations of elliptical {J}osephson junctions with single-domain ferromagnetic layers},}\ }\href {\doibase 10.1063/1.4989392} {\bibfield  {journal} {\bibinfo  {journal} {J. Appl. Phys.}\ }\textbf {\bibinfo {volume} {122}},\ \bibinfo {pages} {133906} (\bibinfo {year} {2017})}\BibitemShut {NoStop}%
\bibitem [{\citenamefont {Yao}\ \emph {et~al.}(2021)\citenamefont {Yao}, \citenamefont {Cai}, \citenamefont {Yang}, \citenamefont {Xing}, \citenamefont {Ma}, \citenamefont {Mori}, \citenamefont {Ji}, \citenamefont {Maekawa}, \citenamefont {Xie},\ and\ \citenamefont {Han}}]{Yao_2021}%
  \BibitemOpen
  \bibfield  {author} {\bibinfo {author} {\bibfnamefont {Y.}~\bibnamefont {Yao}}, \bibinfo {author} {\bibfnamefont {R.}~\bibnamefont {Cai}}, \bibinfo {author} {\bibfnamefont {S.-H.}\ \bibnamefont {Yang}}, \bibinfo {author} {\bibfnamefont {W.}~\bibnamefont {Xing}}, \bibinfo {author} {\bibfnamefont {Y.}~\bibnamefont {Ma}}, \bibinfo {author} {\bibfnamefont {M.}~\bibnamefont {Mori}}, \bibinfo {author} {\bibfnamefont {Y.}~\bibnamefont {Ji}}, \bibinfo {author} {\bibfnamefont {S.}~\bibnamefont {Maekawa}}, \bibinfo {author} {\bibfnamefont {X.-C.}\ \bibnamefont {Xie}}, \ and\ \bibinfo {author} {\bibfnamefont {W.}~\bibnamefont {Han}},\ }\bibfield  {title} {\enquote {\bibinfo {title} {Half-integer {S}hapiro steps in strong ferromagnetic {J}osephson junctions},}\ }\href {\doibase 10.1103/PhysRevB.104.104414} {\bibfield  {journal} {\bibinfo  {journal} {Phys. Rev. B}\ }\textbf {\bibinfo {volume} {104}},\ \bibinfo {pages} {104414} (\bibinfo {year} {2021})}\BibitemShut {NoStop}%
\bibitem [{\citenamefont {Mishra}\ \emph {et~al.}(2022)\citenamefont {Mishra}, \citenamefont {Klaes}, \citenamefont {Willard}, \citenamefont {Loloee},\ and\ \citenamefont {Birge}}]{Mishra_PRB_2022}%
  \BibitemOpen
  \bibfield  {author} {\bibinfo {author} {\bibfnamefont {S.~S.}\ \bibnamefont {Mishra}}, \bibinfo {author} {\bibfnamefont {R.~M.}\ \bibnamefont {Klaes}}, \bibinfo {author} {\bibfnamefont {J.}~\bibnamefont {Willard}}, \bibinfo {author} {\bibfnamefont {R.}~\bibnamefont {Loloee}}, \ and\ \bibinfo {author} {\bibfnamefont {N.~O.}\ \bibnamefont {Birge}},\ }\bibfield  {title} {\enquote {\bibinfo {title} {Enhancement of supercurrent through ferromagnetic materials by interface engineering},}\ }\href {\doibase 10.1103/PhysRevB.106.014519} {\bibfield  {journal} {\bibinfo  {journal} {Phys. Rev. B}\ }\textbf {\bibinfo {volume} {106}},\ \bibinfo {pages} {014519} (\bibinfo {year} {2022})}\BibitemShut {NoStop}%
\bibitem [{\citenamefont {Niedzielski}\ \emph {et~al.}(2015)\citenamefont {Niedzielski}, \citenamefont {Gingrich}, \citenamefont {Loloee}, \citenamefont {Pratt},\ and\ \citenamefont {Birge}}]{Niedzielski_SUST_2015}%
  \BibitemOpen
  \bibfield  {author} {\bibinfo {author} {\bibfnamefont {B.~M.}\ \bibnamefont {Niedzielski}}, \bibinfo {author} {\bibfnamefont {E.~C.}\ \bibnamefont {Gingrich}}, \bibinfo {author} {\bibfnamefont {R.}~\bibnamefont {Loloee}}, \bibinfo {author} {\bibfnamefont {W.~P.}\ \bibnamefont {Pratt}}, \ and\ \bibinfo {author} {\bibfnamefont {N.~O.}\ \bibnamefont {Birge}},\ }\bibfield  {title} {\enquote {\bibinfo {title} {{S/F/S J}osephson junctions with single-domain ferromagnets for memory applications},}\ }\href {\doibase 10.1088/0953-2048/28/8/085012} {\bibfield  {journal} {\bibinfo  {journal} {Supercond. Sci. Technol.}\ }\textbf {\bibinfo {volume} {28}},\ \bibinfo {pages} {085012} (\bibinfo {year} {2015})}\BibitemShut {NoStop}%
\bibitem [{\citenamefont {Boothby}\ and\ \citenamefont {Bozorth}(1947)}]{Boothby_1947}%
  \BibitemOpen
  \bibfield  {author} {\bibinfo {author} {\bibfnamefont {O.}~\bibnamefont {Boothby}}\ and\ \bibinfo {author} {\bibfnamefont {R.}~\bibnamefont {Bozorth}},\ }\bibfield  {title} {\enquote {\bibinfo {title} {A new magnetic material of high permeability},}\ }\href {\doibase 10.1063/1.1697599} {\bibfield  {journal} {\bibinfo  {journal} {J. Appl. Phys.}\ }\textbf {\bibinfo {volume} {18}},\ \bibinfo {pages} {173--176} (\bibinfo {year} {1947})}\BibitemShut {NoStop}%
\bibitem [{\citenamefont {Chen}\ \emph {et~al.}(1991)\citenamefont {Chen}, \citenamefont {Gharsallah}, \citenamefont {Gorman},\ and\ \citenamefont {Latimer}}]{Chen_1991}%
  \BibitemOpen
  \bibfield  {author} {\bibinfo {author} {\bibfnamefont {M.}~\bibnamefont {Chen}}, \bibinfo {author} {\bibfnamefont {N.}~\bibnamefont {Gharsallah}}, \bibinfo {author} {\bibfnamefont {G.~L.}\ \bibnamefont {Gorman}}, \ and\ \bibinfo {author} {\bibfnamefont {J.}~\bibnamefont {Latimer}},\ }\bibfield  {title} {\enquote {\bibinfo {title} {{Ternary {NiFeX} as soft biasing film in a magnetoresistive sensor}},}\ }\href {\doibase 10.1063/1.347919} {\bibfield  {journal} {\bibinfo  {journal} {J. Appl. Phys.}\ }\textbf {\bibinfo {volume} {69}},\ \bibinfo {pages} {5631--5633} (\bibinfo {year} {1991})}\BibitemShut {NoStop}%
\bibitem [{\citenamefont {Qader}\ \emph {et~al.}(2017)\citenamefont {Qader}, \citenamefont {Vishina}, \citenamefont {Yu}, \citenamefont {Garcia}, \citenamefont {Singh}, \citenamefont {Rizzo}, \citenamefont {Huang}, \citenamefont {Chamberlin}, \citenamefont {Belashchenko}, \citenamefont {{van Schilfgaarde}},\ and\ \citenamefont {Newman}}]{Qader_JMMM_2017}%
  \BibitemOpen
  \bibfield  {author} {\bibinfo {author} {\bibfnamefont {M.~A.}\ \bibnamefont {Qader}}, \bibinfo {author} {\bibfnamefont {A.}~\bibnamefont {Vishina}}, \bibinfo {author} {\bibfnamefont {L.}~\bibnamefont {Yu}}, \bibinfo {author} {\bibfnamefont {C.}~\bibnamefont {Garcia}}, \bibinfo {author} {\bibfnamefont {R.}~\bibnamefont {Singh}}, \bibinfo {author} {\bibfnamefont {N.}~\bibnamefont {Rizzo}}, \bibinfo {author} {\bibfnamefont {M.}~\bibnamefont {Huang}}, \bibinfo {author} {\bibfnamefont {R.}~\bibnamefont {Chamberlin}}, \bibinfo {author} {\bibfnamefont {K.}~\bibnamefont {Belashchenko}}, \bibinfo {author} {\bibfnamefont {M.}~\bibnamefont {{van Schilfgaarde}}}, \ and\ \bibinfo {author} {\bibfnamefont {N.}~\bibnamefont {Newman}},\ }\bibfield  {title} {\enquote {\bibinfo {title} {The magnetic, electrical and structural properties of copper-permalloy alloys},}\ }\href {\doibase 10.1016/j.jmmm.2017.06.081} {\bibfield  {journal} {\bibinfo  {journal} {J. Magn. Magn. Mater.}\ }\textbf {\bibinfo {volume} {442}},\ \bibinfo
  {pages} {45--52} (\bibinfo {year} {2017})}\BibitemShut {NoStop}%
\bibitem [{\citenamefont {Sellier}\ \emph {et~al.}(2003)\citenamefont {Sellier}, \citenamefont {Baraduc}, \citenamefont {Lefloch},\ and\ \citenamefont {Calemczuk}}]{Sellier_PRB_2003}%
  \BibitemOpen
  \bibfield  {author} {\bibinfo {author} {\bibfnamefont {H.}~\bibnamefont {Sellier}}, \bibinfo {author} {\bibfnamefont {C.}~\bibnamefont {Baraduc}}, \bibinfo {author} {\bibfnamefont {F.}~\bibnamefont {Lefloch}}, \ and\ \bibinfo {author} {\bibfnamefont {R.}~\bibnamefont {Calemczuk}},\ }\bibfield  {title} {\enquote {\bibinfo {title} {Temperature-induced crossover between $0$ and $\ensuremath{\pi}$ states in {S/F/S} junctions},}\ }\href {\doibase 10.1103/PhysRevB.68.054531} {\bibfield  {journal} {\bibinfo  {journal} {Phys. Rev. B}\ }\textbf {\bibinfo {volume} {68}},\ \bibinfo {pages} {054531} (\bibinfo {year} {2003})}\BibitemShut {NoStop}%
\bibitem [{\citenamefont {Oboznov}\ \emph {et~al.}(2006)\citenamefont {Oboznov}, \citenamefont {Bol'ginov}, \citenamefont {Feofanov}, \citenamefont {Ryazanov},\ and\ \citenamefont {Buzdin}}]{Oboznov_PRL_2006}%
  \BibitemOpen
  \bibfield  {author} {\bibinfo {author} {\bibfnamefont {V.~A.}\ \bibnamefont {Oboznov}}, \bibinfo {author} {\bibfnamefont {V.~V.}\ \bibnamefont {Bol'ginov}}, \bibinfo {author} {\bibfnamefont {A.~K.}\ \bibnamefont {Feofanov}}, \bibinfo {author} {\bibfnamefont {V.~V.}\ \bibnamefont {Ryazanov}}, \ and\ \bibinfo {author} {\bibfnamefont {A.~I.}\ \bibnamefont {Buzdin}},\ }\bibfield  {title} {\enquote {\bibinfo {title} {{Thickness Dependence of the Josephson Ground States of Superconductor-Ferromagnet-Superconductor Junctions}},}\ }\href {\doibase 10.1103/PhysRevLett.96.197003} {\bibfield  {journal} {\bibinfo  {journal} {Phys. Rev. Lett.}\ }\textbf {\bibinfo {volume} {96}},\ \bibinfo {pages} {197003} (\bibinfo {year} {2006})}\BibitemShut {NoStop}%
\bibitem [{\citenamefont {Ruppelt}\ \emph {et~al.}(2015)\citenamefont {Ruppelt}, \citenamefont {Sickinger}, \citenamefont {Menditto}, \citenamefont {Goldobin}, \citenamefont {Koelle}, \citenamefont {Kleiner}, \citenamefont {Vavra},\ and\ \citenamefont {Kohlstedt}}]{Ruppelt_2015}%
  \BibitemOpen
  \bibfield  {author} {\bibinfo {author} {\bibfnamefont {N.}~\bibnamefont {Ruppelt}}, \bibinfo {author} {\bibfnamefont {H.}~\bibnamefont {Sickinger}}, \bibinfo {author} {\bibfnamefont {R.}~\bibnamefont {Menditto}}, \bibinfo {author} {\bibfnamefont {E.}~\bibnamefont {Goldobin}}, \bibinfo {author} {\bibfnamefont {D.}~\bibnamefont {Koelle}}, \bibinfo {author} {\bibfnamefont {R.}~\bibnamefont {Kleiner}}, \bibinfo {author} {\bibfnamefont {O.}~\bibnamefont {Vavra}}, \ and\ \bibinfo {author} {\bibfnamefont {H.}~\bibnamefont {Kohlstedt}},\ }\bibfield  {title} {\enquote {\bibinfo {title} {{Observation of 0–$\pi$ transition in {SIsFS J}osephson junctions}},}\ }\href {\doibase 10.1063/1.4905672} {\bibfield  {journal} {\bibinfo  {journal} {Appl. Phys. Lett.}\ }\textbf {\bibinfo {volume} {106}},\ \bibinfo {pages} {022602} (\bibinfo {year} {2015})}\BibitemShut {NoStop}%
\bibitem [{\citenamefont {Bolginov}\ \emph {et~al.}(2018)\citenamefont {Bolginov}, \citenamefont {Rossolenko}, \citenamefont {Shkarin}, \citenamefont {Oboznov},\ and\ \citenamefont {Ryazanov}}]{Bolginov_JLTP_2018}%
  \BibitemOpen
  \bibfield  {author} {\bibinfo {author} {\bibfnamefont {V.~V.}\ \bibnamefont {Bolginov}}, \bibinfo {author} {\bibfnamefont {A.~N.}\ \bibnamefont {Rossolenko}}, \bibinfo {author} {\bibfnamefont {A.~B.}\ \bibnamefont {Shkarin}}, \bibinfo {author} {\bibfnamefont {V.~A.}\ \bibnamefont {Oboznov}}, \ and\ \bibinfo {author} {\bibfnamefont {V.~V.}\ \bibnamefont {Ryazanov}},\ }\bibfield  {title} {\enquote {\bibinfo {title} {{Fabrication of Optimized Superconducting Phase Inverters Based on Superconductor--Ferromagnet--Superconductor} $\pi$-{J}unctions},}\ }\href {\doibase 10.1007/s10909-017-1843-6} {\bibfield  {journal} {\bibinfo  {journal} {J. Low Temp. Phys.}\ }\textbf {\bibinfo {volume} {190}},\ \bibinfo {pages} {302--314} (\bibinfo {year} {2018})}\BibitemShut {NoStop}%
\bibitem [{\citenamefont {Yamashita}, \citenamefont {Kawakami},\ and\ \citenamefont {Terai}(2017)}]{Yamashita_2017}%
  \BibitemOpen
  \bibfield  {author} {\bibinfo {author} {\bibfnamefont {T.}~\bibnamefont {Yamashita}}, \bibinfo {author} {\bibfnamefont {A.}~\bibnamefont {Kawakami}}, \ and\ \bibinfo {author} {\bibfnamefont {H.}~\bibnamefont {Terai}},\ }\bibfield  {title} {\enquote {\bibinfo {title} {Nb{N-Based F}erromagnetic 0 and $\ensuremath{\pi}$ {Josephson J}unctions},}\ }\href {\doibase 10.1103/PhysRevApplied.8.054028} {\bibfield  {journal} {\bibinfo  {journal} {Phys. Rev. Appl.}\ }\textbf {\bibinfo {volume} {8}},\ \bibinfo {pages} {054028} (\bibinfo {year} {2017})}\BibitemShut {NoStop}%
\bibitem [{\citenamefont {Zeng}\ \emph {et~al.}(2022)\citenamefont {Zeng}, \citenamefont {Chen}, \citenamefont {Zhong}, \citenamefont {Wang}, \citenamefont {Pan}, \citenamefont {Zhang}, \citenamefont {Yu}, \citenamefont {Wu}, \citenamefont {Zhang}, \citenamefont {Peng} \emph {et~al.}}]{zeng_2022}%
  \BibitemOpen
  \bibfield  {author} {\bibinfo {author} {\bibfnamefont {J.}~\bibnamefont {Zeng}}, \bibinfo {author} {\bibfnamefont {L.}~\bibnamefont {Chen}}, \bibinfo {author} {\bibfnamefont {X.}~\bibnamefont {Zhong}}, \bibinfo {author} {\bibfnamefont {Y.}~\bibnamefont {Wang}}, \bibinfo {author} {\bibfnamefont {Y.}~\bibnamefont {Pan}}, \bibinfo {author} {\bibfnamefont {D.}~\bibnamefont {Zhang}}, \bibinfo {author} {\bibfnamefont {S.}~\bibnamefont {Yu}}, \bibinfo {author} {\bibfnamefont {L.}~\bibnamefont {Wu}}, \bibinfo {author} {\bibfnamefont {L.}~\bibnamefont {Zhang}}, \bibinfo {author} {\bibfnamefont {W.}~\bibnamefont {Peng}},  \emph {et~al.},\ }\bibfield  {title} {\enquote {\bibinfo {title} {Nonvolatile memory cell using a superconducting-ferromagnetic $\pi$ {J}osephson junction},}\ }\href {\doibase 10.1088/1361-6668/ac80d9} {\bibfield  {journal} {\bibinfo  {journal} {Supercond. Sci. Technol.}\ }\textbf {\bibinfo {volume} {35}},\ \bibinfo {pages} {105009} (\bibinfo {year} {2022})}\BibitemShut {NoStop}%
\bibitem [{\citenamefont {Pham}\ \emph {et~al.}(2022)\citenamefont {Pham}, \citenamefont {Sugimoto}, \citenamefont {Oba}, \citenamefont {Takeshita}, \citenamefont {Li}, \citenamefont {Tanaka}, \citenamefont {Yamashita},\ and\ \citenamefont {Fujimaki}}]{Pham_2022}%
  \BibitemOpen
  \bibfield  {author} {\bibinfo {author} {\bibfnamefont {D.}~\bibnamefont {Pham}}, \bibinfo {author} {\bibfnamefont {R.}~\bibnamefont {Sugimoto}}, \bibinfo {author} {\bibfnamefont {K.}~\bibnamefont {Oba}}, \bibinfo {author} {\bibfnamefont {Y.}~\bibnamefont {Takeshita}}, \bibinfo {author} {\bibfnamefont {F.}~\bibnamefont {Li}}, \bibinfo {author} {\bibfnamefont {M.}~\bibnamefont {Tanaka}}, \bibinfo {author} {\bibfnamefont {T.}~\bibnamefont {Yamashita}}, \ and\ \bibinfo {author} {\bibfnamefont {A.}~\bibnamefont {Fujimaki}},\ }\bibfield  {title} {\enquote {\bibinfo {title} {Weak spin-flip scattering in {P}d$_{89}${N}i$_{11}$ interlayer of {NbN}-based ferromagnetic {J}osephson junctions},}\ }\href {\doibase 10.1038/s41598-022-10967-6} {\bibfield  {journal} {\bibinfo  {journal} {Sci. Rep.}\ }\textbf {\bibinfo {volume} {12}},\ \bibinfo {pages} {6863} (\bibinfo {year} {2022})}\BibitemShut {NoStop}%
\bibitem [{\citenamefont {Stolyarov}\ \emph {et~al.}(2022)\citenamefont {Stolyarov}, \citenamefont {Oboznov}, \citenamefont {Kasatonov}, \citenamefont {Neilo}, \citenamefont {Bakurskiy}, \citenamefont {Klenov}, \citenamefont {Soloviev}, \citenamefont {Kupriyanov}, \citenamefont {Golubov}, \citenamefont {Cren} \emph {et~al.}}]{Stolyarov_2022}%
  \BibitemOpen
  \bibfield  {author} {\bibinfo {author} {\bibfnamefont {V.}~\bibnamefont {Stolyarov}}, \bibinfo {author} {\bibfnamefont {V.}~\bibnamefont {Oboznov}}, \bibinfo {author} {\bibfnamefont {D.}~\bibnamefont {Kasatonov}}, \bibinfo {author} {\bibfnamefont {A.}~\bibnamefont {Neilo}}, \bibinfo {author} {\bibfnamefont {S.}~\bibnamefont {Bakurskiy}}, \bibinfo {author} {\bibfnamefont {N.}~\bibnamefont {Klenov}}, \bibinfo {author} {\bibfnamefont {I.}~\bibnamefont {Soloviev}}, \bibinfo {author} {\bibfnamefont {M.}~\bibnamefont {Kupriyanov}}, \bibinfo {author} {\bibfnamefont {A.}~\bibnamefont {Golubov}}, \bibinfo {author} {\bibfnamefont {T.}~\bibnamefont {Cren}},  \emph {et~al.},\ }\bibfield  {title} {\enquote {\bibinfo {title} {Effective {Exchange Energy in a Thin, Spatially Inhomogeneous CuNi Layer Proximized by N}b},}\ }\href {\doibase 10.1021/acs.jpclett.2c00978} {\bibfield  {journal} {\bibinfo  {journal} {J. Phys. Chem. Lett.}\ }\textbf {\bibinfo {volume} {13}},\ \bibinfo {pages} {6400--6406} (\bibinfo {year}
  {2022})}\BibitemShut {NoStop}%
\bibitem [{\citenamefont {Weides}\ \emph {et~al.}(2006{\natexlab{b}})\citenamefont {Weides}, \citenamefont {Kemmler}, \citenamefont {Kohlstedt}, \citenamefont {Waser}, \citenamefont {Koelle}, \citenamefont {Kleiner},\ and\ \citenamefont {Goldobin}}]{Weides_PRL_2006}%
  \BibitemOpen
  \bibfield  {author} {\bibinfo {author} {\bibfnamefont {M.}~\bibnamefont {Weides}}, \bibinfo {author} {\bibfnamefont {M.}~\bibnamefont {Kemmler}}, \bibinfo {author} {\bibfnamefont {H.}~\bibnamefont {Kohlstedt}}, \bibinfo {author} {\bibfnamefont {R.}~\bibnamefont {Waser}}, \bibinfo {author} {\bibfnamefont {D.}~\bibnamefont {Koelle}}, \bibinfo {author} {\bibfnamefont {R.}~\bibnamefont {Kleiner}}, \ and\ \bibinfo {author} {\bibfnamefont {E.}~\bibnamefont {Goldobin}},\ }\bibfield  {title} {\enquote {\bibinfo {title} {$0\mathrm{\text{\ensuremath{-}}}\ensuremath{\pi}$ {Josephson Tunnel Junctions with Ferromagnetic Barrier}},}\ }\href {\doibase 10.1103/PhysRevLett.97.247001} {\bibfield  {journal} {\bibinfo  {journal} {Phys. Rev. Lett.}\ }\textbf {\bibinfo {volume} {97}},\ \bibinfo {pages} {247001} (\bibinfo {year} {2006}{\natexlab{b}})}\BibitemShut {NoStop}%
\bibitem [{\citenamefont {Born}\ \emph {et~al.}(2006)\citenamefont {Born}, \citenamefont {Siegel}, \citenamefont {Hollmann}, \citenamefont {Braak}, \citenamefont {Golubov}, \citenamefont {Gusakova},\ and\ \citenamefont {Kupriyanov}}]{Born_PRB_2006}%
  \BibitemOpen
  \bibfield  {author} {\bibinfo {author} {\bibfnamefont {F.}~\bibnamefont {Born}}, \bibinfo {author} {\bibfnamefont {M.}~\bibnamefont {Siegel}}, \bibinfo {author} {\bibfnamefont {E.~K.}\ \bibnamefont {Hollmann}}, \bibinfo {author} {\bibfnamefont {H.}~\bibnamefont {Braak}}, \bibinfo {author} {\bibfnamefont {A.~A.}\ \bibnamefont {Golubov}}, \bibinfo {author} {\bibfnamefont {D.~Y.}\ \bibnamefont {Gusakova}}, \ and\ \bibinfo {author} {\bibfnamefont {M.~Y.}\ \bibnamefont {Kupriyanov}},\ }\bibfield  {title} {\enquote {\bibinfo {title} {Multiple $0\text{\ensuremath{-}}\ensuremath{\pi}$ transitions in superconductor/insulator/ferromagnet/superconductor {J}osephson tunnel junctions},}\ }\href {\doibase 10.1103/PhysRevB.74.140501} {\bibfield  {journal} {\bibinfo  {journal} {Phys. Rev. B}\ }\textbf {\bibinfo {volume} {74}},\ \bibinfo {pages} {140501(R)} (\bibinfo {year} {2006})}\BibitemShut {NoStop}%
\bibitem [{\citenamefont {Sprungmann}\ \emph {et~al.}(2009)\citenamefont {Sprungmann}, \citenamefont {Westerholt}, \citenamefont {Zabel}, \citenamefont {Weides},\ and\ \citenamefont {Kohlstedt}}]{Sprungmann_2009}%
  \BibitemOpen
  \bibfield  {author} {\bibinfo {author} {\bibfnamefont {D.}~\bibnamefont {Sprungmann}}, \bibinfo {author} {\bibfnamefont {K.}~\bibnamefont {Westerholt}}, \bibinfo {author} {\bibfnamefont {H.}~\bibnamefont {Zabel}}, \bibinfo {author} {\bibfnamefont {M.}~\bibnamefont {Weides}}, \ and\ \bibinfo {author} {\bibfnamefont {H.}~\bibnamefont {Kohlstedt}},\ }\bibfield  {title} {\enquote {\bibinfo {title} {Josephson tunnel junctions with ferromagnetic {F}e$_{0.75}${C}o$_{0.25}$ barriers},}\ }\href {\doibase 10.1088/0022-3727/42/7/075005} {\bibfield  {journal} {\bibinfo  {journal} {J. Phys. D: Appl. Phys.}\ }\textbf {\bibinfo {volume} {42}},\ \bibinfo {pages} {075005} (\bibinfo {year} {2009})}\BibitemShut {NoStop}%
\bibitem [{\citenamefont {Satchell}\ \emph {et~al.}(2021)\citenamefont {Satchell}, \citenamefont {Mitchell}, \citenamefont {Shepley}, \citenamefont {Darwin}, \citenamefont {Hickey},\ and\ \citenamefont {Burnell}}]{Satchell_SciRep_2021}%
  \BibitemOpen
  \bibfield  {author} {\bibinfo {author} {\bibfnamefont {N.}~\bibnamefont {Satchell}}, \bibinfo {author} {\bibfnamefont {T.}~\bibnamefont {Mitchell}}, \bibinfo {author} {\bibfnamefont {P.}~\bibnamefont {Shepley}}, \bibinfo {author} {\bibfnamefont {E.}~\bibnamefont {Darwin}}, \bibinfo {author} {\bibfnamefont {B.}~\bibnamefont {Hickey}}, \ and\ \bibinfo {author} {\bibfnamefont {G.}~\bibnamefont {Burnell}},\ }\bibfield  {title} {\enquote {\bibinfo {title} {Pt and {CoB} trilayer {J}osephson $\pi$ junctions with perpendicular magnetic anisotropy},}\ }\href {\doibase 10.1038/s41598-021-90432-y} {\bibfield  {journal} {\bibinfo  {journal} {Sci. Rep.}\ }\textbf {\bibinfo {volume} {11}},\ \bibinfo {pages} {11173} (\bibinfo {year} {2021})}\BibitemShut {NoStop}%
\bibitem [{\citenamefont {Loving}\ \emph {et~al.}(2021)\citenamefont {Loving}, \citenamefont {Ambrose}, \citenamefont {Keebaugh}, \citenamefont {Miller}, \citenamefont {Pownall}, \citenamefont {Rizzo}, \citenamefont {Sidorov},\ and\ \citenamefont {Siwak}}]{Loving_arXiv_2021}%
  \BibitemOpen
  \bibfield  {author} {\bibinfo {author} {\bibfnamefont {M.~G.}\ \bibnamefont {Loving}}, \bibinfo {author} {\bibfnamefont {T.~F.}\ \bibnamefont {Ambrose}}, \bibinfo {author} {\bibfnamefont {S.}~\bibnamefont {Keebaugh}}, \bibinfo {author} {\bibfnamefont {D.~L.}\ \bibnamefont {Miller}}, \bibinfo {author} {\bibfnamefont {R.}~\bibnamefont {Pownall}}, \bibinfo {author} {\bibfnamefont {N.~D.}\ \bibnamefont {Rizzo}}, \bibinfo {author} {\bibfnamefont {A.~N.}\ \bibnamefont {Sidorov}}, \ and\ \bibinfo {author} {\bibfnamefont {N.~P.}\ \bibnamefont {Siwak}},\ }\bibfield  {title} {\enquote {\bibinfo {title} {{Magnetic Reversal and Critical Current Transparency of CoFeB Superconductor-Ferromagnet-Superconductor H}eterostructures},}\ }\href {https://arxiv.org/abs/2104.03188} {\bibfield  {journal} {\bibinfo  {journal} {arXiv:2104.03188}\ } (\bibinfo {year} {2021})}\BibitemShut {NoStop}%
\bibitem [{\citenamefont {Komori}\ \emph {et~al.}(2022)\citenamefont {Komori}, \citenamefont {Thompson}, \citenamefont {Yang}, \citenamefont {Kimbell}, \citenamefont {Stelmashenko}, \citenamefont {Blamire},\ and\ \citenamefont {Robinson}}]{Komori_PRApp_2022}%
  \BibitemOpen
  \bibfield  {author} {\bibinfo {author} {\bibfnamefont {S.}~\bibnamefont {Komori}}, \bibinfo {author} {\bibfnamefont {J.~E.}\ \bibnamefont {Thompson}}, \bibinfo {author} {\bibfnamefont {G.}~\bibnamefont {Yang}}, \bibinfo {author} {\bibfnamefont {G.}~\bibnamefont {Kimbell}}, \bibinfo {author} {\bibfnamefont {N.}~\bibnamefont {Stelmashenko}}, \bibinfo {author} {\bibfnamefont {M.~G.}\ \bibnamefont {Blamire}}, \ and\ \bibinfo {author} {\bibfnamefont {J.~W.~A.}\ \bibnamefont {Robinson}},\ }\bibfield  {title} {\enquote {\bibinfo {title} {Enhancement of {Josephson Critical Currents in F}erromagnetic {C}o$_{40}${F}e$_{40}${B}$_{20}$ by {Thermal A}nnealing},}\ }\href {\doibase 10.1103/PhysRevApplied.17.L021002} {\bibfield  {journal} {\bibinfo  {journal} {Phys. Rev. Appl.}\ }\textbf {\bibinfo {volume} {17}},\ \bibinfo {pages} {L021002} (\bibinfo {year} {2022})}\BibitemShut {NoStop}%
\bibitem [{\citenamefont {Madden}(2022)}]{madden_thesis_2022}%
  \BibitemOpen
  \bibfield  {author} {\bibinfo {author} {\bibfnamefont {A.~E.}\ \bibnamefont {Madden}},\ }\href@noop {} {\emph {\bibinfo {title} {Ferromagnetic and Ferrimagnetic Materials as Josephson Junction Barriers}}}\ (\bibinfo  {publisher} {Michigan State University},\ \bibinfo {year} {2022})\BibitemShut {NoStop}%
\bibitem [{\citenamefont {Caruso}\ \emph {et~al.}(2019)\citenamefont {Caruso}, \citenamefont {Massarotti}, \citenamefont {Campagnano}, \citenamefont {Pal}, \citenamefont {Ahmad}, \citenamefont {Lucignano}, \citenamefont {Eschrig}, \citenamefont {Blamire},\ and\ \citenamefont {Tafuri}}]{caruso_2019}%
  \BibitemOpen
  \bibfield  {author} {\bibinfo {author} {\bibfnamefont {R.}~\bibnamefont {Caruso}}, \bibinfo {author} {\bibfnamefont {D.}~\bibnamefont {Massarotti}}, \bibinfo {author} {\bibfnamefont {G.}~\bibnamefont {Campagnano}}, \bibinfo {author} {\bibfnamefont {A.}~\bibnamefont {Pal}}, \bibinfo {author} {\bibfnamefont {H.~G.}\ \bibnamefont {Ahmad}}, \bibinfo {author} {\bibfnamefont {P.}~\bibnamefont {Lucignano}}, \bibinfo {author} {\bibfnamefont {M.}~\bibnamefont {Eschrig}}, \bibinfo {author} {\bibfnamefont {M.~G.}\ \bibnamefont {Blamire}}, \ and\ \bibinfo {author} {\bibfnamefont {F.}~\bibnamefont {Tafuri}},\ }\bibfield  {title} {\enquote {\bibinfo {title} {Tuning of {Magnetic Activity in S}pin-{F}ilter {Josephson} {Junctions Towards S}pin-{Triplet T}ransport},}\ }\href {\doibase 10.1103/PhysRevLett.122.047002} {\bibfield  {journal} {\bibinfo  {journal} {Phys. Rev. Lett.}\ }\textbf {\bibinfo {volume} {122}},\ \bibinfo {pages} {047002} (\bibinfo {year} {2019})}\BibitemShut {NoStop}%
\bibitem [{\citenamefont {Kang}\ \emph {et~al.}(2022)\citenamefont {Kang}, \citenamefont {Berger}, \citenamefont {Watanabe}, \citenamefont {Taniguchi}, \citenamefont {Forr{\'o}}, \citenamefont {Shan},\ and\ \citenamefont {Mak}}]{kang_2022}%
  \BibitemOpen
  \bibfield  {author} {\bibinfo {author} {\bibfnamefont {K.}~\bibnamefont {Kang}}, \bibinfo {author} {\bibfnamefont {H.}~\bibnamefont {Berger}}, \bibinfo {author} {\bibfnamefont {K.}~\bibnamefont {Watanabe}}, \bibinfo {author} {\bibfnamefont {T.}~\bibnamefont {Taniguchi}}, \bibinfo {author} {\bibfnamefont {L.}~\bibnamefont {Forr{\'o}}}, \bibinfo {author} {\bibfnamefont {J.}~\bibnamefont {Shan}}, \ and\ \bibinfo {author} {\bibfnamefont {K.~F.}\ \bibnamefont {Mak}},\ }\bibfield  {title} {\enquote {\bibinfo {title} {van der {W}aals $\pi$ {Josephson J}unctions},}\ }\href {\doibase 10.1021/acs.nanolett.2c01640} {\bibfield  {journal} {\bibinfo  {journal} {Nano Lett.}\ }\textbf {\bibinfo {volume} {22}},\ \bibinfo {pages} {5510--5515} (\bibinfo {year} {2022})}\BibitemShut {NoStop}%
\bibitem [{\citenamefont {Lavrijsen}\ \emph {et~al.}(2010)\citenamefont {Lavrijsen}, \citenamefont {Malinowski}, \citenamefont {Franken}, \citenamefont {Kohlhepp}, \citenamefont {Swagten}, \citenamefont {Koopmans}, \citenamefont {Czapkiewicz},\ and\ \citenamefont {Stobiecki}}]{Lavrijsen_2010}%
  \BibitemOpen
  \bibfield  {author} {\bibinfo {author} {\bibfnamefont {R.}~\bibnamefont {Lavrijsen}}, \bibinfo {author} {\bibfnamefont {G.}~\bibnamefont {Malinowski}}, \bibinfo {author} {\bibfnamefont {J.~H.}\ \bibnamefont {Franken}}, \bibinfo {author} {\bibfnamefont {J.~T.}\ \bibnamefont {Kohlhepp}}, \bibinfo {author} {\bibfnamefont {H.~J.~M.}\ \bibnamefont {Swagten}}, \bibinfo {author} {\bibfnamefont {B.}~\bibnamefont {Koopmans}}, \bibinfo {author} {\bibfnamefont {M.}~\bibnamefont {Czapkiewicz}}, \ and\ \bibinfo {author} {\bibfnamefont {T.}~\bibnamefont {Stobiecki}},\ }\bibfield  {title} {\enquote {\bibinfo {title} {Reduced domain wall pinning in ultrathin {P}t/{C}o$_{100-x}${B}$_x$/{P}t with perpendicular magnetic anisotropy},}\ }\href {\doibase 10.1063/1.3280373} {\bibfield  {journal} {\bibinfo  {journal} {Appl. Phys. Lett.}\ }\textbf {\bibinfo {volume} {96}},\ \bibinfo {pages} {022501} (\bibinfo {year} {2010})}\BibitemShut {NoStop}%
\bibitem [{\citenamefont {Ikeda}\ \emph {et~al.}(2008)\citenamefont {Ikeda}, \citenamefont {Hayakawa}, \citenamefont {Ashizawa}, \citenamefont {Lee}, \citenamefont {Miura}, \citenamefont {Hasegawa}, \citenamefont {Tsunoda}, \citenamefont {Matsukura},\ and\ \citenamefont {Ohno}}]{Ikeda_2008}%
  \BibitemOpen
  \bibfield  {author} {\bibinfo {author} {\bibfnamefont {S.}~\bibnamefont {Ikeda}}, \bibinfo {author} {\bibfnamefont {J.}~\bibnamefont {Hayakawa}}, \bibinfo {author} {\bibfnamefont {Y.}~\bibnamefont {Ashizawa}}, \bibinfo {author} {\bibfnamefont {Y.~M.}\ \bibnamefont {Lee}}, \bibinfo {author} {\bibfnamefont {K.}~\bibnamefont {Miura}}, \bibinfo {author} {\bibfnamefont {H.}~\bibnamefont {Hasegawa}}, \bibinfo {author} {\bibfnamefont {M.}~\bibnamefont {Tsunoda}}, \bibinfo {author} {\bibfnamefont {F.}~\bibnamefont {Matsukura}}, \ and\ \bibinfo {author} {\bibfnamefont {H.}~\bibnamefont {Ohno}},\ }\bibfield  {title} {\enquote {\bibinfo {title} {{Tunnel magnetoresistance of 604$\%$ at 300K by suppression of Ta diffusion in CoFeB/MgO/CoFeB pseudo-spin-valves annealed at high temperature}},}\ }\href {\doibase 10.1063/1.2976435} {\bibfield  {journal} {\bibinfo  {journal} {Appl. Phys. Lett.}\ }\textbf {\bibinfo {volume} {93}},\ \bibinfo {pages} {082508} (\bibinfo {year} {2008})}\BibitemShut {NoStop}%
\bibitem [{\citenamefont {Naylor}, \citenamefont {Burnell},\ and\ \citenamefont {Hickey}(2012)}]{Naylor2012}%
  \BibitemOpen
  \bibfield  {author} {\bibinfo {author} {\bibfnamefont {A.~D.}\ \bibnamefont {Naylor}}, \bibinfo {author} {\bibfnamefont {G.}~\bibnamefont {Burnell}}, \ and\ \bibinfo {author} {\bibfnamefont {B.~J.}\ \bibnamefont {Hickey}},\ }\bibfield  {title} {\enquote {\bibinfo {title} {Transport spin polarization of the rare-earth transition-metal alloy {C}o${}_{1\ensuremath{-}x}${G}d${}_{x}$},}\ }\href {\doibase 10.1103/PhysRevB.85.064410} {\bibfield  {journal} {\bibinfo  {journal} {Phys. Rev. B}\ }\textbf {\bibinfo {volume} {85}},\ \bibinfo {pages} {064410} (\bibinfo {year} {2012})}\BibitemShut {NoStop}%
\bibitem [{\citenamefont {Jimbo}\ \emph {et~al.}(1997)\citenamefont {Jimbo}, \citenamefont {Komiyama}, \citenamefont {Shirota}, \citenamefont {Fujiwara}, \citenamefont {Tsunashima},\ and\ \citenamefont {Matsuura}}]{Jimbo1997}%
  \BibitemOpen
  \bibfield  {author} {\bibinfo {author} {\bibfnamefont {M.}~\bibnamefont {Jimbo}}, \bibinfo {author} {\bibfnamefont {K.}~\bibnamefont {Komiyama}}, \bibinfo {author} {\bibfnamefont {Y.}~\bibnamefont {Shirota}}, \bibinfo {author} {\bibfnamefont {Y.}~\bibnamefont {Fujiwara}}, \bibinfo {author} {\bibfnamefont {S.}~\bibnamefont {Tsunashima}}, \ and\ \bibinfo {author} {\bibfnamefont {M.}~\bibnamefont {Matsuura}},\ }\bibfield  {title} {\enquote {\bibinfo {title} {Thermal stability of spin valves using amorphous {CoFeB}},}\ }\href {\doibase https://doi.org/10.1016/S0304-8853(96)00538-0} {\bibfield  {journal} {\bibinfo  {journal} {J. Magn. Magn. Mater.}\ }\textbf {\bibinfo {volume} {165}},\ \bibinfo {pages} {308--311} (\bibinfo {year} {1997})},\ \bibinfo {note} {symposium E: Magnetic Ultrathin Films, Multilayers and Surfaces}\BibitemShut {NoStop}%
\bibitem [{\citenamefont {Vecchione}\ \emph {et~al.}(2011)\citenamefont {Vecchione}, \citenamefont {Fittipaldi}, \citenamefont {Cirillo}, \citenamefont {Hesselberth}, \citenamefont {Aarts}, \citenamefont {Prischepa}, \citenamefont {Kushnir}, \citenamefont {Kupriyanov},\ and\ \citenamefont {Attanasio}}]{Vecchione_2011}%
  \BibitemOpen
  \bibfield  {author} {\bibinfo {author} {\bibfnamefont {A.}~\bibnamefont {Vecchione}}, \bibinfo {author} {\bibfnamefont {R.}~\bibnamefont {Fittipaldi}}, \bibinfo {author} {\bibfnamefont {C.}~\bibnamefont {Cirillo}}, \bibinfo {author} {\bibfnamefont {M.}~\bibnamefont {Hesselberth}}, \bibinfo {author} {\bibfnamefont {J.}~\bibnamefont {Aarts}}, \bibinfo {author} {\bibfnamefont {S.}~\bibnamefont {Prischepa}}, \bibinfo {author} {\bibfnamefont {V.}~\bibnamefont {Kushnir}}, \bibinfo {author} {\bibfnamefont {M.~Y.}\ \bibnamefont {Kupriyanov}}, \ and\ \bibinfo {author} {\bibfnamefont {C.}~\bibnamefont {Attanasio}},\ }\bibfield  {title} {\enquote {\bibinfo {title} {X-ray scattering study of interfacial roughness in {N}b/{PdN}i multilayers},}\ }\href {\doibase 10.1016/j.susc.2011.06.013} {\bibfield  {journal} {\bibinfo  {journal} {Surf. Sci.}\ }\textbf {\bibinfo {volume} {605}},\ \bibinfo {pages} {1791--1796} (\bibinfo {year} {2011})}\BibitemShut {NoStop}%
\bibitem [{\citenamefont {Khaydukov}\ \emph {et~al.}(2015)\citenamefont {Khaydukov}, \citenamefont {Morari}, \citenamefont {Soltwedel}, \citenamefont {Keller}, \citenamefont {Christiani}, \citenamefont {Logvenov}, \citenamefont {Kupriyanov}, \citenamefont {Sidorenko},\ and\ \citenamefont {Keimer}}]{Khaydukov_2015}%
  \BibitemOpen
  \bibfield  {author} {\bibinfo {author} {\bibfnamefont {Y.}~\bibnamefont {Khaydukov}}, \bibinfo {author} {\bibfnamefont {R.}~\bibnamefont {Morari}}, \bibinfo {author} {\bibfnamefont {O.}~\bibnamefont {Soltwedel}}, \bibinfo {author} {\bibfnamefont {T.}~\bibnamefont {Keller}}, \bibinfo {author} {\bibfnamefont {G.}~\bibnamefont {Christiani}}, \bibinfo {author} {\bibfnamefont {G.}~\bibnamefont {Logvenov}}, \bibinfo {author} {\bibfnamefont {M.}~\bibnamefont {Kupriyanov}}, \bibinfo {author} {\bibfnamefont {A.}~\bibnamefont {Sidorenko}}, \ and\ \bibinfo {author} {\bibfnamefont {B.}~\bibnamefont {Keimer}},\ }\bibfield  {title} {\enquote {\bibinfo {title} {Interfacial roughness and proximity effects in superconductor/ferromagnet {CuN}i/{N}b heterostructures},}\ }\href {\doibase 10.1063/1.4936789} {\bibfield  {journal} {\bibinfo  {journal} {J. Appl. Phys.}\ }\textbf {\bibinfo {volume} {118}},\ \bibinfo {pages} {213905} (\bibinfo {year} {2015})}\BibitemShut {NoStop}%
\bibitem [{\citenamefont {Garcia}\ \emph {et~al.}(2009)\citenamefont {Garcia}, \citenamefont {Fernandez~Pinel}, \citenamefont {De~la Venta}, \citenamefont {Quesada}, \citenamefont {Bouzas}, \citenamefont {Fern{\'a}ndez}, \citenamefont {Romero}, \citenamefont {Gonz{\'a}lez},\ and\ \citenamefont {Costa-Kr{\"a}mer}}]{Garcia_2009}%
  \BibitemOpen
  \bibfield  {author} {\bibinfo {author} {\bibfnamefont {M.}~\bibnamefont {Garcia}}, \bibinfo {author} {\bibfnamefont {E.}~\bibnamefont {Fernandez~Pinel}}, \bibinfo {author} {\bibfnamefont {J.}~\bibnamefont {De~la Venta}}, \bibinfo {author} {\bibfnamefont {A.}~\bibnamefont {Quesada}}, \bibinfo {author} {\bibfnamefont {V.}~\bibnamefont {Bouzas}}, \bibinfo {author} {\bibfnamefont {J.}~\bibnamefont {Fern{\'a}ndez}}, \bibinfo {author} {\bibfnamefont {J.}~\bibnamefont {Romero}}, \bibinfo {author} {\bibfnamefont {M.}~\bibnamefont {Gonz{\'a}lez}}, \ and\ \bibinfo {author} {\bibfnamefont {J.}~\bibnamefont {Costa-Kr{\"a}mer}},\ }\bibfield  {title} {\enquote {\bibinfo {title} {{Sources of experimental errors in the observation of nanoscale magnetism}},}\ }\href {\doibase 10.1063/1.3060808} {\bibfield  {journal} {\bibinfo  {journal} {J. Appl. Phys.}\ }\textbf {\bibinfo {volume} {105}},\ \bibinfo {pages} {013925} (\bibinfo {year} {2009})}\BibitemShut {NoStop}%
\bibitem [{\citenamefont {Satchell}\ and\ \citenamefont {Birge}(2018)}]{satchellSOC2018}%
  \BibitemOpen
  \bibfield  {author} {\bibinfo {author} {\bibfnamefont {N.}~\bibnamefont {Satchell}}\ and\ \bibinfo {author} {\bibfnamefont {N.~O.}\ \bibnamefont {Birge}},\ }\bibfield  {title} {\enquote {\bibinfo {title} {Supercurrent in ferromagnetic {J}osephson junctions with heavy metal interlayers},}\ }\href {\doibase 10.1103/PhysRevB.97.214509} {\bibfield  {journal} {\bibinfo  {journal} {Phys. Rev. B}\ }\textbf {\bibinfo {volume} {97}},\ \bibinfo {pages} {214509} (\bibinfo {year} {2018})}\BibitemShut {NoStop}%
\bibitem [{\citenamefont {Bass}\ and\ \citenamefont {Pratt}(2007)}]{bass_2007}%
  \BibitemOpen
  \bibfield  {author} {\bibinfo {author} {\bibfnamefont {J.}~\bibnamefont {Bass}}\ and\ \bibinfo {author} {\bibfnamefont {W.~P.}\ \bibnamefont {Pratt}},\ }\bibfield  {title} {\enquote {\bibinfo {title} {Spin-diffusion lengths in metals and alloys, and spin-flipping at metal/metal interfaces: an experimentalist’s critical review},}\ }\href {\doibase 10.1088/0953-8984/19/18/183201} {\bibfield  {journal} {\bibinfo  {journal} {J. Phys. Condens. Matter}\ }\textbf {\bibinfo {volume} {19}},\ \bibinfo {pages} {183201} (\bibinfo {year} {2007})}\BibitemShut {NoStop}%
\bibitem [{\citenamefont {Thomas}, \citenamefont {Ulmer},\ and\ \citenamefont {Ketterson}(1998)}]{Thomas_1998}%
  \BibitemOpen
  \bibfield  {author} {\bibinfo {author} {\bibfnamefont {C.~D.}\ \bibnamefont {Thomas}}, \bibinfo {author} {\bibfnamefont {M.~P.}\ \bibnamefont {Ulmer}}, \ and\ \bibinfo {author} {\bibfnamefont {J.~B.}\ \bibnamefont {Ketterson}},\ }\bibfield  {title} {\enquote {\bibinfo {title} {Superconducting tunnel junction base electrode planarization},}\ }\href {\doibase 10.1063/1.368037} {\bibfield  {journal} {\bibinfo  {journal} {J. Appl. Phys.}\ }\textbf {\bibinfo {volume} {84}},\ \bibinfo {pages} {364--367} (\bibinfo {year} {1998})}\BibitemShut {NoStop}%
\bibitem [{\citenamefont {Wang}, \citenamefont {Pratt},\ and\ \citenamefont {Birge}(2012)}]{Wang_2012}%
  \BibitemOpen
  \bibfield  {author} {\bibinfo {author} {\bibfnamefont {Y.}~\bibnamefont {Wang}}, \bibinfo {author} {\bibfnamefont {W.~P.}\ \bibnamefont {Pratt}}, \ and\ \bibinfo {author} {\bibfnamefont {N.~O.}\ \bibnamefont {Birge}},\ }\bibfield  {title} {\enquote {\bibinfo {title} {Area-dependence of spin-triplet supercurrent in ferromagnetic {Josephson} junctions},}\ }\href {\doibase 10.1103/PhysRevB.85.214522} {\bibfield  {journal} {\bibinfo  {journal} {Phys. Rev. B}\ }\textbf {\bibinfo {volume} {85}},\ \bibinfo {pages} {214522} (\bibinfo {year} {2012})}\BibitemShut {NoStop}%
\bibitem [{\citenamefont {Quarterman}\ \emph {et~al.}(2020)\citenamefont {Quarterman}, \citenamefont {Satchell}, \citenamefont {Kirby}, \citenamefont {Loloee}, \citenamefont {Burnell}, \citenamefont {Birge},\ and\ \citenamefont {Borchers}}]{quarterman2020distortions}%
  \BibitemOpen
  \bibfield  {author} {\bibinfo {author} {\bibfnamefont {P.}~\bibnamefont {Quarterman}}, \bibinfo {author} {\bibfnamefont {N.}~\bibnamefont {Satchell}}, \bibinfo {author} {\bibfnamefont {B.~J.}\ \bibnamefont {Kirby}}, \bibinfo {author} {\bibfnamefont {R.}~\bibnamefont {Loloee}}, \bibinfo {author} {\bibfnamefont {G.}~\bibnamefont {Burnell}}, \bibinfo {author} {\bibfnamefont {N.~O.}\ \bibnamefont {Birge}}, \ and\ \bibinfo {author} {\bibfnamefont {J.~A.}\ \bibnamefont {Borchers}},\ }\bibfield  {title} {\enquote {\bibinfo {title} {Distortions to the penetration depth and coherence length of superconductor/normal-metal superlattices},}\ }\href {\doibase 10.1103/PhysRevMaterials.4.074801} {\bibfield  {journal} {\bibinfo  {journal} {Phys. Rev. Materials}\ }\textbf {\bibinfo {volume} {4}},\ \bibinfo {pages} {074801} (\bibinfo {year} {2020})}\BibitemShut {NoStop}%
\bibitem [{\citenamefont {{Bell}}\ \emph {et~al.}(2005)\citenamefont {{Bell}}, \citenamefont {{Burnell}}, \citenamefont {{C.W. Leung}}, \citenamefont {{Tarte}},\ and\ \citenamefont {{Blamire}}}]{Bell_IEEE_2005}%
  \BibitemOpen
  \bibfield  {author} {\bibinfo {author} {\bibfnamefont {C.}~\bibnamefont {{Bell}}}, \bibinfo {author} {\bibfnamefont {G.}~\bibnamefont {{Burnell}}}, \bibinfo {author} {\bibnamefont {{C.W. Leung}}}, \bibinfo {author} {\bibfnamefont {E.~J.}\ \bibnamefont {{Tarte}}}, \ and\ \bibinfo {author} {\bibfnamefont {M.~G.}\ \bibnamefont {{Blamire}}},\ }\bibfield  {title} {\enquote {\bibinfo {title} {Spin valve {Josephson} junctions},}\ }\href {\doibase 10.1109/TASC.2005.850112} {\bibfield  {journal} {\bibinfo  {journal} {IEEE Trans. Appl. Supercond.}\ }\textbf {\bibinfo {volume} {15}},\ \bibinfo {pages} {908--911} (\bibinfo {year} {2005})}\BibitemShut {NoStop}%
\bibitem [{\citenamefont {Baek}\ \emph {et~al.}(2015)\citenamefont {Baek}, \citenamefont {Rippard}, \citenamefont {Pufall}, \citenamefont {Benz}, \citenamefont {Russek}, \citenamefont {Rogalla},\ and\ \citenamefont {Dresselhaus}}]{Baek_PRApp_2015}%
  \BibitemOpen
  \bibfield  {author} {\bibinfo {author} {\bibfnamefont {B.}~\bibnamefont {Baek}}, \bibinfo {author} {\bibfnamefont {W.~H.}\ \bibnamefont {Rippard}}, \bibinfo {author} {\bibfnamefont {M.~R.}\ \bibnamefont {Pufall}}, \bibinfo {author} {\bibfnamefont {S.~P.}\ \bibnamefont {Benz}}, \bibinfo {author} {\bibfnamefont {S.~E.}\ \bibnamefont {Russek}}, \bibinfo {author} {\bibfnamefont {H.}~\bibnamefont {Rogalla}}, \ and\ \bibinfo {author} {\bibfnamefont {P.~D.}\ \bibnamefont {Dresselhaus}},\ }\bibfield  {title} {\enquote {\bibinfo {title} {{Spin-Transfer Torque Switching in Nanopillar Superconducting-Magnetic Hybrid Josephson Junctions}},}\ }\href {\doibase 10.1103/PhysRevApplied.3.011001} {\bibfield  {journal} {\bibinfo  {journal} {Phys. Rev. Applied}\ }\textbf {\bibinfo {volume} {3}},\ \bibinfo {pages} {011001} (\bibinfo {year} {2015})}\BibitemShut {NoStop}%
\bibitem [{\citenamefont {Niedzielski}\ \emph {et~al.}(2018)\citenamefont {Niedzielski}, \citenamefont {Bertus}, \citenamefont {Glick}, \citenamefont {Loloee}, \citenamefont {Pratt},\ and\ \citenamefont {Birge}}]{Niedzielski_PRB_2018}%
  \BibitemOpen
  \bibfield  {author} {\bibinfo {author} {\bibfnamefont {B.~M.}\ \bibnamefont {Niedzielski}}, \bibinfo {author} {\bibfnamefont {T.~J.}\ \bibnamefont {Bertus}}, \bibinfo {author} {\bibfnamefont {J.~A.}\ \bibnamefont {Glick}}, \bibinfo {author} {\bibfnamefont {R.}~\bibnamefont {Loloee}}, \bibinfo {author} {\bibfnamefont {W.~P.}\ \bibnamefont {Pratt}}, \ and\ \bibinfo {author} {\bibfnamefont {N.~O.}\ \bibnamefont {Birge}},\ }\bibfield  {title} {\enquote {\bibinfo {title} {Spin-valve {J}osephson junctions for cryogenic memory},}\ }\href {\doibase 10.1103/PhysRevB.97.024517} {\bibfield  {journal} {\bibinfo  {journal} {Phys. Rev. B}\ }\textbf {\bibinfo {volume} {97}},\ \bibinfo {pages} {024517} (\bibinfo {year} {2018})}\BibitemShut {NoStop}%
\bibitem [{\citenamefont {Satchell}\ \emph {et~al.}(2020)\citenamefont {Satchell}, \citenamefont {Shepley}, \citenamefont {Algarni}, \citenamefont {Vaughan}, \citenamefont {Darwin}, \citenamefont {Ali}, \citenamefont {Rosamond}, \citenamefont {Chen}, \citenamefont {Linfield}, \citenamefont {Hickey},\ and\ \citenamefont {Burnell}}]{Satchell_APL_2020}%
  \BibitemOpen
  \bibfield  {author} {\bibinfo {author} {\bibfnamefont {N.}~\bibnamefont {Satchell}}, \bibinfo {author} {\bibfnamefont {P.~M.}\ \bibnamefont {Shepley}}, \bibinfo {author} {\bibfnamefont {M.}~\bibnamefont {Algarni}}, \bibinfo {author} {\bibfnamefont {M.}~\bibnamefont {Vaughan}}, \bibinfo {author} {\bibfnamefont {E.}~\bibnamefont {Darwin}}, \bibinfo {author} {\bibfnamefont {M.}~\bibnamefont {Ali}}, \bibinfo {author} {\bibfnamefont {M.~C.}\ \bibnamefont {Rosamond}}, \bibinfo {author} {\bibfnamefont {L.}~\bibnamefont {Chen}}, \bibinfo {author} {\bibfnamefont {E.~H.}\ \bibnamefont {Linfield}}, \bibinfo {author} {\bibfnamefont {B.~J.}\ \bibnamefont {Hickey}}, \ and\ \bibinfo {author} {\bibfnamefont {G.}~\bibnamefont {Burnell}},\ }\bibfield  {title} {\enquote {\bibinfo {title} {Spin-valve {J}osephson junctions with perpendicular magnetic anisotropy for cryogenic memory},}\ }\href {\doibase 10.1063/1.5140095} {\bibfield  {journal} {\bibinfo  {journal} {Appl. Phys. Lett.}\ }\textbf {\bibinfo {volume} {116}},\
  \bibinfo {pages} {022601} (\bibinfo {year} {2020})}\BibitemShut {NoStop}%
\bibitem [{\citenamefont {Kulik}(1966)}]{Kulik_1966}%
  \BibitemOpen
  \bibfield  {author} {\bibinfo {author} {\bibfnamefont {I.~O.}\ \bibnamefont {Kulik}},\ }\bibfield  {title} {\enquote {\bibinfo {title} {Magnitude of the critical {J}osephson tunnel current},}\ }\href@noop {} {\bibfield  {journal} {\bibinfo  {journal} {JETP}\ }\textbf {\bibinfo {volume} {22}},\ \bibinfo {pages} {841--843} (\bibinfo {year} {1966})}\BibitemShut {NoStop}%
\bibitem [{\citenamefont {Massarotti}\ \emph {et~al.}(2015)\citenamefont {Massarotti}, \citenamefont {Pal}, \citenamefont {Rotoli}, \citenamefont {Longobardi}, \citenamefont {Blamire},\ and\ \citenamefont {Tafuri}}]{massarotti_2015}%
  \BibitemOpen
  \bibfield  {author} {\bibinfo {author} {\bibfnamefont {D.}~\bibnamefont {Massarotti}}, \bibinfo {author} {\bibfnamefont {A.}~\bibnamefont {Pal}}, \bibinfo {author} {\bibfnamefont {G.}~\bibnamefont {Rotoli}}, \bibinfo {author} {\bibfnamefont {L.}~\bibnamefont {Longobardi}}, \bibinfo {author} {\bibfnamefont {M.~G.}\ \bibnamefont {Blamire}}, \ and\ \bibinfo {author} {\bibfnamefont {F.}~\bibnamefont {Tafuri}},\ }\bibfield  {title} {\enquote {\bibinfo {title} {Macroscopic quantum tunnelling in spin filter ferromagnetic {Josephson} junctions},}\ }\href {\doibase 10.1038/ncomms8376} {\bibfield  {journal} {\bibinfo  {journal} {Nat. Commun.}\ }\textbf {\bibinfo {volume} {6}},\ \bibinfo {pages} {7376} (\bibinfo {year} {2015})}\BibitemShut {NoStop}%
\bibitem [{\citenamefont {Massarotti}\ \emph {et~al.}(2017)\citenamefont {Massarotti}, \citenamefont {Caruso}, \citenamefont {Pal}, \citenamefont {Rotoli}, \citenamefont {Longobardi}, \citenamefont {Pepe}, \citenamefont {Blamire},\ and\ \citenamefont {Tafuri}}]{massarotti_2017}%
  \BibitemOpen
  \bibfield  {author} {\bibinfo {author} {\bibfnamefont {D.}~\bibnamefont {Massarotti}}, \bibinfo {author} {\bibfnamefont {R.}~\bibnamefont {Caruso}}, \bibinfo {author} {\bibfnamefont {A.}~\bibnamefont {Pal}}, \bibinfo {author} {\bibfnamefont {G.}~\bibnamefont {Rotoli}}, \bibinfo {author} {\bibfnamefont {L.}~\bibnamefont {Longobardi}}, \bibinfo {author} {\bibfnamefont {G.~P.}\ \bibnamefont {Pepe}}, \bibinfo {author} {\bibfnamefont {M.}~\bibnamefont {Blamire}}, \ and\ \bibinfo {author} {\bibfnamefont {F.}~\bibnamefont {Tafuri}},\ }\bibfield  {title} {\enquote {\bibinfo {title} {Low temperature properties of spin filter {NbN/GdN/NbN J}osephson junctions},}\ }\href {\doibase 10.1016/j.physc.2016.07.018} {\bibfield  {journal} {\bibinfo  {journal} {Physica C Supercond.}\ }\textbf {\bibinfo {volume} {533}},\ \bibinfo {pages} {53--58} (\bibinfo {year} {2017})}\BibitemShut {NoStop}%
\bibitem [{\citenamefont {Cascales}\ \emph {et~al.}(2019)\citenamefont {Cascales}, \citenamefont {Takamura}, \citenamefont {Stephen}, \citenamefont {Heiman}, \citenamefont {Bergeret},\ and\ \citenamefont {Moodera}}]{cascales_2019}%
  \BibitemOpen
  \bibfield  {author} {\bibinfo {author} {\bibfnamefont {J.~P.}\ \bibnamefont {Cascales}}, \bibinfo {author} {\bibfnamefont {Y.}~\bibnamefont {Takamura}}, \bibinfo {author} {\bibfnamefont {G.~M.}\ \bibnamefont {Stephen}}, \bibinfo {author} {\bibfnamefont {D.}~\bibnamefont {Heiman}}, \bibinfo {author} {\bibfnamefont {F.~S.}\ \bibnamefont {Bergeret}}, \ and\ \bibinfo {author} {\bibfnamefont {J.~S.}\ \bibnamefont {Moodera}},\ }\bibfield  {title} {\enquote {\bibinfo {title} {Switchable {J}osephson junction based on interfacial exchange field},}\ }\href {\doibase 10.1063/1.5050382} {\bibfield  {journal} {\bibinfo  {journal} {Appl. Phys. Lett.}\ }\textbf {\bibinfo {volume} {114}},\ \bibinfo {pages} {022601} (\bibinfo {year} {2019})}\BibitemShut {NoStop}%
\bibitem [{\citenamefont {Ahmad}\ \emph {et~al.}(2020)\citenamefont {Ahmad}, \citenamefont {Caruso}, \citenamefont {Pal}, \citenamefont {Rotoli}, \citenamefont {Pepe}, \citenamefont {Blamire}, \citenamefont {Tafuri},\ and\ \citenamefont {Massarotti}}]{Ahmad_2020}%
  \BibitemOpen
  \bibfield  {author} {\bibinfo {author} {\bibfnamefont {H.}~\bibnamefont {Ahmad}}, \bibinfo {author} {\bibfnamefont {R.}~\bibnamefont {Caruso}}, \bibinfo {author} {\bibfnamefont {A.}~\bibnamefont {Pal}}, \bibinfo {author} {\bibfnamefont {G.}~\bibnamefont {Rotoli}}, \bibinfo {author} {\bibfnamefont {G.}~\bibnamefont {Pepe}}, \bibinfo {author} {\bibfnamefont {M.}~\bibnamefont {Blamire}}, \bibinfo {author} {\bibfnamefont {F.}~\bibnamefont {Tafuri}}, \ and\ \bibinfo {author} {\bibfnamefont {D.}~\bibnamefont {Massarotti}},\ }\bibfield  {title} {\enquote {\bibinfo {title} {Electrodynamics of {Highly Spin-Polarized Tunnel Josephson J}unctions},}\ }\href {\doibase 10.1103/PhysRevApplied.13.014017} {\bibfield  {journal} {\bibinfo  {journal} {Phys. Rev. Appl.}\ }\textbf {\bibinfo {volume} {13}},\ \bibinfo {pages} {014017} (\bibinfo {year} {2020})}\BibitemShut {NoStop}%
\bibitem [{\citenamefont {Ahmad}\ \emph {et~al.}(2022{\natexlab{b}})\citenamefont {Ahmad}, \citenamefont {Minutillo}, \citenamefont {Capecelatro}, \citenamefont {Pal}, \citenamefont {Caruso}, \citenamefont {Passarelli}, \citenamefont {Blamire}, \citenamefont {Tafuri}, \citenamefont {Lucignano},\ and\ \citenamefont {Massarotti}}]{Ahmad_2022b}%
  \BibitemOpen
  \bibfield  {author} {\bibinfo {author} {\bibfnamefont {H.~G.}\ \bibnamefont {Ahmad}}, \bibinfo {author} {\bibfnamefont {M.}~\bibnamefont {Minutillo}}, \bibinfo {author} {\bibfnamefont {R.}~\bibnamefont {Capecelatro}}, \bibinfo {author} {\bibfnamefont {A.}~\bibnamefont {Pal}}, \bibinfo {author} {\bibfnamefont {R.}~\bibnamefont {Caruso}}, \bibinfo {author} {\bibfnamefont {G.}~\bibnamefont {Passarelli}}, \bibinfo {author} {\bibfnamefont {M.~G.}\ \bibnamefont {Blamire}}, \bibinfo {author} {\bibfnamefont {F.}~\bibnamefont {Tafuri}}, \bibinfo {author} {\bibfnamefont {P.}~\bibnamefont {Lucignano}}, \ and\ \bibinfo {author} {\bibfnamefont {D.}~\bibnamefont {Massarotti}},\ }\bibfield  {title} {\enquote {\bibinfo {title} {Coexistence and tuning of spin-singlet and triplet transport in spin-filter {J}osephson junctions},}\ }\href {\doibase 10.1038/s42005-021-00783-1} {\bibfield  {journal} {\bibinfo  {journal} {Commun. Phys.}\ }\textbf {\bibinfo {volume} {5}},\ \bibinfo {pages} {2} (\bibinfo {year}
  {2022}{\natexlab{b}})}\BibitemShut {NoStop}%
\bibitem [{\citenamefont {Razmadze}\ \emph {et~al.}(2023)\citenamefont {Razmadze}, \citenamefont {Souto}, \citenamefont {Galletti}, \citenamefont {Maiani}, \citenamefont {Liu}, \citenamefont {Krogstrup}, \citenamefont {Schrade}, \citenamefont {Gyenis}, \citenamefont {Marcus},\ and\ \citenamefont {Vaitiekėnas}}]{razmadze_2023}%
  \BibitemOpen
  \bibfield  {author} {\bibinfo {author} {\bibfnamefont {D.}~\bibnamefont {Razmadze}}, \bibinfo {author} {\bibfnamefont {R.~S.}\ \bibnamefont {Souto}}, \bibinfo {author} {\bibfnamefont {L.}~\bibnamefont {Galletti}}, \bibinfo {author} {\bibfnamefont {A.}~\bibnamefont {Maiani}}, \bibinfo {author} {\bibfnamefont {Y.}~\bibnamefont {Liu}}, \bibinfo {author} {\bibfnamefont {P.}~\bibnamefont {Krogstrup}}, \bibinfo {author} {\bibfnamefont {C.}~\bibnamefont {Schrade}}, \bibinfo {author} {\bibfnamefont {A.}~\bibnamefont {Gyenis}}, \bibinfo {author} {\bibfnamefont {C.~M.}\ \bibnamefont {Marcus}}, \ and\ \bibinfo {author} {\bibfnamefont {S.}~\bibnamefont {Vaitiekėnas}},\ }\bibfield  {title} {\enquote {\bibinfo {title} {Supercurrent reversal in ferromagnetic hybrid nanowire {Josephson} junctions},}\ }\href {\doibase 10.1103/PhysRevB.107.L081301} {\bibfield  {journal} {\bibinfo  {journal} {Phys. Rev. B}\ }\textbf {\bibinfo {volume} {107}},\ \bibinfo {pages} {L081301} (\bibinfo {year} {2023})}\BibitemShut {NoStop}%
\bibitem [{\citenamefont {Ai}\ \emph {et~al.}(2021)\citenamefont {Ai}, \citenamefont {Zhang}, \citenamefont {Yang}, \citenamefont {Xie}, \citenamefont {Yang}, \citenamefont {Jia}, \citenamefont {Zhang}, \citenamefont {Liu}, \citenamefont {Li}, \citenamefont {Leng} \emph {et~al.}}]{ai_2021}%
  \BibitemOpen
  \bibfield  {author} {\bibinfo {author} {\bibfnamefont {L.}~\bibnamefont {Ai}}, \bibinfo {author} {\bibfnamefont {E.}~\bibnamefont {Zhang}}, \bibinfo {author} {\bibfnamefont {J.}~\bibnamefont {Yang}}, \bibinfo {author} {\bibfnamefont {X.}~\bibnamefont {Xie}}, \bibinfo {author} {\bibfnamefont {Y.}~\bibnamefont {Yang}}, \bibinfo {author} {\bibfnamefont {Z.}~\bibnamefont {Jia}}, \bibinfo {author} {\bibfnamefont {Y.}~\bibnamefont {Zhang}}, \bibinfo {author} {\bibfnamefont {S.}~\bibnamefont {Liu}}, \bibinfo {author} {\bibfnamefont {Z.}~\bibnamefont {Li}}, \bibinfo {author} {\bibfnamefont {P.}~\bibnamefont {Leng}},  \emph {et~al.},\ }\bibfield  {title} {\enquote {\bibinfo {title} {Van der {W}aals ferromagnetic {J}osephson junctions},}\ }\href {\doibase 10.1038/s41467-021-26946-w} {\bibfield  {journal} {\bibinfo  {journal} {Nat. Commun.}\ }\textbf {\bibinfo {volume} {12}},\ \bibinfo {pages} {6580} (\bibinfo {year} {2021})}\BibitemShut {NoStop}%
\bibitem [{\citenamefont {Idzuchi}\ \emph {et~al.}(2021)\citenamefont {Idzuchi}, \citenamefont {Pientka}, \citenamefont {Huang}, \citenamefont {Harada}, \citenamefont {G{\"u}l}, \citenamefont {Shin}, \citenamefont {Nguyen}, \citenamefont {Jo}, \citenamefont {Shindo}, \citenamefont {Cava} \emph {et~al.}}]{idzuchi_2021}%
  \BibitemOpen
  \bibfield  {author} {\bibinfo {author} {\bibfnamefont {H.}~\bibnamefont {Idzuchi}}, \bibinfo {author} {\bibfnamefont {F.}~\bibnamefont {Pientka}}, \bibinfo {author} {\bibfnamefont {K.-F.}\ \bibnamefont {Huang}}, \bibinfo {author} {\bibfnamefont {K.}~\bibnamefont {Harada}}, \bibinfo {author} {\bibfnamefont {{\"O}.}~\bibnamefont {G{\"u}l}}, \bibinfo {author} {\bibfnamefont {Y.~J.}\ \bibnamefont {Shin}}, \bibinfo {author} {\bibfnamefont {L.}~\bibnamefont {Nguyen}}, \bibinfo {author} {\bibfnamefont {N.}~\bibnamefont {Jo}}, \bibinfo {author} {\bibfnamefont {D.}~\bibnamefont {Shindo}}, \bibinfo {author} {\bibfnamefont {R.}~\bibnamefont {Cava}},  \emph {et~al.},\ }\bibfield  {title} {\enquote {\bibinfo {title} {Unconventional supercurrent phase in {I}sing superconductor {J}osephson junction with atomically thin magnetic insulator},}\ }\href {\doibase 10.1038/s41467-021-25608-1} {\bibfield  {journal} {\bibinfo  {journal} {Nat. Commun.}\ }\textbf {\bibinfo {volume} {12}},\ \bibinfo {pages} {5332} (\bibinfo {year}
  {2021})}\BibitemShut {NoStop}%
\bibitem [{\citenamefont {Hu}\ \emph {et~al.}(2023)\citenamefont {Hu}, \citenamefont {Wang}, \citenamefont {Wang}, \citenamefont {Zhang}, \citenamefont {Feng}, \citenamefont {Wang}, \citenamefont {Niu}, \citenamefont {Zhang},\ and\ \citenamefont {Xiang}}]{hu_2023}%
  \BibitemOpen
  \bibfield  {author} {\bibinfo {author} {\bibfnamefont {G.}~\bibnamefont {Hu}}, \bibinfo {author} {\bibfnamefont {C.}~\bibnamefont {Wang}}, \bibinfo {author} {\bibfnamefont {S.}~\bibnamefont {Wang}}, \bibinfo {author} {\bibfnamefont {Y.}~\bibnamefont {Zhang}}, \bibinfo {author} {\bibfnamefont {Y.}~\bibnamefont {Feng}}, \bibinfo {author} {\bibfnamefont {Z.}~\bibnamefont {Wang}}, \bibinfo {author} {\bibfnamefont {Q.}~\bibnamefont {Niu}}, \bibinfo {author} {\bibfnamefont {Z.}~\bibnamefont {Zhang}}, \ and\ \bibinfo {author} {\bibfnamefont {B.}~\bibnamefont {Xiang}},\ }\bibfield  {title} {\enquote {\bibinfo {title} {Long-range skin {Josephson} supercurrent across a van der {Waals} ferromagnet},}\ }\href {\doibase 10.1038/s41467-023-37603-9} {\bibfield  {journal} {\bibinfo  {journal} {Nat. Commun.}\ }\textbf {\bibinfo {volume} {14}},\ \bibinfo {pages} {1779} (\bibinfo {year} {2023})}\BibitemShut {NoStop}%
\bibitem [{\citenamefont {Heikkil{\"a}}, \citenamefont {Wilhelm},\ and\ \citenamefont {Sch{\"o}n}(2000)}]{heikkila_2000}%
  \BibitemOpen
  \bibfield  {author} {\bibinfo {author} {\bibfnamefont {T.~T.}\ \bibnamefont {Heikkil{\"a}}}, \bibinfo {author} {\bibfnamefont {F.~K.}\ \bibnamefont {Wilhelm}}, \ and\ \bibinfo {author} {\bibfnamefont {G.}~\bibnamefont {Sch{\"o}n}},\ }\bibfield  {title} {\enquote {\bibinfo {title} {Non-equilibrium supercurrent through mesoscopic ferromagnetic weak links},}\ }\href {\doibase 10.1209/epl/i2000-00513-x} {\bibfield  {journal} {\bibinfo  {journal} {EPL}\ }\textbf {\bibinfo {volume} {51}},\ \bibinfo {pages} {434} (\bibinfo {year} {2000})}\BibitemShut {NoStop}%
\bibitem [{\citenamefont {Yip}(2000)}]{Yip_2000}%
  \BibitemOpen
  \bibfield  {author} {\bibinfo {author} {\bibfnamefont {S.-K.}\ \bibnamefont {Yip}},\ }\bibfield  {title} {\enquote {\bibinfo {title} {Magnetic-field effect on the supercurrent of an {SNS} junction},}\ }\href {\doibase 10.1103/PhysRevB.62.R6127} {\bibfield  {journal} {\bibinfo  {journal} {Phys. Rev. B}\ }\textbf {\bibinfo {volume} {62}},\ \bibinfo {pages} {R6127--R6130} (\bibinfo {year} {2000})}\BibitemShut {NoStop}%
\bibitem [{\citenamefont {Li}\ \emph {et~al.}(2019)\citenamefont {Li}, \citenamefont {de~Ronde}, \citenamefont {de~Boer}, \citenamefont {Ridderbos}, \citenamefont {Zwanenburg}, \citenamefont {Huang}, \citenamefont {Golubov},\ and\ \citenamefont {Brinkman}}]{Li_2019}%
  \BibitemOpen
  \bibfield  {author} {\bibinfo {author} {\bibfnamefont {C.}~\bibnamefont {Li}}, \bibinfo {author} {\bibfnamefont {B.}~\bibnamefont {de~Ronde}}, \bibinfo {author} {\bibfnamefont {J.}~\bibnamefont {de~Boer}}, \bibinfo {author} {\bibfnamefont {J.}~\bibnamefont {Ridderbos}}, \bibinfo {author} {\bibfnamefont {F.}~\bibnamefont {Zwanenburg}}, \bibinfo {author} {\bibfnamefont {Y.}~\bibnamefont {Huang}}, \bibinfo {author} {\bibfnamefont {A.}~\bibnamefont {Golubov}}, \ and\ \bibinfo {author} {\bibfnamefont {A.}~\bibnamefont {Brinkman}},\ }\bibfield  {title} {\enquote {\bibinfo {title} {Zeeman-{E}ffect-{I}nduced $0\text{\ensuremath{-}}\ensuremath{\pi}$ {Transitions in Ballistic Dirac Semimetal Josephson J}unctions},}\ }\href {\doibase 10.1103/PhysRevLett.123.026802} {\bibfield  {journal} {\bibinfo  {journal} {Phys. Rev. Lett.}\ }\textbf {\bibinfo {volume} {123}},\ \bibinfo {pages} {026802} (\bibinfo {year} {2019})}\BibitemShut {NoStop}%
\bibitem [{\citenamefont {Dvir}\ \emph {et~al.}(2021)\citenamefont {Dvir}, \citenamefont {Zalic}, \citenamefont {Fyhn}, \citenamefont {Amundsen}, \citenamefont {Taniguchi}, \citenamefont {Watanabe}, \citenamefont {Linder},\ and\ \citenamefont {Steinberg}}]{Dvir_2021}%
  \BibitemOpen
  \bibfield  {author} {\bibinfo {author} {\bibfnamefont {T.}~\bibnamefont {Dvir}}, \bibinfo {author} {\bibfnamefont {A.}~\bibnamefont {Zalic}}, \bibinfo {author} {\bibfnamefont {E.~H.}\ \bibnamefont {Fyhn}}, \bibinfo {author} {\bibfnamefont {M.}~\bibnamefont {Amundsen}}, \bibinfo {author} {\bibfnamefont {T.}~\bibnamefont {Taniguchi}}, \bibinfo {author} {\bibfnamefont {K.}~\bibnamefont {Watanabe}}, \bibinfo {author} {\bibfnamefont {J.}~\bibnamefont {Linder}}, \ and\ \bibinfo {author} {\bibfnamefont {H.}~\bibnamefont {Steinberg}},\ }\bibfield  {title} {\enquote {\bibinfo {title} {Planar graphene-{NbSe}$_{2}$ {J}osephson junctions in a parallel magnetic field},}\ }\href {\doibase 10.1103/PhysRevB.103.115401} {\bibfield  {journal} {\bibinfo  {journal} {Phys. Rev. B}\ }\textbf {\bibinfo {volume} {103}},\ \bibinfo {pages} {115401} (\bibinfo {year} {2021})}\BibitemShut {NoStop}%
\bibitem [{\citenamefont {Schönle}\ \emph {et~al.}(2019)\citenamefont {Schönle}, \citenamefont {Borisov}, \citenamefont {Klett}, \citenamefont {Dyck}, \citenamefont {Balestro}, \citenamefont {Reiss},\ and\ \citenamefont {Wernsdorfer}}]{schnle_2019}%
  \BibitemOpen
  \bibfield  {author} {\bibinfo {author} {\bibfnamefont {J.}~\bibnamefont {Schönle}}, \bibinfo {author} {\bibfnamefont {K.}~\bibnamefont {Borisov}}, \bibinfo {author} {\bibfnamefont {R.}~\bibnamefont {Klett}}, \bibinfo {author} {\bibfnamefont {D.}~\bibnamefont {Dyck}}, \bibinfo {author} {\bibfnamefont {F.}~\bibnamefont {Balestro}}, \bibinfo {author} {\bibfnamefont {G.}~\bibnamefont {Reiss}}, \ and\ \bibinfo {author} {\bibfnamefont {W.}~\bibnamefont {Wernsdorfer}},\ }\bibfield  {title} {\enquote {\bibinfo {title} {Field-tunable 0-$\pi$-transitions in {SnTe} topological crystalline insulator {SQUIDs}},}\ }\href {\doibase 10.1038/s41598-018-38008-1} {\bibfield  {journal} {\bibinfo  {journal} {Sci. Rep.}\ }\textbf {\bibinfo {volume} {9}},\ \bibinfo {pages} {1987} (\bibinfo {year} {2019})}\BibitemShut {NoStop}%
\bibitem [{\citenamefont {Ke}\ \emph {et~al.}(2019)\citenamefont {Ke}, \citenamefont {Moehle}, \citenamefont {de~Vries}, \citenamefont {Thomas}, \citenamefont {Metti}, \citenamefont {Guinn}, \citenamefont {Kallaher}, \citenamefont {Lodari}, \citenamefont {Scappucci}, \citenamefont {Wang}, \citenamefont {Diaz}, \citenamefont {Gardner}, \citenamefont {Manfra},\ and\ \citenamefont {Goswami}}]{ke_2019}%
  \BibitemOpen
  \bibfield  {author} {\bibinfo {author} {\bibfnamefont {C.~T.}\ \bibnamefont {Ke}}, \bibinfo {author} {\bibfnamefont {C.~M.}\ \bibnamefont {Moehle}}, \bibinfo {author} {\bibfnamefont {F.~K.}\ \bibnamefont {de~Vries}}, \bibinfo {author} {\bibfnamefont {C.}~\bibnamefont {Thomas}}, \bibinfo {author} {\bibfnamefont {S.}~\bibnamefont {Metti}}, \bibinfo {author} {\bibfnamefont {C.~R.}\ \bibnamefont {Guinn}}, \bibinfo {author} {\bibfnamefont {R.}~\bibnamefont {Kallaher}}, \bibinfo {author} {\bibfnamefont {M.}~\bibnamefont {Lodari}}, \bibinfo {author} {\bibfnamefont {G.}~\bibnamefont {Scappucci}}, \bibinfo {author} {\bibfnamefont {T.}~\bibnamefont {Wang}}, \bibinfo {author} {\bibfnamefont {R.~E.}\ \bibnamefont {Diaz}}, \bibinfo {author} {\bibfnamefont {G.~C.}\ \bibnamefont {Gardner}}, \bibinfo {author} {\bibfnamefont {M.~J.}\ \bibnamefont {Manfra}}, \ and\ \bibinfo {author} {\bibfnamefont {S.}~\bibnamefont {Goswami}},\ }\bibfield  {title} {\enquote {\bibinfo {title} {Ballistic superconductivity and tunable
  $\pi$-junctions in {InSb} quantum wells},}\ }\href {\doibase 10.1038/s41467-019-11742-4} {\bibfield  {journal} {\bibinfo  {journal} {Nat. Commun.}\ }\textbf {\bibinfo {volume} {10}},\ \bibinfo {pages} {3764} (\bibinfo {year} {2019})}\BibitemShut {NoStop}%
\bibitem [{\citenamefont {Haxell}\ \emph {et~al.}(2023)\citenamefont {Haxell}, \citenamefont {Coraiola}, \citenamefont {Sabonis}, \citenamefont {Hinderling}, \citenamefont {Ten~Kate}, \citenamefont {Cheah}, \citenamefont {Krizek}, \citenamefont {Schott}, \citenamefont {Wegscheider},\ and\ \citenamefont {Nichele}}]{haxell_2023}%
  \BibitemOpen
  \bibfield  {author} {\bibinfo {author} {\bibfnamefont {D.~Z.}\ \bibnamefont {Haxell}}, \bibinfo {author} {\bibfnamefont {M.}~\bibnamefont {Coraiola}}, \bibinfo {author} {\bibfnamefont {D.}~\bibnamefont {Sabonis}}, \bibinfo {author} {\bibfnamefont {M.}~\bibnamefont {Hinderling}}, \bibinfo {author} {\bibfnamefont {S.~C.}\ \bibnamefont {Ten~Kate}}, \bibinfo {author} {\bibfnamefont {E.}~\bibnamefont {Cheah}}, \bibinfo {author} {\bibfnamefont {F.}~\bibnamefont {Krizek}}, \bibinfo {author} {\bibfnamefont {R.}~\bibnamefont {Schott}}, \bibinfo {author} {\bibfnamefont {W.}~\bibnamefont {Wegscheider}}, \ and\ \bibinfo {author} {\bibfnamefont {F.}~\bibnamefont {Nichele}},\ }\bibfield  {title} {\enquote {\bibinfo {title} {Zeeman-and {Orbital-Driven Phase Shifts in Planar Josephson J}unctions},}\ }\href {\doibase 10.1021/acsnano.3c04957} {\bibfield  {journal} {\bibinfo  {journal} {ACS Nano}\ }\textbf {\bibinfo {volume} {17}},\ \bibinfo {pages} {18139--18147} (\bibinfo {year} {2023})}\BibitemShut {NoStop}%
\bibitem [{\citenamefont {Whiticar}\ \emph {et~al.}(2021)\citenamefont {Whiticar}, \citenamefont {Fornieri}, \citenamefont {Banerjee}, \citenamefont {Drachmann}, \citenamefont {Gronin}, \citenamefont {Gardner}, \citenamefont {Lindemann}, \citenamefont {Manfra},\ and\ \citenamefont {Marcus}}]{Whiticar_2021}%
  \BibitemOpen
  \bibfield  {author} {\bibinfo {author} {\bibfnamefont {A.~M.}\ \bibnamefont {Whiticar}}, \bibinfo {author} {\bibfnamefont {A.}~\bibnamefont {Fornieri}}, \bibinfo {author} {\bibfnamefont {A.}~\bibnamefont {Banerjee}}, \bibinfo {author} {\bibfnamefont {A.~C.~C.}\ \bibnamefont {Drachmann}}, \bibinfo {author} {\bibfnamefont {S.}~\bibnamefont {Gronin}}, \bibinfo {author} {\bibfnamefont {G.~C.}\ \bibnamefont {Gardner}}, \bibinfo {author} {\bibfnamefont {T.}~\bibnamefont {Lindemann}}, \bibinfo {author} {\bibfnamefont {M.~J.}\ \bibnamefont {Manfra}}, \ and\ \bibinfo {author} {\bibfnamefont {C.~M.}\ \bibnamefont {Marcus}},\ }\bibfield  {title} {\enquote {\bibinfo {title} {Zeeman-driven parity transitions in an {A}ndreev quantum dot},}\ }\href {\doibase 10.1103/PhysRevB.103.245308} {\bibfield  {journal} {\bibinfo  {journal} {Phys. Rev. B}\ }\textbf {\bibinfo {volume} {103}},\ \bibinfo {pages} {245308} (\bibinfo {year} {2021})}\BibitemShut {NoStop}%
\bibitem [{\citenamefont {Likharev}(1999)}]{likharev_1999}%
  \BibitemOpen
  \bibfield  {author} {\bibinfo {author} {\bibfnamefont {K.~K.}\ \bibnamefont {Likharev}},\ }\bibfield  {title} {\enquote {\bibinfo {title} {Single-electron devices and their applications},}\ }\href {\doibase 10.1109/5.752518} {\bibfield  {journal} {\bibinfo  {journal} {Proceedings of the IEEE}\ }\textbf {\bibinfo {volume} {87}},\ \bibinfo {pages} {606--632} (\bibinfo {year} {1999})}\BibitemShut {NoStop}%
\bibitem [{\citenamefont {Van~Dam}\ \emph {et~al.}(2006)\citenamefont {Van~Dam}, \citenamefont {Nazarov}, \citenamefont {Bakkers}, \citenamefont {De~Franceschi},\ and\ \citenamefont {Kouwenhoven}}]{van_2006}%
  \BibitemOpen
  \bibfield  {author} {\bibinfo {author} {\bibfnamefont {J.~A.}\ \bibnamefont {Van~Dam}}, \bibinfo {author} {\bibfnamefont {Y.~V.}\ \bibnamefont {Nazarov}}, \bibinfo {author} {\bibfnamefont {E.~P.}\ \bibnamefont {Bakkers}}, \bibinfo {author} {\bibfnamefont {S.}~\bibnamefont {De~Franceschi}}, \ and\ \bibinfo {author} {\bibfnamefont {L.~P.}\ \bibnamefont {Kouwenhoven}},\ }\bibfield  {title} {\enquote {\bibinfo {title} {Supercurrent reversal in quantum dots},}\ }\href {\doibase 10.1038/nature05018} {\bibfield  {journal} {\bibinfo  {journal} {Nature}\ }\textbf {\bibinfo {volume} {442}},\ \bibinfo {pages} {667--670} (\bibinfo {year} {2006})}\BibitemShut {NoStop}%
\bibitem [{\citenamefont {J{\o}rgensen}\ \emph {et~al.}(2007)\citenamefont {J{\o}rgensen}, \citenamefont {Novotn{\`y}}, \citenamefont {Grove-Rasmussen}, \citenamefont {Flensberg},\ and\ \citenamefont {Lindelof}}]{jorgensen_2007}%
  \BibitemOpen
  \bibfield  {author} {\bibinfo {author} {\bibfnamefont {H.~I.}\ \bibnamefont {J{\o}rgensen}}, \bibinfo {author} {\bibfnamefont {T.}~\bibnamefont {Novotn{\`y}}}, \bibinfo {author} {\bibfnamefont {K.}~\bibnamefont {Grove-Rasmussen}}, \bibinfo {author} {\bibfnamefont {K.}~\bibnamefont {Flensberg}}, \ and\ \bibinfo {author} {\bibfnamefont {P.}~\bibnamefont {Lindelof}},\ }\bibfield  {title} {\enquote {\bibinfo {title} {Critical current 0- $\pi$ transition in designed {J}osephson quantum dot junctions},}\ }\href {\doibase 10.1021/nl071152w} {\bibfield  {journal} {\bibinfo  {journal} {Nano Lett.}\ }\textbf {\bibinfo {volume} {7}},\ \bibinfo {pages} {2441--2445} (\bibinfo {year} {2007})}\BibitemShut {NoStop}%
\bibitem [{\citenamefont {Maurand}\ \emph {et~al.}(2012)\citenamefont {Maurand}, \citenamefont {Meng}, \citenamefont {Bonet}, \citenamefont {Florens}, \citenamefont {Marty},\ and\ \citenamefont {Wernsdorfer}}]{Maurand_2012}%
  \BibitemOpen
  \bibfield  {author} {\bibinfo {author} {\bibfnamefont {R.}~\bibnamefont {Maurand}}, \bibinfo {author} {\bibfnamefont {T.}~\bibnamefont {Meng}}, \bibinfo {author} {\bibfnamefont {E.}~\bibnamefont {Bonet}}, \bibinfo {author} {\bibfnamefont {S.}~\bibnamefont {Florens}}, \bibinfo {author} {\bibfnamefont {L.}~\bibnamefont {Marty}}, \ and\ \bibinfo {author} {\bibfnamefont {W.}~\bibnamefont {Wernsdorfer}},\ }\bibfield  {title} {\enquote {\bibinfo {title} {First-order $0\mathrm{\text{\ensuremath{-}}}\ensuremath{\pi}$ {Quantum Phase Transition in the Kondo Regime of a Superconducting Carbon-Nanotube Quantum D}ot},}\ }\href {\doibase 10.1103/PhysRevX.2.011009} {\bibfield  {journal} {\bibinfo  {journal} {Phys. Rev. X}\ }\textbf {\bibinfo {volume} {2}},\ \bibinfo {pages} {011009} (\bibinfo {year} {2012})}\BibitemShut {NoStop}%
\bibitem [{\citenamefont {Delagrange}\ \emph {et~al.}(2016)\citenamefont {Delagrange}, \citenamefont {Weil}, \citenamefont {Kasumov}, \citenamefont {Ferrier}, \citenamefont {Bouchiat},\ and\ \citenamefont {Deblock}}]{Delagrange_2016}%
  \BibitemOpen
  \bibfield  {author} {\bibinfo {author} {\bibfnamefont {R.}~\bibnamefont {Delagrange}}, \bibinfo {author} {\bibfnamefont {R.}~\bibnamefont {Weil}}, \bibinfo {author} {\bibfnamefont {A.}~\bibnamefont {Kasumov}}, \bibinfo {author} {\bibfnamefont {M.}~\bibnamefont {Ferrier}}, \bibinfo {author} {\bibfnamefont {H.}~\bibnamefont {Bouchiat}}, \ and\ \bibinfo {author} {\bibfnamefont {R.}~\bibnamefont {Deblock}},\ }\bibfield  {title} {\enquote {\bibinfo {title} {0-$\ensuremath{\pi}$ quantum transition in a carbon nanotube {J}osephson junction: {U}niversal phase dependence and orbital degeneracy},}\ }\href {\doibase 10.1103/PhysRevB.93.195437} {\bibfield  {journal} {\bibinfo  {journal} {Phys. Rev. B}\ }\textbf {\bibinfo {volume} {93}},\ \bibinfo {pages} {195437} (\bibinfo {year} {2016})}\BibitemShut {NoStop}%
\bibitem [{\citenamefont {Salamone}\ \emph {et~al.}(2021)\citenamefont {Salamone}, \citenamefont {Svendsen}, \citenamefont {Amundsen},\ and\ \citenamefont {Jacobsen}}]{Salamone_2021}%
  \BibitemOpen
  \bibfield  {author} {\bibinfo {author} {\bibfnamefont {T.}~\bibnamefont {Salamone}}, \bibinfo {author} {\bibfnamefont {M.~B.~M.}\ \bibnamefont {Svendsen}}, \bibinfo {author} {\bibfnamefont {M.}~\bibnamefont {Amundsen}}, \ and\ \bibinfo {author} {\bibfnamefont {S.}~\bibnamefont {Jacobsen}},\ }\bibfield  {title} {\enquote {\bibinfo {title} {Curvature-induced long-range supercurrents in diffusive superconductor-ferromagnet-superconductor {J}osephson junctions with a dynamic $0\text{\ensuremath{-}}\ensuremath{\pi}$ transition},}\ }\href {\doibase 10.1103/PhysRevB.104.L060505} {\bibfield  {journal} {\bibinfo  {journal} {Phys. Rev. B}\ }\textbf {\bibinfo {volume} {104}},\ \bibinfo {pages} {L060505} (\bibinfo {year} {2021})}\BibitemShut {NoStop}%
\bibitem [{\citenamefont {Skarpeid}\ \emph {et~al.}(2024)\citenamefont {Skarpeid}, \citenamefont {Hugdal}, \citenamefont {Salamone}, \citenamefont {Amundsen},\ and\ \citenamefont {Jacobsen}}]{Skarpeid_2023}%
  \BibitemOpen
  \bibfield  {author} {\bibinfo {author} {\bibfnamefont {A.~J.}\ \bibnamefont {Skarpeid}}, \bibinfo {author} {\bibfnamefont {H.~G.}\ \bibnamefont {Hugdal}}, \bibinfo {author} {\bibfnamefont {T.}~\bibnamefont {Salamone}}, \bibinfo {author} {\bibfnamefont {M.}~\bibnamefont {Amundsen}}, \ and\ \bibinfo {author} {\bibfnamefont {S.~H.}\ \bibnamefont {Jacobsen}},\ }\bibfield  {title} {\enquote {\bibinfo {title} {Non-constant geometric curvature for tailored spin-orbit coupling and chirality in superconductor-magnet heterostructures},}\ }\href {http://iopscience.iop.org/article/10.1088/1361-648X/ad2e23} {\bibfield  {journal} {\bibinfo  {journal} {J. Condens. Matter Phys.}\ ,\ \bibinfo {pages} {10.1088/1361--648X/ad2e23}} (\bibinfo {year} {2024})}\BibitemShut {NoStop}%
\bibitem [{\citenamefont {Giroud}\ \emph {et~al.}(1998)\citenamefont {Giroud}, \citenamefont {Courtois}, \citenamefont {Hasselbach}, \citenamefont {Mailly},\ and\ \citenamefont {Pannetier}}]{giroud_1998}%
  \BibitemOpen
  \bibfield  {author} {\bibinfo {author} {\bibfnamefont {M.}~\bibnamefont {Giroud}}, \bibinfo {author} {\bibfnamefont {H.}~\bibnamefont {Courtois}}, \bibinfo {author} {\bibfnamefont {K.}~\bibnamefont {Hasselbach}}, \bibinfo {author} {\bibfnamefont {D.}~\bibnamefont {Mailly}}, \ and\ \bibinfo {author} {\bibfnamefont {B.}~\bibnamefont {Pannetier}},\ }\bibfield  {title} {\enquote {\bibinfo {title} {Superconducting proximity effect in a mesoscopic ferromagnetic wire},}\ }\href {\doibase 10.1103/PhysRevB.58.R11872} {\bibfield  {journal} {\bibinfo  {journal} {Phys. Rev. B}\ }\textbf {\bibinfo {volume} {58}},\ \bibinfo {pages} {R11872--R11875} (\bibinfo {year} {1998})}\BibitemShut {NoStop}%
\bibitem [{\citenamefont {Lawrence}\ and\ \citenamefont {Giordano}(1999)}]{lawrence_1999}%
  \BibitemOpen
  \bibfield  {author} {\bibinfo {author} {\bibfnamefont {M.~D.}\ \bibnamefont {Lawrence}}\ and\ \bibinfo {author} {\bibfnamefont {N.}~\bibnamefont {Giordano}},\ }\bibfield  {title} {\enquote {\bibinfo {title} {Proximity effects in superconductor-ferromagnet junctions},}\ }\href {\doibase 10.1088/0953-8984/11/4/016} {\bibfield  {journal} {\bibinfo  {journal} {J. Phys.: Condens. Matter}\ }\textbf {\bibinfo {volume} {11}},\ \bibinfo {pages} {1089--1094} (\bibinfo {year} {1999})}\BibitemShut {NoStop}%
\bibitem [{\citenamefont {Petrashov}\ \emph {et~al.}(1999)\citenamefont {Petrashov}, \citenamefont {Sosnin}, \citenamefont {Cox}, \citenamefont {Parsons},\ and\ \citenamefont {Troadec}}]{petrashov_1999}%
  \BibitemOpen
  \bibfield  {author} {\bibinfo {author} {\bibfnamefont {V.~T.}\ \bibnamefont {Petrashov}}, \bibinfo {author} {\bibfnamefont {I.~A.}\ \bibnamefont {Sosnin}}, \bibinfo {author} {\bibfnamefont {I.}~\bibnamefont {Cox}}, \bibinfo {author} {\bibfnamefont {A.}~\bibnamefont {Parsons}}, \ and\ \bibinfo {author} {\bibfnamefont {C.}~\bibnamefont {Troadec}},\ }\bibfield  {title} {\enquote {\bibinfo {title} {Giant mutual proximity effects in ferromagnetic/superconducting nanostructures},}\ }\href {\doibase 10.1103/PhysRevLett.83.3281} {\bibfield  {journal} {\bibinfo  {journal} {Phys. Rev. Lett.}\ }\textbf {\bibinfo {volume} {83}},\ \bibinfo {pages} {3281--3284} (\bibinfo {year} {1999})}\BibitemShut {NoStop}%
\bibitem [{\citenamefont {Keizer}\ \emph {et~al.}(2006)\citenamefont {Keizer}, \citenamefont {Goennenwein}, \citenamefont {Klapwijk}, \citenamefont {Miao}, \citenamefont {Xiao},\ and\ \citenamefont {Gupta}}]{keizer_2006}%
  \BibitemOpen
  \bibfield  {author} {\bibinfo {author} {\bibfnamefont {R.}~\bibnamefont {Keizer}}, \bibinfo {author} {\bibfnamefont {S.}~\bibnamefont {Goennenwein}}, \bibinfo {author} {\bibfnamefont {T.}~\bibnamefont {Klapwijk}}, \bibinfo {author} {\bibfnamefont {G.}~\bibnamefont {Miao}}, \bibinfo {author} {\bibfnamefont {G.}~\bibnamefont {Xiao}}, \ and\ \bibinfo {author} {\bibfnamefont {A.}~\bibnamefont {Gupta}},\ }\bibfield  {title} {\enquote {\bibinfo {title} {A spin triplet supercurrent through the half-metallic ferromagnet {CrO}2},}\ }\href {\doibase 10.1038/nature04499} {\bibfield  {journal} {\bibinfo  {journal} {Nature}\ }\textbf {\bibinfo {volume} {439}},\ \bibinfo {pages} {825--827} (\bibinfo {year} {2006})}\BibitemShut {NoStop}%
\bibitem [{\citenamefont {Sosnin}\ \emph {et~al.}(2006)\citenamefont {Sosnin}, \citenamefont {Cho}, \citenamefont {Petrashov},\ and\ \citenamefont {Volkov}}]{Sosnin_2006}%
  \BibitemOpen
  \bibfield  {author} {\bibinfo {author} {\bibfnamefont {I.}~\bibnamefont {Sosnin}}, \bibinfo {author} {\bibfnamefont {H.}~\bibnamefont {Cho}}, \bibinfo {author} {\bibfnamefont {V.~T.}\ \bibnamefont {Petrashov}}, \ and\ \bibinfo {author} {\bibfnamefont {A.~F.}\ \bibnamefont {Volkov}},\ }\bibfield  {title} {\enquote {\bibinfo {title} {Superconducting {P}hase {C}oherent {Electron Transport in Proximity Conical F}erromagnets},}\ }\href {\doibase 10.1103/PhysRevLett.96.157002} {\bibfield  {journal} {\bibinfo  {journal} {Phys. Rev. Lett.}\ }\textbf {\bibinfo {volume} {96}},\ \bibinfo {pages} {157002} (\bibinfo {year} {2006})}\BibitemShut {NoStop}%
\bibitem [{\citenamefont {Khaire}\ \emph {et~al.}(2010)\citenamefont {Khaire}, \citenamefont {Khasawneh}, \citenamefont {Pratt},\ and\ \citenamefont {Birge}}]{khaire_2010}%
  \BibitemOpen
  \bibfield  {author} {\bibinfo {author} {\bibfnamefont {T.~S.}\ \bibnamefont {Khaire}}, \bibinfo {author} {\bibfnamefont {M.~A.}\ \bibnamefont {Khasawneh}}, \bibinfo {author} {\bibfnamefont {W.~P.}\ \bibnamefont {Pratt}}, \ and\ \bibinfo {author} {\bibfnamefont {N.~O.}\ \bibnamefont {Birge}},\ }\bibfield  {title} {\enquote {\bibinfo {title} {Observation of spin-triplet superconductivity in {Co}-based {Josephson} junctions},}\ }\href {\doibase 10.1103/PhysRevLett.104.137002} {\bibfield  {journal} {\bibinfo  {journal} {Phys. Rev. Lett.}\ }\textbf {\bibinfo {volume} {104}},\ \bibinfo {pages} {137002} (\bibinfo {year} {2010})}\BibitemShut {NoStop}%
\bibitem [{\citenamefont {Sprungmann}\ \emph {et~al.}(2010)\citenamefont {Sprungmann}, \citenamefont {Westerholt}, \citenamefont {Zabel}, \citenamefont {Weides},\ and\ \citenamefont {Kohlstedt}}]{sprungmann_2010}%
  \BibitemOpen
  \bibfield  {author} {\bibinfo {author} {\bibfnamefont {D.}~\bibnamefont {Sprungmann}}, \bibinfo {author} {\bibfnamefont {K.}~\bibnamefont {Westerholt}}, \bibinfo {author} {\bibfnamefont {H.}~\bibnamefont {Zabel}}, \bibinfo {author} {\bibfnamefont {M.}~\bibnamefont {Weides}}, \ and\ \bibinfo {author} {\bibfnamefont {H.}~\bibnamefont {Kohlstedt}},\ }\bibfield  {title} {\enquote {\bibinfo {title} {Evidence for triplet superconductivity in {Josephson} junctions with barriers of the ferromagnetic {Heusler} alloy},}\ }\href {\doibase 10.1103/PhysRevB.82.060505} {\bibfield  {journal} {\bibinfo  {journal} {Phys. Rev. B}\ }\textbf {\bibinfo {volume} {82}},\ \bibinfo {pages} {060505} (\bibinfo {year} {2010})}\BibitemShut {NoStop}%
\bibitem [{\citenamefont {Anwar}\ \emph {et~al.}(2012)\citenamefont {Anwar}, \citenamefont {Veldhorst}, \citenamefont {Brinkman},\ and\ \citenamefont {Aarts}}]{anwar_2012}%
  \BibitemOpen
  \bibfield  {author} {\bibinfo {author} {\bibfnamefont {M.~S.}\ \bibnamefont {Anwar}}, \bibinfo {author} {\bibfnamefont {M.}~\bibnamefont {Veldhorst}}, \bibinfo {author} {\bibfnamefont {A.}~\bibnamefont {Brinkman}}, \ and\ \bibinfo {author} {\bibfnamefont {J.}~\bibnamefont {Aarts}},\ }\bibfield  {title} {\enquote {\bibinfo {title} {Long range supercurrents in ferromagnetic {CrO}$_2$ using a multilayer contact structure},}\ }\href {\doibase 10.1063/1.3681138} {\bibfield  {journal} {\bibinfo  {journal} {Appl. Phys. Lett}\ }\textbf {\bibinfo {volume} {100}},\ \bibinfo {pages} {052602} (\bibinfo {year} {2012})}\BibitemShut {NoStop}%
\bibitem [{\citenamefont {Birge}(2018)}]{Birge_review_2018}%
  \BibitemOpen
  \bibfield  {author} {\bibinfo {author} {\bibfnamefont {N.~O.}\ \bibnamefont {Birge}},\ }\bibfield  {title} {\enquote {\bibinfo {title} {Spin-triplet supercurrents in {J}osephson junctions containing strong ferromagnetic materials},}\ }\href {\doibase 10.1098/rsta.2015.0150} {\bibfield  {journal} {\bibinfo  {journal} {Philos. Trans. Royal Soc. A}\ }\textbf {\bibinfo {volume} {376}},\ \bibinfo {pages} {20150150} (\bibinfo {year} {2018})}\BibitemShut {NoStop}%
\bibitem [{\citenamefont {Houzet}\ and\ \citenamefont {Buzdin}(2007)}]{houzet_2007}%
  \BibitemOpen
  \bibfield  {author} {\bibinfo {author} {\bibfnamefont {M.}~\bibnamefont {Houzet}}\ and\ \bibinfo {author} {\bibfnamefont {A.~I.}\ \bibnamefont {Buzdin}},\ }\bibfield  {title} {\enquote {\bibinfo {title} {Long range triplet {Josephson} effect through a ferromagnetic trilayer},}\ }\href {\doibase 10.1103/PhysRevB.76.060504} {\bibfield  {journal} {\bibinfo  {journal} {Phys. Rev. B}\ }\textbf {\bibinfo {volume} {76}},\ \bibinfo {pages} {060504} (\bibinfo {year} {2007})}\BibitemShut {NoStop}%
\bibitem [{\citenamefont {Volkov}\ and\ \citenamefont {Efetov}(2010)}]{volkov_2010}%
  \BibitemOpen
  \bibfield  {author} {\bibinfo {author} {\bibfnamefont {A.~F.}\ \bibnamefont {Volkov}}\ and\ \bibinfo {author} {\bibfnamefont {K.~B.}\ \bibnamefont {Efetov}},\ }\bibfield  {title} {\enquote {\bibinfo {title} {Odd spin-triplet superconductivity in a multilayered superconductor-ferromagnet {Josephson} junction},}\ }\href {\doibase 10.1103/PhysRevB.81.144522} {\bibfield  {journal} {\bibinfo  {journal} {Phys. Rev. B}\ }\textbf {\bibinfo {volume} {81}},\ \bibinfo {pages} {144522} (\bibinfo {year} {2010})}\BibitemShut {NoStop}%
\bibitem [{\citenamefont {Trifunovic}\ and\ \citenamefont {Radović}(2010)}]{trifunovic_2010}%
  \BibitemOpen
  \bibfield  {author} {\bibinfo {author} {\bibfnamefont {L.}~\bibnamefont {Trifunovic}}\ and\ \bibinfo {author} {\bibfnamefont {Z.}~\bibnamefont {Radović}},\ }\bibfield  {title} {\enquote {\bibinfo {title} {Long-range spin-triplet proximity effect in {Josephson} junctions with multilayered ferromagnets},}\ }\href {\doibase 10.1103/PhysRevB.82.020505} {\bibfield  {journal} {\bibinfo  {journal} {Phys. Rev. B}\ }\textbf {\bibinfo {volume} {82}},\ \bibinfo {pages} {020505} (\bibinfo {year} {2010})}\BibitemShut {NoStop}%
\bibitem [{\citenamefont {Volkov}, \citenamefont {Anishchanka},\ and\ \citenamefont {Efetov}(2006)}]{volkov_2006}%
  \BibitemOpen
  \bibfield  {author} {\bibinfo {author} {\bibfnamefont {A.~F.}\ \bibnamefont {Volkov}}, \bibinfo {author} {\bibfnamefont {A.}~\bibnamefont {Anishchanka}}, \ and\ \bibinfo {author} {\bibfnamefont {K.~B.}\ \bibnamefont {Efetov}},\ }\bibfield  {title} {\enquote {\bibinfo {title} {Odd triplet superconductivity in a superconductor/ferromagnet system with a spiral magnetic structure},}\ }\href {\doibase 10.1103/{PhysRevB}.73.104412} {\bibfield  {journal} {\bibinfo  {journal} {Phys. Rev. B}\ }\textbf {\bibinfo {volume} {73}},\ \bibinfo {pages} {104412} (\bibinfo {year} {2006})}\BibitemShut {NoStop}%
\bibitem [{\citenamefont {Halász}\ \emph {et~al.}(2009)\citenamefont {Halász}, \citenamefont {Robinson}, \citenamefont {Annett},\ and\ \citenamefont {Blamire}}]{halsz_2009}%
  \BibitemOpen
  \bibfield  {author} {\bibinfo {author} {\bibfnamefont {G.~B.}\ \bibnamefont {Halász}}, \bibinfo {author} {\bibfnamefont {J.~W.~A.}\ \bibnamefont {Robinson}}, \bibinfo {author} {\bibfnamefont {J.~F.}\ \bibnamefont {Annett}}, \ and\ \bibinfo {author} {\bibfnamefont {M.~G.}\ \bibnamefont {Blamire}},\ }\bibfield  {title} {\enquote {\bibinfo {title} {Critical current of a {J}osephson junction containing a conical magnet},}\ }\href {\doibase 10.1103/{PhysRevB}.79.224505} {\bibfield  {journal} {\bibinfo  {journal} {Phys. Rev. B}\ }\textbf {\bibinfo {volume} {79}},\ \bibinfo {pages} {224505} (\bibinfo {year} {2009})}\BibitemShut {NoStop}%
\bibitem [{\citenamefont {Glick}\ \emph {et~al.}(2018)\citenamefont {Glick}, \citenamefont {Aguilar}, \citenamefont {Gougam}, \citenamefont {Niedzielski}, \citenamefont {Gingrich}, \citenamefont {Loloee}, \citenamefont {Pratt},\ and\ \citenamefont {Birge}}]{Glick_SciAdv_2018}%
  \BibitemOpen
  \bibfield  {author} {\bibinfo {author} {\bibfnamefont {J.~A.}\ \bibnamefont {Glick}}, \bibinfo {author} {\bibfnamefont {V.}~\bibnamefont {Aguilar}}, \bibinfo {author} {\bibfnamefont {A.~B.}\ \bibnamefont {Gougam}}, \bibinfo {author} {\bibfnamefont {B.~M.}\ \bibnamefont {Niedzielski}}, \bibinfo {author} {\bibfnamefont {E.~C.}\ \bibnamefont {Gingrich}}, \bibinfo {author} {\bibfnamefont {R.}~\bibnamefont {Loloee}}, \bibinfo {author} {\bibfnamefont {W.~P.}\ \bibnamefont {Pratt}}, \ and\ \bibinfo {author} {\bibfnamefont {N.~O.}\ \bibnamefont {Birge}},\ }\bibfield  {title} {\enquote {\bibinfo {title} {Phase control in a spin-triplet {SQUID}},}\ }\href {\doibase 10.1126/sciadv.aat9457} {\bibfield  {journal} {\bibinfo  {journal} {Sci. Adv.}\ }\textbf {\bibinfo {volume} {4}},\ \bibinfo {pages} {eaat9457} (\bibinfo {year} {2018})}\BibitemShut {NoStop}%
\bibitem [{\citenamefont {Birge}\ and\ \citenamefont {Houzet}(2019)}]{Birge_IEEE_2019}%
  \BibitemOpen
  \bibfield  {author} {\bibinfo {author} {\bibfnamefont {N.~O.}\ \bibnamefont {Birge}}\ and\ \bibinfo {author} {\bibfnamefont {M.}~\bibnamefont {Houzet}},\ }\bibfield  {title} {\enquote {\bibinfo {title} {Spin-{Singlet and Spin-Triplet Josephson Junctions for Cryogenic M}emory},}\ }\href {\doibase 10.1109/LMAG.2019.2955419} {\bibfield  {journal} {\bibinfo  {journal} {IEEE Magn. Lett.}\ }\textbf {\bibinfo {volume} {10}},\ \bibinfo {pages} {4509605} (\bibinfo {year} {2019})}\BibitemShut {NoStop}%
\bibitem [{\citenamefont {Buzdin}\ and\ \citenamefont {Koshelev}(2003)}]{buzdin_2003}%
  \BibitemOpen
  \bibfield  {author} {\bibinfo {author} {\bibfnamefont {A.}~\bibnamefont {Buzdin}}\ and\ \bibinfo {author} {\bibfnamefont {A.~E.}\ \bibnamefont {Koshelev}},\ }\bibfield  {title} {\enquote {\bibinfo {title} {Periodic alternating 0- and $\pi$-junction structures as realization of $\phi$-{Josephson} junctions},}\ }\href {\doibase 10.1103/PhysRevB.67.220504} {\bibfield  {journal} {\bibinfo  {journal} {Phys. Rev. B}\ }\textbf {\bibinfo {volume} {67}},\ \bibinfo {pages} {220504} (\bibinfo {year} {2003})}\BibitemShut {NoStop}%
\bibitem [{\citenamefont {Goldobin}\ \emph {et~al.}(2007)\citenamefont {Goldobin}, \citenamefont {Koelle}, \citenamefont {Kleiner},\ and\ \citenamefont {Buzdin}}]{goldobin_2007}%
  \BibitemOpen
  \bibfield  {author} {\bibinfo {author} {\bibfnamefont {E.}~\bibnamefont {Goldobin}}, \bibinfo {author} {\bibfnamefont {D.}~\bibnamefont {Koelle}}, \bibinfo {author} {\bibfnamefont {R.}~\bibnamefont {Kleiner}}, \ and\ \bibinfo {author} {\bibfnamefont {A.}~\bibnamefont {Buzdin}},\ }\bibfield  {title} {\enquote {\bibinfo {title} {Josephson junctions with second harmonic in the current-phase relation: {P}roperties of $\phi$ junctions},}\ }\href {https://journals.aps.org/prb/abstract/10.1103/PhysRevB.76.224523} {\bibfield  {journal} {\bibinfo  {journal} {Phys. Rev. B}\ }\textbf {\bibinfo {volume} {76}},\ \bibinfo {pages} {224523} (\bibinfo {year} {2007})}\BibitemShut {NoStop}%
\bibitem [{\citenamefont {Pugach}\ \emph {et~al.}(2010)\citenamefont {Pugach}, \citenamefont {Goldobin}, \citenamefont {Kleiner},\ and\ \citenamefont {Koelle}}]{pugach_2010}%
  \BibitemOpen
  \bibfield  {author} {\bibinfo {author} {\bibfnamefont {N.~G.}\ \bibnamefont {Pugach}}, \bibinfo {author} {\bibfnamefont {E.}~\bibnamefont {Goldobin}}, \bibinfo {author} {\bibfnamefont {R.}~\bibnamefont {Kleiner}}, \ and\ \bibinfo {author} {\bibfnamefont {D.}~\bibnamefont {Koelle}},\ }\bibfield  {title} {\enquote {\bibinfo {title} {Method for reliable realization of a $\varphi$ {Josephson} junction},}\ }\href {https://journals.aps.org/prb/abstract/10.1103/PhysRevB.81.104513} {\bibfield  {journal} {\bibinfo  {journal} {Phys. Rev. B}\ }\textbf {\bibinfo {volume} {81}},\ \bibinfo {pages} {104513} (\bibinfo {year} {2010})}\BibitemShut {NoStop}%
\bibitem [{\citenamefont {Sickinger}\ \emph {et~al.}(2012)\citenamefont {Sickinger}, \citenamefont {Lipman}, \citenamefont {Weides}, \citenamefont {Mints}, \citenamefont {Kohlstedt}, \citenamefont {Koelle}, \citenamefont {Kleiner},\ and\ \citenamefont {Goldobin}}]{sickinger_2012}%
  \BibitemOpen
  \bibfield  {author} {\bibinfo {author} {\bibfnamefont {H.}~\bibnamefont {Sickinger}}, \bibinfo {author} {\bibfnamefont {A.}~\bibnamefont {Lipman}}, \bibinfo {author} {\bibfnamefont {M.}~\bibnamefont {Weides}}, \bibinfo {author} {\bibfnamefont {R.~G.}\ \bibnamefont {Mints}}, \bibinfo {author} {\bibfnamefont {H.}~\bibnamefont {Kohlstedt}}, \bibinfo {author} {\bibfnamefont {D.}~\bibnamefont {Koelle}}, \bibinfo {author} {\bibfnamefont {R.}~\bibnamefont {Kleiner}}, \ and\ \bibinfo {author} {\bibfnamefont {E.}~\bibnamefont {Goldobin}},\ }\bibfield  {title} {\enquote {\bibinfo {title} {Experimental {E}vidence of a $\varphi$ {Josephson} {J}unction},}\ }\href {\doibase 10.1103/PhysRevLett.109.107002} {\bibfield  {journal} {\bibinfo  {journal} {Phys. Rev. Lett.}\ }\textbf {\bibinfo {volume} {109}},\ \bibinfo {pages} {107002} (\bibinfo {year} {2012})}\BibitemShut {NoStop}%
\bibitem [{\citenamefont {Bakurskiy}\ \emph {et~al.}(2013{\natexlab{b}})\citenamefont {Bakurskiy}, \citenamefont {Klenov}, \citenamefont {Karminskaya}, \citenamefont {Kupriyanov},\ and\ \citenamefont {Golubov}}]{bakurskiy_2013}%
  \BibitemOpen
  \bibfield  {author} {\bibinfo {author} {\bibfnamefont {S.~V.}\ \bibnamefont {Bakurskiy}}, \bibinfo {author} {\bibfnamefont {N.~V.}\ \bibnamefont {Klenov}}, \bibinfo {author} {\bibfnamefont {T.~Y.}\ \bibnamefont {Karminskaya}}, \bibinfo {author} {\bibfnamefont {M.~Y.}\ \bibnamefont {Kupriyanov}}, \ and\ \bibinfo {author} {\bibfnamefont {A.~A.}\ \bibnamefont {Golubov}},\ }\bibfield  {title} {\enquote {\bibinfo {title} {Josephson $\phi$-junctions based on structures with complex normal/ferromagnet bilayer},}\ }\href {\doibase 10.1088/0953-2048/26/1/015005} {\bibfield  {journal} {\bibinfo  {journal} {Supercond. Sci. Technol.}\ }\textbf {\bibinfo {volume} {26}},\ \bibinfo {pages} {015005} (\bibinfo {year} {2013}{\natexlab{b}})}\BibitemShut {NoStop}%
\bibitem [{\citenamefont {Silaev}, \citenamefont {Tokatly},\ and\ \citenamefont {Bergeret}(2017)}]{silaev_2017}%
  \BibitemOpen
  \bibfield  {author} {\bibinfo {author} {\bibfnamefont {M.~A.}\ \bibnamefont {Silaev}}, \bibinfo {author} {\bibfnamefont {I.~V.}\ \bibnamefont {Tokatly}}, \ and\ \bibinfo {author} {\bibfnamefont {F.~S.}\ \bibnamefont {Bergeret}},\ }\bibfield  {title} {\enquote {\bibinfo {title} {Anomalous current in diffusive ferromagnetic {Josephson} junctions},}\ }\href {\doibase 10.1103/PhysRevB.95.184508} {\bibfield  {journal} {\bibinfo  {journal} {Phys. Rev. B}\ }\textbf {\bibinfo {volume} {95}},\ \bibinfo {pages} {184508} (\bibinfo {year} {2017})}\BibitemShut {NoStop}%
\bibitem [{\citenamefont {Braude}\ and\ \citenamefont {Nazarov}(2007)}]{braude_2007}%
  \BibitemOpen
  \bibfield  {author} {\bibinfo {author} {\bibfnamefont {V.}~\bibnamefont {Braude}}\ and\ \bibinfo {author} {\bibfnamefont {Y.~V.}\ \bibnamefont {Nazarov}},\ }\bibfield  {title} {\enquote {\bibinfo {title} {Fully developed triplet proximity effect},}\ }\href {\doibase 10.1103/PhysRevLett.98.077003} {\bibfield  {journal} {\bibinfo  {journal} {Phys. Rev. Lett.}\ }\textbf {\bibinfo {volume} {98}},\ \bibinfo {pages} {077003} (\bibinfo {year} {2007})}\BibitemShut {NoStop}%
\bibitem [{\citenamefont {Buzdin}(2008)}]{buzdin_2008}%
  \BibitemOpen
  \bibfield  {author} {\bibinfo {author} {\bibfnamefont {A.}~\bibnamefont {Buzdin}},\ }\bibfield  {title} {\enquote {\bibinfo {title} {Direct coupling between magnetism and superconducting current in the {Josephson} $\phi_0$ junction},}\ }\href {\doibase 10.1103/PhysRevLett.101.107005} {\bibfield  {journal} {\bibinfo  {journal} {Phys. Rev. Lett.}\ }\textbf {\bibinfo {volume} {101}},\ \bibinfo {pages} {107005} (\bibinfo {year} {2008})}\BibitemShut {NoStop}%
\bibitem [{\citenamefont {Galaktionov}, \citenamefont {Kalenkov},\ and\ \citenamefont {Zaikin}(2008)}]{galaktionov_2008}%
  \BibitemOpen
  \bibfield  {author} {\bibinfo {author} {\bibfnamefont {A.~V.}\ \bibnamefont {Galaktionov}}, \bibinfo {author} {\bibfnamefont {M.~S.}\ \bibnamefont {Kalenkov}}, \ and\ \bibinfo {author} {\bibfnamefont {A.~D.}\ \bibnamefont {Zaikin}},\ }\bibfield  {title} {\enquote {\bibinfo {title} {Josephson current and {Andreev} states in superconductor–half metal–superconductor heterostructures},}\ }\href {\doibase 10.1103/PhysRevB.77.094520} {\bibfield  {journal} {\bibinfo  {journal} {Phys. Rev. B}\ }\textbf {\bibinfo {volume} {77}},\ \bibinfo {pages} {094520} (\bibinfo {year} {2008})}\BibitemShut {NoStop}%
\bibitem [{\citenamefont {Grein}\ \emph {et~al.}(2009)\citenamefont {Grein}, \citenamefont {Eschrig}, \citenamefont {Metalidis},\ and\ \citenamefont {Schön}}]{grein_2009}%
  \BibitemOpen
  \bibfield  {author} {\bibinfo {author} {\bibfnamefont {R.}~\bibnamefont {Grein}}, \bibinfo {author} {\bibfnamefont {M.}~\bibnamefont {Eschrig}}, \bibinfo {author} {\bibfnamefont {G.}~\bibnamefont {Metalidis}}, \ and\ \bibinfo {author} {\bibfnamefont {G.}~\bibnamefont {Schön}},\ }\bibfield  {title} {\enquote {\bibinfo {title} {Spin-dependent {Cooper} pair phase and pure spin supercurrents in strongly polarized ferromagnets.}}\ }\href {\doibase 10.1103/PhysRevLett.102.227005} {\bibfield  {journal} {\bibinfo  {journal} {Phys. Rev. Lett.}\ }\textbf {\bibinfo {volume} {102}},\ \bibinfo {pages} {227005} (\bibinfo {year} {2009})}\BibitemShut {NoStop}%
\bibitem [{\citenamefont {Margaris}, \citenamefont {Paltoglou},\ and\ \citenamefont {Flytzanis}(2010)}]{Margaris_2010}%
  \BibitemOpen
  \bibfield  {author} {\bibinfo {author} {\bibfnamefont {I.}~\bibnamefont {Margaris}}, \bibinfo {author} {\bibfnamefont {V.}~\bibnamefont {Paltoglou}}, \ and\ \bibinfo {author} {\bibfnamefont {N.}~\bibnamefont {Flytzanis}},\ }\bibfield  {title} {\enquote {\bibinfo {title} {Zero phase difference supercurrent in ferromagnetic {J}osephson junctions},}\ }\href {\doibase 10.1088/0953-8984/22/44/445701} {\bibfield  {journal} {\bibinfo  {journal} {J. Phys. Condens. Matter}\ }\textbf {\bibinfo {volume} {22}},\ \bibinfo {pages} {445701} (\bibinfo {year} {2010})}\BibitemShut {NoStop}%
\bibitem [{\citenamefont {Liu}\ and\ \citenamefont {Chan}(2010)}]{liu_2010}%
  \BibitemOpen
  \bibfield  {author} {\bibinfo {author} {\bibfnamefont {J.-F.}\ \bibnamefont {Liu}}\ and\ \bibinfo {author} {\bibfnamefont {K.~S.}\ \bibnamefont {Chan}},\ }\bibfield  {title} {\enquote {\bibinfo {title} {Anomalous {Josephson} current through a ferromagnetic trilayer junction},}\ }\href {\doibase 10.1103/PhysRevB.82.184533} {\bibfield  {journal} {\bibinfo  {journal} {Phys. Rev. B}\ }\textbf {\bibinfo {volume} {82}},\ \bibinfo {pages} {184533} (\bibinfo {year} {2010})}\BibitemShut {NoStop}%
\bibitem [{\citenamefont {Alidoust}\ and\ \citenamefont {Linder}(2013)}]{alidoust_2013}%
  \BibitemOpen
  \bibfield  {author} {\bibinfo {author} {\bibfnamefont {M.}~\bibnamefont {Alidoust}}\ and\ \bibinfo {author} {\bibfnamefont {J.}~\bibnamefont {Linder}},\ }\bibfield  {title} {\enquote {\bibinfo {title} {$\varphi$-state and inverted {F}raunhofer pattern in nonaligned {Josephson} junctions},}\ }\href {\doibase 10.1103/PhysRevB.87.060503} {\bibfield  {journal} {\bibinfo  {journal} {Phys. Rev. B}\ }\textbf {\bibinfo {volume} {87}},\ \bibinfo {pages} {060503} (\bibinfo {year} {2013})}\BibitemShut {NoStop}%
\bibitem [{\citenamefont {Kulagina}\ and\ \citenamefont {Linder}(2014)}]{kulagina_2014}%
  \BibitemOpen
  \bibfield  {author} {\bibinfo {author} {\bibfnamefont {I.}~\bibnamefont {Kulagina}}\ and\ \bibinfo {author} {\bibfnamefont {J.}~\bibnamefont {Linder}},\ }\bibfield  {title} {\enquote {\bibinfo {title} {Spin supercurrent, magnetization dynamics, and $\varphi$-state in spin-textured {Josephson} junctions},}\ }\href {https://journals.aps.org/prb/abstract/10.1103/PhysRevB.90.054504} {\bibfield  {journal} {\bibinfo  {journal} {Phys. Rev. B}\ }\textbf {\bibinfo {volume} {90}},\ \bibinfo {pages} {054504} (\bibinfo {year} {2014})}\BibitemShut {NoStop}%
\bibitem [{\citenamefont {Konschelle}, \citenamefont {Tokatly},\ and\ \citenamefont {Bergeret}(2015)}]{konschelle_2015}%
  \BibitemOpen
  \bibfield  {author} {\bibinfo {author} {\bibfnamefont {F.}~\bibnamefont {Konschelle}}, \bibinfo {author} {\bibfnamefont {I.~V.}\ \bibnamefont {Tokatly}}, \ and\ \bibinfo {author} {\bibfnamefont {F.~S.}\ \bibnamefont {Bergeret}},\ }\bibfield  {title} {\enquote {\bibinfo {title} {Theory of the spin-galvanic effect and the anomalous phase shift in superconductors and {Josephson} junctions with intrinsic spin-orbit coupling},}\ }\href {\doibase 10.1103/PhysRevB.92.125443} {\bibfield  {journal} {\bibinfo  {journal} {Phys. Rev. B}\ }\textbf {\bibinfo {volume} {92}},\ \bibinfo {pages} {125443} (\bibinfo {year} {2015})}\BibitemShut {NoStop}%
\bibitem [{\citenamefont {Pal}\ and\ \citenamefont {Benjamin}(2019)}]{pal_2019}%
  \BibitemOpen
  \bibfield  {author} {\bibinfo {author} {\bibfnamefont {S.}~\bibnamefont {Pal}}\ and\ \bibinfo {author} {\bibfnamefont {C.}~\bibnamefont {Benjamin}},\ }\bibfield  {title} {\enquote {\bibinfo {title} {Quantized {Josephson} phase battery},}\ }\href {\doibase 10.1209/0295-5075/126/57002} {\bibfield  {journal} {\bibinfo  {journal} {EPL}\ }\textbf {\bibinfo {volume} {126}},\ \bibinfo {pages} {57002} (\bibinfo {year} {2019})}\BibitemShut {NoStop}%
\bibitem [{\citenamefont {Shukrinov}(2022)}]{shukrinov_2022}%
  \BibitemOpen
  \bibfield  {author} {\bibinfo {author} {\bibfnamefont {Y.~M.}\ \bibnamefont {Shukrinov}},\ }\bibfield  {title} {\enquote {\bibinfo {title} {Anomalous {Josephson} effect},}\ }\href {\doibase 10.3367/UFNe.2020.11.038894} {\bibfield  {journal} {\bibinfo  {journal} {Phys.-Usp.}\ }\textbf {\bibinfo {volume} {65}},\ \bibinfo {pages} {317--354} (\bibinfo {year} {2022})}\BibitemShut {NoStop}%
\bibitem [{\citenamefont {Bobkova}, \citenamefont {Bobkov},\ and\ \citenamefont {Silaev}(2022)}]{bobkova_2022}%
  \BibitemOpen
  \bibfield  {author} {\bibinfo {author} {\bibfnamefont {I.~V.}\ \bibnamefont {Bobkova}}, \bibinfo {author} {\bibfnamefont {A.~M.}\ \bibnamefont {Bobkov}}, \ and\ \bibinfo {author} {\bibfnamefont {M.~A.}\ \bibnamefont {Silaev}},\ }\bibfield  {title} {\enquote {\bibinfo {title} {Magnetoelectric effects in {Josephson} junctions},}\ }\href {\doibase 10.1088/1361-648X/ac7994} {\bibfield  {journal} {\bibinfo  {journal} {J. Phys. Condens. Matter}\ }\textbf {\bibinfo {volume} {34}},\ \bibinfo {pages} {353001} (\bibinfo {year} {2022})}\BibitemShut {NoStop}%
\bibitem [{\citenamefont {Szombati}\ \emph {et~al.}(2016)\citenamefont {Szombati}, \citenamefont {Nadj-Perge}, \citenamefont {Car}, \citenamefont {Plissard}, \citenamefont {Bakkers},\ and\ \citenamefont {Kouwenhoven}}]{szombati_2016}%
  \BibitemOpen
  \bibfield  {author} {\bibinfo {author} {\bibfnamefont {D.~B.}\ \bibnamefont {Szombati}}, \bibinfo {author} {\bibfnamefont {S.}~\bibnamefont {Nadj-Perge}}, \bibinfo {author} {\bibfnamefont {D.}~\bibnamefont {Car}}, \bibinfo {author} {\bibfnamefont {S.~R.}\ \bibnamefont {Plissard}}, \bibinfo {author} {\bibfnamefont {E.~P. A.~M.}\ \bibnamefont {Bakkers}}, \ and\ \bibinfo {author} {\bibfnamefont {L.~P.}\ \bibnamefont {Kouwenhoven}},\ }\bibfield  {title} {\enquote {\bibinfo {title} {{Josephson} $\phi_0$-junction in nanowire quantum dots},}\ }\href {\doibase 10.1038/nphys3742} {\bibfield  {journal} {\bibinfo  {journal} {Nat. Phys.}\ }\textbf {\bibinfo {volume} {12}},\ \bibinfo {pages} {568--572} (\bibinfo {year} {2016})}\BibitemShut {NoStop}%
\bibitem [{\citenamefont {Murani}\ \emph {et~al.}(2017)\citenamefont {Murani}, \citenamefont {Kasumov}, \citenamefont {Sengupta}, \citenamefont {Kasumov}, \citenamefont {Volkov}, \citenamefont {Khodos}, \citenamefont {Brisset}, \citenamefont {Delagrange}, \citenamefont {Chepelianskii}, \citenamefont {Deblock}, \citenamefont {Bouchiat},\ and\ \citenamefont {Guéron}}]{murani_2017}%
  \BibitemOpen
  \bibfield  {author} {\bibinfo {author} {\bibfnamefont {A.}~\bibnamefont {Murani}}, \bibinfo {author} {\bibfnamefont {A.}~\bibnamefont {Kasumov}}, \bibinfo {author} {\bibfnamefont {S.}~\bibnamefont {Sengupta}}, \bibinfo {author} {\bibfnamefont {Y.~A.}\ \bibnamefont {Kasumov}}, \bibinfo {author} {\bibfnamefont {V.~T.}\ \bibnamefont {Volkov}}, \bibinfo {author} {\bibfnamefont {I.~I.}\ \bibnamefont {Khodos}}, \bibinfo {author} {\bibfnamefont {F.}~\bibnamefont {Brisset}}, \bibinfo {author} {\bibfnamefont {R.}~\bibnamefont {Delagrange}}, \bibinfo {author} {\bibfnamefont {A.}~\bibnamefont {Chepelianskii}}, \bibinfo {author} {\bibfnamefont {R.}~\bibnamefont {Deblock}}, \bibinfo {author} {\bibfnamefont {H.}~\bibnamefont {Bouchiat}}, \ and\ \bibinfo {author} {\bibfnamefont {S.}~\bibnamefont {Guéron}},\ }\bibfield  {title} {\enquote {\bibinfo {title} {Ballistic edge states in {B}ismuth nanowires revealed by {SQUID} interferometry},}\ }\href {\doibase 10.1038/ncomms15941} {\bibfield  {journal} {\bibinfo  {journal}
  {Nat. Commun.}\ }\textbf {\bibinfo {volume} {8}},\ \bibinfo {pages} {15941} (\bibinfo {year} {2017})}\BibitemShut {NoStop}%
\bibitem [{\citenamefont {Assouline}\ \emph {et~al.}(2019)\citenamefont {Assouline}, \citenamefont {Feuillet-Palma}, \citenamefont {Bergeal}, \citenamefont {Zhang}, \citenamefont {Mottaghizadeh}, \citenamefont {Zimmers}, \citenamefont {Lhuillier}, \citenamefont {Eddrie}, \citenamefont {Atkinson}, \citenamefont {Aprili},\ and\ \citenamefont {Aubin}}]{assouline_2019}%
  \BibitemOpen
  \bibfield  {author} {\bibinfo {author} {\bibfnamefont {A.}~\bibnamefont {Assouline}}, \bibinfo {author} {\bibfnamefont {C.}~\bibnamefont {Feuillet-Palma}}, \bibinfo {author} {\bibfnamefont {N.}~\bibnamefont {Bergeal}}, \bibinfo {author} {\bibfnamefont {T.}~\bibnamefont {Zhang}}, \bibinfo {author} {\bibfnamefont {A.}~\bibnamefont {Mottaghizadeh}}, \bibinfo {author} {\bibfnamefont {A.}~\bibnamefont {Zimmers}}, \bibinfo {author} {\bibfnamefont {E.}~\bibnamefont {Lhuillier}}, \bibinfo {author} {\bibfnamefont {M.}~\bibnamefont {Eddrie}}, \bibinfo {author} {\bibfnamefont {P.}~\bibnamefont {Atkinson}}, \bibinfo {author} {\bibfnamefont {M.}~\bibnamefont {Aprili}}, \ and\ \bibinfo {author} {\bibfnamefont {H.}~\bibnamefont {Aubin}},\ }\bibfield  {title} {\enquote {\bibinfo {title} {Spin-{O}rbit induced phase-shift in {Bi}$_2${Se}$_3$ {Josephson} junctions.}}\ }\href {\doibase 10.1038/s41467-018-08022-y} {\bibfield  {journal} {\bibinfo  {journal} {Nat. Commun.}\ }\textbf {\bibinfo {volume} {10}},\ \bibinfo {pages}
  {126} (\bibinfo {year} {2019})}\BibitemShut {NoStop}%
\bibitem [{\citenamefont {Ren}\ \emph {et~al.}(2019)\citenamefont {Ren}, \citenamefont {Pientka}, \citenamefont {Hart}, \citenamefont {Pierce}, \citenamefont {Kosowsky}, \citenamefont {Lunczer}, \citenamefont {Schlereth}, \citenamefont {Scharf}, \citenamefont {Hankiewicz}, \citenamefont {Molenkamp}, \citenamefont {Halperin},\ and\ \citenamefont {Yacoby}}]{ren_2019}%
  \BibitemOpen
  \bibfield  {author} {\bibinfo {author} {\bibfnamefont {H.}~\bibnamefont {Ren}}, \bibinfo {author} {\bibfnamefont {F.}~\bibnamefont {Pientka}}, \bibinfo {author} {\bibfnamefont {S.}~\bibnamefont {Hart}}, \bibinfo {author} {\bibfnamefont {A.~T.}\ \bibnamefont {Pierce}}, \bibinfo {author} {\bibfnamefont {M.}~\bibnamefont {Kosowsky}}, \bibinfo {author} {\bibfnamefont {L.}~\bibnamefont {Lunczer}}, \bibinfo {author} {\bibfnamefont {R.}~\bibnamefont {Schlereth}}, \bibinfo {author} {\bibfnamefont {B.}~\bibnamefont {Scharf}}, \bibinfo {author} {\bibfnamefont {E.~M.}\ \bibnamefont {Hankiewicz}}, \bibinfo {author} {\bibfnamefont {L.~W.}\ \bibnamefont {Molenkamp}}, \bibinfo {author} {\bibfnamefont {B.~I.}\ \bibnamefont {Halperin}}, \ and\ \bibinfo {author} {\bibfnamefont {A.}~\bibnamefont {Yacoby}},\ }\bibfield  {title} {\enquote {\bibinfo {title} {Topological superconductivity in a phase-controlled {Josephson} junction},}\ }\href {\doibase 10.1038/s41586-019-1148-9} {\bibfield  {journal} {\bibinfo  {journal} {Nature}\
  }\textbf {\bibinfo {volume} {569}},\ \bibinfo {pages} {93--98} (\bibinfo {year} {2019})}\BibitemShut {NoStop}%
\bibitem [{\citenamefont {Mayer}\ \emph {et~al.}(2020)\citenamefont {Mayer}, \citenamefont {Dartiailh}, \citenamefont {Yuan}, \citenamefont {Wickramasinghe}, \citenamefont {Rossi},\ and\ \citenamefont {Shabani}}]{Mayer_2020}%
  \BibitemOpen
  \bibfield  {author} {\bibinfo {author} {\bibfnamefont {W.}~\bibnamefont {Mayer}}, \bibinfo {author} {\bibfnamefont {M.~C.}\ \bibnamefont {Dartiailh}}, \bibinfo {author} {\bibfnamefont {J.}~\bibnamefont {Yuan}}, \bibinfo {author} {\bibfnamefont {K.~S.}\ \bibnamefont {Wickramasinghe}}, \bibinfo {author} {\bibfnamefont {E.}~\bibnamefont {Rossi}}, \ and\ \bibinfo {author} {\bibfnamefont {J.}~\bibnamefont {Shabani}},\ }\bibfield  {title} {\enquote {\bibinfo {title} {Gate controlled anomalous phase shift in {Al/InAs} {Josephson} junctions},}\ }\href {\doibase 10.1038/s41467-019-14094-1} {\bibfield  {journal} {\bibinfo  {journal} {Nat. Commun.}\ }\textbf {\bibinfo {volume} {11}},\ \bibinfo {pages} {212} (\bibinfo {year} {2020})}\BibitemShut {NoStop}%
\bibitem [{\citenamefont {Strambini}\ \emph {et~al.}(2020)\citenamefont {Strambini}, \citenamefont {Iorio}, \citenamefont {Durante}, \citenamefont {Citro}, \citenamefont {Sanz-Fernández}, \citenamefont {Guarcello}, \citenamefont {Tokatly}, \citenamefont {Braggio}, \citenamefont {Rocci}, \citenamefont {Ligato}, \citenamefont {Zannier}, \citenamefont {Sorba}, \citenamefont {Bergeret},\ and\ \citenamefont {Giazotto}}]{strambini_2020}%
  \BibitemOpen
  \bibfield  {author} {\bibinfo {author} {\bibfnamefont {E.}~\bibnamefont {Strambini}}, \bibinfo {author} {\bibfnamefont {A.}~\bibnamefont {Iorio}}, \bibinfo {author} {\bibfnamefont {O.}~\bibnamefont {Durante}}, \bibinfo {author} {\bibfnamefont {R.}~\bibnamefont {Citro}}, \bibinfo {author} {\bibfnamefont {C.}~\bibnamefont {Sanz-Fernández}}, \bibinfo {author} {\bibfnamefont {C.}~\bibnamefont {Guarcello}}, \bibinfo {author} {\bibfnamefont {I.~V.}\ \bibnamefont {Tokatly}}, \bibinfo {author} {\bibfnamefont {A.}~\bibnamefont {Braggio}}, \bibinfo {author} {\bibfnamefont {M.}~\bibnamefont {Rocci}}, \bibinfo {author} {\bibfnamefont {N.}~\bibnamefont {Ligato}}, \bibinfo {author} {\bibfnamefont {V.}~\bibnamefont {Zannier}}, \bibinfo {author} {\bibfnamefont {L.}~\bibnamefont {Sorba}}, \bibinfo {author} {\bibfnamefont {F.~S.}\ \bibnamefont {Bergeret}}, \ and\ \bibinfo {author} {\bibfnamefont {F.}~\bibnamefont {Giazotto}},\ }\bibfield  {title} {\enquote {\bibinfo {title} {A {Josephson} phase battery},}\ }\href {\doibase
  10.1038/s41565-020-0712-7} {\bibfield  {journal} {\bibinfo  {journal} {Nat. Nanotechnol.}\ }\textbf {\bibinfo {volume} {15}},\ \bibinfo {pages} {656--660} (\bibinfo {year} {2020})}\BibitemShut {NoStop}%
\bibitem [{\citenamefont {Baumgartner}\ \emph {et~al.}(2022)\citenamefont {Baumgartner}, \citenamefont {Fuchs}, \citenamefont {Costa}, \citenamefont {Picó-Cortés}, \citenamefont {Reinhardt}, \citenamefont {Gronin}, \citenamefont {Gardner}, \citenamefont {Lindemann}, \citenamefont {Manfra}, \citenamefont {Faria~Junior}, \citenamefont {Kochan}, \citenamefont {Fabian}, \citenamefont {Paradiso},\ and\ \citenamefont {Strunk}}]{baumgartner_2022}%
  \BibitemOpen
  \bibfield  {author} {\bibinfo {author} {\bibfnamefont {C.}~\bibnamefont {Baumgartner}}, \bibinfo {author} {\bibfnamefont {L.}~\bibnamefont {Fuchs}}, \bibinfo {author} {\bibfnamefont {A.}~\bibnamefont {Costa}}, \bibinfo {author} {\bibfnamefont {J.}~\bibnamefont {Picó-Cortés}}, \bibinfo {author} {\bibfnamefont {S.}~\bibnamefont {Reinhardt}}, \bibinfo {author} {\bibfnamefont {S.}~\bibnamefont {Gronin}}, \bibinfo {author} {\bibfnamefont {G.~C.}\ \bibnamefont {Gardner}}, \bibinfo {author} {\bibfnamefont {T.}~\bibnamefont {Lindemann}}, \bibinfo {author} {\bibfnamefont {M.~J.}\ \bibnamefont {Manfra}}, \bibinfo {author} {\bibfnamefont {P.~E.}\ \bibnamefont {Faria~Junior}}, \bibinfo {author} {\bibfnamefont {D.}~\bibnamefont {Kochan}}, \bibinfo {author} {\bibfnamefont {J.}~\bibnamefont {Fabian}}, \bibinfo {author} {\bibfnamefont {N.}~\bibnamefont {Paradiso}}, \ and\ \bibinfo {author} {\bibfnamefont {C.}~\bibnamefont {Strunk}},\ }\bibfield  {title} {\enquote {\bibinfo {title} {Effect of {Rashba} and {Dresselhaus}
  spin-orbit coupling on supercurrent rectification and magnetochiral anisotropy of ballistic {Josephson} junctions},}\ }\href {\doibase 10.1088/1361-648X/ac4d5e} {\bibfield  {journal} {\bibinfo  {journal} {J. Phys. Condens. Matter}\ }\textbf {\bibinfo {volume} {34}},\ \bibinfo {pages} {154005} (\bibinfo {year} {2022})}\BibitemShut {NoStop}%
\bibitem [{\citenamefont {Margineda}\ \emph {et~al.}(2023)\citenamefont {Margineda}, \citenamefont {Claydon}, \citenamefont {Qejvanaj},\ and\ \citenamefont {Checkley}}]{Margineda_2023}%
  \BibitemOpen
  \bibfield  {author} {\bibinfo {author} {\bibfnamefont {D.}~\bibnamefont {Margineda}}, \bibinfo {author} {\bibfnamefont {J.~S.}\ \bibnamefont {Claydon}}, \bibinfo {author} {\bibfnamefont {F.}~\bibnamefont {Qejvanaj}}, \ and\ \bibinfo {author} {\bibfnamefont {C.}~\bibnamefont {Checkley}},\ }\bibfield  {title} {\enquote {\bibinfo {title} {Observation of anomalous {J}osephson effect in nonequilibrium {A}ndreev interferometers},}\ }\href {\doibase 10.1103/PhysRevB.107.L100502} {\bibfield  {journal} {\bibinfo  {journal} {Phys. Rev. B}\ }\textbf {\bibinfo {volume} {107}},\ \bibinfo {pages} {L100502} (\bibinfo {year} {2023})}\BibitemShut {NoStop}%
\bibitem [{\citenamefont {Bell}\ \emph {et~al.}(2003{\natexlab{a}})\citenamefont {Bell}, \citenamefont {Burnell}, \citenamefont {Kang}, \citenamefont {Hadfield}, \citenamefont {Kappers},\ and\ \citenamefont {Blamire}}]{Bell_Nano_2003}%
  \BibitemOpen
  \bibfield  {author} {\bibinfo {author} {\bibfnamefont {C.}~\bibnamefont {Bell}}, \bibinfo {author} {\bibfnamefont {G.}~\bibnamefont {Burnell}}, \bibinfo {author} {\bibfnamefont {D.-J.}\ \bibnamefont {Kang}}, \bibinfo {author} {\bibfnamefont {R.~H.}\ \bibnamefont {Hadfield}}, \bibinfo {author} {\bibfnamefont {M.~J.}\ \bibnamefont {Kappers}}, \ and\ \bibinfo {author} {\bibfnamefont {M.~G.}\ \bibnamefont {Blamire}},\ }\bibfield  {title} {\enquote {\bibinfo {title} {Fabrication of nanoscale heterostructure devices with a focused ion beam microscope},}\ }\href {\doibase 10.1088/0957-4484/14/6/312} {\bibfield  {journal} {\bibinfo  {journal} {Nanotechnology}\ }\textbf {\bibinfo {volume} {14}},\ \bibinfo {pages} {630} (\bibinfo {year} {2003}{\natexlab{a}})}\BibitemShut {NoStop}%
\bibitem [{\citenamefont {Meiklejohn}\ and\ \citenamefont {Bean}(1957)}]{Meiklejohn_1957}%
  \BibitemOpen
  \bibfield  {author} {\bibinfo {author} {\bibfnamefont {W.~H.}\ \bibnamefont {Meiklejohn}}\ and\ \bibinfo {author} {\bibfnamefont {C.~P.}\ \bibnamefont {Bean}},\ }\bibfield  {title} {\enquote {\bibinfo {title} {New magnetic anisotropy},}\ }\href {\doibase 10.1103/PhysRev.105.904} {\bibfield  {journal} {\bibinfo  {journal} {Phys. Rev.}\ }\textbf {\bibinfo {volume} {105}},\ \bibinfo {pages} {904--913} (\bibinfo {year} {1957})}\BibitemShut {NoStop}%
\bibitem [{\citenamefont {Bell}\ \emph {et~al.}(2003{\natexlab{b}})\citenamefont {Bell}, \citenamefont {Tarte}, \citenamefont {Burnell}, \citenamefont {Leung}, \citenamefont {Kang},\ and\ \citenamefont {Blamire}}]{Bell_2003}%
  \BibitemOpen
  \bibfield  {author} {\bibinfo {author} {\bibfnamefont {C.}~\bibnamefont {Bell}}, \bibinfo {author} {\bibfnamefont {E.~J.}\ \bibnamefont {Tarte}}, \bibinfo {author} {\bibfnamefont {G.}~\bibnamefont {Burnell}}, \bibinfo {author} {\bibfnamefont {C.~W.}\ \bibnamefont {Leung}}, \bibinfo {author} {\bibfnamefont {D.-J.}\ \bibnamefont {Kang}}, \ and\ \bibinfo {author} {\bibfnamefont {M.~G.}\ \bibnamefont {Blamire}},\ }\bibfield  {title} {\enquote {\bibinfo {title} {Proximity and {Josephson} effects in superconductor/antiferromagnetic {Nb}/$\gamma$ -- {Fe}$_{50}${Mn}$_{50}$ heterostructures},}\ }\href {\doibase 10.1103/PhysRevB.68.144517} {\bibfield  {journal} {\bibinfo  {journal} {Phys. Rev. B}\ }\textbf {\bibinfo {volume} {68}},\ \bibinfo {pages} {144517} (\bibinfo {year} {2003}{\natexlab{b}})}\BibitemShut {NoStop}%
\bibitem [{\citenamefont {Weides}\ \emph {et~al.}(2009)\citenamefont {Weides}, \citenamefont {Disch}, \citenamefont {Kohlstedt},\ and\ \citenamefont {B\"urgler}}]{Weides_PRB_2009}%
  \BibitemOpen
  \bibfield  {author} {\bibinfo {author} {\bibfnamefont {M.}~\bibnamefont {Weides}}, \bibinfo {author} {\bibfnamefont {M.}~\bibnamefont {Disch}}, \bibinfo {author} {\bibfnamefont {H.}~\bibnamefont {Kohlstedt}}, \ and\ \bibinfo {author} {\bibfnamefont {D.~E.}\ \bibnamefont {B\"urgler}},\ }\bibfield  {title} {\enquote {\bibinfo {title} {Observation of {J}osephson coupling through an interlayer of antiferromagnetically ordered chromium},}\ }\href {\doibase 10.1103/PhysRevB.80.064508} {\bibfield  {journal} {\bibinfo  {journal} {Phys. Rev. B}\ }\textbf {\bibinfo {volume} {80}},\ \bibinfo {pages} {064508} (\bibinfo {year} {2009})}\BibitemShut {NoStop}%
\bibitem [{\citenamefont {Klaes}, \citenamefont {Loloee},\ and\ \citenamefont {Birge}(2023)}]{Klaes_2023}%
  \BibitemOpen
  \bibfield  {author} {\bibinfo {author} {\bibfnamefont {R.~M.}\ \bibnamefont {Klaes}}, \bibinfo {author} {\bibfnamefont {R.}~\bibnamefont {Loloee}}, \ and\ \bibinfo {author} {\bibfnamefont {N.~O.}\ \bibnamefont {Birge}},\ }\bibfield  {title} {\enquote {\bibinfo {title} {Critical {C}urrent {D}ecay in {Josephson} {J}unctions {C}ontaining {A}ntiferromagnetic {NiMn}},}\ }\href {\doibase 10.1109/TASC.2023.3257769} {\bibfield  {journal} {\bibinfo  {journal} {IEEE Trans. Appl. Supercond.}\ }\textbf {\bibinfo {volume} {33}},\ \bibinfo {pages} {1800903} (\bibinfo {year} {2023})}\BibitemShut {NoStop}%
\bibitem [{\citenamefont {Vettoliere}\ \emph {et~al.}(2022{\natexlab{a}})\citenamefont {Vettoliere}, \citenamefont {Satariano}, \citenamefont {Ferraiuolo}, \citenamefont {Di~Palma}, \citenamefont {Ahmad}, \citenamefont {Ausanio}, \citenamefont {Pepe}, \citenamefont {Tafuri}, \citenamefont {Massarotti}, \citenamefont {Montemurro} \emph {et~al.}}]{Vettoliere_2022a}%
  \BibitemOpen
  \bibfield  {author} {\bibinfo {author} {\bibfnamefont {A.}~\bibnamefont {Vettoliere}}, \bibinfo {author} {\bibfnamefont {R.}~\bibnamefont {Satariano}}, \bibinfo {author} {\bibfnamefont {R.}~\bibnamefont {Ferraiuolo}}, \bibinfo {author} {\bibfnamefont {L.}~\bibnamefont {Di~Palma}}, \bibinfo {author} {\bibfnamefont {H.~G.}\ \bibnamefont {Ahmad}}, \bibinfo {author} {\bibfnamefont {G.}~\bibnamefont {Ausanio}}, \bibinfo {author} {\bibfnamefont {G.~P.}\ \bibnamefont {Pepe}}, \bibinfo {author} {\bibfnamefont {F.}~\bibnamefont {Tafuri}}, \bibinfo {author} {\bibfnamefont {D.}~\bibnamefont {Massarotti}}, \bibinfo {author} {\bibfnamefont {D.}~\bibnamefont {Montemurro}},  \emph {et~al.},\ }\bibfield  {title} {\enquote {\bibinfo {title} {High-{Quality ferromagnetic J}osephson junctions based on aluminum electrodes},}\ }\href {\doibase 10.3390/nano12234155} {\bibfield  {journal} {\bibinfo  {journal} {Nanomater.}\ }\textbf {\bibinfo {volume} {12}},\ \bibinfo {pages} {4155} (\bibinfo {year}
  {2022}{\natexlab{a}})}\BibitemShut {NoStop}%
\bibitem [{\citenamefont {Vettoliere}\ \emph {et~al.}(2022{\natexlab{b}})\citenamefont {Vettoliere}, \citenamefont {Satariano}, \citenamefont {Ferraiuolo}, \citenamefont {Di~Palma}, \citenamefont {Ahmad}, \citenamefont {Ausanio}, \citenamefont {Pepe}, \citenamefont {Tafuri}, \citenamefont {Montemurro}, \citenamefont {Granata} \emph {et~al.}}]{Vettoliere_2022b}%
  \BibitemOpen
  \bibfield  {author} {\bibinfo {author} {\bibfnamefont {A.}~\bibnamefont {Vettoliere}}, \bibinfo {author} {\bibfnamefont {R.}~\bibnamefont {Satariano}}, \bibinfo {author} {\bibfnamefont {R.}~\bibnamefont {Ferraiuolo}}, \bibinfo {author} {\bibfnamefont {L.}~\bibnamefont {Di~Palma}}, \bibinfo {author} {\bibfnamefont {H.}~\bibnamefont {Ahmad}}, \bibinfo {author} {\bibfnamefont {G.}~\bibnamefont {Ausanio}}, \bibinfo {author} {\bibfnamefont {G.}~\bibnamefont {Pepe}}, \bibinfo {author} {\bibfnamefont {F.}~\bibnamefont {Tafuri}}, \bibinfo {author} {\bibfnamefont {D.}~\bibnamefont {Montemurro}}, \bibinfo {author} {\bibfnamefont {C.}~\bibnamefont {Granata}},  \emph {et~al.},\ }\bibfield  {title} {\enquote {\bibinfo {title} {Aluminum-ferromagnetic {J}osephson tunnel junctions for high quality magnetic switching devices},}\ }\href {\doibase 10.1063/5.0101686} {\bibfield  {journal} {\bibinfo  {journal} {Appl. Phys. Lett.}\ }\textbf {\bibinfo {volume} {120}},\ \bibinfo {pages} {262601} (\bibinfo {year}
  {2022}{\natexlab{b}})}\BibitemShut {NoStop}%
\bibitem [{\citenamefont {Tavares}, \citenamefont {Yang},\ and\ \citenamefont {Meyers}(2023)}]{Tavares_2023}%
  \BibitemOpen
  \bibfield  {author} {\bibinfo {author} {\bibfnamefont {S.}~\bibnamefont {Tavares}}, \bibinfo {author} {\bibfnamefont {K.}~\bibnamefont {Yang}}, \ and\ \bibinfo {author} {\bibfnamefont {M.~A.}\ \bibnamefont {Meyers}},\ }\bibfield  {title} {\enquote {\bibinfo {title} {Heusler alloys: {P}ast, properties, new alloys, and prospects},}\ }\href {\doibase https://doi.org/10.1016/j.pmatsci.2022.101017} {\bibfield  {journal} {\bibinfo  {journal} {Prog. Mater. Sci.}\ }\textbf {\bibinfo {volume} {132}},\ \bibinfo {pages} {101017} (\bibinfo {year} {2023})}\BibitemShut {NoStop}%
\bibitem [{\citenamefont {K\={o}no}(1958)}]{Kono_1958}%
  \BibitemOpen
  \bibfield  {author} {\bibinfo {author} {\bibfnamefont {H.}~\bibnamefont {K\={o}no}},\ }\bibfield  {title} {\enquote {\bibinfo {title} {{On the Ferromagnetic Phase in Manganese-Aluminum S}ystem},}\ }\href {\doibase 10.1143/JPSJ.13.1444} {\bibfield  {journal} {\bibinfo  {journal} {J. Phys. Soc. Japan.}\ }\textbf {\bibinfo {volume} {13}},\ \bibinfo {pages} {1444--1451} (\bibinfo {year} {1958})}\BibitemShut {NoStop}%
\bibitem [{\citenamefont {Bither}\ and\ \citenamefont {Cloud}(2004)}]{Bither_2004}%
  \BibitemOpen
  \bibfield  {author} {\bibinfo {author} {\bibfnamefont {T.~A.}\ \bibnamefont {Bither}}\ and\ \bibinfo {author} {\bibfnamefont {W.~H.}\ \bibnamefont {Cloud}},\ }\bibfield  {title} {\enquote {\bibinfo {title} {{Magnetic {T}etragonal $\delta$ {P}hase in the {Mn–Ga B}inary}},}\ }\href {\doibase 10.1063/1.1714349} {\bibfield  {journal} {\bibinfo  {journal} {J. Appl. Phys.}\ }\textbf {\bibinfo {volume} {36}},\ \bibinfo {pages} {1501--1502} (\bibinfo {year} {2004})}\BibitemShut {NoStop}%
\bibitem [{\citenamefont {Waintal}\ and\ \citenamefont {Brouwer}(2002)}]{Waintal_2002}%
  \BibitemOpen
  \bibfield  {author} {\bibinfo {author} {\bibfnamefont {X.}~\bibnamefont {Waintal}}\ and\ \bibinfo {author} {\bibfnamefont {P.~W.}\ \bibnamefont {Brouwer}},\ }\bibfield  {title} {\enquote {\bibinfo {title} {Magnetic exchange interaction induced by a {J}osephson current},}\ }\href {\doibase 10.1103/PhysRevB.65.054407} {\bibfield  {journal} {\bibinfo  {journal} {Phys. Rev. B}\ }\textbf {\bibinfo {volume} {65}},\ \bibinfo {pages} {054407} (\bibinfo {year} {2002})}\BibitemShut {NoStop}%
\bibitem [{\citenamefont {Zhao}\ and\ \citenamefont {Sauls}(2008)}]{Zhao_2008}%
  \BibitemOpen
  \bibfield  {author} {\bibinfo {author} {\bibfnamefont {E.}~\bibnamefont {Zhao}}\ and\ \bibinfo {author} {\bibfnamefont {J.~A.}\ \bibnamefont {Sauls}},\ }\bibfield  {title} {\enquote {\bibinfo {title} {Theory of nonequilibrium spin transport and spin-transfer torque in superconducting-ferromagnetic nanostructures},}\ }\href {\doibase 10.1103/PhysRevB.78.174511} {\bibfield  {journal} {\bibinfo  {journal} {Phys. Rev. B}\ }\textbf {\bibinfo {volume} {78}},\ \bibinfo {pages} {174511} (\bibinfo {year} {2008})}\BibitemShut {NoStop}%
\bibitem [{\citenamefont {Konschelle}\ and\ \citenamefont {Buzdin}(2009)}]{Konschelle_2009}%
  \BibitemOpen
  \bibfield  {author} {\bibinfo {author} {\bibfnamefont {F.}~\bibnamefont {Konschelle}}\ and\ \bibinfo {author} {\bibfnamefont {A.}~\bibnamefont {Buzdin}},\ }\bibfield  {title} {\enquote {\bibinfo {title} {Magnetic {Moment Manipulation by a Josephson C}urrent},}\ }\href {\doibase 10.1103/PhysRevLett.102.017001} {\bibfield  {journal} {\bibinfo  {journal} {Phys. Rev. Lett.}\ }\textbf {\bibinfo {volume} {102}},\ \bibinfo {pages} {017001} (\bibinfo {year} {2009})}\BibitemShut {NoStop}%
\bibitem [{\citenamefont {Linder}\ and\ \citenamefont {Yokoyama}(2011)}]{Linder_2011}%
  \BibitemOpen
  \bibfield  {author} {\bibinfo {author} {\bibfnamefont {J.}~\bibnamefont {Linder}}\ and\ \bibinfo {author} {\bibfnamefont {T.}~\bibnamefont {Yokoyama}},\ }\bibfield  {title} {\enquote {\bibinfo {title} {Supercurrent-induced magnetization dynamics in a {Josephson} junction with two misaligned ferromagnetic layers},}\ }\href {\doibase 10.1103/PhysRevB.83.012501} {\bibfield  {journal} {\bibinfo  {journal} {Phys. Rev. B}\ }\textbf {\bibinfo {volume} {83}},\ \bibinfo {pages} {012501} (\bibinfo {year} {2011})}\BibitemShut {NoStop}%
\bibitem [{\citenamefont {Shao}\ \emph {et~al.}(2021)\citenamefont {Shao}, \citenamefont {Li}, \citenamefont {Liu}, \citenamefont {Yang}, \citenamefont {Fukami}, \citenamefont {Razavi}, \citenamefont {Wu}, \citenamefont {Wang}, \citenamefont {Freimuth}, \citenamefont {Mokrousov}, \citenamefont {Stiles}, \citenamefont {Emori}, \citenamefont {Hoffmann}, \citenamefont {Åkerman}, \citenamefont {Roy}, \citenamefont {Wang}, \citenamefont {Yang}, \citenamefont {Garello},\ and\ \citenamefont {Zhang}}]{Shao_2021}%
  \BibitemOpen
  \bibfield  {author} {\bibinfo {author} {\bibfnamefont {Q.}~\bibnamefont {Shao}}, \bibinfo {author} {\bibfnamefont {P.}~\bibnamefont {Li}}, \bibinfo {author} {\bibfnamefont {L.}~\bibnamefont {Liu}}, \bibinfo {author} {\bibfnamefont {H.}~\bibnamefont {Yang}}, \bibinfo {author} {\bibfnamefont {S.}~\bibnamefont {Fukami}}, \bibinfo {author} {\bibfnamefont {A.}~\bibnamefont {Razavi}}, \bibinfo {author} {\bibfnamefont {H.}~\bibnamefont {Wu}}, \bibinfo {author} {\bibfnamefont {K.}~\bibnamefont {Wang}}, \bibinfo {author} {\bibfnamefont {F.}~\bibnamefont {Freimuth}}, \bibinfo {author} {\bibfnamefont {Y.}~\bibnamefont {Mokrousov}}, \bibinfo {author} {\bibfnamefont {M.~D.}\ \bibnamefont {Stiles}}, \bibinfo {author} {\bibfnamefont {S.}~\bibnamefont {Emori}}, \bibinfo {author} {\bibfnamefont {A.}~\bibnamefont {Hoffmann}}, \bibinfo {author} {\bibfnamefont {J.}~\bibnamefont {Åkerman}}, \bibinfo {author} {\bibfnamefont {K.}~\bibnamefont {Roy}}, \bibinfo {author} {\bibfnamefont {J.-P.}\ \bibnamefont {Wang}}, \bibinfo
  {author} {\bibfnamefont {S.-H.}\ \bibnamefont {Yang}}, \bibinfo {author} {\bibfnamefont {K.}~\bibnamefont {Garello}}, \ and\ \bibinfo {author} {\bibfnamefont {W.}~\bibnamefont {Zhang}},\ }\bibfield  {title} {\enquote {\bibinfo {title} {Roadmap of {Spin–Orbit T}orques},}\ }\href {\doibase 10.1109/TMAG.2021.3078583} {\bibfield  {journal} {\bibinfo  {journal} {IEEE Trans. Magn.}\ }\textbf {\bibinfo {volume} {57}},\ \bibinfo {pages} {1--39} (\bibinfo {year} {2021})}\BibitemShut {NoStop}%
\bibitem [{\citenamefont {Nguyen}\ \emph {et~al.}(2020)\citenamefont {Nguyen}, \citenamefont {Ribeill}, \citenamefont {Gustafsson}, \citenamefont {Shi}, \citenamefont {Aradhya}, \citenamefont {Wagner}, \citenamefont {Ranzani}, \citenamefont {Zhu}, \citenamefont {Baghdadi}, \citenamefont {Butters} \emph {et~al.}}]{nguyen_2020}%
  \BibitemOpen
  \bibfield  {author} {\bibinfo {author} {\bibfnamefont {M.-H.}\ \bibnamefont {Nguyen}}, \bibinfo {author} {\bibfnamefont {G.~J.}\ \bibnamefont {Ribeill}}, \bibinfo {author} {\bibfnamefont {M.~V.}\ \bibnamefont {Gustafsson}}, \bibinfo {author} {\bibfnamefont {S.}~\bibnamefont {Shi}}, \bibinfo {author} {\bibfnamefont {S.~V.}\ \bibnamefont {Aradhya}}, \bibinfo {author} {\bibfnamefont {A.~P.}\ \bibnamefont {Wagner}}, \bibinfo {author} {\bibfnamefont {L.~M.}\ \bibnamefont {Ranzani}}, \bibinfo {author} {\bibfnamefont {L.}~\bibnamefont {Zhu}}, \bibinfo {author} {\bibfnamefont {R.}~\bibnamefont {Baghdadi}}, \bibinfo {author} {\bibfnamefont {B.}~\bibnamefont {Butters}},  \emph {et~al.},\ }\bibfield  {title} {\enquote {\bibinfo {title} {Cryogenic memory architecture integrating spin {H}all effect based magnetic memory and superconductive cryotron devices},}\ }\href {\doibase 10.1038/s41598-019-57137-9} {\bibfield  {journal} {\bibinfo  {journal} {Sci. Rep.}\ }\textbf {\bibinfo {volume} {10}},\ \bibinfo {pages} {248}
  (\bibinfo {year} {2020})}\BibitemShut {NoStop}%
\bibitem [{\citenamefont {\ifmmode~\check{S}\else \v{S}\fi{}mejkal}, \citenamefont {Sinova},\ and\ \citenamefont {Jungwirth}(2022)}]{smejkal_2022}%
  \BibitemOpen
  \bibfield  {author} {\bibinfo {author} {\bibfnamefont {L.}~\bibnamefont {\ifmmode~\check{S}\else \v{S}\fi{}mejkal}}, \bibinfo {author} {\bibfnamefont {J.}~\bibnamefont {Sinova}}, \ and\ \bibinfo {author} {\bibfnamefont {T.}~\bibnamefont {Jungwirth}},\ }\bibfield  {title} {\enquote {\bibinfo {title} {Emerging {R}esearch {L}andscape of {A}ltermagnetism},}\ }\href {\doibase 10.1103/PhysRevX.12.040501} {\bibfield  {journal} {\bibinfo  {journal} {Phys. Rev. X}\ }\textbf {\bibinfo {volume} {12}},\ \bibinfo {pages} {040501} (\bibinfo {year} {2022})}\BibitemShut {NoStop}%
\bibitem [{\citenamefont {Zhang}, \citenamefont {Hu},\ and\ \citenamefont {Neupert}(2024)}]{zhang_2023}%
  \BibitemOpen
  \bibfield  {author} {\bibinfo {author} {\bibfnamefont {S.-B.}\ \bibnamefont {Zhang}}, \bibinfo {author} {\bibfnamefont {L.-H.}\ \bibnamefont {Hu}}, \ and\ \bibinfo {author} {\bibfnamefont {T.}~\bibnamefont {Neupert}},\ }\bibfield  {title} {\enquote {\bibinfo {title} {Finite-momentum {C}ooper pairing in proximitized altermagnets},}\ }\href {\doibase 10.1038/s41467-024-45951-3} {\bibfield  {journal} {\bibinfo  {journal} {Nat. Commun.}\ }\textbf {\bibinfo {volume} {15}},\ \bibinfo {pages} {1801} (\bibinfo {year} {2024})}\BibitemShut {NoStop}%
\bibitem [{\citenamefont {Ouassou}, \citenamefont {Brataas},\ and\ \citenamefont {Linder}(2023)}]{ouassou_2023}%
  \BibitemOpen
  \bibfield  {author} {\bibinfo {author} {\bibfnamefont {J.~A.}\ \bibnamefont {Ouassou}}, \bibinfo {author} {\bibfnamefont {A.}~\bibnamefont {Brataas}}, \ and\ \bibinfo {author} {\bibfnamefont {J.}~\bibnamefont {Linder}},\ }\bibfield  {title} {\enquote {\bibinfo {title} {dc {J}osephson {E}ffect in {A}ltermagnets},}\ }\href {\doibase 10.1103/PhysRevLett.131.076003} {\bibfield  {journal} {\bibinfo  {journal} {Phys. Rev. Lett.}\ }\textbf {\bibinfo {volume} {131}},\ \bibinfo {pages} {076003} (\bibinfo {year} {2023})}\BibitemShut {NoStop}%
\bibitem [{\citenamefont {Beenakker}\ and\ \citenamefont {Vakhtel}(2023)}]{beenakker_2023}%
  \BibitemOpen
  \bibfield  {author} {\bibinfo {author} {\bibfnamefont {C.~W.~J.}\ \bibnamefont {Beenakker}}\ and\ \bibinfo {author} {\bibfnamefont {T.}~\bibnamefont {Vakhtel}},\ }\bibfield  {title} {\enquote {\bibinfo {title} {Phase-shifted {A}ndreev levels in an altermagnet {J}osephson junction},}\ }\href {\doibase 10.1103/PhysRevB.108.075425} {\bibfield  {journal} {\bibinfo  {journal} {Phys. Rev. B}\ }\textbf {\bibinfo {volume} {108}},\ \bibinfo {pages} {075425} (\bibinfo {year} {2023})}\BibitemShut {NoStop}%
\bibitem [{\citenamefont {Papaj}(2023)}]{papaj_2023}%
  \BibitemOpen
  \bibfield  {author} {\bibinfo {author} {\bibfnamefont {M.}~\bibnamefont {Papaj}},\ }\bibfield  {title} {\enquote {\bibinfo {title} {Andreev reflection at the altermagnet-superconductor interface},}\ }\href {\doibase 10.1103/PhysRevB.108.L060508} {\bibfield  {journal} {\bibinfo  {journal} {Phys. Rev. B}\ }\textbf {\bibinfo {volume} {108}},\ \bibinfo {pages} {L060508} (\bibinfo {year} {2023})}\BibitemShut {NoStop}%
\bibitem [{\citenamefont {Feng}\ \emph {et~al.}(2022)\citenamefont {Feng}, \citenamefont {Zhou}, \citenamefont {{\v{S}}mejkal}, \citenamefont {Wu}, \citenamefont {Zhu}, \citenamefont {Guo}, \citenamefont {Gonz{\'a}lez-Hern{\'a}ndez}, \citenamefont {Wang}, \citenamefont {Yan}, \citenamefont {Qin} \emph {et~al.}}]{feng_2022}%
  \BibitemOpen
  \bibfield  {author} {\bibinfo {author} {\bibfnamefont {Z.}~\bibnamefont {Feng}}, \bibinfo {author} {\bibfnamefont {X.}~\bibnamefont {Zhou}}, \bibinfo {author} {\bibfnamefont {L.}~\bibnamefont {{\v{S}}mejkal}}, \bibinfo {author} {\bibfnamefont {L.}~\bibnamefont {Wu}}, \bibinfo {author} {\bibfnamefont {Z.}~\bibnamefont {Zhu}}, \bibinfo {author} {\bibfnamefont {H.}~\bibnamefont {Guo}}, \bibinfo {author} {\bibfnamefont {R.}~\bibnamefont {Gonz{\'a}lez-Hern{\'a}ndez}}, \bibinfo {author} {\bibfnamefont {X.}~\bibnamefont {Wang}}, \bibinfo {author} {\bibfnamefont {H.}~\bibnamefont {Yan}}, \bibinfo {author} {\bibfnamefont {P.}~\bibnamefont {Qin}},  \emph {et~al.},\ }\bibfield  {title} {\enquote {\bibinfo {title} {An anomalous {H}all effect in altermagnetic ruthenium dioxide},}\ }\href {\doibase 10.1038/s41928-022-00866-z} {\bibfield  {journal} {\bibinfo  {journal} {Nat. Electron.}\ }\textbf {\bibinfo {volume} {5}},\ \bibinfo {pages} {735--743} (\bibinfo {year} {2022})}\BibitemShut {NoStop}%
\bibitem [{\citenamefont {Crotty}, \citenamefont {Schult},\ and\ \citenamefont {Segall}(2010)}]{Crotty_2010}%
  \BibitemOpen
  \bibfield  {author} {\bibinfo {author} {\bibfnamefont {P.}~\bibnamefont {Crotty}}, \bibinfo {author} {\bibfnamefont {D.}~\bibnamefont {Schult}}, \ and\ \bibinfo {author} {\bibfnamefont {K.}~\bibnamefont {Segall}},\ }\bibfield  {title} {\enquote {\bibinfo {title} {Josephson junction simulation of neurons},}\ }\href {\doibase 10.1103/PhysRevE.82.011914} {\bibfield  {journal} {\bibinfo  {journal} {Phys. Rev. E}\ }\textbf {\bibinfo {volume} {82}},\ \bibinfo {pages} {011914} (\bibinfo {year} {2010})}\BibitemShut {NoStop}%
\bibitem [{\citenamefont {Schneider}\ \emph {et~al.}(2022)\citenamefont {Schneider}, \citenamefont {Toomey}, \citenamefont {Rowlands}, \citenamefont {Shainline}, \citenamefont {Tschirhart},\ and\ \citenamefont {Segall}}]{Schneider_2022}%
  \BibitemOpen
  \bibfield  {author} {\bibinfo {author} {\bibfnamefont {M.}~\bibnamefont {Schneider}}, \bibinfo {author} {\bibfnamefont {E.}~\bibnamefont {Toomey}}, \bibinfo {author} {\bibfnamefont {G.}~\bibnamefont {Rowlands}}, \bibinfo {author} {\bibfnamefont {J.}~\bibnamefont {Shainline}}, \bibinfo {author} {\bibfnamefont {P.}~\bibnamefont {Tschirhart}}, \ and\ \bibinfo {author} {\bibfnamefont {K.}~\bibnamefont {Segall}},\ }\bibfield  {title} {\enquote {\bibinfo {title} {Super{M}ind: a survey of the potential of superconducting electronics for neuromorphic computing},}\ }\href {\doibase 10.1088/1361-6668/ac4cd2} {\bibfield  {journal} {\bibinfo  {journal} {Supercond. Sci. Technol.}\ }\textbf {\bibinfo {volume} {35}},\ \bibinfo {pages} {053001} (\bibinfo {year} {2022})}\BibitemShut {NoStop}%
\bibitem [{\citenamefont {Schegolev}\ \emph {et~al.}(2022)\citenamefont {Schegolev}, \citenamefont {Klenov}, \citenamefont {Bakurskiy}, \citenamefont {Soloviev}, \citenamefont {Kupriyanov}, \citenamefont {Tereshonok},\ and\ \citenamefont {Sidorenko}}]{Schegolev_2022}%
  \BibitemOpen
  \bibfield  {author} {\bibinfo {author} {\bibfnamefont {A.~E.}\ \bibnamefont {Schegolev}}, \bibinfo {author} {\bibfnamefont {N.~V.}\ \bibnamefont {Klenov}}, \bibinfo {author} {\bibfnamefont {S.~V.}\ \bibnamefont {Bakurskiy}}, \bibinfo {author} {\bibfnamefont {I.~I.}\ \bibnamefont {Soloviev}}, \bibinfo {author} {\bibfnamefont {M.~Y.}\ \bibnamefont {Kupriyanov}}, \bibinfo {author} {\bibfnamefont {M.~V.}\ \bibnamefont {Tereshonok}}, \ and\ \bibinfo {author} {\bibfnamefont {A.~S.}\ \bibnamefont {Sidorenko}},\ }\bibfield  {title} {\enquote {\bibinfo {title} {Tunable superconducting neurons for networks based on radial basis functions},}\ }\href {\doibase 10.3762/bxiv.2022.16.v1} {\bibfield  {journal} {\bibinfo  {journal} {Beilstein J. Nanotechnol.}\ }\textbf {\bibinfo {volume} {13}},\ \bibinfo {pages} {444--454} (\bibinfo {year} {2022})}\BibitemShut {NoStop}%
\bibitem [{\citenamefont {Schneider}, \citenamefont {Donnelly},\ and\ \citenamefont {Russek}(2018)}]{schneider_2018}%
  \BibitemOpen
  \bibfield  {author} {\bibinfo {author} {\bibfnamefont {M.~L.}\ \bibnamefont {Schneider}}, \bibinfo {author} {\bibfnamefont {C.~A.}\ \bibnamefont {Donnelly}}, \ and\ \bibinfo {author} {\bibfnamefont {S.~E.}\ \bibnamefont {Russek}},\ }\bibfield  {title} {\enquote {\bibinfo {title} {Tutorial: {H}igh-speed low-power neuromorphic systems based on magnetic {J}osephson junctions},}\ }\href {\doibase 10.1063/1.5042425} {\bibfield  {journal} {\bibinfo  {journal} {J. Appl. Phys.}\ }\textbf {\bibinfo {volume} {124}} (\bibinfo {year} {2018}),\ 10.1063/1.5042425}\BibitemShut {NoStop}%
\bibitem [{\citenamefont {Schneider}\ \emph {et~al.}(2018)\citenamefont {Schneider}, \citenamefont {Donnelly}, \citenamefont {Russek}, \citenamefont {Baek}, \citenamefont {Pufall}, \citenamefont {Hopkins}, \citenamefont {Dresselhaus}, \citenamefont {Benz},\ and\ \citenamefont {Rippard}}]{schneider_2018a}%
  \BibitemOpen
  \bibfield  {author} {\bibinfo {author} {\bibfnamefont {M.~L.}\ \bibnamefont {Schneider}}, \bibinfo {author} {\bibfnamefont {C.~A.}\ \bibnamefont {Donnelly}}, \bibinfo {author} {\bibfnamefont {S.~E.}\ \bibnamefont {Russek}}, \bibinfo {author} {\bibfnamefont {B.}~\bibnamefont {Baek}}, \bibinfo {author} {\bibfnamefont {M.~R.}\ \bibnamefont {Pufall}}, \bibinfo {author} {\bibfnamefont {P.~F.}\ \bibnamefont {Hopkins}}, \bibinfo {author} {\bibfnamefont {P.~D.}\ \bibnamefont {Dresselhaus}}, \bibinfo {author} {\bibfnamefont {S.~P.}\ \bibnamefont {Benz}}, \ and\ \bibinfo {author} {\bibfnamefont {W.~H.}\ \bibnamefont {Rippard}},\ }\bibfield  {title} {\enquote {\bibinfo {title} {Ultralow power artificial synapses using nanotextured magnetic {J}osephson junctions},}\ }\href {\doibase 10.1126/sciadv.1701329} {\bibfield  {journal} {\bibinfo  {journal} {Sci. Adv.}\ }\textbf {\bibinfo {volume} {4}},\ \bibinfo {pages} {e1701329} (\bibinfo {year} {2018})}\BibitemShut {NoStop}%
\bibitem [{\citenamefont {Ju{\'e}}\ \emph {et~al.}(2022{\natexlab{a}})\citenamefont {Ju{\'e}}, \citenamefont {Iankevich}, \citenamefont {Reisinger}, \citenamefont {Hahn}, \citenamefont {Provenzano}, \citenamefont {Pufall}, \citenamefont {Haygood}, \citenamefont {Rippard},\ and\ \citenamefont {Schneider}}]{jue_2022}%
  \BibitemOpen
  \bibfield  {author} {\bibinfo {author} {\bibfnamefont {E.}~\bibnamefont {Ju{\'e}}}, \bibinfo {author} {\bibfnamefont {G.}~\bibnamefont {Iankevich}}, \bibinfo {author} {\bibfnamefont {T.}~\bibnamefont {Reisinger}}, \bibinfo {author} {\bibfnamefont {H.}~\bibnamefont {Hahn}}, \bibinfo {author} {\bibfnamefont {V.}~\bibnamefont {Provenzano}}, \bibinfo {author} {\bibfnamefont {M.~R.}\ \bibnamefont {Pufall}}, \bibinfo {author} {\bibfnamefont {I.~W.}\ \bibnamefont {Haygood}}, \bibinfo {author} {\bibfnamefont {W.~H.}\ \bibnamefont {Rippard}}, \ and\ \bibinfo {author} {\bibfnamefont {M.~L.}\ \bibnamefont {Schneider}},\ }\bibfield  {title} {\enquote {\bibinfo {title} {Artificial synapses based on {J}osephson junctions with {F}e nanoclusters in the amorphous {G}e barrier},}\ }\href {\doibase 10.1063/5.0080841} {\bibfield  {journal} {\bibinfo  {journal} {J. Appl. Phys.}\ }\textbf {\bibinfo {volume} {131}},\ \bibinfo {pages} {073902} (\bibinfo {year} {2022}{\natexlab{a}})}\BibitemShut {NoStop}%
\bibitem [{\citenamefont {Ju{\'e}}\ \emph {et~al.}(2022{\natexlab{b}})\citenamefont {Ju{\'e}}, \citenamefont {Pufall}, \citenamefont {Haygood}, \citenamefont {Rippard},\ and\ \citenamefont {Schneider}}]{jue_2022a}%
  \BibitemOpen
  \bibfield  {author} {\bibinfo {author} {\bibfnamefont {E.}~\bibnamefont {Ju{\'e}}}, \bibinfo {author} {\bibfnamefont {M.~R.}\ \bibnamefont {Pufall}}, \bibinfo {author} {\bibfnamefont {I.~W.}\ \bibnamefont {Haygood}}, \bibinfo {author} {\bibfnamefont {W.~H.}\ \bibnamefont {Rippard}}, \ and\ \bibinfo {author} {\bibfnamefont {M.~L.}\ \bibnamefont {Schneider}},\ }\bibfield  {title} {\enquote {\bibinfo {title} {Perspectives on nanoclustered magnetic {J}osephson junctions as artificial synapses},}\ }\href {\doibase 10.1063/5.0118287} {\bibfield  {journal} {\bibinfo  {journal} {Appl. Phys. Lett.}\ }\textbf {\bibinfo {volume} {121}},\ \bibinfo {pages} {240501} (\bibinfo {year} {2022}{\natexlab{b}})}\BibitemShut {NoStop}%
\end{thebibliography}%

\end{document}